\documentclass[ twoside,openright,titlepage,numbers=noenddot,headinclude,
                footinclude=true,cleardoublepage=empty,abstractoff, 
                BCOR=5mm,paper=b5,fontsize=10pt,
                dutch,american%
                ]{scrreprt}

\PassOptionsToPackage{eulerchapternumbers,listings,
				 pdfspacing,
				 beramono,parts,dottedtoc}{classicthesis}										

\usepackage{ifthen}
\newboolean{enable-backrefs} 
\setboolean{enable-backrefs}{true} 

\newcommand{\myTitle}{Supersymmetry \& the Spectral Action\xspace}
\newcommand{\mySubtitle}{On a geometrical interpretation of the MSSM\xspace}

\newcommand{\myName}{Thijs van den Broek\xspace}

\newcommand{\myFaculty}{Put data here\xspace}

\newcommand{\myUni}{Put data here\xspace}

\newcommand{\myVersion}{version 0.1\xspace}

\newcounter{dummy} 
\providecommand{\mLyX}{L\kern-.1667em\lower.25em\hbox{Y}\kern-.125emX\@}

\PassOptionsToPackage{latin9}{inputenc}	
 \usepackage{inputenc}				

\PassOptionsToPackage{american}{babel}   
 \usepackage{babel}					

\PassOptionsToPackage{square,numbers}{natbib}
 \usepackage{natbib}				

\PassOptionsToPackage{fleqn}{amsmath}		
 \usepackage{amsmath}

\PassOptionsToPackage{T1}{fontenc} 
	\usepackage{fontenc}     
\usepackage{textcomp} 
\usepackage{scrhack} 
\usepackage{xspace} 
\usepackage{mparhack} 
\usepackage{fixltx2e} 
\PassOptionsToPackage{printonlyused,smaller}{acronym}
	\usepackage{acronym} 

\usepackage{tabularx} 
\usepackage{caption}
\usepackage{listings} 
\PassOptionsToPackage{hyperfootnotes=false,pdfpagelabels}{hyperref}
	\usepackage{hyperref}  
	\usepackage{graphicx} 

\newcommand{\backrefnotcitedstring}{\relax}
\newcommand{\backrefcitedsinglestring}[1]{(Cited on page~#1.)}
\newcommand{\backrefcitedmultistring}[1]{(Cited on pages~#1.)}
\ifthenelse{\boolean{enable-backrefs}}%
{%
		\PassOptionsToPackage{hyperpageref}{backref}
		\usepackage{backref} 
		   \renewcommand*{\backref}[1]{}  
		   \renewcommand*{\backrefalt}[4]{
		      \ifcase #1 %
		         \backrefnotcitedstring%
		      \or%
		         \backrefcitedsinglestring{#2}%
		      \else%
		         \backrefcitedmultistring{#2}%
		      \fi}%
}{\relax}    

\hypersetup{%
    colorlinks=true, linktocpage=true, pdfstartpage=3, pdfstartview=FitV,%
    breaklinks=true, pdfpagemode=UseNone, pageanchor=true, pdfpagemode=UseOutlines,%
    plainpages=false, bookmarksnumbered, bookmarksopen=true, bookmarksopenlevel=1,%
    hypertexnames=true, pdfhighlight=/O,
    urlcolor=webbrown, linkcolor=RoyalBlue, citecolor=webgreen, 
    pdftitle={\myTitle},%
    pdfauthor={\textcopyright\ \myName, \myUni, \myFaculty},%
    pdfsubject={},%
    pdfkeywords={},%
    pdfcreator={pdfLaTeX},%
    pdfproducer={LaTeX with hyperref and classicthesis}%
}   

\makeatletter
\@ifpackageloaded{babel}%
    {%
       \addto\extrasamerican{%
				}%
       \addto\extrasngerman{%
				}%
    }{\relax}
\makeatother

\usepackage{classicthesis} 

\usepackage{mydefs}
\usepackage{slashed}
\usepackage{mathrsfs}
\usepackage{nicefrac}
\usepackage{amssymb}

\usepackage[b5paper,layout=b5paper]{geometry}

\usepackage[style=long,xindy,toc,numberline,translate=false,numberedsection=autolabel]{glossaries}
\usepackage{glossaries-babel}
\usepackage{glossary-mcols}
\renewcommand*{\glsmcols}{2}

\usepackage{xstring}

\usepackage[final]{showkeys}

\usepackage{parskip}
\titlespacing*{\subsection}{0pt}{*4}{*1.5}
\titlespacing*{\section}{0pt}{*4}{*1.5}
\titlespacing*{\subsubsection}{0pt}{*4}{*1.5}

\usepackage{footmisc}

\newlength{\gloraise}
\def\sfer{\ensuremath{\phi}\xspace}
\def\asfer{\ensuremath{\bar\phi}\xspace}

\newglossary[slg]{symbolslist}{syi}{syg}{List of symbols}

 \GlsAddXdyAttribute{inclp} 
\makeglossaries

\newglossaryentry{CofM}{
	name=\ensuremath{C(M, \com)},
	description={},
	type=symbolslist,
	sort=C
}
\newglossaryentry{Cinfty}{
	name=\ensuremath{C^{\infty}(M, \com)},
	description={},
	type=symbolslist,
	sort=Ci
}
\newglossaryentry{quat}{
	name={\ensuremath{\protect\mathbb{H}}},
	description={},
	type={symbolslist},
	sort={H}
}
\newglossaryentry{BH}{
	name=\ensuremath{B(\H)},
	description={The $C^*$-algebra of bounded operators on the Hilbert space $\H$.},
	type=symbolslist
}
\newglossaryentry{D}{
	name=\ensuremath{D},
	description={},
	type=symbolslist,
	sort=D
}
\newglossaryentry{A}{
	name=\ensuremath{\A},
	description={},
	type=symbolslist,
	sort=A
}
\newglossaryentry{space}{
	name=\ensuremath{\H},
	description={},
	type=symbolslist,
	sort=Hilb
}
\newglossaryentry{g}{
	name=\ensuremath{g},
	description={},
	type=symbolslist,
	sort=g
}
\newglossaryentry{L2MS}{
	name=\ensuremath{L^2(M,S)},
	description={},
	type=symbolslist,
	sort=LM
}
\newglossaryentry{M}{
	name=\ensuremath{M},
	description={},
	type=symbolslist,
	sort=M
}
\newglossaryentry{omegamu}{
	name=\ensuremath{\omega_\mu},
	description={},
	type=symbolslist,
	sort=omega
}
\newglossaryentry{m}{
	name=\ensuremath{m},
	description={},
	type=symbolslist,
	sort=m
}
\newglossaryentry{vierbein}{
	name=\ensuremath{\gamma^a},
	description={},
	type=symbolslist,
	sort=gammai
}
\newglossaryentry{dirac}{
	name=\ensuremath{\dirac},
	description={},
	type=symbolslist,
	sort=dslashm
}
\newglossaryentry{h}{
	name=\ensuremath{h},
	description={},
	type=symbolslist,
	sort=h
}
\newglossaryentry{AF}{
	name=\ensuremath{\A_F},
	description={},
	type=symbolslist,
	sort=AF
}
\newglossaryentry{HF}{
	name=\ensuremath{\H_F},
	description={},
	type=symbolslist,
	sort=HF
}
\newglossaryentry{DF}{
	name=\ensuremath{D_F},
	description={},
	type=symbolslist,
	sort=DF
}
\newglossaryentry{gamma}{
	name=\ensuremath{\gamma},
	description={},
	type=symbolslist,
	sort=gamma
}
\newglossaryentry{J}{
	name=\ensuremath{J},
	description={},
	type=symbolslist,
	sort=J
}
\newglossaryentry{gammaF}{
	name=\ensuremath{\protect\gamma_F},
	description={},
	type=symbolslist,
	sort=gammaF
}
\newglossaryentry{JF}{
	name=\ensuremath{J_F},
	description={},
	type=symbolslist,
	sort=JF
}
\newglossaryentry{aopp}{
	name=\ensuremath{a^o},
	description={},
	type=symbolslist,
	sort=aopp
}

\newglossaryentry{epsilon1}{
	name=\ensuremath{\protect\epsilon},
	description={},
	type=symbolslist,
	sort=epsilon1
}
\newglossaryentry{epsilon2}{
	name=\ensuremath{\protect\epsilon'},
	description={},
	type=symbolslist,
	sort=epsilon2
}
\newglossaryentry{epsilon3}{
	name=\ensuremath{\protect\epsilon''},
	description={},
	type=symbolslist,
	sort=epsilon3
}
\newglossaryentry{JM}{
	name=\ensuremath{J_M},
	description={},
	type=symbolslist,
	sort=JM
}
\newglossaryentry{gammaM}{
	name=\ensuremath{\gamma_M},
	description={},
	type=symbolslist,
	sort=gammaM
}
\newglossaryentry{u}{
	name=\ensuremath{u},
	description={},
	type=symbolslist,
	sort=u
}
\newglossaryentry{U}{
	name=\ensuremath{U},
	description={},
	type=symbolslist,
	sort=U
}
\newglossaryentry{DA}{
	name=\ensuremath{D_A},
	description={},
	type=symbolslist,
	sort=DA
}
\newglossaryentry{canA}{
	name=\ensuremath{\can_A},
	description={},
	type=symbolslist,
	sort=DA2
}
\newglossaryentry{Phi}{
	name=\ensuremath{\Phi},
	description={},
	type=symbolslist,
	sort=fi
}
\newglossaryentry{bbAmu}{
	name=\ensuremath{\protect\mathbb{A}_\mu},
	description={},
	type=symbolslist,
	sort=Amu
}
\newglossaryentry{f}{
	name=\ensuremath{f},
	description={},
	type=symbolslist,
	sort=f
}
\newglossaryentry{Hplus}{
	name=\ensuremath{\H^+},
	description={},
	type=symbolslist,
	sort=Hplus
}
\newglossaryentry{Lambda}{
	name=\ensuremath{\Lambda},
	description={},
	type=symbolslist,
	sort=Lambda
}
\newglossaryentry{f2}{
	name=\ensuremath{f_2},
	description={},
	type=symbolslist,
	sort=f2
}
\newglossaryentry{f4}{
	name=\ensuremath{f_4},
	description={},
	type=symbolslist,
	sort=f4
}
\newglossaryentry{lambda}{
	name=\ensuremath{\protect\lambda},
	description={},
	type=symbolslist,
	sort=lambda
}
\newglossaryentry{UpsilonR}{
	name=\ensuremath{\protect\Upsilon_R},
	description={},
	type=symbolslist,
	sort=UpsilonR
}
\newglossaryentry{AFC}{
	name=\ensuremath{\A_F^\com},
	description={},
	type=symbolslist,
	sort=AFC
}
\newglossaryentry{srepi}{
	name=\ensuremath{\protect\srep{i}},
	description={},
	type=symbolslist,
	sort=Ni
}
\newglossaryentry{srepoj}{
	name=\ensuremath{\protect\srepo{j}},
	description={},
	type=symbolslist,
	sort=Nio
}
\newglossaryentry{MNiNj}{
	name=\ensuremath{M_{N_iN_j}},
	description={},
	type=symbolslist,
	sort=MNiNj
}
\newglossaryentry{Dijkl}{
	name=\ensuremath{\protect\D{ij}{kl}},
	description={},
	type=symbolslist,
	sort=Dijkl
}
\newglossaryentry{M3C}{
	name=\ensuremath{M_3(\com)},
	description={},
	type=symbolslist,
	sort=M3C
}
\newglossaryentry{MNF}{
	name=\ensuremath{\protect M_{N_i}(\mathbb{F}_i)},
	description={},
	type=symbolslist,
	sort=MNF
}
\newglossaryentry{Dmu}{
	name=\ensuremath{\protect D_\mu},
	description={},
	type=symbolslist,
	sort=Dmu
}
\newglossaryentry{bbFmunu}{
	name=\ensuremath{\protect\mathbb{F}_{\mu\nu}},
	description={},
	type=symbolslist,
	sort=Fmunu
}
\newglossaryentry{Riemm}{
	name=\ensuremath{\protect R_{\mu\nu\lambda\sigma}},
	description={},
	type=symbolslist,
	sort=Rmunulambdasigma
}
\newglossaryentry{Phiik}{
	name=\ensuremath{\protect\Phi_{ik}},
	description={},
	type=symbolslist,
	sort=fiik
}
\newglossaryentry{bbA}{
	name=\ensuremath{\protect\mathbb{A}},
	description={},
	type=symbolslist,
	sort=Abb
}
\newglossaryentry{Amu}{
	name=\ensuremath{\protect A_\mu},
	description={},
	type=symbolslist,
	sort=Amu
}
\newglossaryentry{Rp}{
	name=\ensuremath{\protect R},
	description={},
	type=symbolslist,
	sort=Rp
}
\newglossaryentry{R}{
	name=\ensuremath{\protect R},
	description={},
	type=symbolslist,
	sort=R
}
\newglossaryentry{a024}{
	name=\ensuremath{\protect a_{0,2,4}},
	description={},
	type=symbolslist,
	sort=a0
}
\newglossaryentry{Rmunu}{
	name=\ensuremath{\protect R_{\mu\nu}},
	description={},
	type=symbolslist,
	sort=Rmunu
}

\newglossaryentry{grading}{name=grading,description={A grading $\gamma$ is a linear operator on a Hilbert space that satisfies $\gamma^* = \gamma$ and $\gamma^2 = 1$. This makes it possible to divide the space it acts on into subspaces of positive and negative $\gamma$ eigenvalue via the projection operators $\tfrac{1}{2}(1 \pm \gamma)$.}}
\newglossaryentry{norm}{name=norm,description={Let $V$ be a vector space over the field $\mathbb{F}$. A norm $\|.\|$ is a map from $V$ to $\mathbb{R}^{+}$ that satisfies \begin{itemize}\item $\|u + v\| \leq \|u\| + \|v\|$ (triangle inequality), \item $\|v\| = 0 \Leftrightarrow v = 0$, \item $\|\lambda v\| = |\lambda|\|v\|$\end{itemize} for all $u, v \in V$, $\lambda \in \mathbb{F}$.}}
\newglossaryentry{algebra}{name=algebra,description={An (associative) algebra $\A$ is a vector space over a field $\mathbb{F}$ that is equipped with a multiplication (here denoted by juxtaposition) which satisfies \begin{itemize}\item $(\lambda a)(\mu b) = (\lambda\mu) (ab)$,\item $(a + b)c = ac + bc$ and $a(b + c) = ab + ac$,\item $(ab)c = a(bc)$ (associativity)\end{itemize} for all $\lambda, \mu$ in $\mathbb{F}$, $a, b, c$ in $\A$.}}
\newglossaryentry{banachalgebra}{name={Banach algebra},description={A Banach algebra $\A$ is a Banach space (i.e.~a vector space that carries a \gls{norm} $\|.\|$ and is complete with respect to that norm) that in addition is an algebra. The norm on $\A$ and multiplication of its elements must moreover satisfy $\|ab\| \leq \|a\|\|b\|$ for all $a, b$ in $\A$.}}
\newglossaryentry{involutive}{name=involutive,description={An algebra $\A$ is called involutive if for each $a \in \A$, there is an $a^* \in \A$ such that $(a^*)^* = a$, $(ab)^* = b^*a^*$ and $(\lambda a)^* = \bar\lambda a^*$, $\forall a, b \in \A, \lambda \in \mathbb{F}$}.}
\newglossaryentry{unital}{name=unital,description={A unital algebra is an algebra $\A$ that has a multiplicative element $1$, satisfying $1\,a = a\,1 = a$ for all $a$ in $\A$.}}
\newglossaryentry{quaternions}{name=quaternions,description={The algebra over $\mathbb{R}$ that is generated by the elements $1$, $I$, $J$ and $K$, satisfying $I^2 = J^2 = K^2 = -1$ and $IJ = K$. The quaternions can be represented by two-by-two complex matrices in which case they are spanned by the unit matrix and the Pauli spin matrices. 
}}
\newglossaryentry{self-adjoint}{name=self-adjoint,description={An operator $A$ on a Hilbert space $\H$ is said to be self-adjoint, if $A^* = A$ (with respect to the inner product on $\H$) and the domains of $A$ and $A^*$ coincide. If $A$ is a finite-dimensional matrix, then it is also said to be \emph{Hermitian}.}}
\newglossaryentry{Morita equivalence}{name=Morita equivalence,description={Is an equivalence that is a bit weaker than unitary equivalence, for example each algebra $\A$ is Morita equivalent with the algebra of $\A$-valued $N \times N$-matrices.}}
\newglossaryentry{cstaralgebra}{name={\ensuremath{C^*}-algebra}, description={An algebra $\A$ over $\mathbb{C}$ is said to be a $C^*$-algebra when it is an \gls{involutive} \gls{banachalgebra} that satisfies $\|a^*a\| = \|a\|\|a^*\|$ for all $a$ in $\A$.}}
\newglossaryentry{module}{name={module}, description={Let $\A$ be an \gls{algebra} over the field $\mathbb{F}$. A left module $\E$ over $\A$ is a vector space over $\mathbb{F}$ that at the same time carries a left-action of $\A$, satisfying:
\begin{itemize}
	\item $a(b\eta) = (ab)\eta$,
	\item $(a + b)\eta = a\eta + b\eta$,
	\item $a(\eta + \zeta) = a\eta + a\zeta$,
\end{itemize}
for all $a, b$ in $\A$ and $\eta, \zeta$ in $\E$. In a similar fashion we can define right modules.}} 
\newglossaryentry{bimodule}{name=bimodule, description={If $\E$ is a left \gls{module} over an \gls{algebra} $\A$ and a right module over $\mathcal{B}$ and the left and right actions on $\E$ are compatible in the sense that $(a\eta)b = a(\eta b)$ for all $a \in \A$, $b\in \mathcal{B}$ and $\eta \in \E$, then $\E$ is said to be a ($\A$-$\mathcal{B}$-)bimodule.}}

\newglossaryentry{automorphism}{name=automorphism, description={An automorphism is an isomorphism from a certain mathematical object to itself. What the demands for being isomorphic are depends on the object in question. In the case of an algebra $\A$ it is a bijective map $\sigma$ from $\A$ to itself that satisfies $\sigma(\lambda a) = \lambda \sigma(a)$, $\sigma(a + b) = \sigma(a) + \sigma(b)$ and $\sigma(ab) = \sigma(a)\sigma(b)$ for all $\lambda$ in $\mathbb{F}$, $a, b$ in $\A$.}}
\newglossaryentry{order}{name={order},description={nog invulling aan geven}}
\newglossaryentry{elliptic}{name={elliptic},description={nog invulling aan geven}}
\newglossaryentry{connection}{name={connection},description={Let $M$ be a differentiable manifold and $(M, \pi, E)$ a smooth \gls{vector bundle}. For a tangent vector field $v \in \glslink{vector bundle}{TM}$, the connection $\nabla$ defines a covariant derivative $\nabla_v$ along $v$ which is a map from the \glspl{section} $\Gamma(E)$ of the vector bundle to itself. Examples are the Levi-Civita connection $\nabla^g_{\partial_\mu}$, for which $\nabla^g_{\partial_\mu}(\partial_\nu) = \Gamma_{\mu\nu}^{\lambda}\partial_{\lambda}$ (with $\Gamma_{\mu\nu}^{\lambda}$ the \emph{Christoffel symbols}) when $(M, g)$ is a Riemannian manifold and $E = TM$ (e.g.~\cite[\S 7.1]{GVF00}), and the spin connection on a \gls{spinor bundle}, when $M$ is a Riemannian spin manifold (e.g.~\cite[9.3]{GVF00}).}}
\newglossaryentry{endomorphism}{name={endomorphism},description={A homomorphism $\phi: \E \to \E$ (i.e.~a `structure preserving' map) from a certain object to itself. When $\E$ is a left $\A$-\gls{module}, $\phi$ is an endomorphism if $\phi(a\eta) = a\phi(\eta) \in \E$ and $\phi(\eta_1 + \eta_2) = \phi(\eta_1) + \phi(\eta_2) \in \E$ for all $\eta,\eta_{1,2} \in \E$ and $a \in \A$.}}
\newglossaryentry{contragredient}{name={contragredient}, description={For a left-$\A$ \gls{module} $\E$, we define the \emph{contragredient module} $\E^o$ by
\bas
	\E^o := \{\bar\eta, \eta \in \E\}, \quad \text{with}\quad \bar\eta a = \overline{a^*\eta}, \forall a \in \A.
\eas
Then $\E^o$ is seen to be a right-$\A$ \gls{module}, i.e. $(\bar\eta a)b =(\bar\eta)(ab)$, $\bar\eta (a + b) = \bar\eta a + \bar\eta b$ and 
$(\bar\eta + \bar\zeta)a = \bar\eta a + \bar\zeta a$, for all $a, b \in \A$ and $\eta, \zeta \in \E$.}}
\newglossaryentry{antiunitary}{name={antiunitary},description={An antiunitary operator $U$ on a Hilbert space $\H$ is a bijective, antilinear operator such that
	\bas
		\inpr{U\psi_1}{U\psi_2} = \inpr{\psi_2}{\psi_1}
	\eas 
	for all $\psi_{1,2} \in \H$.}}
\newglossaryentry{resolvent}{name={resolvent},description={For a closed operator $A$ on a Hilbert space $\H$ and complex number $z$ not in the spectrum of $A$,
		\bas
			R_z(A) := (z I - A)^{-1}
		\eas
		is called the \emph{resolvent} of $A$ (see e.g.~\cite[\rnum{8}.1]{RS80I}, \cite[\S 7.4]{GVF00}).}}
\newglossaryentry{compact}{name={compact},description={A topological space $X$ is said to be compact (e.g.~\cite[\S 2.3.5]{NAK90}) if for any \emph{cover} $\mathcal{U}$, i.e.
		\bas
				X = \bigcup_{U \in\, \mathcal{U}} U,\qquad \mathcal{U} = \{U, U \subset X\}
		\eas 
		there exists a finite subcover $\{U_1, \ldots, U_n\} \subset \mathcal{U}$, i.e.~$X = \bigcup_i U_i$ also. A compact operator is an operator that maps any bounded set to a relatively compact set. Perhaps a bit more intuitively, a compact operator $K$ on a Hilbert space $\H$ can be written \cite[\rnum{6}]{RS80I} in the form
		\bas
				K = \sum_{i = 1}^{N} \lambda_i \inpr{e_i}{.}f_i,
		\eas
		with orthonormal sets $\{e_i\}_{i = 1}^N$ and $\{f_i\}_{i = 1}^{N}$ of elements of $\H$ and $\{\lambda_i\}_{i = 1}^{N}$ real and positive, such that $\lambda_i \to 0$, which is the only accumulation point of $K$. The number $N$ may be finite or infinite.}}
\newglossaryentry{Hausdorff}{name={Hausdorff},description={A property of a space that is of topological nature. A topological space $X$ is called a Hausdorff space if any two distinct points that it contains are separated by neighbourhoods \cite[\S 2.3.3]{NAK90}. (Practically, all spaces that one encounters in physics are Hausdorff spaces.)}}
\newglossaryentry{square-integrable}{name={square-integrable},description={
The square integrable \glspl{section} of a \gls{spinor bundle} are those $\psi$ for which 
\begin{align*}
	\langle\psi, \psi \rangle \equiv \int_M (\psi,\psi)(x)\sqrt{g} \mathrm{d}^mx < \infty,
\end{align*}
where 
\bas
	(\,.\,,\,.\,) : \Gamma(S) \times \Gamma(S) \to C(M)
\eas
is the Hermitian pairing on $\Gamma(S)$, the sections of $S$ (see e.g.~\cite[\S 9.3]{GVF00})
}}
\newglossaryentry{spinor bundle}{name={spinor bundle},description={An example of a \gls{vector bundle}. The \emph{base space} of a spinor bundle is a Riemannian manifold $(M, g)$ that admits a spin structure. Roughly speaking, the fibers $S_x$ of the spinor bundle $(M, \pi, S)$ are in addition to vector spaces also representations of the spin group, i.e.~the elements of $S_x$ are spinors (see e.g.~\cite[Ch.~2]{LM89}).}}
\newglossaryentry{section}{name={section},description={A section $\sigma$ of a \gls{vector bundle} $(M, \pi, E)$ is a continuous map $\sigma : U \to E$, with $U$ an open subset of $M$, such that $\pi \circ \sigma = \id_{U}$ (see e.g.~\cite[Ch.~2]{GVF00}). The sections of a vector bundle are vector fields (i.e.~$\sigma(x) \in E_x$), those of a \gls{spinor bundle} are spinor fields. The collection of sections with values in $E_x$ is typically denoted by $\Gamma(E)$.\\
{\protect\makebox[\textwidth][c]{
	\def\svgwidth{.45\textwidth}
		\protect
\begingroup%
  \makeatletter%
  \providecommand\color[2][]{%
    \errmessage{(Inkscape) Color is used for the text in Inkscape, but the package 'color.sty' is not loaded}%
    \renewcommand\color[2][]{}%
  }%
  \providecommand\transparent[1]{%
    \errmessage{(Inkscape) Transparency is used (non-zero) for the text in Inkscape, but the package 'transparent.sty' is not loaded}%
    \renewcommand\transparent[1]{}%
  }%
  \providecommand\rotatebox[2]{#2}%
  \ifx\svgwidth\undefined%
    \setlength{\unitlength}{373.231915bp}%
    \ifx\svgscale\undefined%
      \relax%
    \else%
      \setlength{\unitlength}{\unitlength * \real{\svgscale}}%
    \fi%
  \else%
    \setlength{\unitlength}{\svgwidth}%
  \fi%
  \global\let\svgwidth\undefined%
  \global\let\svgscale\undefined%
  \makeatother%
  \begin{picture}(1,0.85977882)%
    \put(0,0){\includegraphics[width=\unitlength]{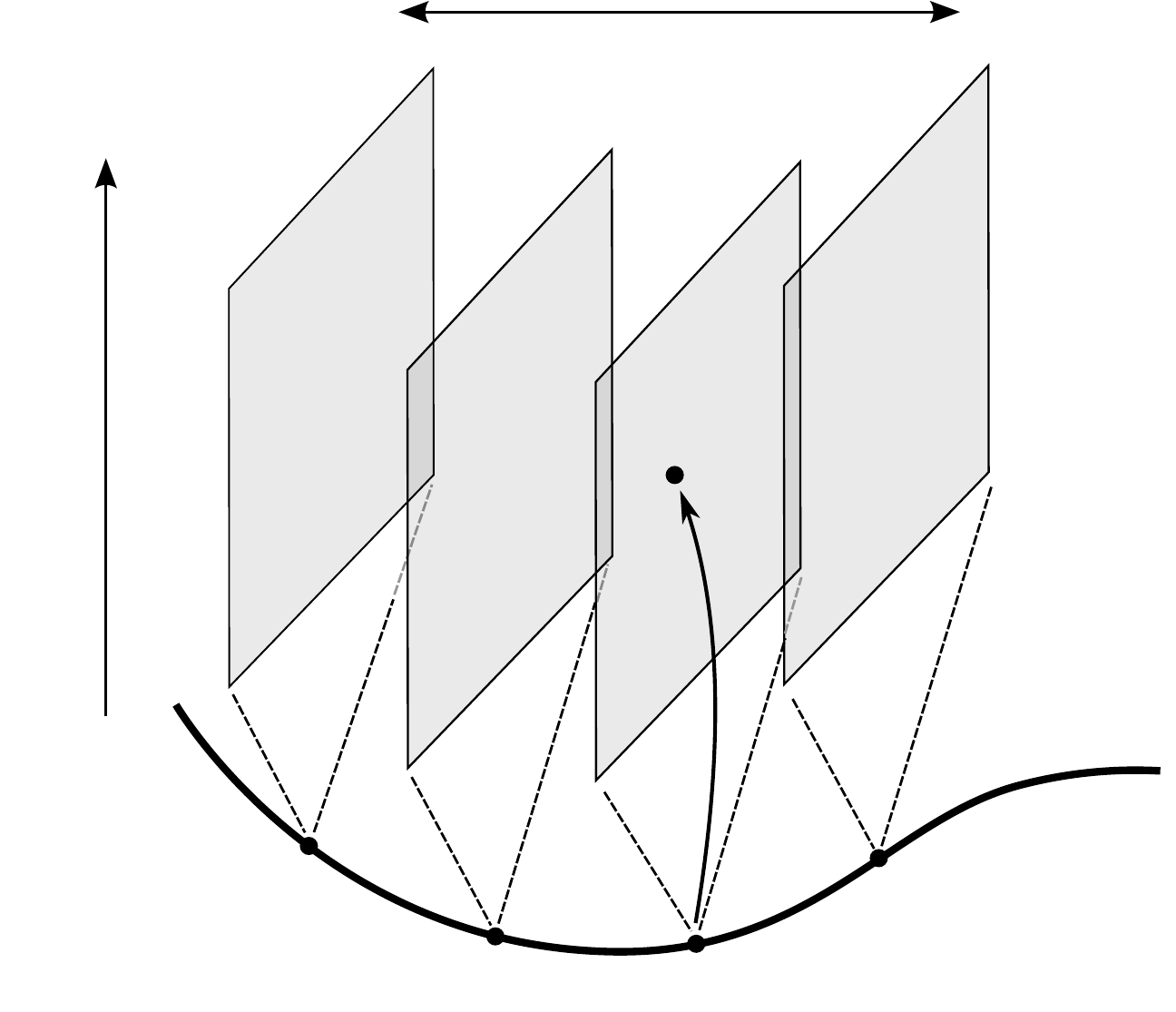}}%
    \put(0.5442842,0.78054658){\color[rgb]{0,0,0}\makebox(0,0)[lb]{\smash{$E$}}}%
    \put(0.204337,0.11053035){\color[rgb]{0,0,0}\makebox(0,0)[lt]{\begin{minipage}{0.0666348\unitlength}\raggedright $x_1$\end{minipage}}}%
    \put(0.39232198,0.03256663){\color[rgb]{0,0,0}\makebox(0,0)[lt]{\begin{minipage}{0.0666348\unitlength}\raggedright $x_2$\end{minipage}}}%
    \put(0.56961816,0.02304738){\color[rgb]{0,0,0}\makebox(0,0)[lt]{\begin{minipage}{0.0666348\unitlength}\raggedright $x_3$\end{minipage}}}%
    \put(0.73263545,0.08730236){\color[rgb]{0,0,0}\makebox(0,0)[lt]{\begin{minipage}{0.0666348\unitlength}\raggedright $x_4$\end{minipage}}}%
    \put(0.95871781,0.17773531){\color[rgb]{0,0,0}\makebox(0,0)[lt]{\begin{minipage}{0.0666348\unitlength}\raggedright $M$\end{minipage}}}%
    \put(0.71803611,0.58201014){\color[rgb]{0,0,0}\makebox(0,0)[lt]{\begin{minipage}{0.15230816\unitlength}\raggedright $E_{x_4}$\end{minipage}}}%
    \put(0.52324821,0.54007082){\color[rgb]{0,0,0}\makebox(0,0)[lt]{\begin{minipage}{0.14992825\unitlength}\raggedright $\sigma(x_3)$\end{minipage}}}%
    \put(0.22095733,0.5930066){\color[rgb]{0,0,0}\makebox(0,0)[lt]{\begin{minipage}{0.15230816\unitlength}\raggedright $E_{x_1}$\end{minipage}}}%
    \put(-0.00158996,0.47437776){\color[rgb]{0,0,0}\makebox(0,0)[lt]{\begin{minipage}{0.14992825\unitlength}\raggedright $\sigma$\end{minipage}}}%
  \end{picture}%
\endgroup%

}}\\ In the Figure there is a very schematic visualization of a section $\sigma$ of a vector bundle $E$.},plural={sections}}
\newglossaryentry{completion}{name={completion},description={nog invulling aan geven}}
\newglossaryentry{vector bundle}{name={vector bundle},description={A vector bundle over a manifold $M$ is the object $(M, \pi, E)$, where $\pi : E \to M$ is a continuous surjection and 
\bas
	E = \bigsqcup_{x \in M} E_x\qquad E_x = \pi^{-1}(x),
\eas 
the disjoint union of spaces $E_x$, each carrying the structure of a vector space (see e.g.~\cite[Ch.~2]{GVF00}). A classical example is $E = TM$ (with $E_x = T_xM$, the tangent space of $M$ at the point $x$).\\
{\protect\makebox[\textwidth][c]{
	\def\svgwidth{.35\textwidth}
		\protect
\begingroup%
  \makeatletter%
  \providecommand\color[2][]{%
    \errmessage{(Inkscape) Color is used for the text in Inkscape, but the package 'color.sty' is not loaded}%
    \renewcommand\color[2][]{}%
  }%
  \providecommand\transparent[1]{%
    \errmessage{(Inkscape) Transparency is used (non-zero) for the text in Inkscape, but the package 'transparent.sty' is not loaded}%
    \renewcommand\transparent[1]{}%
  }%
  \providecommand\rotatebox[2]{#2}%
  \ifx\svgwidth\undefined%
    \setlength{\unitlength}{470.67274871bp}%
    \ifx\svgscale\undefined%
      \relax%
    \else%
      \setlength{\unitlength}{\unitlength * \real{\svgscale}}%
    \fi%
  \else%
    \setlength{\unitlength}{\svgwidth}%
  \fi%
  \global\let\svgwidth\undefined%
  \global\let\svgscale\undefined%
  \makeatother%
  \begin{picture}(1,0.89724587)%
    \put(0,0){\includegraphics[width=\unitlength]{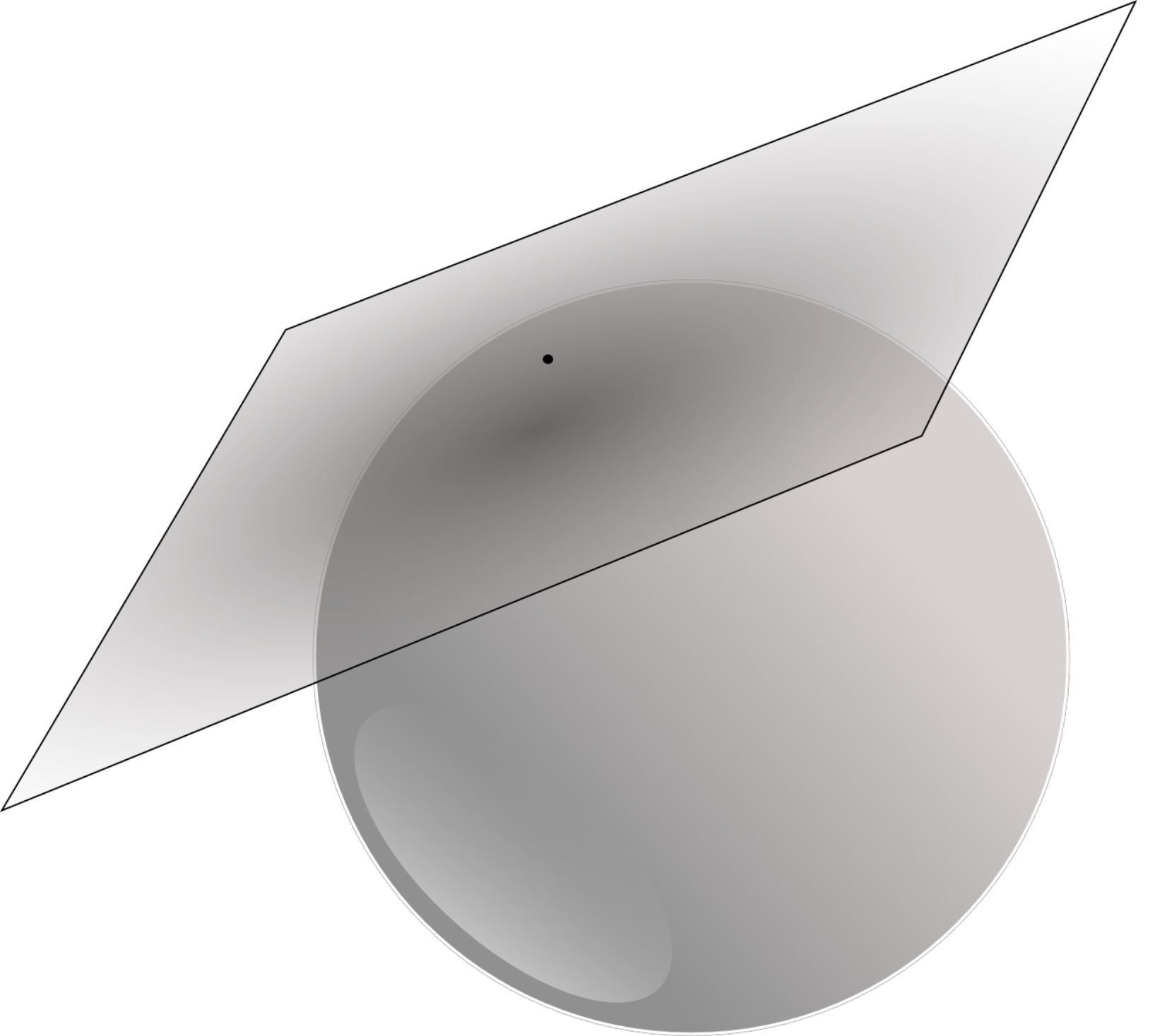}}%
    \put(0.95408815,0.17911114){\color[rgb]{0,0,0}\makebox(0,0)[lt]{\begin{minipage}{0.28250734\unitlength}\raggedright $S^2$\end{minipage}}}%
    \put(0.90806883,0.58177997){\color[rgb]{0,0,0}\makebox(0,0)[lt]{\begin{minipage}{0.20836511\unitlength}\raggedright $T_xS^2$\end{minipage}}}%
    \put(0.4295723,0.56263358){\color[rgb]{0,0,0}\makebox(0,0)[lt]{\begin{minipage}{0.18407715\unitlength}\raggedright $x$\end{minipage}}}%
  \end{picture}%
\endgroup%

}}\\ In the Figure this is demonstrated for the case of the two-sphere $S^2$. For each of its points $x$ we can take the tangent space $T_xS^2$, the collection of all such spaces makes up $TS^2$.}}

\lccode`\(=`\(
\lccode`\)=`\) 
\graphicspath{{./gfx}}

\begin{document}

\frenchspacing
\raggedbottom
\selectlanguage{american} 
\pagenumbering{roman}
\pagestyle{plain}

\refstepcounter{dummy}
\pdfbookmark[0]{}{}
\thispagestyle{empty}

        \begin{center}
\ \\[50pt]
{\LARGE Supersymmetry and the Spectral Action}\\[20pt]
 {\Large On a geometrical interpretation of the MSSM}\\[130pt]

\vspace{\stretch{1}}

Thijs van den Broek

\end{center}

\thispagestyle{empty}

\hfill

\vfill

This work has been funded by Nikhef, the National Institute for Subatomic Physics, in Amsterdam. \\[10pt]
Printed by Ipskamp Drukkers. \\[10pt]
\noindent\myName: \textit{\myTitle,} \mySubtitle. 

%
%
%
%
%

\cleardoublepage
\refstepcounter{dummy}
\pdfbookmark[0]{}{}
\thispagestyle{empty}

        \begin{center}
\ \\[50pt]
{\LARGE Supersymmetry and the Spectral Action}\\[20pt]
 {\Large On a geometrical interpretation of the MSSM}\\[130pt]

 {\large 
Proefschrift\\[20pt]
ter verkrijging van de graad van doctor\\
aan de Radboud Universiteit Nijmegen \\
op gezag van de rector magnificus prof.~mr.~S.C.J.J.~Kortmann,\\
volgens besluit van het college van decanen in het openbaar te verdedigen op vrijdag 5 september 2014
om 12.30 uur precies\\[45pt]
door} \\[45pt]
{\Large Thijs Cornelis Henricus van den Broek}\\[15pt]
geboren op 23 januari 1983 te Eindhoven

\end{center}

\newpage

\begin{tabular}{p{.5cm}p{.8\textwidth}}
\multicolumn{2}{l}{\emph{Promotor}}\\ 
	& Prof.~dr.~R.H.P.~Kleiss\\[10pt]
\multicolumn{2}{l}{\emph{Copromotoren}}\\ 
& Dr.~W.J.P.~Beenakker \\
							& Dr.~W.D.~van Suijlekom\\[10pt]
\multicolumn{2}{l}{\emph{Manuscriptcommissie}}\\ 
				& Prof.~dr.~S.J.~de Jong\\
				& Prof.~dr.~A.~Connes, Institut des Hautes \'Etudes Scientifiques, Bures-sur-Yvette, France\\
				& Prof.~dr.~F.~Lizzi, Universit\`a di Napoli Federico \rnum{2}, Napoli, Italy\\
				& Prof.~dr.~A.N.~Schellekens\\
				& Prof.~dr.~T.~Sch\"ucker, Universit\'e de Provence, Aix-Marseille, France
\end{tabular}

\cleardoublepage
\pdfbookmark[1]{Abstract}{Abstract}
\begingroup
\let\clearpage\relax
\let\cleardoublepage\relax
\let\cleardoublepage\relax

\chapter*{Foreword}

Dear reader,

Apparently you have started reading my thesis on the interplay between noncommutative geometry and supersymmetry, that is a result of the research I did for my PhD. 
I hope you will enjoy it.

Of course this thesis is self-contained, starting with an introduction and finishing with the conclusions, but as a guide for reading I have some suggestions.
\begin{itemize}
	\item To make life a bit easier for the non-physicist: I recommend you to at least start with the English or Dutch summaries in Chapter \ref{ch:summary}.
	\item The (high energy) physicist might skip Sections \ref{sec:introHep} on high energy physics and \ref{sec:introSUSY} on supersymmetry and start with the introduction to noncommutative geometry in Section \ref{sec:introNCG}.
	\item In contrast, the more mathematically inclined might benefit from reading Sections \ref{sec:introHep} and \ref{sec:introSUSY} to get a sense of what I am trying to accomplish in the context of noncommutative geometry.
	\item Those already familiar to noncommutative geometry can even consider just starting with Chapter \ref{ch:constraints}, using the previous chapter as referencing material. 
\end{itemize}

\begin{flushright}
Thijs van den Broek\\
January 2014
\end{flushright}

%

\endgroup			

\vfill

\pagestyle{scrheadings}
\cleardoublepage
\refstepcounter{dummy}
\pdfbookmark[1]{\contentsname}{tableofcontents}
\setcounter{tocdepth}{1} 
\setcounter{secnumdepth}{2} 
\manualmark
\markboth{\spacedlowsmallcaps{\contentsname}}{\spacedlowsmallcaps{\contentsname}}
\tableofcontents 
\automark[section]{chapter}
\renewcommand{\chaptermark}[1]{\markboth{\spacedlowsmallcaps{#1}}{\spacedlowsmallcaps{#1}}}
\renewcommand{\sectionmark}[1]{\markright{\thesection\enspace\spacedlowsmallcaps{#1}}}

\cleardoublepage

\pagenumbering{arabic}
\cleardoublepage





\chapter{Introduction}
\section{High energy physics}\label{sec:introHep}

Man has come a long way in his understanding of the world surrounding him. With the progression of time, fewer and fewer of the natural phenomena that once were mysteries to us have remained so. One of the oldest and most compelling questions in this respect is what our world \emph{is made of}. Aided by technology, wit and a vast amount of curiosity we managed to probe into the structures of matter deeper and deeper, uncovering smaller building blocks along the way.

It is widely recognized that this line of thought started with the hypothesis of \emph{atomos} ---the invisible, basic constituents of matter--- by Demokritos and his teacher Leucippus in ancient Greece. But what we nowadays conceive of as atoms have turned out to be far from indivisible. Work by J.J.~Thomson (1897), E.~Rutherford (1911) and others revealed that these atoms are made of a core, consisting of \emph{protons} and \emph{neutrons}, and one or more surrounding electrons. A big puzzle remained however. Electrodynamics, the theory that describes how charged particles such as the electron move, predicts that atoms will not be stable. The electrons, constantly losing kinetic energy from \emph{bremsstrahlung}, would quickly spiral into and smash against the core. Stable atoms should thus not exist. Electrodynamics turned out to be a theory unsuitable for the atomic realm. Einstein, Bohr, Schr\"odinger, Heisenberg and many others subsequently devised \emph{quantum mechanics} (QM) in which atoms absorb and emit energy only in discrete \emph{quanta}, thereby circumventing the problem of unstable atoms. A key ingredient of quantum mechanics is that the classical phase space coordinates $(\vec{x}, \vec{p})$ should be promoted to \emph{operators} $(\hat{\vec{x}}, \hat{\vec{p}})$ on wave functions $\psi$, satisfying the fundamental commutation relation \cite{BJ25}
\begin{align*}
	[\hat{x}_i, \hat{p}_j] = i\hbar \delta_{ij}.
\end{align*}
For the first time physics faced objects that are \emph{not-commuting}, i.e.~for which $[x, y] := xy - yx \ne 0$. The wave functions are the solutions of the \emph{Schr\"odinger equation}: 
\begin{align*}
	\bigg(\frac{\hat{\vec{p}}\cdot\hat{\vec{p}}}{2m} + V(\hat{\vec{x}})\bigg)\psi = i\hbar\frac{\partial}{\partial t}\psi,\qquad \hat{p}_j = - i\hbar \frac{\partial}{\partial x_j}, i^2 = -1,
\end{align*}
with $V(\hat{\vec{x}})$ an external potential, $m$ the mass of the particle and $\hbar$ Planck's constant. The relevant physical quantity is then $|\psi(\vec{x}, t)|^2$, the probability density of the particle at $(\vec{x}, t)$. As marvelous and widely applicable quantum mechanics turned out to be, its foremost shortcoming is that it is a non-relativistic theory, only valid for speeds much smaller than that of light $c$. Already soon after the advent of QM, Paul Dirac translated \cite{D28} the Schr\"odinger equation into what is nowadays called the \emph{Dirac equation}:
\ba
	(i\hbar\slashed{\partial} - mc)\psi = 0,\qquad \slashed{\partial} \equiv \gamma^\mu \partial_\mu,\label{eq:intro_dirac}
\ea
where the $\gamma^\mu$ are the $4 \times 4$ \emph{gamma matrices} acting on the \emph{spinor} $\psi$. Treating space and time on equal footing, this does yield a relativistic theory, valid for all physically allowed speeds. It was found to describe particles that possess a spin of $\tfrac{1}{2}$, such as the electrons revolving around an atom's core. This equation, amongst others, allowed not only stable atoms, but could predict the spectrum of the hydrogen atom to an unprecedented precision. 

Soon after, in 1929, Werner Heisenberg and Wolfgang Pauli proposed \cite{HP30} that not only the energy of a system can be quantized, but also the wavefunctions themselves are quantized objects. This insight paved the way for the framework that is now known as Quantum Field Theory (QFT). It is considered to be \emph{the} theory, applicable both for arbitrarily small scales, and for speeds arbitrarily close to the speed of light. Put differently, it takes two of the three fundamental constants of nature ($\hbar$, $c$ and the gravitational constant $G$) into account, see Figure \ref{fig:intro_theory_cube}. 

\fig{ht!}{.6\textwidth}{intro_theory_cube_color}{.8\textwidth}{If each of the three fundamental constants of nature is visualized as a dimension, each of which can be taken into account or not, then all major theories in physics (Classical Mechanics, Newtonian Gravity, the theories of Special Relativity and General Relativity, Quantum Mechanics and Quantum Field Theory) are seen to correspond to corners of the cube. The holy grail in physics, Quantum Gravity, corresponds to the front top corner.}{intro_theory_cube}

\subsection{Quantum Field Theory}\label{sec:QFT}

The subject of QFT is vast and intricate, and consequently I can only scratch the surface when it comes to introducing it. For any details I refer the reader to one of the many textbooks that have already been written on QFT, such as \cite{PS95,W05-1,W05,ZJ02}, each with its particular traits. From here on we will switch to natural units, setting $\hbar = 1$, $c = 1$.

A basic ingredient of almost any physical theory is the space(-time) in which physical events take place. In QFT it is usually taken to be Minkowski space $\mathbb{R}^{1,3}$, equipped with a (non-positive definite) inner product $\langle .,.\rangle : \mathbb{R}^{1,3} \times \mathbb{R}^{1,3} \to \mathbb{R}$. The Poincar\'e group $\mathbb{R}^{1,3} \rtimes O(1, 3)$ is defined as the group of isometries of Minkowski space, i.e.~all operations $O$ that preserve the inner product $\langle x - y, x- y\rangle$ for any two space-time events $x$ and $y$. 
Besides the discrete space- and time-inversions, it consists of the \emph{Lorentz boosts}, \emph{rotations} and \emph{translations}. Wigner classified \cite{W39} all unitary (and physically acceptable) representations of the Poincar\'e group and found that they are characterized by two numbers $(s, m)$, where $s \in \frac{1}{2}\mathbb{N}$ is the \emph{spin} of the representation and $m \in \mathbb{R}_+$ its \emph{mass}.  

The most common representations in QFT are the ones having spin $0$, $\tfrac{1}{2}$ and $1$. The Lorentz group $O(1, 3)$, the subgroup of the Poincar\'e group that excludes the translations, acts in the following way on these representations:
	\begin{align*}
		\phi(x) &\to U(\Lambda)\phi(x) := \phi(\Lambda^{-1} x)&& \text{(spin 0)}\\
		\psi(x) &\to U(\Lambda)\psi(x) := S(\Lambda)\psi(\Lambda^{-1} x)&&\text{(spin }\tfrac{1}{2}{)}\\
		A^\mu(x) &\to U(\Lambda)A^\mu(x) := (\Lambda^\mu_{\phantom{\mu}\nu} A^\nu)(\Lambda^{-1} x)&&\text{(spin 1)}
	\end{align*}
for any $\Lambda \in O(1, 3)$. Here $S(\Lambda) = \exp(-\tfrac{i}{2}\omega_{\mu\nu}S^{\mu\nu})$ with $S^{\mu\nu} = \tfrac{i}{4}[\gamma^\mu, \gamma^\nu])$ and $\omega_{\mu\nu}$ a real, antisymmetric tensor that represents $\Lambda$. The components $\omega_{ij}$ ($i,j = 1,2,3$) correspond to rotations, $\omega_{0i}$ to boosts. 

\begin{figure}[ht]
	\centering
	\includegraphics[width=.6\textwidth]{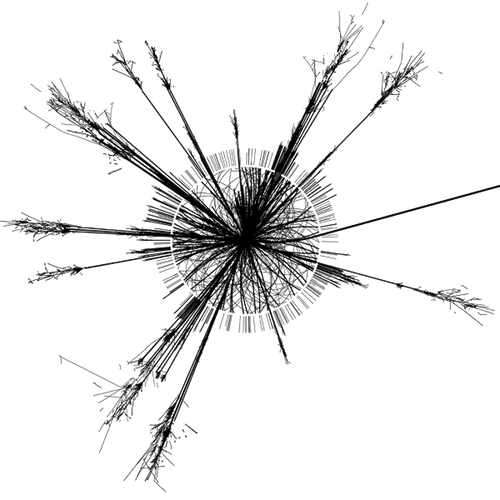}
	\captionsetup{width=.6\textwidth}
	\caption{An event that results from two colliding high energy protons, as is seen in the ATLAS detector at CERN. Image was taken from the CERN document server, at \url{cds.cern.ch/record/1096078}.}
	\label{fig:event}
\end{figure}

Given a set of fields we ultimately want to determine how they interact, i.e.~to calculate cross sections of the various possible interactions and to confront the results with experimental data. But how to go about this? The starting point for this is to specify an \emph{action} functional
\ba
	S[\{\Phi\}] \in \mathbb{R}\label{eq:intro_action},
\ea
where we have written $\{\Phi\} = \{\phi_1, \ldots \phi_l, \psi_1, \ldots \psi_m, A_1, \ldots A_n\}$ for the collection of fields present in the theory.
Often this is written in terms of a \emph{Lagrangian density} $\mathcal{L}$ via
\begin{align}\label{eq:intro_def_action}
	S[\{\Phi\}] = \int_{\mathbb{R}^{1,3}} \mathcal{L}(\{\Phi\}, \partial_\mu\{\Phi\}) \sqrt{g}\mathrm{d}^4 x,\quad g = |\det g_{\mu\nu}|,
\end{align}
where $g_{\mu\nu}$ is the metric and with $\partial_\mu\{\Phi\}$ we have indicated that the Lagrangian (density) typically also depends on (the first) derivatives of the fields, see \eqref{eq:electron_kin} for an example. The machinery of QFT is then aimed at calculating (amongst others) cross sections from this. This is typically done in three steps, as shown schematically:
\begin{align}\label{eq:action_to_crosssection}
	\text{Action } \to \text{ Feynman rules } \to  \text{ Matrix elements } \to \text{ Cross sections}.
\end{align}
As powerful a framework as this might be, it is insufficient to describe what we observe in experiments (e.g.~Figure \ref{fig:event}). What exactly determines which particles interact with each other and how? 

\subsection{The Standard Model}\label{sec:intro_SM}

Via the Poincar\'e group we have so far only described the `external' degrees of freedom (i.e.~related to space-time), but particles are also seen to have `internal' properties, such as the charge of an electron, or the colour of a quark. As for the first example, electrons are seen to interact with light, e.g.~photons. How can such an interaction be understood in QFT? These internal properties cannot be accounted for by the Poincar\'e group, and so we have to turn to the \emph{gauge principle}.

There is the notion of a \emph{gauge group}; a Lie group $\G$ of which the finite part of the fields are a representation:
\begin{align*}
	(\pi(g) \phi)(x) &= g(x)\cdot \phi(x),\qquad \forall g \in \G,
\end{align*}
for any representation $\phi(x)$ of the Poincar\'e group. This is called a \emph{gauge transformation}. Notice the difference with the representation of the Poincar\'e group: even though $g \in \G$ varies with the space-time coordinate $x$, it does not \emph{act} on it.

The only permissible terms in the action \eqref{eq:intro_def_action} are those that keep it gauge invariant:
\bas
S[\pi(g)\phi, \pi(g)\psi, \ldots] = S[\phi, \psi, \ldots].
\eas
In the case of the electron, for example, this means that the kinetic term
\ba
	\mathcal{L}(\bar\psi, \psi) = \bar\psi(i\slashed{\partial} - m)\psi\label{eq:electron_kin},\qquad \bar\psi = \psi^*\gamma^0,
\ea 
whose equation of motion is \eqref{eq:intro_dirac}, as such is not permitted if we require local phase-invariance. The reason for this is that upon letting $\psi \to g\psi$, $g = e^{i\alpha} \in C^{\infty}(\mathbb{R}^{1,3}, U(1))$ we have
\bas
	\bar\psi i\slashed{\partial} \psi \to \bar\psi e^{- i\alpha} i\slashed{\partial}e^{i\alpha} \psi = 
	\bar\psi i\slashed{\partial} \psi - \bar\psi\gamma^\mu\psi \partial_\mu(\alpha).
\eas
Since the result is not equal to $\bar\psi i \slashed{\partial}\psi$, \eqref{eq:electron_kin} is not gauge invariant. The standard solution (see e.g.~\cite[Ch 4.1.]{PS95}) is to \emph{make} the expression gauge invariant again by means of \emph{minimal coupling}:
\ba\label{eq:min_coupling}
	\partial_\mu \to D_\mu := (\partial_\mu - i A_\mu),
\ea
where $A_\mu$ is the (Hermitian) photon field. Thus, we must add to the Lagrangian density \eqref{eq:electron_kin} an interaction $\bar\psi \gamma^\mu A_\mu\psi$, the latter transforming as 
\bas 
	A_\mu \to A_\mu + ie^{i\alpha}\partial_\mu e^{-i\alpha} 
\eas 
under the gauge group. To have propagating photons, the action then has to feature a kinetic term for the photon field as well:
\ba\label{eq:gauge_kin}
	- \frac{1}{4}\int F_{\mu\nu}F^{\mu\nu}\sqrt{g}\mathrm{d}^4x,\qquad F_{\mu\nu} = \partial_\mu A_\nu - \partial_\nu A_\mu,
\ea
which is also gauge invariant. The photon field $A_\mu$, whose interaction strength is proportional to the charge of the particles involved, mediates the electromagnetic force and is thus said to be the `carrier' of that force. Using essentially the same argument all such force carriers, called \emph{gauge bosons}, are introduced into the theory this way. 

The Standard Model (SM) can be obtained by taking the gauge group to be 
\ba\label{eq:SM_gauge_group}
	\G = U(1) \times SU(2) \times SU(3).
\ea 
Taking the representations that are listed in the upper part of Table \ref{tab:SM}, then via the gauge principle the gauge bosons listed in the middle part ---corresponding to the hypercharge, weak and strong forces--- appear. The corresponding Lagrangian density is 
\bas
	\mathcal{L}_{\mathrm{kin}} = \mathcal{L}_{\mathrm{Dirac}} + \mathcal{L}_{\textrm{gauge}}.
\eas
Here, $\mathcal{L}_{\mathrm{Dirac}}$ consists of the kinetic terms for each of the fermionic representations in Table \ref{tab:SM} (such as \eqref{eq:electron_kin}, but coupled to the appropriate gauge bosons via \eqref{eq:min_coupling}) and $\mathcal{L}_{\textrm{gauge}}$ denotes the kinetic terms for each of the gauge bosons in Table \ref{tab:SM} (such as \eqref{eq:gauge_kin}). But this is not the full story: experimentally we know that the fermions possess a mass, but concocting a Lagrangian that describes these masses is not trivial. The Dirac representation is reducible and $\psi= (\psi_L, \psi_R)$, where the labels $L$ and $R$ denote the $\pm$--eigenstates of the operator $\gamma^5$, to which we refer as \emph{left- and right-handed} particles. The density \eqref{eq:electron_kin} might feature a mass term, but since in the SM the left- and right-handed particles are in different representations of the gauge group, a term like
\ba
	m\bar\psi\psi = m (\bar \psi_L \psi_R + \bar\psi_R \psi_L) ,\qquad \psi_{L,R} = e_{L,R}, u_{L,R}, d_{L,R}, \ldots\label{eq:mass_wrong}
\ea
would not be gauge invariant and hence cannot be part of a viable Lagrangian density. On the other hand, 
\ba
	m (\bar\psi_L \psi_L + \bar\psi_R \psi_R), \qquad \psi_{L} = e_L, u_L, d_L, \ldots \label{eq:mass_wrong2}
\ea 
is gauge invariant but would vanish altogether, since the fields are of opposite handedness. This is where the Higgs boson, listed in the bottom part of Table \ref{tab:SM}, comes in. The \emph{Yukawa interactions}
\bas
	- \mathcal{L}_{\textrm{Yukawa}} &= \Upsilon_u \bar q_L \widetilde{\varphi} u_R + \Upsilon_d \bar q_L \varphi d_R + \Upsilon_\nu \bar l_L \widetilde{\varphi} \nu_R + \Upsilon_e \bar l_L \varphi e_R + h.c.,
\eas
that feature the Higgs $SU(2)$-doublet field $\varphi$ also feature the left- and right-handed fermions in a way similar as in \eqref{eq:mass_wrong} but now in a gauge invariant fashion. In the above expression 
\bas
	\widetilde{\varphi} := \begin{pmatrix} 0 & 1 \\ -1 & 0 \end{pmatrix}\overline{\varphi},
\eas
with $\overline{\varphi}$ the complex conjugate of $\varphi$, the symbols $\Upsilon_{u,d, e, \nu}$ denote the $3\times 3$ Yukawa-matrices on \emph{family-space} and a sum over the $3$ generations is implied.\footnote{Here we have tacitly assumed the existence of a right-handed neutrino, which is strictly speaking not part of the SM.} We add to the Lagrangian density a kinetic term for the Higgs and the Higgs potential:
\ba\label{eq:intro_higgs_potential}
	 \mathcal{L}_{\textrm{Higgs}} &= |D_\mu\varphi|^2 - V(\varphi), \qquad V(\varphi) = - \mu^2 |\varphi|^2 + \lambda |\varphi|^4\qquad \lambda, \mu^2 > 0.
\ea
Because of the opposite signs in the potential $V(\varphi)$ one finds that its minimum is not at $|\varphi| = 0$, but rather at $\mu /\sqrt{2\lambda} =: v/\sqrt{2}$. This minimum will serve as the ground state around which the quantum fluctuations (e.g.~the particles) should be considered. Expanding the field around the minimum, we can parametrize it in the \emph{unitary gauge} as 
\bas
	\varphi = \begin{pmatrix}0\\ \tfrac{1}{\sqrt{2}}v + H\end{pmatrix}\qquad\text{where } H(x) \in \mathbb{R},
\eas
which yields from $\mathcal{L}_{\mathrm{Yukawa}}$ the desired fermion mass terms:
\bas
	- \mathcal{L}_{\textrm{Yukawa}} &= \frac{v}{\sqrt{2}} \Big[\Upsilon_u \bar u_L u_R + \Upsilon_d \bar d_L d_R + \Upsilon_\nu \bar \nu_L \nu_R + \Upsilon_e \bar e_L e_R + h.c.\Big]
\eas
This is an example of \emph{spontaneous symmetry breaking}, resulting in 
\bas
	U(1)_{Y} \times SU(2)_L \times SU(3) \to U(1)_{\textrm{EM}} \times SU(3),
\eas
for the gauge group. Here $U(1)_{\textrm{EM}}$ is the group whose representations are labeled by the \emph{electromagnetic charges}. Apart from giving mass to the fermions, this mechanism also provides mass terms for the weak gauge bosons, arising from $|D_\mu \varphi|^2$.

 Since the right-handed neutrino is completely neutral under the gauge group, we can add (e.g.~\cite{D13}) to the Lagrangian density a \emph{Majorana mass term}
\ba
 	\mathcal{L}_{\textrm{Maj}} &= -\frac{1}{2} (\overline{\nu_R^c} M_R\nu_R + h.c.),
\ea
with $M_R$ a symmetric $3\times 3$--matrix in family space and $\nu_R^c := C\bar\psi^t$, the charge conjugate of $\nu_R$, where $\psi^t$ denotes the \emph{transpose} of $\psi$. When the entries of $M_R$ (after diagonalizing it) are large enough, a \emph{seesaw mechanism} \cite{MP04, Y79, GRS79} might explain why the upper bounds for the observed neutrino masses are so low. 

To summarize things, the most general Lagrangian density compatible\footnote{Modulo some terms like the $\theta$-term \cite[\S 23.6]{W05}.} with the representations that are present in the theory and the rules of the game (gauge invariance, etc.), is given by\footnote{See e.g.~\cite{PL03}, but in a slightly different notation. See also \cite{SSZ02}.}:
\ba\label{eq:intro_lag_sm}
	\mathcal{L}_{\textrm{SM}} = \mathcal{L}_{\mathrm{kin}} + \mathcal{L}_{\textrm{Higgs}}+ \mathcal{L}_{\textrm{Yukawa}} + \mathcal{L}_{\textrm{Maj}}. 
\ea
From this expression, all the tools that QFT provides us and some experimental input we can determine what we expect to see in particle colliders.

\begin{table}[ht]
\begin{tabularx}{\textwidth}{XllllX}
	\toprule	
		  & \textbf{Name} & \textbf{Symbol} & \textbf{Representation} &\\
	\midrule 
		\multicolumn{5}{l}{\hspace{.5cm}Fermions (per generation):}\\
	\midrule
	& right-handed up quarks & $u_R$ & $(2/3, 1, 3)$   &\\
	& right-handed down quarks & $d_R$ & $(-1/3, 1, 3)$ &  \\
	& left-handed quarks & $q_L = (u_L, d_L)$ & $(1/6, 2, 3)$&   \\
	& right-handed neutrinos & $\nu_R$ & $(0, 1, 1)$  & \\
	& right-handed electrons & $e_R$ & $(-1, 1, 1)$  & \\
	& left-handed leptons & $l_L = (\nu_L, e_L)$ & $(-1/2, 2, 1)$ & \\
	\midrule
		\multicolumn{5}{l}{\hspace{.5cm}Gauge bosons:}\\
	\midrule
&	 gluons & $\gluon_\mu$ & $(0, 1, 8)$  &\\
&	 weak-force bosons & $\vec{W}_\mu$ & $(0, 3, 1)$ & \\
&	 hypercharge field& $B_\mu$ & $(0, 1, 1)$ & \\
	\midrule
		\multicolumn{5}{l}{\hspace{.5cm}Scalar(s):}\\
	\midrule
&	 Higgs boson & $\varphi$ & $(1/2, 2, 1)$ & \\
	\bottomrule
\end{tabularx}
	\caption[The particles of the Standard Model]{The particles of the Standard Model. The last column gives the representation of the gauge group \eqref{eq:SM_gauge_group} the particle is in. The first number in that column denotes the hypercharge of the $U(1)$-representation. The second denotes the dimension of the $SU(2)$-representation: $1$ for trivial / singlet, $2$ for fundamental / defining and $3$ for adjoint. The third number is the dimension of the $SU(3)$ representation, $1$, $3$ or $8$. }
	\label{tab:SM}
\end{table}


On the 4$^\mathrm{th}$ of July 2012, the LHC collaboration claimed to have found a new particle whose mass was around $125$ GeV. At the time of writing of this thesis it is becoming more and more probable \cite{ATLAS-CONF-2013-034, ATLAS-CONF-2013-040} that this is indeed the long sought for Higgs boson. This discovery is the crown on the Standard Model and a tremendous landmark for physics.

\subsection*{The limits of the Standard Model}\label{sec:intro_break}

As successful the SM may be in describing the world at the scale of elementary particles, many believe that it cannot be the final theory of nature. At some point (with which is meant 'for some energy') new phenomena should occur that the SM cannot account for. To date, particle accelerator experiments have given us no reason to doubt the validity of the SM\footnote{See e.g.~\cite{ATL-PHYS-PROC-2013-339} concerning the search for supersymmetry (\S \ref{sec:introSUSY} ahead) in particular.}, but there is a number of experimental and theoretical arguments in favour of this view. We list the main ones.

\begin{itemize}
	\item No incorporation of gravity. We have four fundamental forces, only three of which are described with the SM. The remaining force, gravity, is very successfully described by means of Einstein's theory of General Relativity. Despite tremendous efforts no way of combining both, corresponding to the front top corner of the cube in Figure \ref{fig:intro_theory_cube}, has yet been found.\footnote{With this I mean theories that not only obtain the SM as a low energy limit, but also make one or more falsifiable predictions.}
	\item Large corrections to the Higgs mass. Due to interactions such as the one depicted in Figure \ref{fig:higgs_self} the Higgs receives extra contributions to its mass squared that are proportional to the square of the mass of the heaviest fermion it couples to (see e.g.~\cite[Ch. 1.2]{DGR04}). If we believe in a Grand Unified Theory (GUT) at a mass scale on the order of $10^{15}-10^{16}$ GeV and assume that there are fermions of this mass coupling to the Higgs, these contributions are huge compared to the \emph{bare} Higgs mass. This is not a sign of a robust theory! 

\fig{ht}{.5\textwidth}{higgs_self}{.5\textwidth}{A contribution to the Higgs self-energy due to its interaction with a fermion / anti-fermion pair $f_L, \bar f_R$.}{higgs_self}

	\item Dark matter (DM). There are various experimental signs\footnote{See e.g.~\cite{Colafrancesco2010} for an overview.} showing that we do not know what the vast majority of our universe consists of.\footnote{Recently, the Planck Collaboration (\cite{PlanckCollaboration2013a} Table 9, \cite{PlanckCollaboration2013}) fitted their cosomological data to the $\Lambda$CDM model (see e.g.~\cite{BCNO12}), implying Cold Dark Matter (CDM) to account for about 26\% of the total energy in the universe and baryonic matter making up only about 5\%.} The amount of matter that is directly visible to us on the one hand and the amount that we indirectly measure (e.g.~via gravity) on the other, are far apart. One of the most viable scenarios\footnote{See e.g.~\cite{Feng2010} for a review on the DM candidates.} is that dark matter consists of one or more types of \emph{weakly interacting massive particles} (or WIMPs), something the SM cannot account for.\footnote{The extension of the SM with right-handed neutrinos does provide WIMPs (the `sterile neutrinos'), but for them to serve as DM candidates means giving up the \emph{seesaw mechanism}, explaining the smallness of the neutrino masses \cite[\S\rnum{7}]{Feng2010}.}
\end{itemize}
 
The search for a viable extension of the SM is ongoing and of massive scale.

\section{Supersymmetry}\label{sec:introSUSY}


The past decades have witnessed the birth of a plethora of `Beyond the Standard Model' theories that try to remedy one or more of the aforementioned shortcomings. \emph{Supersymmetry} (SUSY) is a particular example of such a theory. The purpose of this section is to very briefly discuss its basis notions, apply it to the SM and review some relevant properties of the result. Good introductions to supersymmetry are \cite{DGR04,Lykken2007,Martin2011,Bilal2007}. A more mathematical approach can be found in \cite{freed2006}.

%


In the 1960s the question was raised whether there might be extensions of the Poincar\'e algebra (Section \ref{sec:QFT}), incorporating a symmetry that would prove to be valuable for physics. Coleman and Mandula \cite{CoMa67} proved that ---given certain conditions--- the Poincar\'e algebra constitutes all the symmetries of the $S$-matrix.

Several years later however, Haag et al.~\cite{Haag1975} showed that extending the Poincar\'e algebra \emph{can} possibly lead to new physics, if one extends the notion of a Lie algebra (as is the Poincar\'e algebra) to that of a graded Lie algebra. Elements of such an algebra have a specific degree which determines whether they satisfy commutator or anti-commutator relations. The Poincar\'e algebra (having only zero-degree elements) is then extended with a set of variables $Q^i_{a}$ and $\bar Q^i_{a}$ ($i = 1,\ldots,N$,\footnote{The possible values for $N$, the number of supersymmetry generators, depend on the space-time dimension. For example, for $d = 4$, $N = 1, 2, 4$ or $8$.} $a = 1,2$) of degree $1$ (i.e.~they satisfy anti-commutation relations), transforming in the $(\tfrac{1}{2}, 0)$ and $(0, \tfrac{1}{2})$ representations of the Lorentz group respectively. This extended algebra is called the \emph{supersymmetry algebra}. 
%

\emph{Throughout this thesis we will be considering the case $N=1$ only.}

The nature of these `fermionic' generators $Q, \bar Q$ is then that they relate bosons and fermions. Schematically:
\bas
	Q|\textrm{boson}\rangle &= |\textrm{fermion}\rangle,&
	Q|\textrm{fermion}\rangle &= |\textrm{boson}\rangle.
\eas
To be a bit more precise:

%
%

\begin{defin}[Supersymmetry transformation]
For a constant, two component spinor $\epsilon$, we define (cf.~\cite[p.~21]{wessbagger1992}) a \emph{supersymmetry transformation} on any representation $\zeta$ of the Poincar\'e algebra as 
\begin{align}
 \delta_{\epsilon}\zeta := [(\epsilon Q) + (\bar\epsilon\bar Q)]\zeta. 
\end{align}
\end{defin}


If we define such a $\delta_\epsilon\zeta_i(x)$ for each of the fields $\zeta_1, \ldots, \zeta_n$ appearing in a theory, we can talk about whether or not its action is invariant under supersymmetry. If
\ba
  \delta S[\zeta_1, \ldots, \zeta_n] := \frac{\mathrm{d}}{\mathrm{d}t} S[\zeta_1 + t\delta_\epsilon\zeta_1, \ldots, \zeta_n + t\delta_\epsilon\zeta_n]\Big|_{t=0}\label{eq:invariant_action}
\ea
equals $0$, we call the system \emph{supersymmetric}. A particularly simple example of a supersymmetric system is the following.

\begin{exmpl}[Wess-Zumino \cite{WZ74}]\label{ex:intro-susy-wz}
 The action of a system containing a free Weyl fermion $\xi$ and complex scalar field $\phi$, is (in the notation of \cite{DGR04}) given by 
\ba
 S[\phi, \xi, \bar \xi] = \int\Big(|\partial_{\mu}\phi|^2 + i\xi\sigma^{\mu}[\partial_{\mu}]\bar\xi\Big)\mathrm{d}^4x,\label{eq:intro-susy-wz-action}
\ea
where $\sigma^\mu = (I_2, \sigma^a)$ with $\sigma^a$, $a = 1,2,3$ the Pauli matrices, $\bar\xi$ is the Hermitian conjugate of $\xi$ and $X[\partial_\mu]Y := \tfrac{1}{2}X\partial_\mu Y - \tfrac{1}{2}(\partial_\mu X)Y$. This action is seen to be invariant under the transformations
\ba
  \delta_{\epsilon}\phi &:= \sqrt{2}\epsilon\cdot\xi, & 
  \delta_{\bar\epsilon}\xi &:= -\sqrt{2}i\sigma^{\mu}\bar\epsilon\partial_{\mu}\phi,\label{eq:intro-susy-wz-trans}
\ea
see \cite[\S 4.2]{DGR04}. Fields such as $\phi$ and $\xi$ are called each other's \emph{superpartners}.
\end{exmpl}
 
Actually, \eqref{eq:intro-susy-wz-action} is only supersymmetric \emph{on shell}, i.e.~to prove supersymmetry one has to invoke the equations of motion for $\xi$. This is caused by the fields having the same number of degrees of freedom on shell, but not off shell. We can make this work \text{off shell} as well by introducing a complex scalar (\emph{auxiliary}) field $F$ that appears in the Lagrangian through $\mathcal{L}_F = |F(x)|^2$. Modifying the transformations \eqref{eq:intro-susy-wz-trans} slightly to contain $F$, supersymmetry is seen to hold both on shell and off shell. The example above is a nice illustration of the necessary condition that 
 the total number of fermionic and bosonic degrees of freedom has to be the same
 in order for a system to exhibit supersymmetry at all.


\begin{exmpl}[Wess-Zumino \cite{WZ74}]\label{ex:intro-susy-sym}
	Another important example of a supersymmetric model is the \emph{super Yang-Mills system}, whose action is given by
	\ba
		\int \mathrm{d}^4x \Big(- \frac{1}{4}F_{\mu\nu} F^{\mu\nu} + i \lambda \sigma^\mu  [\partial_\mu] \bar\lambda + \frac{1}{2}D^2\Big).\label{eq:intro-susy-sym-action}
	\ea
	Here $F_{\mu\nu} = \partial_\mu A_\nu - \partial_\nu A_\mu$ is the field strength (curvature) of a $u(1)$ gauge field $A_\mu$, 
$\lambda$ a Weyl spinor and $D$ is a real $u(1)$ auxiliary field. The latter must again be added to ensure an equal number of bosonic and fermionic degrees of freedom both on and off shell. This action is seen to be invariant under the transformations
\bas
	\delta A_\mu &= \epsilon \sigma_\mu \bar \lambda + \lambda\sigma^\mu \bar\epsilon\nn\\
	\delta \lambda &= - \frac{i}{4} \sigma^\mu\sigma^\nu F_{\mu\nu} \epsilon + D\epsilon, \nn\\
	\delta D &= i \partial_\mu (\lambda \sigma^\mu \bar \epsilon + \bar\lambda\bar\sigma^\mu\epsilon),
\eas
where $\bar\sigma^\mu = (I_2, -\sigma^a)$ (see \cite{DGR04}, Chapters 4.1 and 4.4).
\end{exmpl}

In Table \ref{tab:susy_dofs} the role of the auxiliary fields is explicated for the Wess-Zumino and the super Yang-Mills models. For both the bosonic degrees of freedom are seen to be equal to the fermionic ones. 

\begin{table}[ht]
	\begin{tabularx}{\textwidth}{XllllXllllX}
	\toprule
		&	\textbf{Wess-Zumino:} &	$\phi$ & F & $\xi$ && \textbf{Super Yang-Mills:} &	$A_\mu$ & D & $\lambda$ &\\
	\midrule
		&	Off shell: & 2 & 2 & 4 && Off shell: & 3 & 1 & 4 &\\
		&	On shell: &  2 & 0 & 2 && On shell: & 2 & 0 & 2 &\\
	\bottomrule
	\end{tabularx}
	\caption{The number of real degrees of freedom both on and off shell for the Wess-Zumino and Super Yang-Mills models. In all cases the bosonic and fermionic number of degrees of freedom coincide.}
	\label{tab:susy_dofs}
\end{table}


In many of the more advanced treatments of supersymmetry (e.g.~\cite{wessbagger1992}), ordinary space is extended to a \emph{superspace} $(x^\mu, \theta, \overline{\theta})$ (where $\theta$ and $\overline{\theta}$ are two-component Grassmann variables). The particle content of a certain model is then described in terms of \emph{superfields} (fields depending on all coordinates of superspace and containing the particles that are each other's superpartners). Two key examples are the \emph{chiral superfield} $\Phi$, with the particle content of Example \ref{ex:intro-susy-wz}, and the \emph{vector superfield} $V$, whose particle content is that of Example \ref{ex:intro-susy-sym}.  
The action is recovered by integrating certain combinations of the superfields $\Phi$ and $V$ over superspace by means of a \emph{Berezin integral}. In this way the actions \eqref{eq:intro-susy-wz-action} for the chiral superfield and \eqref{eq:intro-susy-sym-action} for the vector superfield can be recovered.

\subsection{The supersymmetric version of the Standard Model}\label{sec:intro_MSSM}

When considering gauge theories, superpartners need to be in the same representation of the gauge group. A glance at Table \ref{tab:SM} makes it clear that the SM is not supersymmetric by itself. We have to introduce its superpartners to \emph{make} it supersymmetric however:
\begin{exmpl}[MSSM]\label{ex:intro_MSSM}
The \emph{Minimally Supersymmetric Standard Model} (MSSM) is the supersymmetric theory that is obtained by adding to the particle content a superpartner\footnote{This makes it an example of $N=1$ supersymmetry.} for each type of SM particles. In addition an extra Higgs doublet and its superpartner are introduced with hypercharge opposite to that of the other pair. One of the two pairs will give mass to the up-type particles, the other to the down-type ones. The adjective 'minimally' is justified by the fact that the MSSM is the smallest (i.e.~with the least number of additional superpartners) viable supersymmetric extension of the SM. See Table \ref{tab:MSSM} and e.g. \cite{DGR04,CEKKLW05} for details.
\end{exmpl}
The following nomenclature is used. The name of superpartners of the fermions get a prefix `s' (i.e.~selectron, stop, etc.). The superpartners of the bosons get the suffix 'ino' (i.e.~gluino, higgsino, etc.). 

Having two higgsino doublets with opposite hypercharge is necessary because adding only one higgsino doublet to the fermionic content of the SM will generate a chiral anomaly. The second higgsino is needed to cancel this anomaly again \cite[\S 8.2]{DGR04}.

The various superpartners are not only distinguished by their spin, but also by their \emph{$R$-parity}. This is a $\mathbb{Z}_2$-grading (or `discrete gauge symmetry') that for the MSSM is equal to
\ba
	R_p = (-1)^{2S + 3B + L}\label{eq:Rpar},
\ea
where $S$ is the spin of the particle, $B$ is its baryon number and $L$ its lepton number. It follows that all SM particles (including the extra Higgses) have $R$-parity $+1$, whereas all superpartners have $R$-parity $-1$.

\begin{table}[ht]
\begin{tabularx}{\textwidth}{XllllllX}
	\toprule	
  & \textbf{Superfield} 					& 			& \multicolumn{3}{c}{\textbf{Spin}} 	& \textbf{Representation} &\\
	\midrule 
	&																&				& $0$ & $\tfrac{1}{2}$ & $1$ 					& 								& \\
	\midrule
	& Left-handed (s)quark 					& $Q_L$ & \sq & $q_L$ 		& -- 							& $(1/6, 2, 3)$  	& \\
	& Up-type (s)quark 							& $U_R$ & \su & $u_R$ 		& -- 							& $(2/3, 1, 3)$  	& \\
	& Down-type (s)quark 						& $D_R$ & \sd & $d_R$ 		& -- 							& $(-1/3, 1, 3)$ 	&  \\
	& Left-handed (s)lepton 				& $L_L$ & \sle& $l_L$ 		& -- 							& $(-1/2, 2, 1)$  	& \\
	& Up-type (s)lepton 						& $N_R$	&\snu	& $\nu_R$ 	& -- 							& $(0, 1, 1)$  	& \\
	& Down-type (s)lepton 					& $E_R$ & \se & $e_R$			& -- 							& $(-1, 1, 1)$ 	&  \\
	\midrule 
	&	Gluon, gluino 								& $V$ 	& --	& \gluino{} & $\gluon_\mu$ 			& $(0, 1, 8)$ 		& \\
	&	$SU(2)$ g.~bosons, gauginos	& $W$ 	& --	& \wino{} 	& $\vec{\weak}_\mu$ & $(0, 3, 1)$ & \\
	& $B$-boson, bino								& $B$ 	& --	& \bino{}		&	$\photon_\mu$ 		& $(0, 1, 1)$ 		& \\
	&	Up-type Higgs(ino) 						& $H_u$ & \hu	& \shu 			& -- 							& $(1/2, 2, 1)$  	&\\
	&	Down-type Higgs(ino) 					& $H_d$	& \hd	& \shd 			& -- 							& $(-1/2, 2, 1)$  	&\\
	\bottomrule
\end{tabularx}
	\caption{The particle content of the $\nu$MSSM, the minimal supersymmetric extension of the Standard Model featuring a right-handed neutrino. Each line represents one superfield, with particle content as indicated. All superpartners are in the same representation of the gauge group. }
	\label{tab:MSSM}
\end{table}

The list of the MSSM's merits is quite impressive. See \cite[ch.~1]{CEKKLW05} for a short overview. Here we will pick out two:
\begin{enumerate}
	\item The MSSM makes the Higgs mass more stable. Roughly speaking, for each of the loop-interactions contributing to the mass of the Higgs there is a second such interaction that features a superpartner. This second contribution compensates for the first one.

	\item If $R$-parity is conserved in the MSSM, the lightest particle that has $R_p = -1$ cannot decay and thus provides a cold Dark Matter candidate.

	\item The additional particle content of the MSSM makes it possible for the three coupling constants $g_1$, $g_2$ and $g_3$ to evolve via the Renormalization Group Equations in such a way that they exactly meet at one energy scale. This hints at the existence of a Grand Unified Theory, that is hoped for by many theorists. See also Sections \ref{sec:intro_NCSM} and \ref{sec:motivation}.
\end{enumerate}

Despite the theoretical arguments in favour of the MSSM, so far no experimental hints for its existence have been detected \cite{ATL-PHYS-PROC-2013-339}.

\section{Noncommutative geometry}\label{sec:introNCG}
Although noncommutative geometry (NCG, \cite{C94}) is a branch of mathematics, there is a number of applications in physics. The aim of this section is to provide a bird's eye view of NCG in relation with its foremost such application. This is the interpretation of the Standard Model as a geometrical theory, a line of thought that started with the Connes-Lott model \cite{CL89} and culminated in \cite{CCM07} with the full SM, including a prediction of the Higgs boson mass. As much as possible I will focus on ideas and concepts and avoid the use of rigorous but technical statements, refering to the literature instead. Good general introductions to the field are e.g.~\cite{GVF00,Landi2008,V06}. The glossary in Appendix \ref{main} provides explanations to some of the terms that are used. In Appendix \ref{symbolslist} a list of symbols can be found. 

Physics and geometry in fact have a rather long and fruitful joint history. Think of the theory of General Relativity and the geometrical interpretation of gauge theories (see e.g.~\cite{EGH80}). NCG may be considered as a generalization of the former and incorporates the latter.

The field is rooted in an idea \cite{GelNai43} that dates back to the 1940s stating that any \gls{compact} \gls{Hausdorff} space $M$ and the commutative, \gls{unital} \glsadd{norm}\glsadd{algebra}\glsadd{banachalgebra}\gls{cstaralgebra} of continous functions on that space,
\begin{align*}
	\gls{CofM} =\{f : M \to \mathbb{C}, f\text{ is continuous}\}
\end{align*} 
are each other's category theoretical dual (loosely stated: they contain the same information). So instead of talking about spaces (which are a topological concept) we might equally well talk about commutative $C^*$-algebras (an algebraic concept). Building upon the above correspondence, various properties of the space $M$ can be translated into properties of the algebra $C(M)$ \cite[\S 1.11]{WO93}. This indicates a deep connection between two different fields of mathematics. 

The essential idea behind NCG is to generalize this correspondence in the sense that also noncommutative algebras are allowed. It provides concepts and techniques in order to work with these noncommutative algebras.

\subsection{Spectral triples}

One such concept lies at the very heart of NCG, namely that of a \emph{spectral triple}, describing a \emph{noncommutative manifold}.

\begin{defin}{(Spectral triple \cite{C94})}\label{def:spectral_triple} A \emph{spectral triple} is a triple $(\A, \H, D)$, where \gls{A} is a \gls{unital}, \gls{involutive} algebra that is represented as bounded operators on a Hilbert space \gls{space} on which also a \emph{Dirac operator} \gls{D} acts. The latter is an (unbounded) \gls{self-adjoint} operator that has \gls{compact} \gls{resolvent} and in addition $[D, a]$ is bounded for all $a \in \A$. 
\end{defin}
We will write $\langle . , . \rangle :\H \times \H \to \mathbb{C}$ for the inner product that $\H$ by definition is equipped with.

This is a rather abstract object. To make it a bit more tangible, we turn to the case of the space \gls{M} again. To make it more interesting for us, we require it to be enriched with extra structures. We will restrict ourselves to \emph{Riemannian spin manifolds}, spaces that \emph{locally} look like the Euclidean space $\mathbb{R}^n$ (for some $n$) on which a Riemannian metric \gls{g} (locally: $g_{\mu\nu}$) exists and that admit spinors. (For a comparison between results for Riemannian / Euclidean and pseudo-Riemannian / Minkowskian backgrounds, see the Appendix \ref{sec:mink_eucl}).
\footnote{Keep in mind though that Minkowski space is not an example of a Riemannian manifold. Rather it is pseudo-Riemannian since its metric is diagonal with negative entries. See Appendix \ref{sec:mink_eucl} for some comments on this.}
\begin{itemize}
\item The algebra \gls{Cinfty} is the subalgebra of $C(M, \com)$ containing only \emph{smooth} (i.e.~infinitely differentiable) functions. It can be made involutive (just as $C(M)$ itself) by defining $f^*: M \to \com$ through $(f^*)(x) := \overline{f(x)} \in \com$ for all $x \in M$.

\item The Hilbert space that is compatible with this algebra 
 is \gls{L2MS} --- or $L^2(S)$ for short. It consists of smooth, spinor-valued functions $\psi$ (i.e.~for each $x \in M$, $\psi(x) \in S_x$ is a spinor). The number of components of that spinor depends on the dimension \gls{m} of the manifold $M$: $\dim S_x = 2^n$, with $m = 2n$ or $m = 2n + 1$, according to whether $m$ is even or odd.\footnote{Technically, $L^2(M, S)$ is the completion of $\Gamma^{\infty}(S)$ ---the smooth \glspl{section} of a \gls{spinor bundle} $S$--- with respect to the inner product \bas\inpr{\psi_1}{\psi_2} := \int_M \sum_{i=1}^{2^n} \psi_1^*(x)_i\psi_2(x)_i \sqrt{g} \mathrm{d}^mx,\eas where $m = 2n$ or $m = 2n + 1$ and $\sqrt{g}$ ($g \equiv \det g$) is the Riemannian volume form. Here we have assumed the orientability of $M$. See e.g.~\cite{GVF00}, Chapter 9, for details.} Note that for a given manifold $M$, $L^2(M, S)$ need not even exist; its existence relies heavily on the properties of $M$ \cite{P86}.

\item The Levi-Civita \gls{connection} ---the unique \gls{connection} on $M$ that is compatible with the metric $g$--- can be lifted to act on spinor-valued functions. This leads to the operator 
\ba
	\gls{dirac} := i \gamma^\mu (\partial_\mu + \gls{omegamu})\label{eq:dirac_canon},
\ea
that is familiar from the Dirac equation \eqref{eq:intro_dirac}, save for the term 
\bas
	\omega_\mu= - \frac{1}{4}\widetilde{\Gamma}^b_{\mu a} \gls{vierbein}\gamma_b
\eas 
which accounts for the manifold $M$ being curved \cite[\S 9.3]{GVF00}. Here the latin indices $a,b$ indicate the use of a frame field \gls{h}, diagonalising the metric 
$
	g^{\mu\nu} = h^\mu_a h^\nu_b \delta ^{ab} 
$ 
and $\gamma$-matrices
\ba\label{eq:intro-gamma-diag}
	\{\gamma^a, \gamma^b\} = 2\delta^{ab},\qquad \gamma^{\mu} = h^\mu_a \gamma^a,
\ea
and $\widetilde{\Gamma}_{\mu a}^b := \Gamma_{\mu\nu}^{\lambda} h^{\nu}_ah_{\lambda}^{b}$, with $\Gamma^{\lambda}_{\mu\nu}$ the Christoffel symbols of the Levi-Civita connection. 
From the metric $g$ thus a Dirac operator is derived and conversely \cite{C89,V06} the metric is completely determined by the Dirac operator.
\end{itemize}
Together these three objects form the \emph{canonical spectral triple}:
\begin{exmpl}{(Canonical spectral triple \cite[Ch.~6.1]{C94})}\label{ex:canon}
	The triple 
	\bas
		(\A, \H, D) = (C^{\infty}(M), L^2(M, S), \dirac = i\gamma^\mu(\partial_\mu + \omega_\mu))
	\eas
is called the canonical spectral triple. Here $M$ is a compact Riemannian spin-manifold and $L^2(M, S)$ denotes the \gls{square-integrable} \glspl{section} of the corresponding \gls{spinor bundle}. The Dirac operator $\dirac$ is associated to the unique spin \gls{connection}, which in turn is derived from the Levi-Civita \gls{connection} on $M$. 
\end{exmpl}
The canonical spectral triple may be said to have served as the motivating example of the field; NCG is more or less modelled to be a generalization of it. 

In the physics parlance the canonical spectral triple roughly speaking determines a physical \emph{system}: the algebra encodes space(-time), the Hilbert space contains spinors `living' on that space(-time) and $\dirac$ determines how these spinors propagate.

A second important example is that of a \emph{finite spectral triple}:

\begin{exmpl}{(Finite spectral triple \cite{PS96, KR97})}\label{ex:finite}
	For a finite-dimensional algebra \gls{AF}, a finite-dimensional left \gls{module} \gls{HF} of $\A_F$ and a Hermitian matrix $\gls{DF} : \H_F \to \H_F$, we call $(\A_F, \H_F, D_F)$ a \emph{finite spectral triple}. 
\end{exmpl}
We will go into (much) more detail on finite spectral triples in Section \ref{sec:finite_krajewski}.

Given a spectral triple one can enrich it with two operators. The first of these, indicated by \gls{J}, has a role similar to that of charge conjugation, whereas the other, indicated by $\gamma$, allows you to make a distinction between positive (`left-handed') and negative (`right-handed') chirality elements of a (reducible) Hilbert space:
\begin{itemize}
\item We call a spectral triple \emph{even} if there exists a \gls{grading} $\gls{gamma} : \H\to\H$, with $[\gamma, a] = 0$ for all $a \in \A$ such that \ba\label{eq:Dg-anti-comm}\gamma D = - D\gamma.\ea
\item We call a spectral triple \emph{real} if there exists an \gls{antiunitary} operator (\emph{real structure}) $\gls{J} : \H \to \H$, satisfying
\ba\label{eq:JD-anti-comm}
	J^2 &= \gls{epsilon1} \id_{\H}, & JD &= \gls{epsilon2} DJ,\qquad \epsilon, \epsilon' \in \{\pm\}. 
\ea
The real structure implements a right action $a^o$ of $a \in \A$ on $\H$, via $\gls{aopp} := Ja^*J^*$ that is required to be compatible with the left action:
\ba\label{eq:left-right}
	[a, Jb^*J^*] = 0\quad,
\ea
i.e.~$(a\psi) b = a(\psi b)$ for all $a, b \in \A, \psi \in \H$. The Dirac operator and real structure are required to be compatible via the \emph{first-order condition}:
	\ba\label{eq:order_one}
		[[D, a], Jb^*J^*] = 0\quad \forall\ a, b \in \A.
	\ea
\item If a spectral triple is both real and even there is the additional compatibility relation 
\ba\label{eq:Jg-anti-comm}
	J\gamma = \gls{epsilon3} \gamma J,\qquad \epsilon'' \in \{\pm\}.
\ea
\end{itemize}
We denote such an enriched spectral triple by $(\A, \H, D; J, \gamma)$ and call it a \emph{real, even spectral triple}. The eight different combinations for the three signs above determine the \emph{KO-dimension} of the spectral triple, cf.~Table \ref{tab:ko_dimensions}. For more details we refer to \cite{GVF00, CM07}.

\begin{table}[ht]
\begin{tabularx}{\textwidth}{XlllllllllX}
\toprule
&	KO-dimension: 								& 0 & 1 & 2 & 3 & 4 & 5 & 6 & 7 &\\
\midrule
& $J^2 = \epsilon \id_{\H}$ 		& $+$ & $+$ & $-$ & $-$ & $-$ & $-$ & $+$ & $+$ & \\
& $JD = \epsilon' DJ$ 					& $+$ & $-$ & $+$ & $+$ & $+$ & $-$ & $+$ & $+$ & \\
& $J\gamma = \epsilon'' \gamma J$& $+$ &     & $-$ &     & $+$ &     & $-$ &     & \\
\bottomrule
\end{tabularx}
\caption{The various possible KO-dimensions and the corresponding values for the signs $J^2 = \epsilon \id_{\H}$, $JD = \epsilon' DJ$ and $J\gamma = \epsilon'' \gamma J$ \cite[\S 9.5]{GVF00}.}
\label{tab:ko_dimensions}
\end{table}

\fig{ht!}{\textwidth}{spectral_triple}{.8\textwidth}{A pictorial overview of the various relations that hold between the constituents of a real and even spectral triple. Not depicted here is the first order condition \eqref{eq:order_one}.}{spectral_triple}

\begin{exmpl}\label{ex:canon_real_even}
The canonical spectral triple (Example \ref{def:spectral_triple}) can be extended by a real structure \gls{JM} (`charge conjugation'). When $\dim M$ is even it can also be extended by a grading $\gls{gammaM} := (-i)^{\dim M/2}\gamma^1\ldots\gamma^M$ (`chirality', often denoted as $\gamma^{\dim M + 1}$). The KO-dimension of a canonical spectral triple always equals the dimension of the manifold $M$ (e.g.~\cite[\S 9.5]{GVF00}).	
\end{exmpl}

For $\dim M = 4$, the case we will be focussing on, we have\footnote{See Appendix \ref{sec:mink_eucl} for the relevant differences with the Minkowski background.}
\bas
	\gamma^5 := - \gamma^1\gamma^2\gamma^3\gamma^4,
\eas
which, using that $\{\gamma^i,\gamma^j\} = 2\delta^{ij}$ (cf.~\eqref{eq:intro-gamma-diag}), indeed satisfies $(\gamma^5)^2 = \id_{L^2(S)}$ and $(\gamma^5)^* = \gamma^5$. This enables us to reduce the space $L^2(M, S)$ into eigenspaces of $\gamma^5$:
\bas
	L^2(S) = L^2(S)_+ \oplus L^2(S)_-,\quad L^2(S)_{\pm} = \{\psi \in L^2(S), \gamma^5\psi = \pm \psi\} .
\eas
Also, $\gamma^5$ is seen to anticommute with $\dirac$. As for the real structure $J$, it is given \cite[\S 5.7]{Landi2008} pointwise as $(J\psi)(x) := C(x)\bar\psi(x)$ with $C(x)$ a \emph{charge conjugation matrix} and the bar denotes complex conjugation. 
One obtains \cite[\S 9.4]{GVF00} a charge conjugation operator that 
satisfies
\bas
	C^2 &= -1, & C\dirac &= \dirac C, & \gamma^5 C &= C\gamma^5.
\eas
Table \ref{tab:ko_dimensions} shows that the KO-dimension indeed equals $\dim M$. Note the difference with the charge conjugation operator in Minkowski space-time, where $C^2 = 1$ and $C\gamma^5 = - \gamma^5 C$ (for signature +\,-\,-\,-, e.g.~\cite[p.97/98]{Z03}) or $C^2 = -1$, $C\gamma^5 = - \gamma^5 C$ (for signature -+++, e.g.~\cite[\S 36]{SR07}).

\begin{exmpl}\label{ex:real_finite}
As in the general case a finite spectral triple (Example \ref{ex:finite}) is called real if there exists a \gls{JF} (implementing a \gls{bimodule} structure of $\H_F$) and even when there exists a \gls{grading} \gls{gammaF} on $\H_F$. 
\end{exmpl}

Given any two spectral triples $(\A_{1,2}, \H_{1,2}, D_{1,2}; J_{1,2}, \gamma_{1,2})$ their tensor product 
\begin{align*}
 (\A_1 \otimes \A_2, \H_1 \otimes \H_2, D_1\otimes 1 + \gamma_1 \otimes D_2, J_{\otimes}, \gamma_1 \otimes \gamma_2),
\end{align*}
is again a spectral triple. Here generally $J_{\otimes} = J_1\otimes J_2$, but with the following exceptions: $J_{\otimes} = J_1\gamma_1\otimes J_2$ when the sum of the respective KO-dimensions is $1$ or $5$ and $J_{\otimes} = J_1 \otimes J_2\gamma_2$ when the KO-dimension of the first spectral triple is $2$ or $6$ and that of the other one is even \cite{Dabrowski2010, Vanhecke2007}. The form of the Dirac operator of the tensor product is necessary to ensure that it anti-commutes with $\gamma_1 \otimes \gamma_2$ and that the resolvent remains compact. It follows that the KO-dimension of this tensor product is the sum of the KO-dimensions of the separate spectral triples. In the canonical spectral triple the algebra encodes space(-time), in a finite spectral triple it will seen to be intimately connected to the gauge group (see \eqref{eq:gauge_group} ahead). In describing particle models we need both. We therefore take the tensor product of a canonical and a finite spectral triple. In the case that $\dim M = 4$ this reads 
\ba\label{eq:acg}
 (C^{\infty}(M, \A_F), L^2(M, S \otimes \H_F), \dirac\otimes 1 + \gamma^5 \otimes D_F, J_M \otimes J_F, \gamma^5 \otimes \gamma_F),
\ea
with $C^{\infty}(M) \otimes \A_F \simeq C^{\infty}(M, \A_F)$. Spectral triples of this form are generally referred to as \emph{almost-commutative geometries} \cite{ISS03}. Noncommutative geometry can thus be said to put the external and internal degrees of freedom of particles on similar footing. To obtain your favourite particle physics model (in four dimensions) the key is to construct the right finite spectral triple that accounts for the gauge group and all internal degrees of freedom and interactions. 

\subsection{Gauge fields and the action functional}\label{sec:gauge_action}

Two more concepts need to be introduced. The first arises from the mathematically natural question ``to what extent is a given spectral triple $(\A, \H, D)$ unique?''. To this end we define the notion of \emph{unitarily equivalent} spectral triples:

\begin{defin}[Unitarily equivalent spin geometries \cite{V06}, \S 7.1]\label{def:unit_equiv}
  Two (real and even) spectral triples $(\A, \H, D; J, \gamma)$ and $(\A, \H, D'; J', \gamma')$ are said to be \emph{unitarily equivalent}, if there exists a unitary operator $U$ on $\H$ such that
  \begin{itemize}
    \item $Ua U^* = \sigma(a)\ \forall\ a \in \A$,
    \item $D' = UDU^*$,
    \item $J' = UJU^*$,
    \item $\gamma' = U\gamma U^*$.
  \end{itemize}
 Here $\sigma$ denotes an \emph{\gls{automorphism}} of the algebra $\A$. 
\end{defin}

Given an algebra $\A$ we can form the group of unitary elements of $\A$: \bas U(\A) := \{\gls{u} \in \A, uu^* = u^*u = 1\}\eas and construct unitary operators $\gls{U} := uJuJ^*$: 
\ba
	U : \H \to \H,\quad \psi \to u\psi u^*. \label{eq:unit_elts}
\ea
Using this group we can construct a particular kind of unitary equivalence for spectral triples, where the automorphism $\sigma$ is seen to be an \emph{inner automorphism}, i.e.~$UaU^* = uau^*$, where we have used \eqref{eq:left-right} and that $J^2 = \epsilon \id$. 

\begin{lem}\label{lem:unit_equiv}
	For $U = uJuJ^*$ with $u \in U(\A)$, the real and even spectral triples $(\A, \H, D; \gamma, J)$ and \ba (\A, \H, D + A + \epsilon' JAJ^*; J, \gamma)\quad\text with \quad A = u[D, u^*], u \in U(\A)\label{eq:unit_equiv},\ea are unitarily equivalent. 
\end{lem}
\begin{proof}
First, since $\gamma$ commutes with all the elements of $\A$ and $J\gamma = \epsilon'' \gamma J$ we have
\bas
	U\gamma U^* \equiv uJuJ^* \gamma Ju^* J^* u^* = (\epsilon'')^2 uJu J^* J u^* J^* u^* \gamma.
\eas
Using that $J^*J = JJ^* = \id$ and $uu^* = 1$, this reduces to $\gamma$.
 Second, we have for the real structure
\bas
	UJU^* \equiv uJuJ^*\, J\, Ju^*J^* u^* = 	uJu Ju^*J^* u^*.
\eas
If we employ \eqref{eq:left-right}, this becomes
\bas
	UJU^* = 	uJ Ju^*J^* u u^* = J, 
\eas 
where we have used that $uu^* = 1$ and twice that $J^*J = \id$ and $J^2 = \epsilon \id$. Finally, we have for the Dirac operator:
\bas
	UDU^* &\equiv uJuJ^* D Ju^*J^* u^* = \epsilon' uJu D u^*J^* u^*, 
\eas
where we have employed $JD = \epsilon' DJ$ and that $JJ^* = \id$. Now, using $uu^* = 1$, write $u D u^* = D + u[D, u^*]$ to yield
\bas
	UDU^* &= \epsilon' uJ(D + u [D, u^*])J^* u^*.
\eas
For the first term of this expression, apply these steps again to obtain $D + u[D, u^*]$. 
 For the second term we use that $J^* = \epsilon J$, insert $1 = JJ^* $ at the very end of this expression and employ the order one condition \eqref{eq:order_one} and the commutant property \eqref{eq:left-right} to get 
\bas
\epsilon'\epsilon^2 uJu [D, u^*](J u^* J^*) J^* = 
\epsilon' uJ (J u^* J^*)u[D, u^*] J^* = \epsilon' Ju[D, u^*]J^*,
\eas
where we have used again twice that $J^* = \epsilon J$.
 Taking terms together, we arrive at the result.
\end{proof}

This result implies that the class of unitarily equivalent spectral triples for $U = uJuJ^*$, $u \in U(\A)$ differ only by the \emph{inner fluctuations} of the Dirac operator. A more general ---but also a somewhat more involved--- way to look at this is by using the notion of \emph{Morita equivalence} of spectral triples (e.g.~\cite[\S \MakeUppercase{\romannumeral 11}]{C00}). In this way the inner fluctuations $A$ of
\bas
	D \to \gls{DA} := D + A + \epsilon' JAJ^*
\eas
are seen to be the self-adjoint elements of
\ba\label{eq:inner_flucts}
	\Omega^1_D(\A) := \Big\{ \sum_n a_n [D, b_n],\ a_n, b_n \in \A\Big\}.
\ea
The action of $U$ (Lemma \ref{lem:unit_equiv}) on $D_A$ (i.e.~$D_A \mapsto UD_AU^*$) induces one on the inner fluctuations: 
\begin{align}
	A \mapsto A^u := uAu^* + u[D, u^*]\label{eq:A_gauge_trans},
\end{align}
an expression that is reminiscent of the way gauge fields transform in quantum field theory. Note that the inner fluctuations that arise using the argument of unitary equivalence in fact only correspond to \emph{pure gauges}.
%
%

In the case of a canonical spectral triple ---for which the left and right actions coincide--- that has $JD = DJ$, the inner fluctuations vanish \cite[\S 8.3]{Landi2008}. In the case of an almost-commutative geometry both components $\dirac$ and $D_F$ of the Dirac operator generate inner fluctuations. 
For these we will write 
\ba\label{eq:fluctDfull}
	D_A &:= \gls{canA} + \gamma_M \otimes \gls{Phi},
\ea	
where $\can_A = i\gamma^\mu (\partial_\mu + \omega_\mu \otimes \id_{\H_F} + \gls{bbA}_\mu)$, with 
\ba
	\mathbb{A}_\mu &= \sum_n \Big(a_n[\partial_\mu, b_n] -\epsilon' J a_n[\partial_\mu, b_n] J^*\Big),\qquad a_n, b_n \in C^{\infty}(M, \A_F),\label{eq:param_A}
\intertext{skew-Hermitian and}
	 \Phi &= D_F + \sum_{n}\Big( a_n [D_F, b_n] + \epsilon' Ja_n[D_F, b_n]J^*\Big),\qquad a_n, b_n \in C^{\infty}(M, \A_F). \nn
\ea
The relative minus sign between the two terms in $\mathbb{A}_\mu$ comes from the identity $J_M \gamma^\mu J_M^* = - \gamma^\mu$ for even-dimensional $\dim M$. The terms will later be seen to contain all gauge fields of the theory \cite[\S \MakeUppercase{\romannumeral 11}]{C00}. The inner fluctuations of the finite Dirac operator $D_F$ (see also \eqref{eq:F-innerfl}) are seen to parametrize all scalar fields, such as the Higgs field. Interestingly, this view places gauge and scalar fields on the same footing, something that is not the case in QFT. See Table \ref{tab:origin_fields} for an overview of the origin of the various fields.
 
%
\begin{table}[ht]
	\begin{tabularx}{\textwidth}{XllX}
	\toprule
			& Type of field & NCG-object & \\
	\midrule
			& Fermions 			& $L^2(M, S) \otimes \H_F$ & \\
			& Scalar bosons				& $\Omega^1_{D_F}(\A)$ & \\
			& Gauge bosons 	& $\Omega^1_{\dirac}(\A)$ & \\
	\bottomrule 
	\end{tabularx}
	\caption{The various possible fields that are ingredients of physical theories and the NCG-objects they originate from in the case of an almost-commutative geometry.}
	\label{tab:origin_fields}
\end{table}

The second and last ingredient that we will need here is a natural, gauge invariant, functional that can serve as the equivalent of the action \eqref{eq:intro_action} we know from high energy physics. For that we want something which as much as possible depends on the data that are present in the spectral triple. \emph{The} choice \cite{CC96, CCM07} for that turns out to be
\begin{align}
	S[\zeta, A] := \frac{1}{2}\langle J\zeta, D_A \zeta \rangle + \tr f(D_A/\Lambda),\qquad \zeta \in \frac{1}{2}(1 + \gamma_M \otimes \gamma_F)\H \equiv \gls{Hplus},\label{eq:totalaction}
\end{align}	
consisting of the \emph{fermionic action} and the \emph{spectral action} respectively. Here \gls{f} is a positive, even function, \gls{Lambda} is a (unknown) mass scale\footnote{The parameter $\Lambda$ more or less serves as a cut-off, and will in the derivation of the SM (Section \ref{sec:intro_NCSM} ahead) be interpreted as the GUT-scale.} and the trace of the second term is over the entire Hilbert space. 
In its original \cite{CL89,CC96} form, the expression for the fermionic action did not feature the real structure (nor the factor $\tfrac{1}{2}$) and did not have elements of only $\H^+$ as input. It was shown \cite{CC97} that for a suitable choice of a spectral triple it does yield the full fermionic part of the Standard Model Lagrangian (see Section \ref{sec:intro_NCSM}), including the Yukawa interactions, but suffered from the fact that the fermionic degrees of freedom were twice what they should be, as pointed out in \cite{LMMS97}. Furthermore it does not allow a theory with massive right-handed neutrinos. Adding $J$ to the expression for the fermionic action and requiring $\{J, \gamma\} = 0$ allows restricting its input to $\H^+$ without vanishing altogether. This expression is seen to solve both problems at the same time \cite{CCM07} (see also \cite{CM07}). We will not further go into details but refer to the mentioned literature instead. 
 
Despite its deceivingly simple form, the second term of \eqref{eq:totalaction} is a rather complicated object and in practice one has to resort to approximations for calculating it explicitly. Most often this is done \cite{CC97} via a \emph{heat kernel expansion} \cite{GIL84}. In four dimensions and for a suitable Dirac operator (see Appendix \ref{sec:spectral_action}) this reads: 
\ba\label{eq:spectral_action}
	\tr f(D_A/\Lambda) \sim 2\Lambda^4f_4\glslink{a024}{a_0}(D_A^2) + 2\Lambda^2f_2\glslink{a024}{a_2}(D_A^2) + f(0)\glslink{a024}{a_4}(D_A^2) + \mathcal{O}(\Lambda^{-2}),
\ea
where \gls{f2}, \gls{f4} are the second and fourth \emph{moments} of $f$
 and the (\emph{Seeley-DeWitt}) coefficients $a_{0, 2, 4}(D_A^2)$ only depend on the square of the Dirac operator. In Appendix \ref{sec:spectral_action} we will cover this expansion in more detail.

Note the resemblance of the kinetic term \eqref{eq:electron_kin} of the electron with the first term of \eqref{eq:totalaction}. This first term of \eqref{eq:totalaction} will in fact yield all fermionic interactions of the theory. 

In the beginning I mentioned that NCG was in some sense a generalization of the theory of General Relativity. With this I mean that if you would compute the spectral action \eqref{eq:totalaction} for the canonical spectral triple (Example \ref{ex:canon}) you get the Einstein-Hilbert action of GR, including a cosmological constant. The first term of \eqref{eq:totalaction} serves as the `matter'-term curving space(-time). See the Appendix \ref{sec:sp_act_canon} for details.

Note that ---in contrast to `normal' high energy physics--- there is no question of adding some terms to the action by hand in order to make something work. The action \eqref{eq:totalaction} is simply fixed by the spectral triple. In some sense it thus puts an additional entry before the first step in the schematical procedure \eqref{eq:action_to_crosssection}.

\subsection{The noncommutative Standard Model (NCSM)}\label{sec:intro_NCSM}

We now have all the essential ingredients to obtain the Standard Model \cite{CCM07}. We take a compact, $4$-dimensional Riemannian spin manifold $M$ without boundary\footnote{This corresponds to the fields to vanish at infinity.} and the corresponding canonical spectral triple. We take the tensor product with a finite spectral triple whose algebra is
\begin{align*}
	\A_F = \com \oplus \mathbb{H} \oplus M_3(\com),
\end{align*}
where with \gls{quat} we mean the \gls{quaternions} and \gls{M3C} the complex $3\!\times\!3$-matrices. Note that it is this finite algebra that makes the resulting spectral triple actually noncommutative. We denote the irreducible representations of its components with $\mathbf{1}$, $\mathbf{2}$ and $\mathbf{3}$ respectively. In addition, we will need the anti-linear representation $\overline{\mathbf{1}}$, on which $\gls{lambda} \in \com$ acts as $\bar\lambda$. With $\mathbf{1}^o$, $\mathbf{2}^o$, etc.~we denote the \gls{contragredient} module. A natural \gls{bimodule} of this algebra\footnote{To be explicit, the element $(\lambda, q, m)\in \A_F$ acts on ---say--- $\repl{2}{3} \ni v\otimes \bar w$ as $qv \otimes \bar w m = qv \otimes \overline{m^*w}$.} (i.e.~the finite Hilbert space),
\ba\label{eq:intro_SM_reps}
	(\repl{2}{1}) \oplus (\repl{1}{1}) \oplus (\repl{\bar 1}{1}) \oplus (\repl{2}{3}) \oplus (\repl{1}{3}) \oplus (\repl{\bar 1}{3}),
\ea
turns out to exactly describe the particle content of the Standard Model; $l_L$, $\nu_R$, $e_R$, $q_L$, $u_R$ and $d_R$ respectively (c.f.~Table \ref{tab:SM}). From the noncommutative point of view having a right-handed neutrino is a desirable feature \cite{CCM07} (see also Section \ref{sec:sm}). If we want to introduce a real structure $J_F$ we also need \repl{1}{2}, etc. (describing the antiparticles). We can construct a grading $\gamma_F$ that distinguishes left- from right-handed particles and that anticommutes with the real structure. This makes the KO-dimension of the finite spectral triple equal to $6$ and consequently that of the almost-commutative geometry equal to $2$. This makes it possible to reduce the fermionic degrees of freedom \cite[\S 4.4.1]{CCM07}. This Hilbert space describes only one generation of particles so we need to take three copies (or \emph{generations}) of it. 

We can check that not only $SU(\A_F)$ (from \eqref{eq:gauge_group}) equals the gauge group of the Standard Model $SU(3) \times SU(2) \times U(1)$ (modulo a finite group) but also that the resulting hypercharges of the representations match those of the particles of the Standard Model (Table \ref{tab:SM}). 

Then there is the Dirac operator $D_F$ for the finite spectral triple. Employing all the demands on the Dirac operator (self-adjointness, anticommutativity \eqref{eq:Dg-anti-comm} with $\gamma$, (anti)commutativity \eqref{eq:JD-anti-comm} with $J$ and the condition \eqref{eq:order_one}), this actually leaves not that much freedom for it. Requiring in addition that the photon remains massless by demanding that
\ba\label{eq:NCSM_extra_demand}
	[D_F, (\lambda, \diag(\lambda, \bar\lambda), 0)] = 0\qquad \forall\ (\lambda, \diag(\lambda, \bar\lambda), 0) \in \A_F, \lambda \in \com
\ea
results in a Dirac operator whose non-zero components are fully determined \cite[\S 2.6]{CCM07} by $3\!\times\!3$-matrices $\Upsilon_\nu, \Upsilon_e, \Upsilon_u, \Upsilon_d$ and a symmetric $3\!\times\!3$-matrix \gls{UpsilonR}, that mix generations. The $\Upsilon_{\nu,e,u,d}$ map between the representations in $\H_F$ that describe the left- and right-handed (anti)leptons and (anti)quarks and are interpreted as the fermion mass mixing matrices. The component $\Upsilon_R$ maps between the representations that describe the right-handed neutrinos and their antiparticles and serves as a Majorana mass matrix.

A second step is to calculate the inner fluctuations of both Dirac operators. For \dirac, the inner fluctuations acting on $\mathbf{1}$ and $\overline{\mathbf{1}}$ are both seen to describe the same $U(1)$ gauge field. To also let the quarks interact with this field in the way they do in the SM, an additional constraint is imposed. This constraint asserts that the total inner fluctuations be traceless: 
	\ba
		\tr_{\H_F} A_\mu  = 0\label{eq:unimod}.
	\ea
This is called the \emph{unimodularity condition} \cite{C94,Alvarez1995}.\footnote{Note that this is not the same as demanding that $\tr_F(A_\mu + JA^*_\mu J^*) = 0$.} In addition it reduces the degrees of freedom of the gauge bosons to the right number. After applying this condition, the inner fluctuations of \dirac turn out to exactly describe the gauge bosons of the Standard Model; the hypercharge field $B_\mu$, the weak-force bosons $\vec{W}_\mu$ and gluons $g_\mu$. The inner fluctuations of $D_F$ on the other hand are seen to describe a scalar field that ---via the action--- interacts with a left-handed and a right-handed lepton or quark: it is the famous Higgs field \cite[\S 3.5]{CCM07}. Since the finite part of the right-handed neutrinos is in $\repl{1}{1} \simeq \com$, the component $\Upsilon_R$ that describes their Majorana masses does not generate a field via the inner fluctuations \eqref{eq:inner_flucts}.

\begin{figure}[ht]
	\centering
	\includegraphics[width=.5\textwidth]{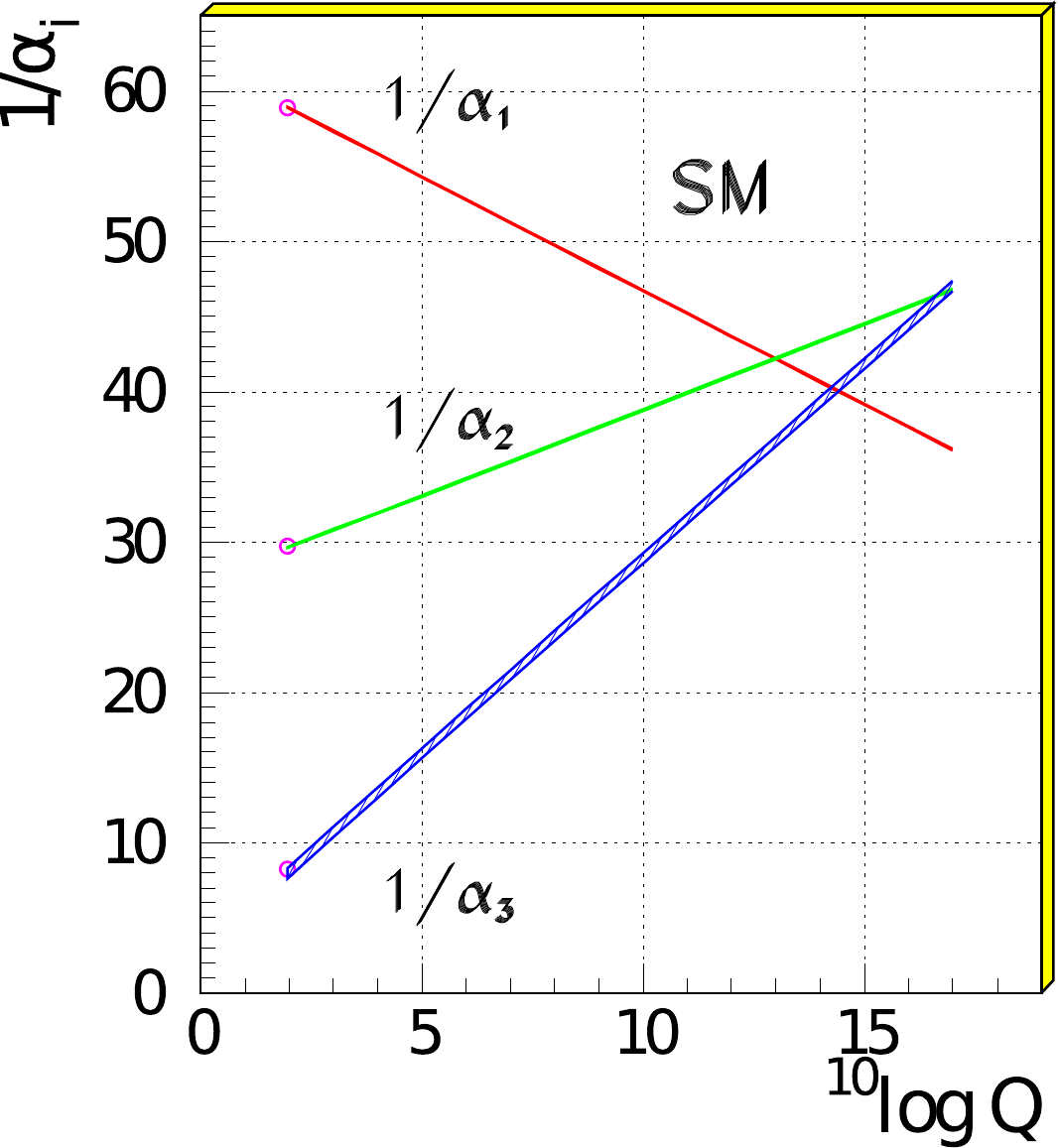}
	\caption{The three (inverse) 'coupling constants' $\alpha_1 = \frac{5}{3}g_1^2/4\pi$, $\alpha_2 = g_2^2/4\pi$ and $\alpha_3 = g_3^2/4\pi$ as a function of the energy. At high energy they are seen to nearly meet in one point. The figure is taken from \cite{Kazakov2009}.}
	\label{fig:gut_scale}
\end{figure}

If we calculate the spectral action for this spectral triple \cite[\S 3.7]{CCM07}, not only do we get the action of the full Standard Model but again the Einstein-Hilbert action of General Relativity too. Various coefficients of terms in the action are determined by variables that are characteristic for NCG (e.g.~the moments $f_{n}$, $\Lambda$, etc.). This gives rise to relations between SM-parameters that are not present in the Standard Model. For example,  if we normalize the kinetic terms of the gauge bosons we automatically find the relation
\begin{align}
	g_3^2 = g_2^2 = \frac{5}{3}g_1^2\label{eq:intro_gut}
\end{align}
between the coupling constants of the strong, weak and hypercharge forces respectively \cite[\S 4.2]{CCM07}. This relation suggests that the interpretation of the so far unknown value of \gls{Lambda} is that of the energy scale at which our theory `lives' and at which the three forces (electromagnetic, weak and strong) are of the same strength. Looking at Figure \ref{fig:gut_scale}, this corresponds to the order of $10^{13} - 10^{17}$ GeV. There is also an additional relation 
\bas
	\lambda &= 4g_2^2\frac{b}{a^2},& b &= \tr[(\yuks{\nu}{}\yuk{\nu}{})^2 + (\yuks{e}{}\yuk{e}{})^2 + 3(\yuks{u}{}\yuk{u}{})^2 + 3(\yuks{d}{}\yuk{d}{})^2],\nn\\
			&& a &= \tr(\yuks{\nu}{}\yuk{\nu} + \yuks{e}{}\yuk{e}{} + 3\yuks{u}{}\yuk{u}{} + 3\yuks{d}{}\yuk{d}{})
\eas
for the coefficient of the Higgs boson self-coupling, as in \eqref{eq:intro_higgs_potential}. Using the value we find for $g_2^2$ from Figure \ref{fig:gut_scale} and approximating the coefficients $a$, $b$ we can infer \cite[\S 5.2]{CCM07} that $\lambda(\Lambda) \approx 0.356$. Inserting this boundary condition into the renormalization group equation for $\lambda$ we obtain a value for the Higgs boson mass at the electroweak scale in the order of $170$ GeV.\footnote{In \cite{DS12} a more detailed analysis is conducted. There the results for the Higgs mass are between $167$ and $176$ GeV \cite[\S 8.2.3]{DS12}.} In addition, this scheme allows a retrodiction of the top quark mass. It is found to be $\lesssim 180$ GeV \cite[\S 5.4]{CCM07}.

This would be a perfect end to the story, if it was not for two things. First of all, the observed Higgs mass ($125.9 \pm 0.4$ GeV/$c^2$ \cite{B12}
) is distinctly different from the above mass range. Second, though we pretended that the three forces are of equal strength at one specific energy-scale $\Lambda$, we know from experiment that ---at least for the SM--- they are in fact not completely, see Figure \ref{fig:gut_scale}. Nonetheless, the fact that NCG allows you to come up with a robust prediction of the Higgs mass in the first place (and that this prediction depends on the particle content, as illustrated by \cite{CC12}) is a promising sign of NCG saying something about reality.

\subsection{Finite spectral triples and Krajewski diagrams}\label{sec:finite_krajewski}

Since we will be using real finite spectral triples (cf.~Examples \ref{ex:finite} and \ref{ex:real_finite}) extensively later on, we cover them in more detail. They are characterized by the following properties:
\begin{itemize}

		\item The finite-dimensional algebra is (by Wedderburn's Theorem) a direct sum of matrix algebras:
	\begin{align}
		\gls{AF} = \bigoplus_{i}^K \gls{MNF}\qquad \mathbb{F}_i = \mathbb{R}, \mathbb{C}, \mathbb{H}\label{eq:finite_algebra}.
	\end{align}	

%
	\item The finite Hilbert space is an $\A_F^\com$-\gls{bimodule}, where \gls{AFC} is the \emph{complexification} of $\A_F$. More specifically, it is a direct sum of tensor products of irreducible representations $\gls{srepi} \equiv \com^{N_i}$ of $M_{N_i}(\mathbb{F}_i)$, for $\mathbb{F}_i = \com, \mathbb{R}$ and\footnote{For the case $\mathbb{F}_i = \mathbb{H}$, the irreducible representation of $M_{N_i}(\mathbb{F}_i)^\com$ is $\com^{2N_i}$.} a \gls{contragredient} representation \gls{srepoj}. 
The latter can be identified with the dual of \srep{j} (by using the canonical inner product on the latter). Thus $\H_F$ is generically of the form
	\begin{align}
	\mkern-36mu	\gls{HF} = \bigoplus_{i \leq j\leq K} \big(\rep{i}{j}\big)^{\oplus M_{N_iN_j}} \oplus \big(\rep{j}{i}\big)^{\oplus M_{N_jN_i}} \oplus \big(\rep{i}{i}\big)^{M_{N_iN_i}}
\label{eq:Hilbertspace}.
	\end{align}
The non-negative integers \gls{MNiNj} denote the \emph{multiplicity} of the representation \rep{i}{j}. When various multiplicities all have one particular value $M$, we speak of ($M$) \emph{generations} that are part of a \emph{family}.

In the rest of this thesis we will not consider representations such as the last part of \eqref{eq:Hilbertspace}, since these are incompatible with $J_F\gamma_F = - \gamma_FJ_F$, necessary for avoiding the fermion doubling problem.

%
	\item The right $\A_F$-module structure is implemented by a real structure 
	\ba\label{eq:fin_real} 
		\gls{JF} :\rep{i}{j} \to \rep{j}{i}
	\ea 
	that takes the adjoint: $J_F(\eta \otimes \bar\zeta) = \zeta \otimes \bar\eta$, for $\eta \in \srep{i}$ and $\zeta \in \srep{j}$. To be explicit: let $a := (a_1, \ldots, a_K) \in \A_F$ and $\eta \otimes \bar\zeta \in \rep{i}{j}$, then
\ba\label{eq:def_right_mult}
	a^o := J_Fa^*J^*_F (\eta \otimes \bar\zeta) = J_Fa^*\zeta \otimes \bar\eta = J_F(a^*_j\zeta \otimes \bar\eta) = \eta \otimes \overline{a_j^*\zeta} \equiv \eta \otimes \bar\zeta a_j.
\ea
From this it is clear that \eqref{eq:left-right} entails the compatibility of the left and right action. For the Hilbert space the existence of a real structure \eqref{eq:fin_real} implies that $M_{N_iN_j} = M_{N_jN_i}$.


	\item For each component of the algebra for which $\mathbb{F}_i = \mathbb{C}$ we will a priori allow both the (complex) linear representation \srep{i} and the anti-linear representation $\overline{\mathbf{N}}_i$, given by: 
	\begin{align*}
		\pi(m)v &:= \overline{m}v,\qquad m \in M_{N_i}(\com), v \in \mathbb{C}^{N_i}.
	\end{align*}

%
\item The finite Dirac operator \gls{DF} consists of components
\ba
	\gls{Dijkl} : \rep{k}{l} \to \rep{i}{j}\label{eq:order_one_finite}.
\ea
The first order condition \eqref{eq:order_one} implies that any component is either left- or right-linear with respect to the algebra \cite{KR97}. This means that $i = k$ or $j = l$.\footnote{An exception to this rule is when one component of the algebra acts in the same way on more than one different representations in $\H_F$.} In both cases it is parametrized by a matrix; in the first case it constitutes of right multiplication with some $\eta_{lj} \in \rep{l}{j}$, in the second case of left multiplication with some $\eta_{ik}\in \rep{i}{k}$. 

\end{itemize}


There exists a very useful graphical representation for finite spectral triples, called \emph{Krajewski diagrams} \cite{KR97}. Such a diagram consists of a two-dimensional grid, labeled by the various $N_i$ and $N_i^o$, representing (the irreducible representations of) the algebra. Any representation \rep{i}{j} that occurs in $\H_F$ then can be represented as a \emph{vertex} on the point $(i, j)$ in this grid. If the finite spectral triple is even, each such representation has a value $\pm$ for the grading $\gamma_F$. We represent it by putting the sign in the corresponding vertex. For real spectral triples, a diagram has to be symmetric with respect to reflection around the diagonal from the upper left to the lower right corner. This is due to the role of $J_F$. The reflection of a particular vertex has the same or an opposite value for the grading, depending on whether $J_F$ commutes or anticommutes with $\gamma_F$. 

\begin{figure}[ht]
\begin{center}
	\def\svgwidth{.4\textwidth}
		\subimport{\the\svgpath}{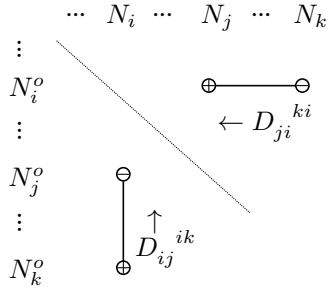}
		\caption{An example of a Krajewski diagram. Each circle in the grid stands for a representation in $\H_F$. A solid line represents a component of the Dirac operator. As can be seen from the signs, $\{J_F, \gamma_F\} = 0$ here.}
		\label{fig:kraj}
\end{center}
\end{figure}

We can represent the component $\D{ij}{kl}$ of the Dirac operator in a Krajewski diagram by an \emph{edge} from $(k,l)$ to $(i, j)$. Since the Dirac operator is self-adjoint, this means that there is also an edge from $(i, j)$ to $(k,l)$ and since it (anti)commutes with $J_F$, this means that there must also be an edge from $(l, k)$ to $(j, i)$. From the first order condition it follows \cite{KR97} that these lines can only be horizontal or vertical. We provide a particularly simple example of a Krajewski diagram in Figure~\ref{fig:kraj}, in which there are two vertices (and their conjugates) between which there is an edge.

\begin{figure}
	\centering
		\def\svgwidth{.6\textwidth}
		\subimport{\the\svgpath}{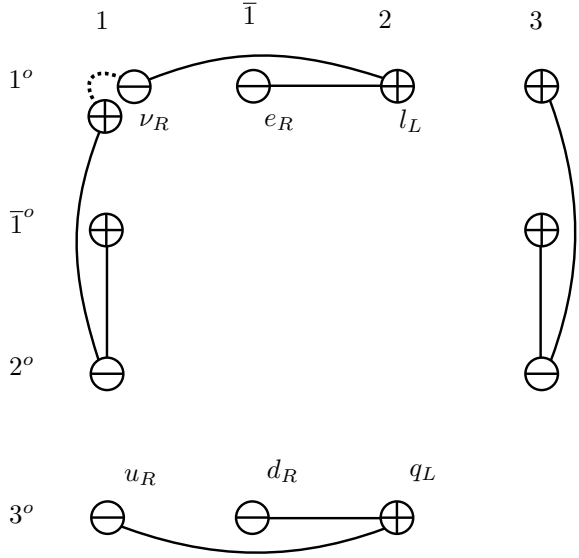}
		\caption{The Krajewski diagram representing the Standard Model. The color of the edges denotes its parametrization. The $0$ components follow from \eqref{eq:NCSM_extra_demand}.}
		\label{fig:KrajSM}	
\end{figure}

Both as an example of the power of Krajewski diagrams and for future reference Figure \ref{fig:KrajSM} shows the diagram that fully determines (the internal structure of) the Standard Model (c.f.~Section \ref{sec:intro_NCSM}). On each point there are in fact three vertices, corresponding to the three generations of particles. The finite Dirac operator was seen to be parametrized by the fermion mass mixing matrices $\Upsilon_{\nu,e,u,d} \in M_{3}(\com)$. Their inner fluctuations generate scalars that are interpreted as the Higgs boson doublet (solid lines), connecting the left- and right-handed representations. Furthermore we have the possibility of adding a Majorana mass $\Upsilon_R$ for the right handed neutrino (dotted line). Note that there are in principle extra components of $D_F$ possible (e.g.~from $\repl{\bar 1}{1}$ to $\repl{3}{1}$) but they are all forbidden by the additional demand \eqref{eq:NCSM_extra_demand}.

The important result of \cite{KR97} is that all properties of a finite spectral triple can be read off from a Krajewski diagram. Although Krajewski diagrams were thus developed as a tool to characterize or classify finite spectral triples, they have turned out to have an applicability beyond that, e.g.~\cite{S12}. Here, we will use them also to determine the value of the trace of the second and fourth powers of the finite Dirac operator $D_F$ (or $\Phi$, including its fluctuations), appearing in the action functional \eqref{eq:spectral_action_acg}. We notice \cite[\S 5.4]{KR97} that 
\begin{itemize}
\item all contributions to the trace of the $n$th power of $D_F$ are given by continuous, closed paths that are comprised of $n$ edges in the Krajewski diagram. 
\item such paths can go back and forth along an edge.
\item a step in the horizontal direction corresponds to a component $\D{ij}{kl}$ of $D_F$ acting on the left of the bimodule $\H_F$, whereas a vertical step corresponds to a component $\D{ij}{kl}$ acting on the right via $J(\D{ij}{kl})^*J^*$. Due to the tensor product structure, the trace that corresponds to a certain closed path is therefore the product of the horizontal and vertical contributions.
\item if a closed path extends in only one direction, this means that the operator acts trivially on either the right or the left of the representation \rep{i}{j} at which the path started. The trace then yields an extra factor $N_i$ or $N_j$, depending on the direction of the path.
\end{itemize}

\begin{figure}
\centering
	\def\svgwidth{\textwidth}
		\subimport{\the\svgpath}{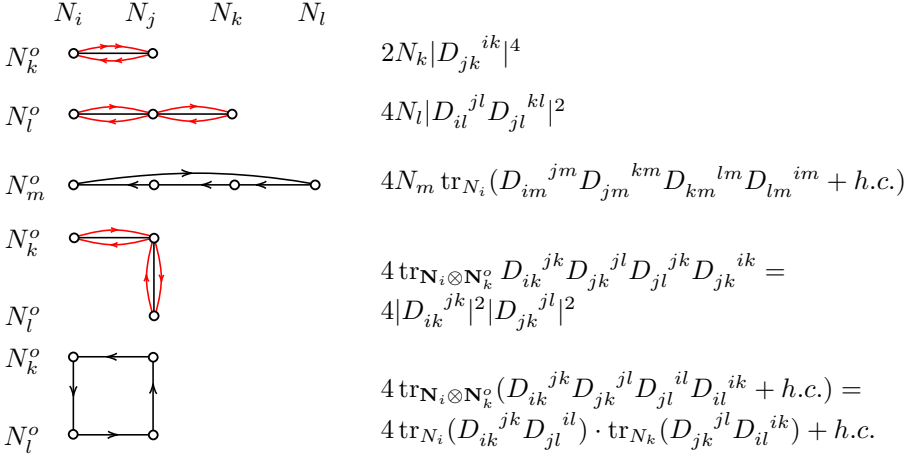}
		\caption{All types of paths contributing to the fourth power of a finite Dirac operator. The last two only occur when it is part of a real spectral triple.}
		\label{fig:KrajPaths}	
\end{figure}

As an example we have depicted in Figure \ref{fig:KrajPaths} all possible contributions to the trace of the fourth power of a $D_F$. This is the highest power that we shall encounter, as we are interested in the action \eqref{eq:spectral_action_acg}. We introduce the notation $|X|^2 := \tr_N X^*X$, for $X^*X \in M_{N}(\com)$. As an illustration of the factors appearing; in the second case a path can start at any of the three vertices, but when it starts in the middle one, it can either go first to the left or to the right. In addition, for a real spectral triple, each path appears in the same way in both directions, giving an extra factor $2$. This last argument does not hold for the last case when $k = i$ and $l = j$, however.

A component $\D{ij}{kj}$ of the finite Dirac operator will develop inner fluctuations \eqref{eq:inner_flucts} that are of the form
\begin{align}
	\D{ij}{kj} &\to \D{ij}{kj} + \sum_n a_n[\D{ij}{kj}, b_n]\nonumber\\
							&=  \D{ij}{kj} + \sum_n (a_n)_i(\D{ij}{kj}(b_n)_k - (b_n)_i\D{ij}{kj}),\qquad a_n, b_n \in \A,\label{eq:F-innerfl}
\end{align}
where $(a_n)_i$ denotes the $i$th component of the algebra element $a_n$. It describes a scalar \gls{Phiik} in the representation \rep{i}{k}. In the expansion \eqref{eq:spectral_action} of the action for an almost commutative geometry (Appendix \ref{sec:spectral_action}) the kinetic terms for the components of $\Phi$ appear via 
\bas
	\{\can_A, \gamma^5 \otimes \Phi\} = i\gamma^\mu\gamma^5[(\partial_A)_\mu, \id_{L^2(S)} \otimes \Phi]
\eas
cf.~\eqref{eq:spectral-action-acg-E}. We determine it for a component $\D{ij}{kj}$ of $\Phi$ in particular by applying it to an element $\zeta_{kj} \in L^2(M, S\otimes \rep{k}{j})$ and find that
\begin{align}
	[(\partial_A)_\mu, \D{ij}{kj}] \zeta_{kj} &= (\partial_\mu + \omega_\mu)(\Phi_{ik}\zeta_{kj}) -i g_iA_{i \mu}\glsadd{Amu} \Phi_{ik}\zeta_{kj} + ig_j\Phi_{ik}\zeta_{kj}A_{j \mu} \nn\\
		&\qquad -\Phi_{ik}(\partial_\mu + \omega_\mu)(\zeta_{kj}) +i g_k \Phi_{ik}A_{k \mu}\zeta_{kj} - ig_j\Phi_{ik}\zeta_{kj}A_{j \mu}\nonumber\\
	&= \big( \partial_\mu(\Phi_{ik}) -i g_i A_{i\mu} \Phi_{ik}  + i g_k \Phi_{ik}A_{k\mu} \big)\zeta_{kj}\nonumber\\
	&\equiv D_\mu(\Phi_{ik})\zeta_{kj},\label{eq:commutatorExpr}
\end{align}
where we have introduced the \emph{covariant derivative} \gls{Dmu} from which the operator $\omega_\mu$ has dropped out completely. We have preliminarily introduced coupling constants $g_{i,k} \in \mathbb{R}$ and wrote $\mathbb{A}_\mu = - i g_iA_{i\mu} + ig_k A_{k\mu}^o$ (with $A_{i\mu}, A_{k\mu}$ Hermitian) to connect with the physics notation.

The \emph{gauge group} that is associated to an algebra of the form \eqref{eq:finite_algebra} is given by
\begin{align}
	SU(\A_F) := \{ u \equiv (u_1, \ldots, u_K) \in U(\A_F), \det{}_{\H_F}(u) = 1\}\label{eq:gauge_group},
\end{align}
where $U(\A_F)$ was defined in \eqref{eq:unit_elts} and with $\det{}_{\H_F}(u)$ we mean the determinant of the entire representation of $u$ on $\H_F$. Applying $U = uJuJ^*$ to an element $\psi_{ij} \in \rep{i}{j} \subset \H_F$ and typical component $\D{ij}{kj}$ of the finite Dirac operator yields
\begin{subequations}\label{eq:fermion_transf}
\ba
	\psi_{ij} &\to uJuJ^*\psi_{ij} = u_i\psi_{ij} u_j^*\\
\intertext{cf.~\eqref{eq:unit_elts} and}
	\D{ij}{kj} &\to uJuJ^* \D{ij}{kj} u^*Ju^*J^* = u_iu_j^{*o}\D{ij}{kj}u_k^*u_j^o = u_i\D{ij}{kj}u_k^*,
\ea
\end{subequations}
respectively.

For future reference we give the spectral action (the second term of \eqref{eq:totalaction}) in the heat kernel expansion (Appendix \ref{sec:spectral_action}) for a general almost-commutative geometry on a 4-dimensional Riemannian spin-manifold without boundary:
\begin{align}
\mkern-18mu \tr f\bigg(\frac{D_A}{\Lambda}\bigg) &\sim \int_M \bigg[\frac{f(0)}{8\pi^2}\Big( - \frac{1}{3}\tr_F\mathbb{F}_{\mu\nu}\mathbb{F}^{\mu\nu} + \tr_F \Phi^4 + \tr_F [D_\mu, \Phi]^2 - \frac{\gls{R}}{6}\tr_F\Phi^2 \nn\\
	&\qquad + \frac{1}{720}(5R^2 - 8\gls{Rmunu}R^{\mu\nu} + 7R_{\mu\nu\sigma\lambda}R^{\mu\nu\sigma\lambda})\mathcal{N}(F)\Big) \label{eq:spectral_action_acg}	\\
		&\quad\qquad - \frac{\Lambda^2}{2\pi^2}f_2\Big(\tr_F\Phi^2 - \frac{1}{12}\mathcal{N}(F)R\Big) + \frac{\Lambda^4}{2\pi^2} f_4\mathcal{N}(F) \bigg] + \mathcal{O}(\Lambda^{-2}),\nn
\end{align}
where $\tr_F$ denotes the trace over the finite Hilbert space, $\mathcal{N}(F) = \dim(\H_F)$ and $\mathbb{F}_{\mu\nu}$ is the (skew-Hermitian) curvature (or field strength) of $\mathbb{A}_\mu$, i.e.
\ba\label{eq:gauge_field_strength}
	\gls{bbFmunu} = [\partial_\mu + \mathbb{A}_\mu, \partial_\nu + \mathbb{A}_\nu],
\ea
while the Riemann tensor \gls{Riemm} is that of the Levi-Civita connection. For a \emph{flat} manifold $M$ in particular, the spectral action \eqref{eq:spectral_action_acg} reduces to
\ba
\tr f\bigg(\frac{D_A}{\Lambda}\bigg) &\sim \int_M \bigg[\frac{f(0)}{8\pi^2}\Big( - \frac{1}{3}\tr_F\mathbb{F}_{\mu\nu}\mathbb{F}^{\mu\nu} + \tr_F \Phi^4 + \tr_F [D_\mu, \Phi]^2\Big) \nn\\
		&\qquad\qquad - \frac{\Lambda^2}{2\pi^2}f_2\tr_F\Phi^2 + \frac{\Lambda^4}{2\pi^2} f_4\mathcal{N}(F) \bigg] + \mathcal{O}(\Lambda^{-2})\label{eq:spectral_action_acg_flat}.
\ea

We have now covered the most important ingredients for particle physics using almost-commutative geometries. We proceed by motivating the choice to search for supersymmetric theories that arise from noncommutative geometry.

\section{Motivation}\label{sec:motivation}

The Standard Model is a tremendously successful theory, but one that at some point will meet its bounds (\S \ref{sec:intro_break}). We are therefore in need of a new theory, respecting the various constraints from both experiment and theory, from which the SM emerges as a low energy limit. 

From the NCG point of view the SM can be beautifully derived from geometrical principles. On top of that the Higgs mass could be predicted, but its value turned out to be off (\S \ref{sec:intro_NCSM}). At the same time any prediction of this sort depends on the contents of the spectral triple (e.g.~\cite{CC12}). Application of noncommutative geometry thus gives us new ways to understand the structure of gauge theories in general and the SM in particular. The question is whether it in addition can teach us more about reality ---via the correct prediction or \emph{retrodiction} of a mass (spectrum)--- than QFT does. In particular, the hope is that there is a theory that can be considered an extension of the NCSM and that, on top of being phenomenologically viable, yields a sufficiently lower value for the Higgs mass. As a guiding principle to such extensions we will in Chapter \ref{ch:constraints} single out a particular set of constraints from physics and look for extensions satisfying them.

The MSSM (\S \ref{sec:intro_MSSM}) is a particularly prominent example of physics beyond the Standard Model. Although the question whether supersymmetry is a real symmetry of nature is still open, the merits of the MSSM and models akin alone make them worthwhile to analyze in full detail. 


\emph{This is the main motivation to search for a theory from noncommutative geometry that describes the MSSM (or something alike), which is the main subject of this PhD thesis.}

To this aim, we will first study the more general question if the action \eqref{eq:totalaction} that stems from NCG can exhibit supersymmetry.\footnote{In a sense the subject fits seamlessly into a broader programme that is out to understand what NCG and its action functional in particular have in store for us.} We do this in Chapter \ref{ch:sst}. If one is after phenomenologically viable theories of supersymmetry, the question on how to break it again is an unavoidable one. We therefore turn to this matter in Chapter \ref{ch:breaking}. Finally, we apply the framework developed in Chapter \ref{ch:sst} to the almost-commutative geometry that is to give the MSSM in this context in Chapter \ref{ch:NCMSSM}.

Previous attempts to reconcile supersymmetry with noncommutative geometry have been made, see e.g.~\cite{HT91a,HT91b,KW96}, but have not led to conclusive answers. We distinguish ourselves from these approaches in the following ways:
\begin{itemize}
	\item We try to stay as close as possible to the framework of NCG/the NCSM, not digressing into superspace and superfields and the likes.
	\item All attempts were made prior to the introduction of the spectral action \eqref{eq:totalaction}.
\end{itemize}
Since the latter has proven itself so well in obtaining the Standard Model and since the (predictive) power of the noncommutative method relies heavily on it, we \emph{choose} it to be our action functional and will ask ourselves in Chapter \ref{ch:sst} ``for what noncommutative geometries is the action supersymmetric?'', or ``what are supersymmetric noncommutative geometries?''. This is in contrast to the question ``what actions are supersymmetric?'' that one typically tries to answer using the superfield formalism. Note the crucial difference here; the intimate connection between an almost-commutative geometry and its associated action forbids us to manually add terms to the latter. We will make an exception to this for \emph{non-physical} fields; the action functional \eqref{eq:totalaction} is an on shell one, not providing the auxiliary fields that are characteristic for supersymmetric theories. We stress that this feature also entails that we have to deal with the terms $\propto \Lambda^2$, that appear to softly break any supersymmetric action. When we are talking about a supersymmetric action we will therefore always mean the part $\propto \Lambda^0$.

One of the merits of noncommutative geometry is that it leads to particle physics models on a curved background. Intuitively this would call for a \emph{local} approach to supersymmetry, e.g.~supergravity \cite{FNF76, wessbagger1992}. We will focus on \emph{global} supersymmetry here, taking the transformation parameters $\epsilon$ to be constants, since we are after the MSSM in particular.\footnote{Besides, the calculations will already be lengthy enough for global supersymmetry.} This requires us to restrict ourselves to a canonical spectral triple on a flat background (i.e. one on which all Christoffel symbols and consequently the Riemann tensor vanish) from Chapter \ref{ch:sst} onward. In this context we will use \eqref{eq:spectral_action_acg_flat}, rather than \eqref{eq:spectral_action_acg}, for the bosonic action. We will thus sacrifice the aforementioned virtue of NCG. It would nonetheless be valuable to extend this to a local approach.

%
%
%

\chapter[Going beyond the Standard Model]{Going beyond the Standard Model\NoCaseChange{\footnote{The contents of this chapter are based on \cite{BS12}.}}}\label{ch:constraints}

Before turning to the description of supersymmetric theories in noncommutative geometry, we first turn our attention to possible extensions of the Standard Model in general. The central question of this chapter is to what extent demands inspired by physics can help us selecting viable noncommutative theories.	
Considerable effort has already been spent on classifying all possible models using the demands that various mathematical and physical arguments put on noncommutative geometries, such as \cite{ISS03,JS05,JSS05,JS08,JS09} and also \cite{CC08}. The approach we take in this chapter is to exploit some of the more recent developments (see \cite{CCM07}) to put constraints on all possible SM extensions. We will demand from any model that
	\begin{enumerate}
		\item the gauge Lie algebra associated to the noncommutative geometry is that of the Standard Model (\S \ref{sec:alg});
		\item the particle content contains at least one copy of each of the particles that the Standard Model features (Table \ref{tab:SM}, \S \ref{sec:alg});
		\item the hypercharges of the particles are such that there is no chiral gauge anomaly (\S \ref{sec:anomalies});
		\item the values of the coupling constants of the electromagnetic, weak and strong forces are such that they satisfy the GUT-relation\footnote{This relation not only appears in the context of a $SU(5)$ Grand Unified Theory, but is also a feature of $SO(10)$ and $E_6$ theories \cite{PL81}, \cite{CEG77}.} \eqref{eq:intro_gut}
			(\S \ref{sec:gut}). Since this relation allows one to determine a scale at which the theory `lives' (c.f.~Section \ref{sec:intro_NCSM}), it plays a vital role in the prediction for the value of the Higgs mass. 
	\end{enumerate}
Together these four demands lead to a number of relations between the multiplicities of the particles, which can be used to constrain the number of viable models. We recover the Standard Model (plus a right-handed neutrino in each generation) using these relations in \S \ref{sec:sm} and finally turn our attention to supersymmetric variants (\S \ref{sec:susy}). 

We must add however, that the scope of the application of these results is much broader than supersymmetry alone.

\section{Constraints on the finite algebra}\label{sec:alg}
Let $\A$ be a $*$-algebra that is represented on a Hilbert space $\H$. Corresponding to the pair $(\A, \H)$ we have defined its \emph{gauge group} in \eqref{eq:gauge_group}.
\begin{lem}\label{lem:possible_algs}
	Suppose that $\A$ is such that $su(\A) \simeq su(3) \oplus su(2) \oplus u(1)$. Then $\A$ must be in the following list:
	\begin{enumerate}
		\item $M_3(\com) \oplus \Qu \oplus M_2(\R)$,
		\item $M_3(\com) \oplus M_3(\mathbb{R}) \oplus M_2(\mathbb{R})$,
		\item $M_3(\com) \oplus \Qu \oplus \C$,
		\item $M_3(\com) \oplus M_2(\C)$ or
		\item $M_3(\com) \oplus M_3(\mathbb{R}) \oplus \com$,
	\end{enumerate}
	modulo extra summands $\mathbb{R}$.
\end{lem}
\begin{proof}
Let $\A$ be of the form \eqref{eq:finite_algebra}, represented on a Hilbert space $\H$ of the form \eqref{eq:Hilbertspace}. We define two Lie algebras
\begin{align}
u(\A) &=  \{ X \in \A : X^* = - X \},\nonumber\\
su(\A) &= \{ X \in u(\A) : \tr_\H X = 0 \}\label{eq:lie-alg}
\end{align}
Note that thus $u(\A)$ is a direct sum of simple Lie algebras $o(N_i), u(N_i), sp(N_i)$ according to $\F_i = \R,\C,\Qu$, respectively. All these matrix Lie algebras have a trace, and we observe that those of the matrices in $o(N_i)$ and $sp(N_i)$ are already zero. For the complex case, we can write $X_i \in u(N_i)$ as $X_i= Y_i+z_i$ where $z_i = \tr X_i$, showing that:
$$
u(N_i) = su(N_i) \oplus u(1).
$$
The unimodularity condition $\tr_\H X = 0$ translates to
$$
\sum_i \alpha_i \tr (X_i) = 0
$$
where $\alpha_i$ are the multiplicities of the fundamental representations of $M_{N_i}(\F_i)$ appearing in $\H$. Using the above property for the traces on simple matrix Lie algebras, we find that unimodularity is equivalent to
$$
\sum_{l=1}^C \alpha_{i_l} z_{i_l} = 0,
$$
where the sum is over the complex factors in $\A$, i.e.~over the indices $i_l \in \{1, \ldots, K\}$ (with $K$ defined in \eqref{eq:finite_algebra}) for which $\F_{i_l} = \C$. Since the number of $u(1)$-factors decreases by one as a result of the condition in \eqref{eq:lie-alg}, we conclude that
$$
su(\A) \simeq \bigoplus_{i=1}^K su(N_i) \oplus u(1) ^{\oplus(C-1)},
$$
where $su(N_i)$ generically denotes $o(N_i), su(N_i)$ or $sp(N_i)$ depending on whether $\F_i = \R,\C$ or $\Qu$, respectively.

In order to get $su(\A) \simeq u(1) \oplus su(2) \oplus su(3)$, we either need that $C = 1$ (with the $u(1) \simeq so(2)$ coming from $M_2(\mathbb{R})$) or $C = 2$. In the first case $su(2)$ must come from either $u(\Qu)$ or from $u(M_3(\mathbb{R})) = o(3)$, using $so(3) \simeq su(2)$, i.e.
 \begin{itemize}
				\item $N_1 = 3_\C, N_2 = 1_\Qu, N_3 = 2_\R$ or 
				\item $N_1 = 3_\C, N_2 = 3_\R, N_3 = 2_\R$. 
\end{itemize}
In the second case we have the following options:
\begin{itemize}
				\item $N_1 = 3_\C, N_2 = 1_\Qu, N_3 = 1_\C$,
				\item $N_1 = 3_\C, N_2 = 2_\C$ or
				\item $N_1 = 3_\C, N_2 = 3_\R, N_3 = 1_\C$.
\end{itemize}
Modulo extra summands $\R$ these are the five options for the algebra $\A$.
\end{proof}

If the Hilbert space is to contain at least one copy of all the SM representations, then the algebra should allow for at least one-, two- and three-dimensional representations. Only the third of these options satisfies this demand\footnote{Although the possible extra summands $\R$ do provide singlets too, the corresponding particles would lack any gauge interactions and are thus unsuitable.}.

\section{Constraints on the finite Hilbert space}

\subsection{Gauge group}\label{sec:gauge}

The gauge group of the Standard Model is known (see e.g.~\cite[\S 3.1]{BH10}) to be
\begin{align}
	\mathcal{G}_{SM} = U(1) \times SU(2) \times SU(3)/\mathbb{Z}_6,\label{eq:gsm}
\end{align}
where the finite abelian subgroup $\mathbb{Z}_6$ stems from the fact that certain elements of $U(1) \times SU(2) \times SU(3)$ act trivially on the Standard Model fermions. 

In the previous paragraph we have demanded that extensions of the SM Hilbert space still have a similar gauge group as that of the SM. Let us explicate this a bit. If we write $u = (u_1, \ldots, u_M)$ for a generic element of $U(\A)$, then the demand $\det_{\H}(u) = 1$ ---that is part of the definition of $SU(\A)$--- applied to the Hilbert space of \eqref{eq:Hilbertspace} translates to
	\begin{align}
		\det{}_\H(u) &= \prod_{i \leq j}^K\det{}_{N_i}(u_i)^{M_{N_iN_j}N_j}\det{}_{N_j}(u_j)^{M_{N_jN_i}N_i} = 1.\label{eq:det}
	\end{align}
	Applying this to the algebra of the SM ---$\com \oplus \mathbb{H} \oplus M_3(\com)$ and correspondingly $U(\A) = U(1) \times SU(2) \times U(3) \ni u = (\lambda, q, m) $--- we have five possibilities for the representations: $\mathbf{N}_i = \mathbf{1}, \overline{\mathbf{1}}, \mathbf{2}$, $\mathbf{3}$ and $\overline{\mathbf{3}}$ on which the determinant equals $\lambda, \lambda^{-1}, 1, \det m$ and $(\det m)^{-1}$ respectively. The relation \eqref{eq:det} then becomes
	\begin{align*}
		\det{}_\H(u) = \lambda^a \det(m)^b = 1,
	\end{align*}
	where 
\begin{subequations}
\label{eq:ab}
\begin{align}
	a  &= 2M_{11} - 2M_{\bar1\bar1} + 2(M_{12}  - M_{\bar1 2}) + 3(M_{13} + M_{1\bar3} - M_{\bar13} - M_{\bar1\bar3})\label{eq:a},\\
	b  &= M_{31} + M_{3\bar1} - M_{\bar31} - M_{\bar3\bar1} + 2(M_{32} - M_{\bar32}) + 6(M_{33} - M_{\bar3\bar3}).\label{eq:b}
\end{align}
\end{subequations}
Here we have used that $M_{N_iN_j} = M_{N_jN_i}$ for the multiplicities of a real spectral triple. The multiplicity $M_{1\bar1}$ does not enter in the expression for $a$ above, for it actually appears twice but with opposite sign.

\begin{lem}
If $a$ divides $b$ then we have for the gauge group
$$
SU(\A) \simeq \left( U(1) \times SU(3)\right)/\Z_3 \times SU(2) \times \Z_a. 
$$
\end{lem}
\proof
We show this in two steps: 
\begin{align}
\tag{I}
&SU(\A) \simeq  G \times SU(2) \times SU(3) / \Z_{3}, \intertext{where $G =\left \{ (\lambda,\mu) \in U(1) \times U(1): \lambda^a \mu^{3b} = 1 \right\}$, containing $\Z_{3}$ as the subgroup $\{e \} \times \Z_{3}$, and}
\tag{II} 
&G \simeq \Z_a \times U(1).
\end{align}
For (I), consider the following map
$$
(\lambda,\mu, q, m) \in G \times SU(2) \times SU(3) \mapsto (\lambda, q, \mu m) \in SU(\A).
$$
We claim that this map is surjective and has kernel $\Z_{3}$. 
If $(\lambda,q, m) \in SU(\A)$ then there exists a $\mu \in U(1)$ such that $\mu^3 = \det m \in U(1)$. Since $\lambda^a \mu^{3b} = \lambda^a \det m^b = 1$ the element $(\lambda,\mu, q, m)$ lies in the preimage of $(\lambda,q, m)$. 
The kernel of the above map consists of pairs $(\lambda,\mu,q, m) \in G \times SU(2) \times SU(3)$ such that $\lambda =1$, $q = 1$ and $m = \mu^{-1} 1_3$. Since $m \in SU(3)$ this $\mu$ satisfies $\mu^{3} = 1$. So we have established (I).

For II we show that the following sequence is split-exact:
$$
1 \to U(1) \to G \to \Z_a \to 1,
$$ 
where the group homomorphisms are given by $\lambda \in U(1) \mapsto (\lambda^{3b/a}, \lambda^{-1}) \in G$ and $(\lambda,\mu) \in G \to \lambda \mu^{3b/a} \in \Z_a$. Exactness can be easily checked, and the splitting map is given by $\lambda \in \Z_a \to (\lambda, 1) \in G$. In this abelian case, the corresponding action of $\Z_a$ on $U(1)$ is trivial so that the resulting semidirect product $G \simeq U(1) \rtimes \Z_a \simeq U(1) \times \Z_a$.
\endproof

\begin{rmk}
In the case that we only allow for representations that already enter in the $\nu$SM (i.e.~$\mathbf{1}, \overline{\mathbf{1}}, \mathbf{2}$ and $\mathbf{3}$), $a$ and $b$ are given by:
\begin{subequations}
\label{eq:ab2}
\begin{align}
	a  &= 2M_{11} - 2M_{\bar1\bar1} + 2(M_{12} - M_{\bar1 2}) + 3(M_{13} - M_{\bar13}) \label{eq:a2},\\
	b  &= M_{31} + M_{3\bar1} + 2M_{32} + 6M_{33}.\label{eq:b2}
\end{align}
\end{subequations}
The Standard Model itself (see also \S\ref{sec:sm}) is given by $M_{11} = M_{1\bar 1} = M_{12} = M_{13}=M_{\bar 1 3} =M_{23} = 3$ (for three families), $M_{ij} = M_{ji}$ and all other multiplicities zero. In this case, $a=b=12$ so that the above Lemma yields $SU(\A) \simeq SU(2) \times (U(1)  \times SU(3))/ {\Z_{3}} \times \Z_{12}$ in concordance with what was found in \cite[Prop. 2.16]{CCM07}. The representation of $SU(\A)$ on $\H$ is as $u \mapsto u JuJ^{-1}$, and the kernel of this representation is $\Z_2$ \cite[Prop. 6.3]{DS11}. In turn, we find that $SU(\A)/\Z_{2} \simeq \mathcal{G}_{SM} \times \Z_{12}$.
\end{rmk}

Note that from the definition of $SU(\A)$ we can determine the hypercharges of the particles from the $U(1)$ factor of $SU(\A)$ (cf. \cite[Prop. 2.16]{CCM07}):
\begin{align}
	\{(\lambda, 1, \lambda^{-a/3b}1_3), \lambda \in \C, |\lambda| = 1\} \subset SU(\A)\label{eq:u1charge},
\end{align}
where the value of $a/b$ must be determined via additional constraints.\footnote{Note however, that these hypercharges are determined up to an overall factor.} The corresponding \emph{hypercharge generator} $Y = (1,0,-\tfrac{a}{3b} 1_3)$ then acts on $\H_F$ as $- i(Y \otimes 1 - 1 \otimes Y^o)$. The hypercharges for the example of the Standard Model then come out right, as illustrated in Table \ref{tab:hypercharges-SM}.

\begin{table}
\begin{tabularx}{\textwidth}{XccrlX}
\toprule
&\textbf{Representation}  & \textbf{M} & \textbf{Hypercharge} & \textbf{SM-particles} &\\
\midrule
&${\bf 1} \otimes {\bf 1}^o$ & $3$ & $1-1=0$& right-handed neutrino &\\
&${\bf \bar 1} \otimes {\bf 1}^o$ & $3$ & $-1-1=-2$& right-handed electron &\\
&${\bf 2} \otimes {\bf 1}^o$ & $3$ & $-1$& left-handed leptons &\\
&${\bf 1} \otimes {\bf 3}^o$ & $3$ & $1+1/3=4/3$& right up-type quarks & \\
&${\bf \bar 1} \otimes {\bf 3}^o$ & $3$ & $-1+1/3=-2/3$& right down-type quarks & \\
&${\bf 2} \otimes {\bf 3}^o$ & $3$ & $1/3$ & left-handed quarks & \\
\bottomrule
\end{tabularx}
\caption{Hypercharges as derived from the finite spectral triple describing the Standard Model. The second column denotes the multiplicity $M$.}
\label{tab:hypercharges-SM}
\end{table}

\subsection{Anomalies and anomaly cancellation}\label{sec:anomalies}
 
In short, a \emph{quantum anomaly} is said to arise when a certain local (gauge) symmetry of a classical theory gets broken upon quantization. Here, we focus attention on the \emph{non-abelian chiral gauge anomaly}: even if the action is invariant under the transformation $\H \ni \zeta \to \exp(\gamma^5 T)\zeta$ (with $T$ anti-Hermitian), the path integral corresponding to this action might not be. The demand of having an anomaly free theory (i.e.~a theory that can be quantized, while preserving its gauge symmetries) can be cast in an expression that depends on the fermionic content of the theory. We will use it to put constraints on the particle content described by the spectral triple. 

In what follows, we assume a finite Hilbert space of the form $\H_F = \H_L \oplus \H_R \oplus \H_{L}^o \oplus \H_R^o$ (containing the left- and right-handed particles and antiparticles respectively) and KO-dimension $6$ (i.e.~$\gamma_F$ anti-commutes with $J_F$) implying that a generic element $\zeta \in \H^+$ ---the physical Hilbert space--- is of the form 
\begin{align}
	\zeta = \xi_L \otimes e_L + \xi_R \otimes e_R + \eta_R \otimes \bar e_L + \eta_L \otimes \bar e_R.\label{eq:spinor_decomp}
\end{align}
Furthermore we take the finite Dirac operator $D_F$ to be zero.\footnote{This corresponds to having massless fermions, which is crucial when considering the chiral gauge anomaly.} $T^a$ will denote a \emph{fixed} generator of the gauge Lie algebra $su(\A_F)$, i.e.~$(T^a)^* = T^a$.
	
\begin{lem}
	Let $\zeta \in \H^+$ and let $\alpha : M \to \mathbb{R}$ be a real function, then the non-abelian chiral gauge transformation
	\begin{align}
	\zeta &\to  U_J\zeta,\text{ with } U_J := \exp(\alpha  \gamma \mathbb{T}^a),\quad \mathbb{T}^a = - i[\pi(T^a) - J\pi( T^a)^* J^*]\label{eq:anomtrans}
	\end{align}
is an on-shell symmetry of the fermionic action in \eqref{eq:totalaction}.
\end{lem}
\begin{proof}
We only have to consider the fermionic part of \eqref{eq:totalaction}, since the gauge fields do not transform. For $U_J$ as in \eqref{eq:anomtrans} with $ \gamma = \gamma^5  \otimes \gamma_F$ one can easily prove that 
\begin{align*}
	J U_J  &= U_J^* J &&\text{and}& U_J\gamma^\mu &= \gamma^\mu U_J^*,
\end{align*} 
by using that left and right actions of the algebra on the Hilbert space must commute and that $\{\gamma^5, \gamma^\mu\} = \{J, \gamma_F\} = 0$. Then for the inner product we have, upon transforming $\zeta$:
\begin{align*}
	\frac{1}{2}\langle JU_J\zeta, D_AU_J\zeta\rangle &= \frac{1}{2}\langle J\zeta, i\gamma^\mu U_J^*(\nabla^S_\mu + \mathbb{A}_\mu)U_J\zeta\rangle \\
&= \frac{1}{2}\langle J\zeta, \dirac\zeta\rangle + \frac{1}{2}\langle J\zeta, i \partial_\mu(\alpha) \gamma^\mu \gamma\,\mathbb{T}^a\zeta\rangle \\
		&\qquad + \frac{1}{2}\langle J\zeta, i\gamma^\mu \exp[-\alpha \gamma\, \ad\mathbb{T}^a]\mathbb{A}_\mu\zeta\rangle, 
\end{align*}
where we have used the identity $\exp(A)B\exp(- A) = \exp(\ad A)B$ for complex $n \times n$ matrices $A$ and $B$. If we expand the exponential in the last term of the previous expression, and use partial integration for the second term, this becomes
\begin{align*}
 & \frac{1}{2}\langle J\zeta, D_A\zeta\rangle - \frac{1}{2}\int_M \alpha \partial_\mu \big[(J\zeta, i \gamma^\mu \gamma \mathbb{T}^a\zeta)\sqrt{g}\big]\mathrm{d}^4x \\
		&\qquad - \frac{1}{2}\langle J\zeta, i\alpha \gamma^\mu\gamma [\mathbb{T}^a,\mathbb{A}_\mu]\zeta\rangle + \mathcal{O}(\alpha^2).
\end{align*}
Writing $\mathbb{A}_\mu = {A}_{\mu\,b} \mathbb{T}^b$ with $A_{\mu, b}$ real (cf.~\eqref{eq:commutatorExpr} and the text below it), and using 
\bas
	[\mathbb{T}^a, \mathbb{T}^b] = \ad[- iT^a, - i T^b] = - if^{ab}_{\phantom{ab}c}\ad(T^c) = f^{ab}_{\phantom{ab}c}\mathbb{T}^c
\eas
this implies
\begin{align*}
 \frac{1}{2} \langle JU_J\zeta, D_AU_J\zeta\rangle &= \frac{1}{2}\langle J\zeta, D_A\zeta\rangle - \frac{1}{2}\int_M \alpha (D_{\mu\,c}^{a} j^{\mu\,c})\mathrm{d}^4x + \mathcal{O}(\alpha^2)
\end{align*}
where 
\begin{align*}
D_{\mu\,c}^{a} &= \partial_\mu \delta^{a}_{\phantom{a}c} + f^{ab}_{\phantom{ab}c}A_{\mu\, b},& j^{\mu\, a} &= (J\zeta, i\gamma^\mu\gamma \mathbb{T}^a\zeta)\sqrt{g}.
\end{align*}
On the other hand we have the following:
\begin{align*}
		\partial_\mu j^{\mu\,a} = [(J\nabla^S_\mu\zeta, i\gamma^\mu\gamma \mathbb{T}^a\zeta) + (J\zeta, i\nabla^S_\mu\gamma^\mu\gamma \mathbb{T}^a\zeta)]\sqrt{g}+ (J\zeta, i\gamma^\mu\gamma \mathbb{T}^a\zeta)\partial_\mu\sqrt{g},
\end{align*}
where we have used that the spin-connection is Hermitian and commutes with $J$. Using that for the tensor density $\sqrt{g}$ we have
$
	\partial_\mu \sqrt{g} = \Gamma^{\lambda}_{\lambda\mu}\sqrt{g}
$ 
(e.g.~\cite[\S 21.2]{misner}) and that for the spin-connection $[\nabla^S_\mu, \gamma^\mu] = -\Gamma^\mu_{\mu\lambda}\gamma^\lambda$ this yields
\begin{align*}
\partial_\mu j^{\mu\,a} = (J\nabla^S_\mu\zeta, i\gamma^\mu\gamma \mathbb{T}^a\zeta)\sqrt{g} + (J\zeta, i\gamma^\mu\gamma \mathbb{T}^a\nabla^S_\mu\zeta)\sqrt{g}.
\end{align*}
Employing the Dirac equation for $\zeta$ and the skew-adjointness of $\mathbb{A}_\mu$ this gives
\begin{align*}
\partial_\mu j^{\mu\,a} = - (J\zeta, i\gamma^\mu\gamma [\mathbb{T}^a,\mathbb{A}_\mu]\zeta)\sqrt{g}.
\end{align*}
i.e.
$
	D_{\mu\,c}^{a} j^{\mu\, c} = 0,
$ 
establishing the result.
\end{proof} 

In order to progress, we need to go a bit more into detail. For a representation $\rep{i}{j}$ of a given pair $(i, j)$, $i \leq j$, the representation $\pi(T^a)$ can be decomposed as $\pi_L(T^a) + \pi_R(T^a)$, where one of the two is trivial depending on the chirality of $\rep{i}{j}$. The representation on the conjugate of $\rep{i}{j}$ is denoted by $\bar\pi_L(T^a) + \bar\pi_R(T^a)$. Hence we write for the full representation $\pi(T^a)$ on $\rep{i}{j} \oplus \rep{j}{i}$:
\begin{align}
\pi(T^a) = \pi_L(T^a) + \pi_R(T^a) +\bar\pi_L(T^a) + \bar\pi_R(T^a).\label{eq:rep_decomp}
\end{align}
\begin{exmpl}
	For the case $\H_F = \mathbf{2}_L \otimes \mathbf{3}^o \oplus \mathbf{3} \otimes \mathbf{2}^o_L$ and $T^a = \tau^a$, the Gell-Mann matrices, we have
	\begin{align*}
		\pi_R(\tau^a) &= \bar\pi_R(\tau^a) = \pi_L(\tau^a) = 0,& \bar\pi_L(\tau^a) &= \tau^a.
	\end{align*}
	For the case $\H_F = \mathbf{1}_R \otimes \mathbf{3}^o \oplus \mathbf{3} \otimes \mathbf{1}^o_R$ and $T^a \equiv Y = (1, 0, -\tfrac{a}{3b}1_3)$, the hypercharge generator, we have
	\begin{align*}
		\pi_R(Y) &= 1, & \bar\pi_R(Y) &= -\frac{a}{3b}1_3, & \pi_L(\lambda) &= \bar\pi_L(\lambda) = 0.
	\end{align*}
\end{exmpl}
Since $\langle J \xi,D_A \xi \rangle$ defines an anti-symmetric bilinear form \cite[Prop. 4.1]{CCM07} we can also write the chiral gauge transformation in terms of the component spinors \eqref{eq:spinor_decomp}, reading:
\begin{align*}
	\frac{1}{2}\langle JU_J\zeta, D_A U_J\zeta\rangle = \langle J_M(U^o)^*\eta, D_AU\xi\rangle 
\end{align*}
i.e.~under the transformation \eqref{eq:anomtrans} we have 
\begin{align}
	\xi &\mapsto \exp(- i \alpha \gamma\, \diag\{\pi_L - \bar\pi^o_L, \pi_R - \bar \pi_R^o\}(T^a)) \xi \equiv U\xi, & \eta &\mapsto (U^*)^o\eta \label{eq:anomtrans_spinors}
\end{align}
in the notation of \eqref{eq:rep_decomp}. 

Now considering the path integral
\begin{align}
	\int \mathcal{D}\eta\mathcal{D}\xi \mathcal{D}\mathbb{A}_\mu \exp(S[\eta,\xi, \mathbb{A}_\mu]),\label{eq:path_integral}
\end{align}
(where in a Euclidean set up the fields $\xi$ and $\eta$ should be considered as independent) there is a second effect from transforming the fermionic fields \eqref{eq:anomtrans_spinors}, which is from the transformation of its measure. The following derivation is primarily based on \cite[\S 22.2]{W05}, \cite[\S 5.6]{BERT96} and \cite[\S 13.2]{NAK90}. We first consider the effect of that transformation on the Dirac spinor $\xi = (\xi_L, \xi_R)$ with finite component. Let 
\[
\xi = \sum_I a_I\psi_I
\]
be its decomposition into the eigenfunctions of $D_A$. Here $I$ is a generic index describing both continuous and finite indices: if, for example, the particle sector of $\H_F$ would equal $(\mathbf{N} \otimes \mathbf{M}^o)_L \oplus (\mathbf{K} \otimes \mathbf{L}^o)_R$ then
\begin{align*}
 	\sum_I a_I\psi_I &\equiv \sum_{j}\Big(\sum_{n,m} a_{jnm} \psi_{jL} \otimes e_n \otimes \bar e_m + \sum_{k,l} a_{jkl} \psi_{jR} \otimes e_k \otimes \bar e_l\Big),\\&\quad a_{jnm} = \langle \psi_{jL} \otimes e_n \otimes \bar e_m, \xi\rangle,
\end{align*}
where $\psi_{jL,R}$ is a left-/right-handed eigenspinor of \dirac and $\{e_n \otimes \bar e_m, 1 \leq n \leq N, 1\leq m \leq M\}$ denotes the basis of the left-handed finite part. Then the transformation
\begin{align*}
	\xi \mapsto \xi' :=  U\xi 
\end{align*}
sets
$
	\xi' = \sum_{I} a_{I}'\psi_I $ with $a_{I}' = \sum_{J} C_{IJ}a_{J}
$
where 
\[ C_{IJ} = \delta_{IJ} + \langle \psi_I, - i \alpha \gamma \diag\{ (\pi_L  - \bar\pi_L^o)(T^a), (\pi_R - \bar\pi_R^o)(T^a)\}\psi_J\rangle + \mathcal{O}(\alpha^2). \]
The effect of the transformation on the measure $\mathcal{D}\xi$ in \eqref{eq:path_integral} is then
\begin{align*}
\mathcal{D}\xi \to \det(C)^{-1}\mathcal{D}\xi,
\end{align*}
and similarly for $\eta$. Writing $X := - i \diag\{ (\pi_L  - \bar\pi_L^o)(T^a), (\pi_R - \bar\pi_R^o)(T^a)\}$, and calculating the determinant by using $\det A = \exp\tr \ln A$ gives
\begin{align*}
	\det(C_{IJ}) &= \exp\tr \ln (\delta_{IJ} + \langle \psi_I , \alpha \gamma X\psi_J \rangle + \mathcal{O}(\alpha^2)) \\
				&\approx \exp\tr \langle \psi_I , \alpha \gamma X\psi_J\rangle\\
				&= \exp \sum_{I}\langle \psi_I , \alpha \gamma X\psi_I \rangle,
\end{align*}
where in going from the first to the second line we have used that $\ln(1 + z) = z + \mathcal{O}(z^2)$, and that $\alpha$ is infinitesimal. The anomaly corresponding to the transformation of $\xi$ is thus
\begin{align*}
	\exp(- \mathscr{A}_\xi),\qquad \mathscr{A}_\xi = \int_M  \alpha\,\mathscr{A}_\xi(x)\,\sqrt{g}\mathrm{d}^4x,\qquad \mathscr{A}_\xi(x) = \sum_{I}(\psi_I, \gamma\,X\psi_I).
\end{align*}
Even though this is an ill-defined quantity, we can make sense of it using the following regularization scheme: 
\begin{align*}
\sum_{I}\langle \psi_I, \alpha\gamma\,X\psi_I\rangle &:= \lim_{\Lambda \to \infty} \sum_{I}\langle \psi_I, \alpha\gamma\,X h(\lambda_{I}^2/\Lambda^2)\psi_I\rangle
\end{align*}
where $h$ can be any function that satisfies $h(0) = 1$ and $\lim_{x \to \infty} h(x) = 0$, and $\lambda_{I}$ is the eigenvalue of $D_A$ with eigenvector $\psi_I$:
\begin{align*}
						\mathscr{A}_\xi &= \lim_{\Lambda \to \infty} \sum_{I}\langle \psi_I, \alpha\gamma\,X h(D_A^2/\Lambda^2)\psi_I\rangle\nn\\
						&= \lim_{\Lambda \to \infty}\int_M \tr_{\H_f}[\alpha \gamma\,X h(D_A^2/\Lambda^2)] \sqrt{g}\mathrm{d}^4x, 
\end{align*}
where with $\H_f = L^2(M, S)_L \otimes \H_L \oplus L^2(M, S)_R \otimes \H_R$ we mean the \emph{particle sector} of the total Hilbert space $\H$, as opposed to the \emph{anti-particle sector} $\H_{\bar f} = L^2(M, S)_R \otimes \H_L^o \oplus L^2(M, S)_L \otimes \H_R^o$. Using a heat kernel expansion \eqref{eq:total_action_exp} this expression equals
\begin{align}
\mathscr{A}_\xi = \lim_{\Lambda \to \infty} \int_M \tr_{\H_f} \alpha\gamma\,X&\Big[2h_4\Lambda^2e_0(x, D_A^2) + 2h_2\Lambda^2 e_2(x, D_A^2)\nn\\&\qquad + h(0)e_4(x, D_A^2) + \mathcal{O}(\Lambda^{-2}) \Big]\sqrt{g}\mathrm{d}^4x\label{eq:anom-xi}.
\end{align}
In an analogous fashion we can determine the anomaly that is caused by the spinor $\eta$, yielding
\begin{align}
  \mathscr{A}_{\eta} = - \lim_{\Lambda \to \infty} \int_M \tr_{\H_{\bar f}} \alpha\gamma\,X^o&\Big[2h_4\Lambda^2e_0(x, D_A^2) + 2h_2\Lambda^2 e_2(x, D_A^2)\nn \\&\qquad+ h(0)e_4(x, D_A^2) + \mathcal{O}(\Lambda^{-2}) \Big]\sqrt{g}\mathrm{d}^4x\label{eq:anom-eta}.
\end{align}
Then we have the following result:
\begin{lem}
	The total anomaly in the path integral due to \eqref{eq:anomtrans} is given by:
	\begin{align*}
	 \mathscr{A} = \int_M \alpha \mathscr{A}(x) \sqrt{g}\mathrm{d}^4x
	\end{align*}
	with 
	\begin{align}
\mathscr{A}(x) &=	\frac{i}{32\pi^2}\epsilon^{\mu\nu\lambda\sigma}F_{\mu\nu}^bF_{\lambda\sigma}^c\Big[ \tr_{\H_L}  (\pi_L - \bar\pi_L^o)(T^a) \{\mathbb{T}^b, \mathbb{T}^c\} - (L \rightarrow R) 
\Big]\nonumber\\
	&\qquad + \frac{i}{384\pi^2}\epsilon^{bcde}R^{\mu\nu}_{\phantom{\mu\nu}bc}R_{\mu\nu\,de}\Big[ \tr_{\H_L}  (\pi_L - \bar\pi_L^o)(T^a)  - (L \rightarrow R)
\Big]\label{eq:anomresult}.
	\end{align}
\end{lem}
\begin{proof}
We start with \eqref{eq:anom-xi}, taking the expressions for $e_{0,2,4}(x, D_A^2)$ from \eqref{eq:gilkey}, 
where, for an almost-commutative geometry, $E$ and $\Omega_{\mu\nu}$ are determined by the field strengths of the gauge fields and the Riemann tensor of $M$:
\begin{align*}
	E &= \frac{1}{4} R - \frac{1}{2}\gamma^\mu\gamma^\nu \mathbb{F}_{\mu\nu},& \Omega_{\mu\nu} = \frac{1}{4}R^{ab}_{\mu\nu}\gamma_{a}\gamma_{b}\otimes 1 + 1\otimes \mathbb{F}_{\mu\nu}, \end{align*}
cf.~\eqref{eq:spectral-action-acg-E} and \eqref{eq:spectral-action-acg-Omega}, but with $\Phi = 0$. Since $\tr \gamma^5 = \tr\gamma^5\gamma^\mu\gamma^\nu = 0$ we only retain, after taking the limit in $\Lambda$, 
\begin{align}
\mathscr{A}_\xi&= h(0)\int_M \tr_{\H_f} \alpha \gamma\,X e_4(x, D_A^2) \nonumber\\ 
&= \frac{h(0)}{(4\pi)^2}\frac{1}{360}\tr_{\H_f} \alpha \gamma\,X \big[180E^2 + 30\Omega_{\mu\nu}\Omega^{\mu\nu}\big]\sqrt{g}\mathrm{d}^4x.
\end{align}
In addition there are boundary terms, but since they will vanish upon integration (the manifold $M$ is taken without boundary), we discard them. Inserting the expressions for $E$ and $\Omega^{\mu\nu}$, performing the trace over Dirac indices and setting $h(0) = 1$ this becomes
\begin{align*}
	\mathscr{A}_\xi = - \int_M \alpha \tr_{\H_L \oplus \H_R}&\bigg[\frac{1}{32\pi^2}\epsilon^{\mu\nu\lambda\sigma}\gamma_F X \mathbb{F}_{\mu\nu}\mathbb{F}_{\lambda\sigma} \\&\qquad+ \frac{1}{768\pi^2}\epsilon^{bcde}R^{\mu\nu}_{\phantom{\mu\nu}bc}R_{\mu\nu\,de}\gamma_F X\bigg]\sqrt{g}\mathrm{d^4}x,
\end{align*}
where we have used that \cite[\S A5.8]{ZJ02}
\bas
	\tr \gamma^5 \gamma^\mu\gamma^\nu\gamma^\lambda\gamma^\sigma = -4 \epsilon^{\mu\nu\lambda\sigma},
\eas
with $\epsilon^{\mu\nu\lambda\sigma}$ the fully anti-symmetric tensor, determined by $\epsilon^{1234} = 1$. The derivation of $\mathscr{A}_\eta$ can be found using the same arguments, reading 
\begin{align*}
	\mathscr{A}_\eta =  \int_M \alpha\tr_{\H_L^o \oplus \H_R^o}&\bigg[ \frac{1}{32\pi^2}\epsilon^{\mu\nu\lambda\sigma}  \gamma_F X^o \mathbb{F}_{\mu\nu}\mathbb{F}_{\lambda\sigma} \\&\qquad + \frac{1}{768\pi^2}\epsilon^{bcde}R^{\mu\nu}_{\phantom{\mu\nu}bc}R_{\mu\nu\,de} \gamma_F X^o\bigg]\sqrt{g}\mathrm{d}^4x.
\end{align*}
Adding $\mathscr{A}_\xi$ and $\mathscr{A}_\eta$ and inserting the expression for $X$ as defined above, using that $\{\gamma_F, J_F\} = 0$ and $\tr_{\H^o}X^o = \tr_{\H}X$ (so contributions from the particle and antiparticle sectors add up), the total anomaly reads
\begin{align*}
	\mathscr{A} &= \int_M \alpha\mathscr{A}(x)\sqrt{g} \mathrm{d}^4x,\quad\text{with}\\
	&\mathscr{A}(x) =  \frac{i}{16\pi^2}\epsilon^{\mu\nu\lambda\sigma}\Big[ \tr_{\H_L}  (\pi_L - \bar\pi_L^o)(T^a) \mathbb{F}_{\mu\nu}\mathbb{F}_{\lambda\sigma} - (L \rightarrow R)
\Big]\\
	&\qquad\qquad  \frac{i}{384\pi^2}\epsilon^{bcde}R^{\mu\nu}_{\phantom{\mu\nu}bc}R_{\mu\nu\,de}\Big[ \tr_{\H_L}  (\pi_L - \bar\pi_L^o)(T^a)  - (L \rightarrow R)
\Big],
\end{align*}
where we have used that $\gamma_F\big|_{\H_L} = - \gamma_F\big|_{\H_R} = 1$. Writing $\mathbb{F}_{\mu\nu} = F_{\mu\nu}^a \mathbb{T}^a$ and exploiting the (anti)symmetries of $\epsilon^{\mu\nu\lambda\sigma}$ and the field strength tensor $F_{\mu\nu}$ we obtain \eqref{eq:anomresult}.
\end{proof}

This result should hold for any generator $T^a$ of the Lie algebra $su(\A)$ of the gauge group (cf.~\eqref{eq:lie-alg}). So, if we want the theory to be anomaly free, \eqref{eq:anomresult} should be zero. 

We apply this general result to the models for which the Lie algebra is $su(\A) = u(1) \oplus su(2) \oplus su(3)$; the $T^a$ can separately be $Y \equiv (1, 0, -\frac{a}{3b}1_3)$, $(0, \sigma^a, 0)$ and $(0, 0, \tau^a)$, the generators of the Lie algebra of the SM. Here $\sigma^a$ and $\tau^a$ denote the Pauli and Gell-Mann matrices respectively. Since $\tr\sigma^i = \tr \tau^a = 0$, the gravitational term of \eqref{eq:anomresult} only gets a contribution from the $u(1)$-part: 
\begin{align*}
	\tr_{\H_L} (Y \otimes 1 - 1 \otimes Y^o) - \tr_{\H_R} (Y \otimes 1 - 1\otimes Y^o),\quad Y = (1, 0, -a/3b\,1_3),
\end{align*}
where although $Y$ denotes the hypercharge \emph{generator} of $u(1)$, it is $Y \otimes 1 -1\otimes Y^o$ that represents the actual hypercharge. For the representations that appear in the Standard Model, this gives
\begin{align}
	&2(-1)M_{21} + 2\cdot 3\frac{a}{3b} M_{23} - 3\bigg(1 + \frac{a}{3b}\bigg)M_{13} - (-2) M_{\bar11} - 3\bigg(-1 + \frac{a}{3b}\bigg)M_{\bar13}\nonumber\\
	&= - 2M_{21} + 2\frac{a}{b} M_{23} - \bigg(3 + \frac{a}{b}\bigg)M_{13} + 2 M_{\bar11} + \bigg(3 - \frac{a}{b}\bigg)M_{\bar13} = 0\label{eq:SManom2}.
\end{align}
Now for the non-gravitational term in \eqref{eq:anomresult}. If we use the cyclicity of the trace, $\tr(\sigma^a) = \tr(\tau^a) = 0$ and $\tr\sigma^a\sigma^b = \tr\tau^a\tau^b = 2\delta^{ab}$, $\{\sigma^a,\sigma^b\} = 2\delta^{ab}$ and $\{\tau^a,\tau^b\} = \tfrac{4}{3}\delta^{ab} + 2d^{ab}_{\phantom{ab}c} \tau^c$, we find ---when restricting to the representations that already appear in the Standard Model--- it to be equivalent to the following relations:
\begin{subequations}
\label{eq:SManom}
\begin{align}
	\intertext{$T^a$ the hypercharge generator, $F_{\mu\nu}$ the $B$-boson field:}
	6\bigg(\frac{a}{3b}\bigg)^3M_{23} - 3\bigg(1 + \frac{a}{3b}\bigg)^3M_{13} - 3\bigg(-1 + \frac{a}{3b}\bigg)^3M_{\bar 13}\qquad& \nn\\
	 + 2(-1)^3M_{21} - (-2)^3M_{\bar11} &= 0\label{eq:SManoma},\\
	\intertext{$T^a$ the hypercharge generator, $F_{\mu\nu}$ the gluon field:}
	2\frac{a}{b}M_{23} - 3\bigg(1 + \frac{a}{3b}\bigg)M_{13} - 3\bigg(- 1 + \frac{a}{3b}\bigg)M_{\bar 13} &= 0\label{eq:SManomb},\\
	\intertext{$T^a$ the hypercharge generator, $F_{\mu\nu}$ the $SU(2)$ boson field:}
	-2M_{21} + 2\frac{a}{b}M_{23} &= 0\label{eq:SManomc},\\
	\intertext{$T^a$ a Gell-Mann matrix, $F_{\mu\nu}$ the gluon field:}
	2M_{23} - M_{13} - M_{\bar13} &= 0.\label{eq:SManomd}
\end{align}
\end{subequations}
All other combinations are seen to vanish. It is evident that the demand of the cancellation of anomalies puts rather stringent constraints on the multiplicities. One well-known result can now be re-confirmed.

\begin{prop}
The minimal Standard Model is anomaly free. 
\end{prop}
\proof
A glance back at the remark just below \eqref{eq:ab2} shows that the non-zero multiplicities of the representation of $\A_F =\C \oplus \Qu \oplus M_3(\C)$ in $\H_F$ are $M_{11} = M_{\bar 11} = M_{21} = M_{13}=M_{\bar 1 3} = M_{23} = 3$ and $a=b=12$, for which the above equations are readily seen to hold. 
\endproof

Two comments are in order here. In using the demand of anomaly cancellation for validating extensions of the Standard Model, we need to know what the value of the grading (chirality) $\gamma$ is on each of the representations. However, there is no such thing as a canonical expression for a grading\footnote{A rather strict demand on the grading would be that of \emph{orientability} \cite{C96, KR97}: for a set $I\ni i$ there exist $a_i, b_i \in \A$ such that $\gamma = \sum_i a_i \otimes b_i^o$. But since already the grading of the SM (with a right-handed neutrino included) is not of this form \cite{ST06}, demanding it for $\gamma$ would in fact be too strict.}, which in principle limits the scope of these constraints to the particles we know the chirality of, i.e.~the SM particles. For non-SM representations we can only try both possible gradings separately. Note, secondly, that the demand for a theory to be anomaly free is most often used for determining the hypercharges of the particles involved. Here, however, these are already determined by the constraint concerning the gauge group, causing the role of anomaly cancellation to be different; it may be used to put constraints on the multiplicities of the representations.

\subsection{GUT-relation}\label{sec:gut}

In the heat kernel expansion \eqref{eq:spectral_action_acg} of an almost-commutative geometry, the kinetic terms of the gauge bosons come from the term
\begin{align*}
	- \frac{f(0)}{24\pi^2}\int_M \tr_{\H_F} \mathbb{F}_{\mu\nu}\mathbb{F}^{\mu\nu}.
\end{align*}
Here the trace runs over the entire (finite) Hilbert space $\H_F$, and $\mathbb{F}_{\mu\nu}$ is a generic symbol denoting the field strength associated to the various gauge fields. Calculating it explicitly ---assuming a Hilbert space of the form \eqref{eq:Hilbertspace}, but for a real spectral triple--- gives 
\begin{align*}
-	\frac{f(0)}{24\pi^2}\tr_{\H_F} \mathbb{F}_{\mu\nu}\mathbb{F}^{\mu\nu} &= \frac{1}{4}\sum_{k}\frac{\K_k}{n_k}\tr_{N_k} F_{\mu\nu}^{k}F^{k,\mu\nu}
\end{align*}
with the coefficients $\K_k$ given by
\begin{align}
 	\K_k  = \frac{f(0)}{3\pi^2}n_kg_k^2\sum_{i, j \geq i} M_{N_i N_j}c^{k}_{ij}\big(q_{ij}^{k}\big)^2.\label{eq:gut1}
\end{align}
Here the index $k$ denotes the \emph{type} of gauge field and $q^{k}_{ij} = -q^{k}_{ji}$ is the charge of the representation $\rep{i}{j}$ associated to the gauge field $A^{k}_\mu$. In the expression for $\K_k$ there is a factor $2$ from summing over both particles and antiparticles. For the number $c^{k}_{ij}$ we have in the case that the only representations are $\mathbf{1}$, $\overline{\mathbf{1}}$, $\mathbf{2}$ and $\mathbf{3}$:
\begin{align*}
	c^{k}_{ij} &= \begin{cases} N_i & \text{if\ } k = j,\\ N_j & \text{if } k= i, \\ N_iN_j & \text{else}. \end{cases}
\end{align*}
Finally the factor $n_k$ stems from the normalization of the generators $T^a_k$ of the gauge group, $\tr T^a_kT^b_k = n_k\delta^{ab}$, and the $n_k^{-1}$ in front of the gauge bosons' kinetic term anticipates the factor $n_k$ that arises when performing the trace over the generators of the gauge group.

In the description of the SM from NCG, the three coefficients are precisely such that upon equating them to the normalisation constant $-1/4$ (by \emph{setting} all $\K_k$ equal to one) in front of the kinetic term ---as is customary--- they automatically satisfy the GUT-relation \eqref{eq:intro_gut} \cite[\S 16.1]{CCM07}. Now certainly in reality, with the particle content of the Standard Model alone, the coupling constants do not meet at a single energy scale. But first of all this feature is too specific to disregard it as a mere coincidence and secondly the entire predictive power of NCG relies \cite[\S 8]{DS11} on it: if it has to say anything more about reality than does the conventional approach to the Standard Model (or any of its extensions), we should take this feature seriously. Furthermore much of the `beyond the Standard Model' research has been conducted in a setting that is characterized by coupling constant unification. \emph{To this end we promote the property that the coupling constants satisfy the GUT relation from a feature to a demand.}\footnote{Certainly at some point one should check that a Hilbert space that satisfies the GUT-relation is compatible with a crossing of the coupling constants as obtained using the Renormalization Group Equations and the very same particle content.} The nature of the constants $\K_k$ is then such that it allows one to put constraints on the Hilbert space.

\begin{table}[h]
\begin{tabularx}{\textwidth}{ccccXccccXcccc}
\toprule
  	 		& $\K_{1}$ 		&$\K_{2}$&$\K_{3}$&& 				& $\K_{1}$ 		&$\K_{2}$&$\K_{3}$\\
\midrule
  \repl{1}{1}		& 0 			& 0 & 0 & & \repl{\bar 1}{\bar 1}	& 0 			& 0 & 0  \\ 
  \repl{1}{\bar 1}	& 4 			& 0 & 0 & & \repl{\bar 1}{2}		& 2 			& 1 & 0 & \\ 
  \repl{1}{2}		& 2 			& 1 & 0 & & \repl{\bar 1}{3}		&$3(-1+\frac{a}{3b})^2$	& 0 & 1  \\ 
  \repl{1}{3}		&$3(1+\frac{a}{3b})^2$	& 0 & 1 & & \repl{\bar 1}{\bar 3}	&$3(-1-\frac{a}{3b})^2$ & 0 & 1  \\
  \repl{1}{\bar 3}	&$3(1-\frac{a}{2b})^2$	& 0 & 1 & & 				&   			&   &    \\
 			&			&   &   & & \repl{3}{3} 			& 0  			& 0 & 6  \\
  \repl{2}{2}		& 0          & 4  & 0 & & \repl{3}{\bar3}		& $9(-\frac{2a}{3b})^2$ & 0 & 6 \\
  \repl{2}{3}		&6$(\frac{a}{3b})^2$	& 3 & 2 & & \repl{\bar 3}{\bar3}		& 0			& 0 & 6  \\
  \repl{2}{\bar 3}	&6$(-\frac{a}{3b})^2$	& 3 & 2 & &				&   			&   & 	\\
\bottomrule
\end{tabularx}
\caption{The contributions to $c^{k}_{ij}(q_{ij}^{k})^2$ to $\K_k$ for each $k = 1, 2, 3$ and each of the representations \rep{i}{j}, with $\srep{i}, \srep{j} = \mathbf{1}, \overline{\mathbf{1}}, \mathbf{2}, \mathbf{3}$, the four representations that occur in the SM.}
\label{tab:coeff}
\end{table}

In the case of the Standard Model algebra  there are three different gauge fields, since the gauge field acting on $\overline{\mathbf{1}}$ is the same as the one acting on $\mathbf{1}$. The $SU(2)$ charge equals $1$ on the representation $\mathbf{2}$ and $0$ on everything else and similarly the $SU(3)$ charge equals $1$ on the representation $\mathbf{3}$ and $0$ on everything else. The charges for the $U(1)$ gauge field are determined by \eqref{eq:u1charge}. The coefficients with which a certain representation contributes to any of the three coupling constants can be found in Table \ref{tab:coeff}. If we only allow the representations that already appear in the Standard Model ---$\mathbf{1}, \overline{\mathbf{1}}, \mathbf{2}, \mathbf{3}$--- we get for the three different coefficients in \eqref{eq:gut1}:
\begin{align*}
	 \K_1 &= \frac{f(0)}{12\pi^2}g_1^2\Big[4M_{1\bar 1} + 2M_{12} + 2M_{\bar 12} + 3\Big(1 + \frac{a}{3b}\Big)^2M_{13} + 3\Big(1 - \frac{a}{3b}\Big)^2M_{\bar 1 3} \\
				&\qquad + 6\Big(\frac{a}{3b}\Big)^2M_{32}\Big],\\
	 \K_{2} &= \frac{f(0)}{3\pi^2}g_2^2n_2[M_{21} + M_{2\bar 1} + 3M_{23} + 4M_{22}],\\
	 \K_{3} &= \frac{f(0)}{3\pi^2}g_3^2n_3[M_{31} + M_{3\bar 1} + 2M_{32} + 6M_{33}],
\end{align*}
with $n_2 = n_3 = \tfrac{1}{2}$ because of the normalisation of $\sigma^a/2$ and $\lambda^a/2$, where $\sigma^a$ and $\lambda^a$ are the Pauli and Gell-Mann matrices respectively. In the expression for $\K_1$ above we have inserted an additional factor $\tfrac{1}{4}$, since we must divide the hypercharges by two to compare with \cite{CCM07}, that has a different parametrization of the gauge fields. Then setting
\begin{align*}
	\K_1 &= \K_2 = \K_3 = 1
\end{align*}
to normalize the gauge bosons, the demand for the GUT-relation \eqref{eq:intro_gut} reads in terms of the multiplicities:
\begin{align}
&\phantom{= 10}\Big[12M_{1\bar1} + 6M_{12} + 6M_{\bar 12} + 9\Big(1 + \frac{a}{3b}\Big)^2M_{13} + 9\Big(- 1 + \frac{a}{3b}\Big)^2M_{\bar 1 3}\nn\\
			&\qquad + 2\frac{a^2}{b^2}M_{32}\Big]\nonumber\\
&= 10[M_{21} + M_{2\bar 1} + 3M_{23} + 4M_{22}]\nonumber\\
&= 10[M_{31} + M_{3\bar 1} + 2M_{32} + 6M_{33}]\label{eq:gut2}.
\end{align} 

\subsection{Bringing it all together}

We summarize Sections \ref{sec:alg}, \ref{sec:gauge}, \ref{sec:anomalies} and \ref{sec:gut} by saying that for any finite spectral triple whose
\begin{itemize}
	\item[C.1] Lie algebra corresponding to the gauge group is the one of the Standard Model;
	\item[C.2] particle content contains at least one copy of each of the Standard Model particles; 
	\item[C.3] particle content is (chiral gauge) anomaly free;
	\item[C.4] three coupling constants satisfy the GUT-relation \eqref{eq:intro_gut};
\end{itemize}
the algebra is $\A = M_3(\com) \oplus \Qu \oplus \com$ and the multiplicities of the fermions are constrained by relations \eqref{eq:SManom} and \eqref{eq:gut2} (with $a$ and $b$ appearing in this last relation determined by \eqref{eq:ab}). 

The reader may have asked himself how and to what extent this approach distinguishes itself from the conventional one, i.e.~the non-NCG approach to (beyond the) SM physics; what more does the former offer compared to the latter? In our opinion there are two main differences. The first is the link between the value for the coupling constants and the Hilbert space ---making the existence of the GUT-relation a consequence of the particle content. Secondly, the demand for the gauge group to be that of the Standard Model made it possible to determine the charges of the featured particles in terms of the powers $a$ and $b$, changing the role of the demand of anomaly cancellations to determining multiplicities. (Both differences are a fruit of the meticulous path from the principles of NCG to the particle content of the Standard Model.)

It might be worthwhile to explicate how this chapter relates to several other analyses having a similar approach, most notably \cite{ISS03,JS05,JSS05,JS08,JS09}. The main differences are that this is the first time that the GUT-relation is explicitly used for constraining the multiplicities, that the demand for the gauge group had not been articulated before in terms of the content of the Hilbert space. Furthermore, this analysis does not regard the question what finite Dirac operators are allowed, so we have no demands related to this.

\section{Solutions of the constraints}

In the following sections we investigate what the above constraints can tell us about some extensions of the Standard Model. But first we will employ these constraints to recover the latter.

\subsection{SM and extensions thereof}\label{sec:sm}

Let $\A = \mathbb{C} \oplus \mathbb{H} \oplus M_3(\mathbb{C})$ and let $\H$ be such that it only contains representations that are present in the Standard Model: $\mathbf{2}_L \otimes \mathbf{1}^o$, $\overline{\mathbf{1}}_R\otimes \mathbf{1}^o$, $\mathbf{2}_L \otimes \mathbf{3}^o$, $\mathbf{1}_R \otimes \mathbf{3}^o$, and $\overline{\mathbf{1}}_R\otimes \mathbf{3}^o$ (representing $(e_L, \nu_L)$, $e_R$, $(u_L, d_L)$, $u_R$ and $d_R$ respectively) and their conjugates. We leave the possibility open for a right-handed neutrino $\mathbf{1}_R \otimes \mathbf{1}^o$. This implies that $\H$ is characterised by a $6$-tuple 
\begin{align*}
 	(M_{11}, M_{13}, M_{\bar 11}, M_{\bar 13}, M_{21}, M_{23})\in \mathbb{N}^6.
\end{align*}
The subscript $L$ ($R$) refers to the value $1$ ($-1$) of $\gamma_F$ on the particular representation.

\begin{lem}\label{lem:solSM}
	Upon demanding C.1 -- C.4 for this spectral triple, the only solution for the multiplicities is:	
	\begin{align*}
		(M_{11}, M_{13}, M_{\bar 11}, M_{\bar 13}, M_{21}, M_{23}) = (M, M, M, M, M, M) 
	\end{align*}
	with $M \in \mathbb{N}$, the number of \emph{generations} or \emph{families}.
\end{lem}
\begin{proof}
From combining \eqref{eq:SManoma} to \eqref{eq:SManomd} we already infer that
\begin{align}
	M_{13} &= M_{\bar13} = M_{23},& M_{\bar 1 1} &= M_{21} = \frac{a}{b}M_{13} \label{eq:smanom2}.
\end{align}
Inserting these equations into the demand for the GUT-relation \eqref{eq:gut2} (but discarding all non-SM representations), gives:
\begin{align*}
2\bigg(9 + 9 \frac{a}{b} + 2\frac{a^2}{b^2}\bigg)M_{13} &= 40M_{13} = \bigg(30 + 10\frac{a}{b}\bigg)M_{13}
\end{align*}
yielding $a = b$, i.e.~the hypercharges are those of the SM and all multiplicities equal each other, except for $M_{11}$ (the right-handed neutrino), which is still unrestricted. Now from the expressions for $a$ and $b$ we find:
\begin{align*}
	2M_{11} + 2M_{12} = 4M_{21},
\end{align*}
establishing the result. 
\end{proof}
Note that according to the previous Lemma, a right-handed neutrino in all generations is necessary from our point of view. That the Standard Model with three generations and an equal number of right-handed neutrinos is at odds with the orientability axiom, was already noted by C.~Stephan \cite{ST06}. For future convenience we succinctly write the solution to the previous Lemma as:
\begin{align*}
	\H_{\mathrm{SM}}' &:= \big(\mathcal{E} \oplus \mathcal{E}^o\big)^{\oplus M},\qquad \mathcal{E} = (\mathbf{2}_L \oplus \mathbf{1}_R \oplus \overline{\mathbf{1}}_R) \otimes (\mathbf{1} \oplus \mathbf{3})^o,
\end{align*}
where we added the prime because the bare SM in fact has $M = 3$ and does not contain any right-handed neutrinos.

We can try to get the most out of the constraints we have derived. We will first focus our attention on the only non-adjoint representation that is absent in the SM --- given that we only use $\mathbf{1}$, $\overline{\mathbf{1}}$, $\mathbf{2}$ and $\mathbf{3}$:
\begin{lem}
	An extension of the SM with only a certain (non-zero) number of copies of $\mathbf{2} \otimes \overline{\mathbf{1}}^o$ (and its conjugate) is possible, provided it is of negative chirality. In that case we have for the multiplicities:
	\begin{align*}
		M_{21} + M_{2\bar1} &= M_{23} = M_{1\bar1} = M_{13} = M_{\bar 13}, & 
		M_{11} &= 3M_{2\bar 1} + M_{21}.
	\end{align*}
\end{lem}
\begin{proof}
	Since the representation under consideration is hypothetical, we do not know whether it is right- or left-handed, but we can try either possibility. We get a $\pm 2M_{2\bar1}$ extra in \eqref{eq:SManoma}, \eqref{eq:SManomc} and \eqref{eq:SManom2}, where the sign depends on the chirality. 

Taking the latter to be positive (i.e.~it has the same chirality as $M_{21}$), we find from anomaly cancellation that
$
	M_{23} = M_{13} = M_{\bar13}, \frac{a}{b}M_{13} = M_{\bar 11} = M_{21} - M_{2\bar1}.
$
 Together with the second equality of \eqref{eq:gut2} this gives 
\begin{align*}\bigg(1 - \frac{a}{b}\bigg)M_{21} = \bigg(-1 + \frac{a}{b}\bigg)M_{2\bar1}.\end{align*}The first equality of \eqref{eq:gut2} solves $a/b = -4 \lor 1$. Using all relations between multiplicities, the first solution demands all multiplicities to vanish, the second solution only sets $M_{2\bar1}$ to be zero and we are back at the SM with right-handed neutrinos. 

Using the other value for the chirality, $M_{21}$ and $M_{2\bar1}$ enter all relations in the same way, except in \eqref{eq:ab} ---whose previous use was to determine $M_{11}$ since that is the only constraint it appears in. This means that we cannot exactly solve all multiplicities. Instead we get the results as stated in the Lemma above.
\end{proof}
Looking at what we observe in particle experiments, the above Lemma suggests that either $\mathbf{2} \otimes \overline{\mathbf{1}}^o$ is absent after all, or that there is a non-zero $M_{2\bar1}$ implying that all other particles ---except for $\mathbf{2} \otimes \mathbf{1}^o$--- come in at least one more generation than is currently observed. The representation $\mathbf{1} \otimes \mathbf{1}^o$ (the right-handed neutrino) on the other hand, needs to have an even higher multiplicity than the others.

\subsection{Supersymmetric extensions}\label{sec:susy}

In this section we focus on supersymmetric extensions of the SM. 
Thus, we want to \emph{at least} extend the finite Hilbert space by the \emph{gauginos}, the superpartners of the gauge bosons (the latter corresponding to the components of the algebra), and the higgsinos (the superpartners of the Higgses \cite[\S 2.2]{CEKKLW05}). The former are in the representations \repl{1}{1}, \repl{2}{2} and \repl{3}{3} and the latter are in \repl{2}{1} and $\mathbf{2}\otimes \overline{\mathbf{1}}^o$. In order for the result to have the right number of degrees of freedom, we need to take two copies of the gaugino representations, both having a different value of the grading. This allows us not only to project onto the physical states of $\H^+$ [c.f.~the first term of \eqref{eq:totalaction}], halving the number of degrees of freedom, but it also allows for the possibility of defining gaugino masses. Having a real structure $J$ makes the higgsinos automatically come with their (charge) conjugates. Since we already have particles in the representation \repl{1}{1} and \repl{2}{1} in the SM, we will distinguish between the SM and supersymmetric versions of this representation by putting a tilde above the latter. In the notation introduced above Lemma \ref{lem:solSM} we write
\begin{align}
	(M_{11} + \widetilde M_{11}, M_{13}, M_{\bar 11}, M_{\bar 13}, M_{21} + \widetilde M_{21}, M_{23}, \widetilde M_{2\bar 1}, \widetilde M_{22}, \widetilde M_{33}) \in\mathbb{N}^{9}\label{eq:solmult2}
\end{align}
with $\widetilde M_{21} = \widetilde M_{2\bar 1} = \widetilde M_{11} = \widetilde M_{22} = \widetilde M_{33} = 1$. Written differently:
\begin{align*}
	\H_{\mathrm{MSSM}}' &:=\H_{\mathrm{SM}}' \oplus \H_{\mathrm{gauginos}} \oplus \H_{\mathrm{higgsinos}}
\end{align*}
with
\begin{align*}
		 \H_{\mathrm{gauginos}} &= \big(\mathbf{1} \otimes \mathbf{1}^o \oplus \mathbf{2} \otimes \mathbf{2}^o \oplus \mathbf{3} \otimes \mathbf{3}^o\big)^{\oplus 2} \simeq \big(\com \oplus M_2(\com) \oplus M_3(\com)\big)^{\oplus 2}\\
		 \H_{\mathrm{higgsinos}} &= \mathbf{2} \otimes \overline{\mathbf{1}}^o \oplus \mathbf{2} \otimes \mathbf{1}^o \oplus \mathbf{1} \otimes \mathbf{2}^o \oplus \overline{\mathbf{1}} \otimes \mathbf{2}^o.
	\end{align*}
We then have:
\begin{lem}
	There is no solution \eqref{eq:solmult2} for the finite Hilbert space that satisfies our demands after extending the Standard Model by two copies of the gauginos and a single copy of the higgsinos.
\end{lem}
\begin{proof}
We can proceed in exactly the same way as in Lemma \ref{lem:solSM}, using the demands C.1 -- C.4. Since the Standard Model representations together satisfied the demands, so should separately do the newly added representations. First of all, the gauginos do not cause an anomaly, since each representation appears both left-handed and right-handed [c.f.~\eqref{eq:anomresult}]. The higgsino in $\mathbf{2} \otimes \mathbf{1}^o$ does cause an anomaly (via the first and third relations in \eqref{eq:SManom}), but the other one in $\mathbf{2} \otimes \overline{\mathbf{1}}^o$ ---having the same grading but an opposite hypercharge--- cancels this anomaly again. So the relations \eqref{eq:smanom2} stay intact, reducing the a priori $6$ unknown SM-multiplicities to only one, $M_{13}$. Next, we find for the three GUT-coefficients:
\begin{align*}
 \mkern-18mu&12 \frac{a}{b}M_{13} + 6\Big(\frac{a}{b}M_{13} + 1\Big) + 6 + 9\Big(1 + \frac{a}{3b}\Big)^2M_{13} \\
	\mkern-18mu&\qquad + 9\Big(-1 + \frac{a}{3b}\Big)^2M_{13} + 2\frac{a^2}{b^2}M_{13}
 =  \bigg[18 + 18\frac{a}{b} + 4\frac{a^2}{b^2}\bigg]M_{13} + 12, \\
 \mkern-18mu& 10\Big[ \Big(\frac{a}{b}M_{13} + 1\Big) + 1 + 3M_{13} + 4\Big] =  10\bigg[\bigg(\frac{a}{b} + 3\bigg)M_{13} + 6 \bigg]\\
\mkern-18mu&\quad\text{and}\\
\mkern-18mu & 10[ M_{13} + M_{13} + 2M_{13} + 6] = 10[4M_{13} + 6],
\end{align*}
respectively. From equating the second and third coefficients one can infer that $a = b$. Inserting this solution into the first coefficient, one gets
\begin{align*}
	 \mkern-18mu 12M_{13} + 6M_{13} + 6 + 6 + 9\bigg(\frac{4}{3}\bigg)^2M_{13} + 9\bigg(-\frac{2}{3}\bigg)^2M_{13} + 2M_{13} = 40M_{13} + 12
\end{align*} 
i.e.~the GUT-relation cannot be satisfied. Moreover, inserting the extra multiplicities in \eqref{eq:ab2} shows that 
\bas
	\mkern-18mua &= 2(M_{13} + 1) - 0 + 2(M_{12} +1 - 1) + 3M_{13}, & b &= M_{31} + M_{31} + 2M_{31} + 6,
\eas
from which we find that $a = b$ only for $M_{13} = \tfrac{4}{3}$, a non-integer number of generations.
%
\end{proof}

Now the previous lemma suggests that the MSSM and NCG are at odds, but there might be models which are not that different from the MSSM that do satisfy all our constraints. We could in principle restore all constraints by adding extra representations compared to the MSSM. In the light of supersymmetry these should all be a superpartner of a scalar particle that enters through a finite Dirac operator. To show that such models exists, we have

\begin{theorem}\label{thm:constraints_mssm_ext}
The smallest possible extension (in the sense of lowest number of extra representations) of the MSSM that satisfies all four constraints, has six additional representations in $\H_F$. Namely, it is one of the following two possibilities for the Hilbert space $\H_{\mathrm{MSSM}}' \oplus \mathcal{F} \oplus \mathcal{F}^o$:
\begin{align}
\mathcal{F} &= (\mathbf{1} \otimes \overline{\mathbf{1}}^o)^{\oplus 4}\, \oplus\, (\mathbf{1} \otimes \mathbf{1}^o)^{\oplus 2}\label{eq:NCMSSM},
\intertext{where two of the copies of $\mathbf{1} \otimes \overline{\mathbf{1}}^o$ should have a grading opposite to the other two copies; or}
\mathcal{F} &= (\mathbf{1} \otimes \overline{\mathbf{1}}^o)^{\oplus 2}\, \oplus\, (\mathbf{1} \otimes \mathbf{3}^o)^{\oplus 2}\, \oplus\, ({\bf 1} \otimes {\bf 2}^o)\, \oplus\, ({\bf{\bar 1}} \otimes {\bf 2}^o)
\label{eq:NCMSSM-2}.
\end{align}
\end{theorem}
\begin{proof}
This can be done with a routine computer check on the equations \eqref{eq:SManom} and \eqref{eq:gut2} while letting the multiplicities $M_{11}, M_{1 \bar 1}, M_{13}, M_{\bar 1 3}, M_{12}, M_{\bar 1 2}, M_{23}, M_{22}$ and $M_{33}$ increase.
\end{proof}

We will for now pause with building specific models, and focus on constructing possibly supersymmetric theories in general. For that we will introduce the notion of $R$-parity into the context of ACGs in Chaper \ref{sec:r-parity}. This will resolve some of the problems with the particle content of the MSSM, as we will see when we return to it in Chapter \ref{ch:NCMSSM}.

\svgpath={./gfx/}

\chapter[Supersymmetric noncommutative geometries]{Supersymmetric noncommutative geometries\NoCaseChange{\footnote{The contents of this chapter are based on \cite{BS13I}.}}}\label{ch:sst}
	

This chapter forms the heart of the thesis. We will digress from the construction of concrete models, turning to the MSSM in particular again in Chapter \ref{ch:NCMSSM}. In contrast, this chapter is devoted to a classification of all almost-commutative geometries (see Section \ref{sec:finite_krajewski}) whose particle content are supersymmetric and the sufficient demands for also the action to be supersymmetric. Throughout this chapter we characterize the finite spectral triples / almost-commutative geometries by their Krajewski diagrams as presented in Section \ref{sec:finite_krajewski}. The canonical part (Example \ref{ex:canon}) of the almost-commutative geometries is sometimes only implicitly there. Already mentioned in Section \ref{sec:motivation} we will restrict ourselves from here to a canonical spectral triple on a flat background, i.e.~all Christoffel symbols and consequently the Riemann tensor vanish. We thus take the expression \eqref{eq:spectral_action_acg_flat} as our bosonic action functional, but will in addition only consider terms $\propto \Lambda^0$ for now. The terms $\propto \Lambda^2$ (and $\Lambda^4$) will be covered in Chapter \ref{ch:breaking}. 

Unless stated otherwise we will restrict ourselves to finite algebras $\A_F$ whose components are matrix algebras over $\com$:
	\begin{align}
		\A_F = \bigoplus_i^K M_{N_i}(\com),\label{eq:alg}
	\end{align}
	cf.~\eqref{eq:finite_algebra}. 
\emph{For a given algebra of this form}, we look for supersymmetric `building blocks' ---made out of representations \rep{i}{j} ($i, j \in \{1, \ldots, K\}$) in the Hilbert space (fermions) and components of the finite Dirac operator (scalars)--- that give a particle content and interactions eligible for supersymmetry. In particular, these building blocks should be `irreducible'; they are the smallest extensions to a spectral triple that are necessary to retain a supersymmetric action. We underline that we do not require that the extra action associated to a building block is supersymmetric in itself. Rather, the building blocks will be defined such that the total action can remain supersymmetric, or can become it again.

We will start by considering all possibilities for a finite algebra consisting of one component in Section \ref{sec:bb1}, step by step extending the algebra in the next sections. But before getting to that, we will introduce the notion of $R$-parity to the context of almost-commutative spectral triples and use it to refine some of the expressions we have been using previously. In Chapter \ref{ch:NCMSSM} these refinements will also be seen to solve some of the problems we encountered with the MSSM in the previous chapter.

\section{NCG and R-parity}\label{sec:r-parity}

One of the key features of many supersymmetric theories is the notion of \emph{$R$-parity}; particles and their superpartners are not only characterized by the fact that they are in the same representation of the gauge group and differ in spin by $\tfrac{1}{2}$, but in addition they have opposite $R$-parity values (cf.~\cite[\S 4.5]{DGR04}). As an illustration of this fact for the MSSM, see Table \ref{tab:rpar}.

\begin{table}[h!]
\begin{tabularx}{\textwidth}{X lllll X}
  \toprule 
& \textbf{Fermions}& \textbf{R-parity} & \textbf{Bosons}& \textbf{R-parity} & \textbf{Multiplicity} &\\
  \midrule
& gauginos			& $-1$ 							& gauge bosons 	& $+1$ & 1 &\\
& SM fermions 	& $+1$							& sfermions 		& $-1$ & 3 &\\
& higgsinos 		& $-1$							&	Higgs(es) 		& $+1$ & 1 &\\
    \bottomrule
\end{tabularx}
\caption{The $R$-parity values for the various particles in the MSSM. In the left column are the fermions, in the right column the bosons. The SM fermions and their superpartners come in three generations each, whereas there is only one copy of the other particles. This statement presupposes that we view the up- and downtype Higgses and higgsinos as being distinct.}
\label{tab:rpar}
\end{table} 

In this section we try to mimic such properties, providing an implementation of this concept in the language of noncommutative geometry: 

\begin{defin}
	\emph{An $R$-extended, real, even spectral triple} is a real and even spectral triple $(\A, \H, D; \gamma, J)$ that is dressed with a grading 
$\gls{Rp} : \H \to \H$ satisfying
\begin{align*}
	[R, \gamma] = [R, J] = [R, a] = 0\ \forall\ a \in \A.
\end{align*}
We will simply write $(\A, \H, D; \gamma, J, R)$ for such an $R$-extended spectral triple.
\end{defin}
Note that, as with any grading, $R$ allows us to split the Hilbert space into an \emph{$R$-even} and \emph{$R$-odd} part:
\begin{align}
	\H = \H_{R = +} \oplus \H_{R = -}, \qquad \H_{R = \pm} = \frac{1}{2}(1 \pm R)\H.\label{eq:R-parity-grading}
\end{align}
Consequently the Dirac operator splits in parts that (anti-)commute with $R$: $D = D_+ + D_-$ with $\{D_-, R\} = [D_+, R] = 0$. We anticipate what is coming in the next section by mentioning that in applying this notion to (the Hilbert space of) the MSSM, elements of $\H_{R = +}$ should coincide with the SM particles and those of $\H_{R=-1}$ with the gauginos and higgsinos. 


\begin{rmk}
	In Krajewski diagrams we will distinguish between objects on which $R = 1$ and on which $R = -1$ in the following way:
	\begin{itemize}
		\item Representations in $\H_F$ on which $R = -1$ get a black fill, whereas those on which $R = +1$ get a white fill with a black stroke.
		\item Scalars (i.e.~components of the Dirac operator) that commute with $R$ are represented by a dashed line, whereas scalars that anti-commute with $R$ get a solid line.	\end{itemize} 
\end{rmk}

We immediately use the $R$-parity operator to make a refinement to the unimodularity condition \eqref{eq:unimod}. Instead of taking the trace over the full (finite) Hilbert space, we only take it over the part on which $R$ equals 1, i.e.~it now reads
\ba
	\tr_{\H_{R =+}}A_\mu &= 0.\label{eq:unimod_new}
\ea 
Analogously, the definition \eqref{eq:gauge_group} of the gauge group must then be modified to 
\begin{align}
	SU(\A) := \{ u \in U(\A), \det{}_{\H_{R = +}}(u) = 1\}.\label{eq:gauge_group_new}
\end{align}
We will justify this choice later, after Lemma \ref{lem:gobinogo}.

Note that adjusting the unimodularity condition has no effect when applying it to the case of the NCSM, since all SM-fermions have $R$-parity $+1$ (Table \ref{tab:rpar}). 

	\section{First building block: the adjoint representation}\label{sec:bb1}

For a finite algebra $\A_F = M_{N_j}(\com)$ that consists of one component, the finite Hilbert space can be taken to be $\rep{j}{j} \simeq M_{N_j}(\com)$, the bimodule of the component $M_{N_j}(\com)$ of the algebra. In order to reduce the fermionic degrees of freedom in the same way as in the NCSM, we need a finite spectral triple of KO-dimension $6$, i.e.~one that satisfies $\{J, \gamma\} = 0$. This requires at least two copies of this bimodule, both having a different value of the finite grading\footnote{We will distinguish the copies by giving them subscripts $L$ and $R$.} and a finite real structure $J_F$ that interchanges these copies (and simultaneously takes their adjoint):
\begin{align*}
	J_F(m, n) := (n^*, m^*).
\end{align*}
We call this 
\begin{defin}\label{def:bb1}
A \emph{building block of the first type} \B{j} ($j \in \{1, \ldots, K\}$) consists of two copies of an adjoint representation $M_{N_j}(\com)$ in the finite Hilbert space, having opposite values for the grading. It is denoted by
\begin{align*}
	\mathcal{B}_j &= { (m, m', 0) \in M_{N_j}(\com)_L \oplus M_{N_j}(\com)_R \oplus \End(\H_F) } \subset H_F \oplus \End(\H_F). 
\end{align*}
\end{defin}

\begin{figure}[ht]
		\begin{center}
		\def\svgwidth{.3\textwidth}
		\subimport{\the\svgpath}{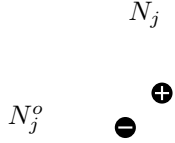}
	\captionsetup{width=.7\textwidth}
		\caption{The first building block consists of two copies in the adjoint representation $M_{N_j}(\com)$, having opposite grading. The solid fill means that they have $R = -1$.}
		\label{fig:bb1}
		\end{center}
	\end{figure}

As for the $R$-parity operator, we put $R|_{M_{N_j}(\com)} = - 1$. Since $D_A$ maps between $R = -1$ representations the gauge field has $R = 1$, indeed opposite to the fermions. The Krajewski diagram that corresponds to this spectral triple is depicted in Figure \ref{fig:bb1}. 

Via the inner fluctuations \eqref{eq:inner_flucts} of the canonical Dirac operator \dirac \eqref{eq:param_A} we obtain gauge fields that act on the $M_{N_j}(\com)$ in the adjoint representation. If we write 
\begin{align*}
	(\gau{jL}', \gau{jR}') \in \H^+ = L^2(S_+ \otimes M_{N_j}(\com)_L) \oplus  L^2(S_- \otimes M_{N_j}(\com)_R)
\end{align*}
for the elements of the Hilbert space as they would appear in the inner product, we find for the fluctuated canonical Dirac operator \eqref{eq:param_A} that:
\begin{align*}
	\can_A(\gau{jL}', \gau{jR}') = i \gamma^\mu (\partial_\mu + \mathbb{A}_\mu)(\gau{jL}', \gau{jR}'),
\end{align*}
with $\mathbb{A}_\mu = - i g_j \ad A_{\mu j}'$. Here we have written $\ad(A_{\mu j}')\gau{L,R}' := A_{\mu j}'\gau{L,R}' - \gau{L,R}'A_{\mu j}'$ with $A_{\mu j}' \in \End(\Gamma(\cS) \otimes u(N_j))$ self-adjoint and we have introduced a coupling constant $g_j$.

\subsection{Matching degrees of freedom}\label{sec:equalizing}

In order for the gauginos to have the same number of finite degrees of freedom as the gauge bosons ---an absolute necessity for supersymmetry--- we can simply reduce their finite part $\gau{jL,R}'$ to $u(N_j)$, as described in \cite[\S 4]{BS10}. However, as is also explained in loc.~cit., even though the finite part of the gauge field $A_{\mu j}'$ is initially also in $u(N_j)$, the trace part is invisible in the action since it acts on the fermions in the adjoint representation. To be explicit, writing $A_{\mu j}' = A_{\mu j} + \tfrac{1}{N_j}B_{\mu j}\id_{N_j}$, with $A_{\mu j}(x) \in su(N_j)$, $B_{\mu j}(x) \in u(1)$ (for conciseness we have left out coupling constants for the moment), we have
\bas
	\ad(A_{\mu j}') = \ad(A_{\mu j}).
\eas
This fact spoils the equality between the number of fermionic and bosonic degrees of freedom again. We observe however that upon splitting the fermions into a traceless and trace part, i.e.~
\ba
	\gau{jL,R}' = \gau{jL,R} + \gau{jL,R}^0 \id_{N_j},\label{eq:bb1-gaugino}
\ea
the latter part is seen to fully decouple from the rest in the fermionic part of the action \eqref{eq:totalaction}:
\bas
	\inpr{J_M\gau{jL}'}{D_A\gau{jR}'} = \inpr{J_M\gau{jL}}{\can_A\gau{jR}} + \inpr{J_M\gau{jL}^0}{\dirac \gau{jR}^0}.
\eas
We discard the trace part from the theory.

\begin{rmk}\label{rmk:bb1-obstr}
	In particular, a building block of the first type with $N_j = 1$ does not yield an action since the bosonic interactions automatically vanish and all fermionic ones are discarded. This is remedied again in a set up such as in the next section. 
\end{rmk} 

Note that applying the unimodularity condition \eqref{eq:unimod_new} does not teach us anything here, for $\H_{R = +}$ is trivial.

One last aspect is hampering a theory with equal fermionic and bosonic degrees of freedom. There is a mismatch between the number of degrees of freedom for the theory \emph{off shell}; the equations of motion for the gauge field and gaugino constrain a different number of degrees of freedom. This is a common issue in supersymmetry and is fixed by means of a non-propagating \emph{auxiliary field}. We mimic this procedure by introducing a variable $G_j := G^a_j T^a_j \in C^{\infty}(M, su(N_j))$ ---with $T^a_j$ the generators of $su(N_j)$--- which appears in the action via:\footnote{This auxiliary field is commonly denoted by $D$. Since this letter already appears frequently in NCG, we instead take $G$ to avoid confusion.}
\begin{align}
- \frac{1}{2n_j}\int_M \tr_{N_j} G^2_j \sqrt{g}\mathrm{d}^4x.\label{eq:auxfieldG}
\end{align}
The factor $n_j$ stems from the normalization of the $T^a_j$, $\tr T^a_jT^b_j = n_j\delta^{ab}$, and is introduced so that in the action $(G^a)^2$ has coefficient $1/2$, as is customary. Typically $n_j = \tfrac{1}{2}$. Using the Euler-Lagrange equations we obtain $G_j = 0$, i.e.~the auxiliary field does not propagate. This means that on shell the action corresponds to what the spectral action yields us. In proving the supersymmetry of the action, however, we will work with the off shell counterpart of the spectral action.

The action of the spectral triple associated to \B{j} has been determined before (e.g.~\cite{CC96}, \cite{CC97}, \cite{Cha94}) and is given by
	\begin{align}\label{eq:SYM-0}
				\act{j}[\gau{}, \mathbb{A}] := \inpr{J_M\gau{jR}'}{\can_A \gau{jL}'} - \frac{f(0)}{24\pi^2}\int_M \tr_{\H_F} \mathbb{F}^j_{\mu\nu} \mathbb{F}^{j,\mu\nu}
 + \mathcal{O}(\Lambda^{-2}),
	\end{align}
where we have written the fermionic terms as they would appear in the path integral (cf.~\cite[\S 16.3]{CM07}).\footnote{It might seem that there are too many independent spinor degrees of freedom, but this is a characteristic feature for a theory on a Euclidean background, see e.g.~\cite{OS1,OS2,NW96} for details.} Using the notation introduced in \eqref{eq:def_right_mult} we write $\mathbb{A}_\mu = - ig_j (A_{\mu j} - A_{\mu j}^o)$ and find for the corresponding field strength \eqref{eq:gauge_field_strength}
\bas
	\mathbb{F}_{\mu\nu} &= - i g_j \big( F_{\mu\nu}^j - (F_{\mu\nu}^j)^o\big),\\
	&\qquad \text{with}\quad F_{\mu\nu}^j = \partial_\mu(A_{\nu j}) - \partial_\nu (A_{\mu j}) - ig_j [A_{\mu j}, A_{\nu j}]
\eas
Hermitian. Consequently we have in the action
\begin{align}
- \frac{f(0)}{24\pi^2}\int_M \tr_{\H_F} \mathbb{F}^j_{\mu\nu} \mathbb{F}^{j,\mu\nu} &=  \frac{1}{4}\frac{\mathcal{K}_j}{n_j}\int_M \tr_{N_j} F^j_{\mu\nu} F^{j,\mu\nu}, \nn\\
		&\qquad\qquad \text{with } \mathcal{K}_j = \frac{f(0)}{3\pi^2}n_jg_j^2 (2N_j)\label{eq:expressionK1},
\end{align}
cf.~\eqref{eq:gut1}. Here we have used that for $X \in M_{N_j}(\com)$ traceless, $\tr_{M_{N_j}(\com)}(X - X^o)^2 = 2N_j\tr_{N_j}X^2$ and there is an additional factor $2$ since there are two copies of $M_{N_j}(\com)$ in $\H_F$. The expression for $\K_j$ gets a contribution from each representation on which the gauge field $A_{\mu j}$ acts, see Remark \ref{rmk:bb2-rmk} ahead. The factor $n_j^{-1}$ in front of the gauge bosons' kinetic term anticipates the same factor arising when performing the trace over the generators of the gauge group. The same thing happens for the gauginos and since we want $\gau{j}^{a}$, rather than $\gau{j}$, to have a normalized kinetic term, we scale these according to
\ba\label{eq:bb1-scaling}
	\gau{j} \to \frac{1}{\sqrt{n_{j}}}\gau{j}, \quad\text{where } \tr T^a_{j}T^b_{j} = n_{j}\delta_{ab}.
\ea
Discarding the trace part of the fermion, scaling the gauginos, introducing the auxiliary field $G_j$ and working out the second term of \eqref{eq:SYM-0} then gives us for the action
\begin{align}
	\act{j}[\gau{}, \mathbb{A}, G_j] &:= \frac{1}{n_j}\inpr{J_M\gau{jL}}{\can_A\gau{jR}} + \frac{1}{4}\frac{\K_j}{n_j}\int_M \tr_{N_j} F^j_{\mu\nu} F^{j,\mu\nu}\nn\\
			&\qquad  - \frac{1}{2n_j}\int_M \tr_{N_j} G^2_j \label{eq:SYM}
\end{align}
with $\gau{jL,R} \in L^2(M, S_{\pm} \otimes su(N_j)_{L,R})$, $A_j \in \End(\Gamma(S) \otimes su(N_j))$ and $G_j \in C^{\infty}(M, su(N_j))$.

For this action we have:
\begin{theorem}\label{thm:bb1}
	The action \eqref{eq:SYM} of an $R$-extended almost-commutative geometry that consists of a building block \B{j} of the first type (Definition \ref{def:bb1}, with $N_j \geq 2$) is supersymmetric under the transformations 
\begin{subequations}\label{eq:bb1-transforms}
\begin{align}
	\delta A_j &= c_{j}\gamma^\mu \big[(J_M\eR, \gamma_\mu\gau{jL})_\cS + (J_M\eL, \gamma_\mu\gau{jR})_\cS\big],  \label{eq:bb1-transforms1}\\
	\delta\gau{jL,R} &= c_{j}'\gamma^\mu\gamma^\nu F^j_{\mu\nu}\eLR + c_{G_j}'G_j\eLR, \label{eq:bb1-transforms2}\\
	\delta G_j &= c_{G_j}\big[(J_M\eR, \can_A\gau{jL})_{\cS} + (J_M\eL, \can_A\gau{jR})_{\cS}\big]\label{eq:bb1-transforms3}
\end{align}
\end{subequations}
	with $c_{j}, c_{j}', c_{G_j}, c_{G_j}' \in \mathbb{C}$ iff
\ba
			2ic_{j}' &= - c_j\K_j, & c_{G_j} = - c_{G_j}'.\label{eq:bb1-constr-final}
\ea
\end{theorem}
\begin{proof}
	The entire proof, together with the explanation of the notation, is given in the Appendix \ref{sec:SYM}.
\end{proof}

We have now established that the building block of Definition \ref{def:bb1} 
 gives the super Yang-Mills action, which is supersymmetric under the transformations \eqref{eq:bb1-transforms}.\footnote{A similar result, without taking two copies of the adjoint representation, was obtained in \cite{BS10}.} This building block is the NCG-analogue of a single vector superfield in the superfield formalism, see Example \ref{ex:intro-susy-sym}.

Note that we cannot define multiple copies of the same building block of the first type without explicitly breaking supersymmetry, since this would add new fermionic degrees of freedom but not bosonic ones. This exhausts all possibilities for a finite algebra that consists of one component. 

\section{Second building block: adding non-adjoint representations}\label{sec:bb2}

If the algebra \eqref{eq:alg} contains two summands, we can first of all have two \emph{different} building blocks of the first type and find that the action is simply the sum of actions of the form \eqref{eq:SYM} and thus still supersymmetric.

 We have a second go at supersymmetry by adding the representation \rep{i}{j} to the finite Hilbert space, corresponding to an off-diagonal vertex in a Krajewski diagram. This introduces non-gaugino fermions to the theory. A real spectral triple then requires us to also add its conjugate \rep{j}{i}. To keep the spectral triple of KO-dimension $6$, both representations should have opposite values of the finite grading $\gamma_F$. For concreteness we choose \rep{i}{j} to have value $+$ in this section, but the opposite sign works equally well with only minor changes in the various expressions. With only this content, the action corresponding to this spectral triple can never be supersymmetric for two reasons. First, it lacks the degrees of freedom of a bosonic (scalar) superpartner. Second, it exhibits interactions with gauge fields (via the inner fluctuations of \dirac) without having the necessary gaugino degrees to make the particle content supersymmetric. However, if we also add the building blocks \B{i} and \B{j} of the first type to the spectral triple, both the gauginos are present and a finite Dirac operator is possible, that might remedy this.

\begin{lem}
	For a finite Hilbert space consisting of two building blocks \B{i} and \B{j} together with the representation \rep{i}{j} and its conjugate the most general finite Dirac operator on the basis 
		\begin{align}
			\rep{i}{j} \oplus M_{N_i}(\com)_L \oplus M_{N_i}(\com)_R \oplus M_{N_j}(\com)_L \oplus M_{N_j}(\com)_R \oplus \rep{j}{i}\label{eq:bb2-basis}. 
	\end{align}
is given by
	\begin{align}
	D_F &= 
	\begin{pmatrix}
			0 & 0 & A & 0 & B & 0\\
				0 & 0 & M_i & 0 & 0 & JA^*J^*\\
			A^* & M^*_i & 0 & 0 & 0 & 0 \\
				0 & 0 & 0 & 0 & M_j & JB^*J^* \\
			B^* & 0 & 0 & M_j^* & 0 & 0 \\
				0 & JAJ^* & 0 & JBJ^* & 0 & 0
		\end{pmatrix}\label{eq:bb2-DF}
	\end{align}
with $A : M_{N_i}(\com)_R \to \rep{i}{j}$ and $B : M_{N_j}(\com)_R \to \rep{i}{j}$.	
\end{lem}
\begin{proof}
	We start with a general $6\times 6$ matrix for $D_F$. Demanding that $\{D_F, \gamma_F\} = 0$ already sets half of its components to zero, leaving $18$ to fill. The first order condition \eqref{eq:order_one} requires all components on the upper-right to lower-left diagonal of \eqref{eq:bb2-DF} to be zero, so $12$ components are left. Furthermore, $D_F$ must be self-adjoint, reducing the degrees of freedom by a factor two. The last demand $J_FD_F = D_FJ_F$ links the remaining half components to the other half, but not for the components that map between the gauginos: because of the particular set up they were already linked via the demand of self-adjointness. This leaves the four independent components $A$, $B$, $M_i$ and $M_j$.
\end{proof}
In this chapter we will set $M_i = M_j = 0$ since these components describe supersymmetry breaking gaugino masses. This will be the subject of Chapter \ref{ch:breaking}. 

\begin{lem}\label{lem:samescalar}
	If the components $A$ and $B$ of \eqref{eq:bb2-DF} differ by only a complex number, then they generate a scalar field $\sfer_{ij}$ in the same representation of the gauge group as the fermion.
\end{lem}
\begin{proof}
We write $\D{ij}{ii} \equiv A$ and $\D{ij}{jj} \equiv B$ in the notation of \eqref{eq:order_one_finite}. First of all, recall that 
$
	\D{ij}{jj} : M_{N_j}(\com) \to \rep{i}{j}
$ 
is given by \emph{left} multiplication with an element $C_{ijj}\, \eta_{ij}$, where $\eta_{ij} \in \rep{i}{j}$ and $C_{ijj} \in \com$. Similarly,
$
	\D{ij}{ii} : M_{N_i}(\com) \to \rep{i}{j}
$ 
is given by \emph{right} multiplication with an element in \rep{i}{j}. If this differs from $\D{ij}{jj}$ by only a complex factor, it is of the form $C_{iij}\eta_{ij}$, with $C_{iij} \in \com$.

Then the inner fluctuations \eqref{eq:F-innerfl} that $\D{ij}{jj}$ develops, are of the form
\begin{align}
	\D{ij}{jj}&\to \D{ij}{jj}	+ \sum_n (a_n)_i\big(\D{ij}{jj}(b_n)_j	- (b_n)_i	\D{ij}{jj}\big) \equiv C_{ijj}\sfer_{ij}\label{eq:bb2-innfs2},
\end{align}
with which we mean left multiplication by the element 
\begin{align*}
\sfer_{ij} \equiv \eta_{ij} + \sum_n (a_n)_i[\eta_{ij} (b_n)_j - (b_n)_i \eta_{ij}] 
\end{align*}
 times the coupling constant $C_{ijj}$. The demand $JD_F = D_FJ$ (cf.~Table \ref{tab:ko_dimensions}) on $D_F$ means that $\D{ki}{ji} = J\D{ik}{ij}J^* = J(\D{ij}{ik})^*J^*$, from which we infer that the component $\D{ii}{ji}$ constitutes of left multiplication with $C_{iij}\eta_{ij}$. Its inner fluctuations are of the form
\begin{align*}
	\D{ii}{ji}&\to \D{ii}{ji}	+ \sum_n (a_n)_i\big(\D{ii}{ji}(b_n)_j	- (b_n)_i	\D{ii}{ji}\big) \equiv C_{iij}\sfer_{ij},
\end{align*}
which coincides with \eqref{eq:bb2-innfs2}. Furthermore, for $U = uJuJ^*$ with $u \in U(\A)$ we find for these components (together with the inner fluctuations) that
\begin{align*}
	U\D{ij}{ii}U &= u_i\D{ij}{ii}u_j^*, & U\D{ij}{jj}U &= u_i\D{ij}{jj}u_j^*,
\end{align*} 
establishing the result.
\end{proof}

\begin{figure}[ht]
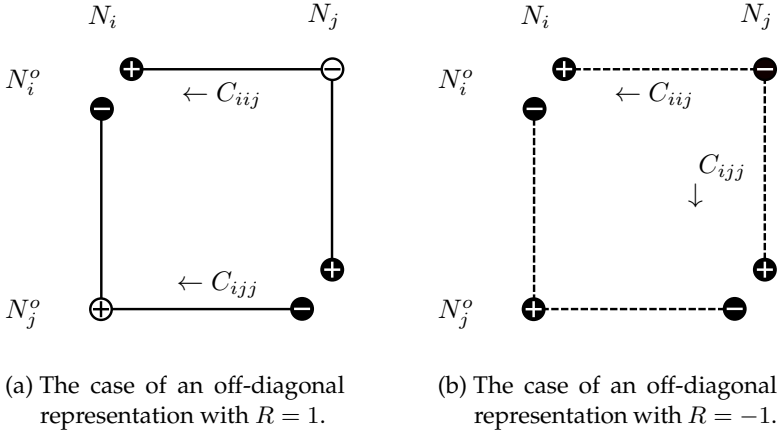

		\captionsetup{width=.9\textwidth}
	\centering
	\begin{subfigure}{.36\textwidth}
		\centering
		\def\svgwidth{\textwidth}
		\subimport{\the\svgpath}{bb2_Rplus.pdf_tex}
		\caption{The case of an off-diagonal representation with $R = 1$.}
		\label{fig:bb2_Rplus}
	\end{subfigure}
	\hspace{30pt}
	\begin{subfigure}{.36\textwidth}
		\centering
		\def\svgwidth{\textwidth}
		\subimport{\the\svgpath}{bb2_Rminus.pdf_tex}	
		\caption{The case of an off-diagonal representation with $R = -1$.}
		\label{fig:bb2_Rminus}
	\end{subfigure}
\caption{After allowing for \emph{off diagonal} representations we need a finite Dirac operator in order to have a chance at supersymmetry. The component $A$ of \eqref{eq:bb2-DF} corresponds to the upper and left lines, whereas the component $B$ corresponds to the lower and right lines. The off-diagonal vertex can have either $R = 1$ (left image) or $R = -1$ (right image). The $R$-value of the components of the finite Dirac operator changes accordingly, as is represented by the (solid/dashed) stroke of the edges.}
\label{fig:bb2}
\end{figure}

Since the diagonal vertices have an $R$-value of $-1$, the scalar field $\sfer_{ij}$ generated by $D_F$ will always have an eigenvalue of $R$ opposite to that of the representation $\rep{i}{j} \in \H_F$. This makes the off-diagonal vertices and these scalars indeed each other's superpartners, hence allowing us to call $\sfer_{ij}$ a sfermion. The Dirac operator \eqref{eq:bb2-DF} (together with the finite Hilbert space) is visualized by means of a Krajewski diagram in Figure \ref{fig:bb2}. Note that we can easily find explicit constructions for $R \in \A_F \otimes \A^o_F$. Requiring that the diagonal representations have an $R$-value of $-1$, we have the implementations $(1_{N_i}, - 1_{N_j}) \otimes (- 1_{N_i}, 1_{N_j})^o$ and $(1_{N_i}, 1_{N_j}) \otimes (-1_{N_i}, - 1_{N_j})^o \in \A_F \otimes \A_F^o$, corresponding to the two possibilities of Figure \ref{fig:bb2}. 

\begin{figure}[ht]
\centering
	\def\svgwidth{.4\textwidth}
		\subimport{\the\svgpath}{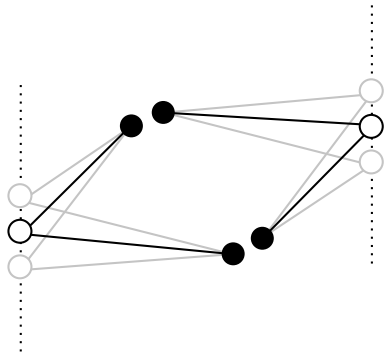}
\captionsetup{width=.9\textwidth}
\caption{An example of a building block of the second type for which the fermion has $R = 1$ and multiple generations.}
\label{fig:bb2-gen}
\end{figure}

We capture this set up with the following definition:
\begin{defin}\label{def:bb2}
The \emph{building block of the second type} \Bc{ij}{\pm} consists of adding the representation \rep{i}{j} (having $\gamma_F$-eigenvalue $\pm$) and its conjugate to a finite Hilbert space containing \B{i} and \B{j}, together with maps between the representations \rep{i}{j} and \rep{j}{i} and the adjoint representations that satisfy the prerequisites of Lemma \ref{lem:samescalar}. Symbolically it is denoted by 
\begin{align*}
	\Bc{ij}{\pm} = (e_i\otimes \bar e_j, e_j'\otimes \bar e_i', \D{ii}{ji} + \D{ij}{jj}) &\in \rep{i}{j} \oplus \rep{j}{i} \oplus \End(\H_F) \\	
		&\qquad \subset \H_F \oplus \End(\H_F) 
\end{align*}
\end{defin}
When necessary, we will denote the chirality of the representation \rep{i}{j} with a subscript $L,R$. Note that such a building block is always characterized by two indices and it can only be defined when \B{i} and \B{j} have previously been defined. In analogy with the building blocks of the first type and with the Higgses/higgsinos of the MSSM in the back of our minds we will require building blocks of the second type whose off-diagonal representation in $\H_F$ has $R = -1$ to have a maximal multiplicity of $1$. In contrast, when the off-diagonal representation in the Hilbert space has $R = 1$ we can take multiple copies (`generations') of the same representation in $\H_F$, all having the \emph{same} value of the grading $\gamma_F$. This also gives rise to an equal number of sfermions, keeping the number of fermionic and scalar degrees of freedom the same, which effectively entails giving the fermion/sfermion-pair a family structure. The $C_{iij}$ and $C_{ijj}$ are then promoted to $M\times M$ matrices acting on these copies. This situation is depicted in Figure \ref{fig:bb2-gen}. We will always allow such a family structure when the fermion has $R = 1$, unless explicitly stated otherwise. There can also be two copies of a building block \B{ij} that have \emph{opposite} values for the grading. We come back to this situation in Section \ref{sec:bb5}. 

Next, we compute the action corresponding to \B{ij}. For a generic element $\zeta$ on the finite basis \eqref{eq:bb2-basis} we will write
\begin{align*}
	\zeta = (\fer{ijL}, \gau{iL}', \gau{iR}', \gau{jL}', \gau{jR}', \afer{ijR}) \in \H^+,
\end{align*}
where the prime on the gauginos suggests that they still contain a trace-part (cf.~\eqref{eq:bb1-gaugino}). To avoid notational clutter, we will write $\fer{L} \equiv \fer{ijL}$, $\afer{R} \equiv \afer{ijR}$ and $\sfer\equiv \sfer_{ijL}$ throughout the rest of this section. The \emph{extra} action as a result of adding a building block \Bc{ij}{+} of the second type (i.e.~additional to that of \eqref{eq:SYM-0} for \B{i} and \B{j}) is given by
\begin{align}
\act{ij}[\gau{i}', \gau{j}', \fer{L}, \afer{R}, \mathbb{A}_i, \mathbb{A}_j, \sfer_{}, \asfer_{}] \equiv \act{ij}[\zeta,\mathbb{A}, \szeta] = \act{f, ij}[\zeta, \mathbb{A}, \szeta] + \act{b, ij}[\mathbb{A}, \szeta]\label{eq:bb2-action}.
\end{align}
The fermionic part of this action reads
\begin{align}
	&\act{f, ij}[\zeta, \mathbb{A}, \szeta] \nn\\
	&= \tfrac{1}{2}\inpr{J(\fer{L}, \afer{R})}{\can_A(\fer{L}, \afer{R})} \nonumber\\
	&\qquad + \tfrac{1}{2}\inpr{J(\fer{L}, \gau{iL}', \gau{iR}', \gau{jL}', \gau{jR}', \afer{R})}{\gamma^5\Phi(\fer{L}, \gau{i,L}', \gau{iR}', \gau{jL}', \gau{jR}', \afer{R})}\nn\\
&= \inpr{J_M\afer{R}}{D_A\fer{L}} + \inpr{J_M\afer{R}}{\gamma^5\gau{iR}'C_{iij}\sfer} + \inpr{J_M\afer{R}}{\gamma^5C_{ijj}\sfer\gau{jR}'}\nonumber \\
	&\qquad + \inpr{J_M\fer{L}}{\gamma^5\asfer C_{iij}^*\gau{iL}'} + \inpr{J_M\fer{L}}{\gamma^5\gau{jL}'\asfer C_{ijj}^*)},\label{eq:bb2-action-ferm}
\end{align}
prior to scaling the gauginos according to \eqref{eq:bb1-scaling}. 
Here we have employed \eqref{eq:bb2-innfs2} and Corollary \ref{cor:symmInnerProd} of the inner product. 
The bosonic part of \eqref{eq:bb2-action} is given by 
\begin{align}
	\act{b, ij}[\mathbb{A}, \szeta] &= \int_M  |\n_{ij} D_\mu \sfer|^2 + \mathcal{M}_{ij}(\sfer, \asfer)\label{eq:bb2-action-term2}
\end{align}
(cf.~\eqref{eq:spectral_action_acg_flat}) with $\n_{ij} = \n_{ij}^*$ the square root of the positive semi-definite $M \times M$--matrix 
\begin{align}
\mathcal{N}_{ij}^2 &= \frac{f(0)}{2\pi^2}(N_i C_{iij}^*C_{iij} + N_jC_{ijj}^* C_{ijj}) \label{eq:exprN},\\
\intertext{where $M$ is the number of particle generations, and} 
\mathcal{M}_{ij}(\sfer, \asfer) &= \frac{f(0)}{2\pi^2}\Big[N_i|C_{iij}\sfer\asfer C_{iij}^*|^2 + N_j|\asfer C_{ijj}^*C_{ijj}\sfer|^2 \nn\\
		&\qquad\qquad + 2|C_{iij}\sfer|^2|C_{ijj}\sfer|^2\Big].\label{eq:exprM1}
\end{align}
The first term of this last equation corresponds to paths in the Krajewski diagram such as in the first example of Figure \ref{fig:KrajPaths}, involving the vertex at $(i, i)$. The second term corresponds to the same type of path but involving $(j, j)$ and the third term consists of paths going in two directions such as the fourth example of Figure \ref{fig:KrajPaths}.

\subsection{Matching degrees of freedom}

As far as the gauginos are concerned, there is a difference compared to the previous section; there the trace parts of the action fully decoupled from the rest of the action, but here this is not the case due to the fermion-sfermion-gaugino interactions in \eqref{eq:bb2-action}. At the same time, the gauge fields $A_{\mu i}'$ and $A_{\mu j}'$ do not act on \rep{i}{j} and \rep{j}{i} in the adjoint representation, causing their trace parts not to vanish either. We thus have fermionic and bosonic $u(1)$ fields, that are each other's potential superpartners.

We distinguish between two cases:
\begin{itemize}
\item In the left image of Figure \ref{fig:bb2} $\H_{R = +} = \rep{i}{j} \oplus \rep{j}{i}$ and thus we can employ the unimodularity condition \eqref{eq:unimod_new}. This yields\footnote{When having multiple copies of the representations \rep{i}{j} and \rep{j}{i} all expressions will be multiplied by the number of copies, since the gauge bosons act on each copy in the same way. This leaves the results unaffected, however.}

\bas
	0 &= \tr_{\rep{i}{j}}g_i'A_{i\mu}' + \tr_{\rep{j}{i}}g_j'A_{j\mu}' \\
		&= N_jg_{B_i}B_{i\mu} + N_ig_{B_j}B_{j\mu} \quad\Longrightarrow \quad B_{j\mu} = - (N_jg_{B_i}/N_ig_{B_j}) B_{i\mu}, 
\eas
	where we have first identified the independent gauge fields before introducing the coupling constants $g_{i,j}$, $g_{B_{i,j}}$ (cf.~\cite[\S 3.5.2]{CCM07}). Consequently the covariant derivative acting on the fermion $\fer{}$ and scalar $\sfer$ and their conjugates is equal to 
$\can_A = i\gamma^\mu D_\mu$ with
\bas
	 D_\mu &= \nabla^S_\mu - i\Big( g_iA_{i\mu} + \frac{g_{B_i}}{N_i}B_i\Big)  + i \Big(g_jA_{j\mu} + \frac{g_{B_j}}{N_j}B_{j}\Big)^o\\
	  	&= \nabla^S_\mu - ig_i A_{i\mu} + ig_j A_{j\mu}^o - 2ig_{B_i}\frac{B_i}{N_i}.
\eas
This also means that the kinetic terms of the $u(1)$ gauge field now appear in the action. After applying the unimodularity condition, the kinetic terms of the gauge bosons, as acting on \rep{i}{j}, are given by
\ba
	& - \tr_{\rep{i}{j}} \mathbb{F}_{\mu\nu}'\mathbb{F}'^{\mu\nu} \nn\\
	&\qquad = \tr_{\rep{i}{j}}\Big(g_iF_{\mu\nu}^i - g_jF_{\mu\nu}^{j\,o} + g_{B_i}\frac{2}{N_i}B^{i}_{\mu\nu}\Big)\nn\\
		&\qquad\qquad\qquad\qquad \times \Big(g_iF^{\mu\nu}_i - g_jF^{\mu\nu\,o}_{j} + g_{B_i}\frac{2}{N_i}B_{i}^{\mu\nu}\Big)\nn\\
			&\qquad = N_jg_i^2\tr_{N_i} F_{\mu\nu}^iF^{\mu\nu}_i + N_ig_j^2\tr_{N_j} F_{\mu\nu}^j F^{\mu\nu}_j + 4\frac{N_j}{N_i}g_{B_i}^2B_{\mu\nu}^iB^{\mu\nu}_i,\label{eq:bb2-gaugekinterms}
\ea
with $B_{i\mu\nu} = \partial_{[\mu} B_{i\nu]}$. The contribution from \rep{j}{i} is the same and those from \rep{i}{i} and \rep{j}{j} have been given in the previous section.

We can use the supersymmetry transformations to also reduce the fermionic degrees of freedom:
\begin{lem}\label{lem:gobinogo}
	Requiring the unimodularity condition \eqref{eq:unimod_new} also for the supersymmetry transformations of the gauge fields, makes the traces of the gauginos proportional to each other.
\end{lem}
\begin{proof}
	We introduce the notation $\gau{iL,R} = \gau{iL,R}^a \otimes T^a_i$, where $T^a_i$, $a = 0, 1, \ldots,\linebreak N_i^2 - 1$ are the generators of $u(N_i) \simeq u(1) \oplus su(N_i)$. Writing out the unimodularity condition \eqref{eq:unimod_new} for the transformation \eqref{eq:bb1-transforms1} of the gauge field reads in this case
	\bas
		0 &=  N_j (g_i\tr \delta A_{i\mu} + g_{B_i}\delta B_{i\mu}) + N_i (g_j\tr \delta A_{j\mu} + g_{B_j}\delta B_{j\mu})
	\eas
	Putting in the expressions for the transformations and using that the $su(N_{i,j})$-parts of the gauginos are automatically traceless, we only retain the trace parts:
	\ba
	 0 &= N_jg_{B_i}\big[(J_M\eR, \gamma_\mu\gau{iL}^0) + (J_M\eL, \gamma_\mu\gau{iR}^0)\big] \nn\\
			&\qquad + N_ig_{B_j}\big[(J_M\eR, \gamma_\mu\gau{jL}^0) + (J_M\eL, \gamma_\mu\gau{jR}^0)\big]\nn\\
			&=\big(J_M\eR, \gamma_\mu(N_jg_{B_i}\gau{iL}^0 + N_ig_{B_j}\gau{jL}^0)\big) + (L \leftrightarrow R)\label{eq:bb2-terms-indep},
	\ea	
where with `$(L\leftrightarrow R)$' we mean the expression preceding it, but everywhere with $L$ and $R$ interchanged. Since $\epsilon = (\eL, \eR)$ can be any covariantly vanishing spinor, $(0, \eR)$ with $\nabla^S\eR = 0$ and $(\eL, 0)$ with $\nabla^S\eL = 0$ are valid solutions for which one of the terms in \eqref{eq:bb2-terms-indep} vanishes, but the other does not. The term with left-handed gauginos is thus independent from that of the right-handed gauginos. Hence, for any $\eR$, 
\bas
\big(J_M\eR, \gamma_\mu(N_jg_{B_i}\gau{iL}^0 + N_ig_{B_j}\gau{jL}^0)\big)
\eas
must vanish, establishing the result.
\end{proof}
Via the transformation \eqref{eq:bb1-transforms2} for the gaugino, we can also reduce one of the $u(1)$ parts of $G_{i,j}' = G_{i,j}^aT^a_{i,j} + H_{i,j} \in C^{\infty}(M, u(N_{i,j}))$.

This provides us a justification for the choice to take the trace in \eqref{eq:unimod_new} only over $\H_F$. For if we had not, we would have been in a bootstrap-like situation in which the gaugino degrees of freedom would have contributed to the relation that we have employed to reduce them by. 

\item In the right image of Figure \ref{fig:bb2} no constraint occurs due to the unimodularity condition because $\H_{R = +} = 0$ and the kinetic terms of the gauge bosons are given by:
\ba
&	- \tr_{\rep{i}{j}} \mathbb{F}_{\mu\nu}'\mathbb{F}'^{\mu\nu}\nn\\ 
&= \tr_{\rep{i}{j}}\Big(g_iF_{\mu\nu}^i - g_jF_{\mu\nu}^{j\,o} + \frac{g_{B_i}}{N_i}B_{i\mu\nu} - \frac{g_{B_j}}{N_j} B_{j\mu\nu}\Big)^2\nn\\
&= N_jg_i^2\tr_{N_i} F_{\mu\nu}^iF^{\mu\nu}_i + N_ig_j^2\tr_{N_j} F_{\mu\nu}^j F^{\mu\nu}_j \nn\\
	&\qquad + N_iN_j\Big(\frac{g_{B_i}B_i}{N_i} - \frac{g_{B_j}B_j}{N_j}\Big)_{\mu\nu}\Big(\frac{g_{B_i}B_i}{N_i} - \frac{g_{B_j}B_j}{N_j}\Big)^{\mu\nu}.\label{eq:bb2-gaugekinterms_right}
\ea
\end{itemize}

Here for the second time we stumble upon problems with the fact that the spectral action gives us an on shell action only. The problem is twofold.  First, there is ---as in the case of \B{i} and \B{j}--- a mismatch in the degrees of freedom off shell between $\fer{} \equiv \fer{ij}$ and $\sfer \equiv \sfer_{ij}$. We compensate for this by introducing a bosonic auxiliary field $F_{ij} \in C^{\infty}(M, \rep{i}{j})$ and its conjugate. They appear in the action via
\begin{align}\label{eq:bb2-auxfields}
	S[F_{ij}, F^*_{ij}] &= - \int_M \tr_{N_j} F_{ij}^* F_{ij}\sqrt{g}\mathrm{d}^4x.
\end{align}
From the Euler-Lagrange equations, it follows that $F_{ij} = F_{ij}^* = 0$, i.e.~$F_{ij}$ and its conjugate only have degrees of freedom off shell. Secondly, the four-scalar self-interaction of $\sfer$ poses an obstacle for a supersymmetric action; regardless of its specific form, a supersymmetry transformation of such a term must involve three scalars and one fermion, a term that cannot be canceled by any other. The standard solution is to rewrite these terms using the auxiliary fields $G_i'$, $G_j'$ that the building blocks of the first type provide us, such that we recover \eqref{eq:bb2-action-term2} on shell. The next lemma tells us that we can do this.

\begin{lem}\label{lem:bb2-offshell}
If $\H_{F,R = +} \ne 0$ then the four-scalar terms \eqref{eq:exprM1} of an almost-commutative geometry that consists of a single building block \B{ij} of the second type can be written in terms of auxiliary fields $G_{i,j} \in C^{\infty}(M, su(N_{i,j}))$ and $H \in C^{\infty}(M, u(1))$, as follows:
\ba
	\mathcal{L}(G_{i,j}, H, \sfer, \asfer) &= -\frac{1}{2n_i}\tr G_i^2 -\frac{1}{2n_j}\tr G_j^2 - \frac{1}{2}H^2 - \tr G_i\P_i' \sfer\asfer \nn\\
		&\qquad - \tr G_j\asfer\P_j'\sfer -  H\tr\Q'\sfer\asfer,\label{eq:bb2-auxterms}
\ea
where in the terms featuring $G_{i,j}$ the trace is over the $N_{i,j}\times N_{i,j}$-matrices and with
\begin{align}
	\P_i' &= \sqrt{\frac{f(0)}{\pi^2n_i} N_i} C_{iij}^*C_{iij},&
	\P_j' &= \sqrt{\frac{f(0)}{\pi^2n_j} N_j} C_{ijj}^*C_{ijj},\nn\\
	\Q' &= \sqrt{\frac{f(0)}{\pi^2}} (C_{iij}^*C_{iij} + C_{ijj}^*C_{ijj}), &&\label{eq:solP}
\end{align}
matrices on $M$-dimensional family space.
\end{lem}
\begin{proof}
	Required for any building block \B{ij} of the second type are the building blocks \B{i} and \B{j} of the first type, initially providing auxiliary fields $G_{i,j} \equiv G^{a}_{i,j} T^a_{i,j} \in C^{\infty}(M, su(N_{i,j}))$ and $H_{i,j} \in C^{\infty}(M, u(1))$. Here the $T^a_{i,j}$ denote the generators of $su(N_{i,j})$ in the fundamental (defining) representation and are normalized according to $\tr T^a_{i,j}T^b_{i,j} = n_{i,j}\delta_{ab}$, where $n_{i,j}$ is the \emph{constant of the representation}. After applying the unimodularity condition \eqref{eq:unimod_new} in the case that $\H_{R = +} \ne 0$ (the left image of Figure \ref{fig:bb2}) for the gauge field and its transformation, only one $u(1)$ auxiliary field $H$ remains. We thus consider the Lagrangian \eqref{eq:bb2-auxterms} with $\P_{i,j}', \Q'$ self-adjoint. (These coefficients are written inside the trace since they may have family indices. However, the combinations $\P_i'\sfer_{}\asfer{}$ and $\asfer_{}\P_j'\sfer_{}$ cannot have family-indices anymore, since $G_i$ and $G_j$ do not.) Applying the Euler-Lagrange equations to this Lagrangian yields
\bas
	G_i^a &= - \tr T^a_i\P_i'\sfer\asfer, &
	G_j^a &= - \tr T^a_j\asfer\P_j'\sfer, &
	H &= - \tr\Q'\sfer\asfer
\eas
and consequently \eqref{eq:bb2-auxterms} equals \emph{on shell} 
\bas
&\mathcal{L}(G_{i,j}, H, \sfer, \asfer) \nn\\
&\qquad= \frac{1}{2}\tr (T_i^a\P_i'\sfer_{ij}\asfer_{ij})^2 + \frac{1}{2}\tr (T_j^a\asfer_{ij}\P_j'\sfer_{ij})^2 + \frac{1}{2}\tr (\Q'\sfer_{ij}\asfer_{ij})^2\nn\\
&\qquad = \frac{n_i}{2}\Big(|\P_i'\sfer\asfer|^2 - \frac{1}{N_i}|\P'^{1/2}_i\sfer|^4\Big) + \frac{n_j}{2}\Big(|\asfer\P_j'\sfer|^2 - \frac{1}{N_j}|\P_j'^{1/2}\sfer|^4\Big)\nn\\
&\qquad\qquad\nn\\
&\qquad\qquad + \frac{1}{2}|\Q'^{1/2}\sfer|^4.
\eas
Here we have employed the identity
\ba\label{eq:idn-sun-gens}
	(T^a_{i,j})_{mn} (T^a_{i,j})_{kl} = n_{i,j}\Big(\delta_{ml}\delta_{kn} - \frac{1}{N_{i,j}}\delta_{mn}\delta_{kl}\Big).
\ea
With the choices \eqref{eq:solP}
we indeed recover the four-scalar terms \eqref{eq:exprM1} of the spectral action. 
\end{proof}

Even though in the case that $\H_{F,R = +} = 0$ (the right image of Figure \ref{fig:bb2}) the unimodularity condition cannot be used to relate the $u(1)$ fields $H_i$ and $H_j$ to each other, a similar solution is possible:
\begin{cor}
If $\H_{R = +} = 0$ then the four-scalar terms \eqref{eq:exprM1} of a building block \B{ij} of the second type can be written off shell using the Lagrangian
\ba
	\mathcal{L}(G_{i,j}, H_{i,j}, \sfer, \asfer) &= -\frac{1}{2n_i}\tr G_i^2 -\frac{1}{2n_j}\tr G_j^2 - \frac{1}{2}H_i^2 - \frac{1}{2}H_j^2 -  \tr G_i\P_i'\sfer\asfer \nn\\&\qquad- \tr G_j\asfer \P_j'\sfer - H_i \tr\Q'_i\sfer\asfer - H_j\tr\Q'_j\sfer\asfer,\label{eq:bb2-auxterms2}
\ea
with
\begin{align*}
	\P'_i &= \sqrt{\frac{f(0)}{\pi^2n_i} N_i} C_{iij}^*C_{iij},&
	\P'_j &= \sqrt{\frac{f(0)}{\pi^2n_j} N_j} C_{ijj}^*C_{ijj},\nn\\
	\Q'_i &= \Q'_j = \sqrt{\frac{f(0)}{2\pi^2}}(C_{iij}^*C_{iij} + C_{ijj}^*C_{ijj}),&&
\end{align*}
not carrying a family-index.
\end{cor}

In both cases we have obtained a system that has equal bosonic and fermionic degrees of freedom, both on shell and off shell.

\subsection{The final action and supersymmetry}

 We first turn to the case that $\H_{R = + }\ne 0$. Reducing the degrees of freedom by identifying half of the $u(1)$ fields with the other half and rewriting \eqref{eq:bb2-action} to an off shell action 
we find the \emph{extra} contributions
\begin{align*}
	& \inpr{J_M\afer{R}}{\gamma^5(\gau{iR}'C_{iij}\sfer + C_{ijj}\sfer\gau{jR}')} + \inpr{J_M\fer{L}}{\gamma^5(\asfer C_{iij}^*\gau{iL}' + \gau{jL}'\asfer C_{ijj}^*)}\nonumber\\
	 & \qquad + \inpr{J_M\afer{R}}{D_A\fer{L}} + \int_M\Big[|\n_{ij}D_\mu \sfer|^2 - \tr_{N_i}\big( \P_i'\sfer\asfer G_i\big) \nn\\
	&\qquad\qquad - \tr_{N_j}\big(\asfer \P_j'\sfer G_j\big) - H\tr_{N_i} \Q'\sfer\asfer - \tr_{N_j^{\oplus M}}F_{ij}^*F_{ij}\Big]
\end{align*}
to the total action, with 
\bas
	\gau{i}' &= \gau{i} + \gau{i}^0 \id_{N_i},&
	\gau{j}' &= \gau{j} - N_{j/i} \gau{i}^0 \id_{N_j}
\eas
and $G_{i,j} \in C^{\infty}(M, su(N_{i,j}))$, $H \in C^{\infty}(M, u(1))$. For notational convenience we will suppress the subscripts in the traces when no confusion is likely to arise. In addition, adding a building block \B{ij} slightly changes the expressions for the pre-factors of the kinetic terms of $A_{i\mu}$ and $A_{j\mu}$ (cf.~Remark \ref{rmk:bb2-rmk} below). 

As a final step we scale the sfermion $\sfer_{ij}$ according to 
\ba\label{eq:bb3-scalingfields}
	\sfer_{ij}&\to \n_{ij}^{-1} \sfer_{ij},&
	\asfer_{ij}&\to  \asfer_{ij}\n_{ij}^{-1},
\ea
and the gauginos according to \eqref{eq:bb1-scaling} to give us the correctly normalized kinetic terms for both:
\begin{align}
	&\inpr{J_M\afer{R}}{\gamma^5[\gau{iR}'\Cw{i,j}\sfer + \Cw{j,i}\sfer\gau{jR}']} + \inpr{J_M\fer{L}}{\gamma^5[\asfer \Cw{i,j}^*\gau{iL}' + \gau{jL}'\asfer \Cw{j,i}^*]}\nonumber\\
	 & \qquad + \inpr{J_M\afer{R}}{D_A\fer{L}} + \int_M\Big[|D_\mu \sfer|^2 - \tr\big( \P_i\sfer\asfer G_i\big) - \tr\big(\asfer \P_j\sfer G_j\big) \nn\\
	&\qquad\qquad - \tr HQ\sfer\asfer - \tr_{N_j^{\oplus M}}F_{ij}^*F_{ij}\Big]\label{eq:bb2-action-offshell}.
\end{align}
Here we have written
\ba	\label{eq:bb2-scalingG}
\Cw{i,j} &:= \frac{C_{iij}}{\sqrt{n_i}}\n_{ij}^{-1},& \Cw{j,i} &:= \frac{C_{ijj}}{\sqrt{n_j}}\n_{ij}^{-1}, \nn\\
	\P_{i,j} &:= \n_{ij}^{-1} \P_{i,j}' \n_{ij}^{-1} &
	\Q &:= \n_{ij}^{-1} \Q' \n_{ij}^{-1}
\ea	
for the scaled versions of the parameters. For this action we have:

\begin{theorem}\label{prop:bb2}
		The total action that is associated to $\B{i} \oplus \B{j} \oplus \B{ij}$, given by \eqref{eq:SYM} and \eqref{eq:bb2-action-offshell}, is supersymmetric under the transformations \eqref{eq:bb1-transforms},
\begin{subequations}\label{eq:susytransforms4}
\begin{align}
				\delta \sfer &= c_{ij}(J_M\epsilon_L, \gamma^5 \fer{L})_{\cS},  &\delta \asfer &=  c_{ij}^*(J_M\epsilon_R, \gamma^5 \afer{R})_{\cS}\label{eq:transforms4.1},\\
				\delta \fer{L} &= c_{ij}' \gamma^5 [\can_A, \sfer]\epsilon_R + d_{ij}'F_{ij}\epsilon_L,  & \delta \afer{R} &=  c_{ij}'^*\gamma^5 [\can_A, \asfer]\epsilon_L + d_{ij}'^* F_{ij}^*\epsilon_R\label{eq:transforms4.2}
\
\end{align}
\end{subequations}
and
\begin{subequations}\label{eq:susytransforms5}
\begin{align}
	\delta F_{ij} &= d_{ij}(J_M\eR, \can_A\fer{L})_{\cS} + d_{ij,i}(J_M\eR, \gamma^5\gau{iR}\sfer)_{\cS} - d_{ij,j}(J_M\eR, \gamma^5\sfer\gau{jR})_{\cS},\label{eq:transforms4.4a}\\
	 \delta F_{ij}^* &= d_{ij}^*(J_M\eL, \can_A\afer{R})_{\cS} + d_{ij,i}^*(J_M\eL, \gamma^5\asfer\gau{iL})_{\cS} - d_{ij,j}^*(J_M\eL, \gamma^5\gau{jL}\asfer)_{\cS}\label{eq:transforms4.4b},
\end{align}
\end{subequations}
with $c_{ij}, c_{ij}', d_{ij}, d_{ij}', d_{ij,i}$ and $d_{ij,j}$ complex numbers, if and only if 
\ba
	\Cw{i,j} &= \sgnc_{i,j}\sqrt{\frac{2}{\K_i}}g_i\id_M,& \Cw{j,i} &=  \sgnc_{j,i}\sqrt{\frac{2}{\K_j}}g_j\id_M,\nn\\ 
\P_{i}^2 &= \frac{g_{i}^2}{\K_{i}}\id_M, & \P_{j}^2 &= \frac{g_{j}^2}{\K_{j}}\id_M,\label{eq:bb2-resultCiij}
\ea
for the unknown parameters of the finite Dirac operator (where $\id_M$ is the identity on family-space, which equals unity of $\fer{ij}$ has no family index) and 
	\bas
		c_{ij}' &=  c_{ij}^* = \sgnc_{i,j} \sqrt{2\K_i}c_i = - \sgnc_{j,i} \sqrt{2\K_j}c_j,\nn\\ 
		 d_{ij} &= d_{ij}'^* = \sgnc_{i,j} \sqrt{\frac{\K_i}{2}} \frac{d_{ij,i}}{g_i}  = - \sgnc_{j,i} \sqrt{\frac{\K_j}{2}} \frac{d_{ij,j}}{g_j},&
		c_{G_i} &= \sgnc_i \sqrt{\K_i} c_i,
	\eas	
	 with $\sgnc_{i}, \sgnc_{i,j}, \sgnc_{j,i} \in \{\pm 1\}$ for the transformation constants.
	\end{theorem}
	\begin{proof}
			Since the action \eqref{eq:SYM} is already supersymmetric by virtue of Theorem \ref{thm:bb1}, we only have to prove that the same holds for the contribution \eqref{eq:bb2-action-offshell} to the action from \B{ij}. The detailed proof of this fact can be found in Appendix \ref{sec:bb2-proof}.
	\end{proof}

Then for $C_{iij}$ and $\P_{i,j}$ that satisfy these relations (setting $\K_{i,j} = 1$), the supersymmetric action (but omitting the $u(1)$-terms for conciseness now) reads:
\begin{align}
	&\inpr{J_M\afer{R}}{\can_A\fer{L}} + \sqrt{2}\inpr{J_M\afer{R}}{\gamma^5(\sgnc_{i,j} g_i\gau{iR}\sfer + \sgnc_{j,i} \sfer g_j\gau{jR})}\nonumber \\
	&\quad + \sqrt{2}\inpr{J_M\fer{L}}{\gamma^5(\sgnc_{i,j} \asfer g_i\gau{iL} + \sgnc_{j,i} g_j\gau{jL}\asfer )}\label{eq:bb2-action-susy}\\
	 & \qquad + \int_M\Big[|D_\mu \sfer|^2 - g_i\tr_{N_i}\big(\sfer\asfer G_i\big) - g_j\tr_{N_j}\big(\asfer \sfer G_j\big) - \tr_{N_j^{\oplus M}} F_{ij}^*F_{ij}\Big]\nn,
\end{align}
i.e.
we recover the pre-factors for the fermion-sfermion-gaugino and four-scalar interactions that are familiar for supersymmetry. The signs $\sgnc_{i,j}$ and $\sgnc_{j,i}$ above can be chosen freely.

\begin{rmk}\label{rmk:bb2-obstr}In the case that $\H_{R = +} = 0$, there is an interaction 
\ba
	\propto \int_M B_{i\mu\nu}B^{\mu\nu}_j\label{eq:u1cross}
\ea
present (see the last term of \eqref{eq:bb2-gaugekinterms_right}). Transforming the gauge fields appearing in that interaction shows that the supersymmetry of the total action requires an interaction 
\bas
	\propto \inpr{J_M\gau{i}^0}{\dirac \gau{j}^0}, 
\eas
a term that the fermionic action does not provide. Thus, a situation in which there are two different $u(1)$ fields that both act on the same representation \rep{i}{j} is an obstruction for supersymmetry. This is also the reason that a supersymmetric action with gauge groups $U(N_{i,j})$ is not possible in the presence of a representation \rep{i}{j}, since 
\bas
&- \tr_{\rep{i}{j}}\mathbb{F}_{\mu\nu} \mathbb{F}^{\mu\nu}	\nn\\
	&= \tr_{\rep{i}{j}}(g_iF^i_{\mu\nu} - g_jF^{j\,o}_{\mu\nu})(g_iF_i^{\mu\nu} - g_jF_{j}^{\mu\nu\,o}) \\
			&= N_jg_i^2\tr F^i_{\mu\nu}F_i^{\mu\nu} + N_ig_j^2\tr F^j_{\mu\nu}F_j^{\mu\nu} - 2g_ig_j\tr F^i_{\mu\nu}\tr F_j^{\mu\nu}, 
\eas
of which the last term spoils supersymmetry. Averting a theory in which two independent $u(1)$ gauge fields act on the same representation will be seen to put an important constraint on realistic supersymmetric models from noncommutative geometry.
\end{rmk}

Note that it is not per se the presence of an $R = -1$ off-diagonal fermion in the first place that is causing this; in a spectral triple that contains at least one $R = +1$ fermion the interaction \eqref{eq:u1cross} vanishes due to the unimodularity condition \eqref{eq:unimod_new}.

\begin{rmk}\label{rmk:bb2-rmk}
	In the previous section we have compactly written
	\begin{align*}
		\K_i = \frac{f(0)}{3\pi^2}2N_ig_i^2n_i
	\end{align*}
	only partly for notational convenience. There are two other reasons. The first is that since the kinetic terms for the gauge bosons are normalized to $-1/4$, $\mathcal{K}_i$ must in the end have the value of $1$. This puts a relation between $f(0)$ and $g_i$. This is the same as in the Standard Model \cite[\S 17.1]{CCM07}. Secondly, the expression for $\mathcal{K}_i$ depends on the contents of the spectral triple. As \eqref{eq:bb2-gaugekinterms} shows, when the Hilbert space is extended with \rep{i}{j} and its opposite (both having $R = 1$), then \eqref{eq:expressionK1} changes to
\begin{align}
	\K_i &= \frac{f(0)}{3\pi^2}g_i^2n_i(2N_i + MN_j), & \K_j &= \frac{f(0)}{3\pi^2}g_j^2n_j(MN_i + 2N_j), \nn\\
	 \K_{B} &= \frac{4f(0)}{3\pi^2}\frac{N_j}{N_i}M g_B^2.&&\label{eq:expressionK2}
\end{align}
Here $M$ denotes the number of generations that the fermion--sfermion pair comes in. In fact, the relation between the coupling constant(s) $g_i$ and the function $f$ should be evaluated only for the full spectral triple. In this case however, setting all three terms equal to one, implies the GUT-like relation 
		\bas
			n_i(2N_i + MN_j) g_i^2 = n_j(2N_j + MN_i) g_j^2 = 4\frac{N_j}{N_i}M g_B^2.
		\eas 
\end{rmk}

What remains, is to check whether there exist solutions for $C_{iij}$ and $C_{ijj}$ that satisfy the supersymmetry constraints \eqref{eq:bb2-resultCiij}. 
\begin{prop}\label{lem:bb2-nosol}
	Consider an almost-commutative geometry whose finite algebra is of the form $M_{N_i}(\com) \oplus M_{N_j}(\com)$. The particle content and action associated to this almost-commutative geometry are both supersymmetric off shell if and only if it consists of two disjoint building blocks \B{i,j} of the first type, for which $N_{i}, N_{j} > 1$.
\end{prop}
\begin{proof}
We will prove this by showing that the action of a single building block \B{ij} of the second type is not supersymmetric, falling back to Theorem \ref{thm:bb1} for a positive result. For the action of a \B{ij} of the second type to be supersymmetric requires the existence of parameters $C_{iij}$ and $C_{ijj}$ that ---after scaling according to \eqref{eq:bb2-scalingG}--- satisfy \eqref{eq:bb2-resultCiij} both directly and indirectly via $\P_{i,j}$ of the form \eqref{eq:solP}. To check whether they directly satisfy \eqref{eq:bb2-resultCiij} we note that the pre-factor $\n_{ij}^2$ for the kinetic term of the sfermion $\sfer_{ij}$ appearing in \eqref{eq:bb2-scalingG} itself is an expression in terms of $C_{iij}$ and $C_{ijj}$. We multiply the first relation of \eqref{eq:bb2-resultCiij} with its conjugate and multiply with $\n_{ij}$ on both sides to get
\bas
	C_{iij}^*C_{iij} = \frac{2}{\K_i}n_ig_i^2 \n_{ij}^2.
\eas
Inserting the expression \eqref{eq:exprN} for $\n_{ij}^2$, we obtain
\bas
	C_{iij}^*C_{iij} &= g_i^2n_i\frac{f(0)}{\pi^2}\frac{1}{\K_i}\Big[N_iC_{iij}^*C_{iij} + N_jC_{ijj}^*C_{ijj}\Big].
\eas
From \eqref{eq:bb2-scalingG} and \eqref{eq:bb2-resultCiij} we infer that $C_{ijj}^*C_{ijj} = (n_jg_j^2/n_ig_i^2)C_{iij}^*C_{iij}$, i.e.~we require:
\bas
	\K_i &= \frac{f(0)}{\pi^2}\Big[g_i^2n_iN_i + n_jg_j^2N_j\Big].
\eas
If we use the expressions \eqref{eq:expressionK2} for the pre-factors of the gauge bosons' kinetic terms to express the combinations $f(0)n_{i,j}g_{i,j}^2/\pi^2$ in terms of $N_{i,j}$ and $M$, the requirement for consistency reads
\bas
		1 &= \bigg(\frac{3N_i}{2N_i + MN_j} + \frac{3N_j}{MN_i + 2N_j}\bigg).
\eas
The only solutions to this equation are given by $M = 4$ and $N_i = N_j$. However, inserting the solution \eqref{eq:bb2-resultCiij} for $C_{iij}^*C_{iij}$ into the expression \eqref{eq:solP} for $\P_{i}, \P_j$ (necessary to write the action off shell) gives
\bas
	\P_{i}^2 &= 4\frac{f(0)}{\pi^2}N_{i}g_{i}^4\frac{n_{i}}{\K_i^2},&
	\P_{j}^2 &= 4\frac{f(0)}{\pi^2}N_{j}g_{j}^4\frac{n_{j}}{\K_j^2},
\eas
with an $\id_M$ where appropriate. We again use Remark \ref{rmk:bb2-rmk} to replace $f(0)g_i^2/(\pi^2\K_i)$ by an expression featuring $N_{i,j}$, $M$ and $n_{i,j}$. This yields
\bas
	\P_{i}^2 &= \frac{12N_i}{2N_i + MN_j}\frac{g_i^2}{\K_i} = 2\frac{g_i^2}{\K_i},&
	\P_{j}^2 &= \frac{12N_j}{2N_j + MN_i}\frac{g_j^2}{\K_j} = 2\frac{g_j^2}{\K_j}
\eas
for the values $M = 4$, $N_i = N_j$ that gave the correct fermion-sfermion-gaugino interactions. We thus have a contradiction with the demand on $\P_{i,j}^2$ from \eqref{eq:bb2-resultCiij}, necessary for supersymmetry.
\end{proof}

We shortly pay attention to a case that is of similar nature but lies outside the scope of the above Proposition. 

\begin{rmk}
	For $\A_F = \com \oplus \com$, a building block \B{ij} of the second type does not have a supersymmetric action either. In this case there are only $u(1)$ fields present in the theory and $G_i$, $G_j$ and $H$ are seen to coincide. It is possible to rewrite the four-scalar interaction of the spectral action off shell, but 
this set up also suffers from a similar problem as in Proposition \ref{lem:bb2-nosol}.
\end{rmk}

We can extend the result of Proposition \ref{lem:bb2-nosol} to components of the finite algebra that are defined over other fields than $\com$. For this, we first need the following lemma.

\begin{lem}\label{lem:bb2-otherfields}
	The inner fluctuations \eqref{eq:inner_flucts} of $\dirac$ caused by a component of the finite algebra that is defined over $\mathbb{R}$ or $\mathbb{H}$, are traceless.
\end{lem}
\begin{proof}
	The inner fluctuations are of the form
	\bas
		i \gamma^\mu A_\mu^{\mathbb{F}},\quad  A_\mu^{\mathbb{F}} = \sum_i a_i \partial_\mu (b_i),
	\eas
 with $a_i, b_i \in C^{\infty}(M, M_{N}(\mathbb{F}))$, $\mathbb{F} = \mathbb{R}, \mathbb{H}$. This implies that $A_\mu^\mathbb{F}$ is itself an $M_{N}(\mathbb{F})$-valued function. For the inner fluctuations to be self-adjoint, $A_\mu^\mathbb{F}$ must be skew-Hermitian. In the case that $\mathbb{F} = \mathbb{R}$ this implies that all components on the diagonal vanish and consequently so does the trace. In the case that $\mathbb{F} = \mathbb{H}$, all elements on the diagonal must themselves be skew-Hermitian. Since all quaternions are of the form
	\bas
	\begin{pmatrix}
		\alpha & \beta\\
		- \bar \beta & \bar \alpha
	\end{pmatrix} \quad \alpha, \beta \in \com,
	\eas
	this means that the diagonal of $A_\mu^\mathbb{H}$ consists of purely imaginary numbers that vanish pairwise. Its trace is thus also $0$.
\end{proof}

Then we have 
\begin{theorem}
	Consider an almost-commutative geometry whose finite algebra is of the form $M_{N_i}(\mathbb{F}_i) \oplus M_{N_j}(\mathbb{F}_j)$ with $\mathbb{F}_i, \mathbb{F}_j = \mathbb{R}, \com, \mathbb{H}$. The particle content and action associated to this almost-commutative geometry are both supersymmetric off shell if and only if it consists of two disjoint building blocks \B{i,j} of the first type, for which $N_{i}, N_{j} > 1$.
\end{theorem}
\begin{proof}
Not only do we have different possibilities for the fields $\mathbb{F}_{i,j}$ over which the components are defined, but we can also have various combinations for the values of the $R$-parity. We cover all possible cases one by one.
 
If $R = +1$ on the representations in the finite Hilbert space that describe the gauginos, then the gauginos and gauge bosons have the same $R$-parity and the particle content is not supersymmetric. 

If $R = -1$ for these representations, and $R = +1$ on the off-diagonal representations, suppose at least one of the $\mathbb{F}_i$, $\mathbb{F}_j$ is equal to $\mathbb{R}$ or $\mathbb{H}$. Then using Lemma \ref{lem:bb2-otherfields} we see that after application of the unimodularity condition \eqref{eq:unimod_new} there is no $u(1)$-valued gauge field left. Lemma \ref{lem:gobinogo} then also causes the absence of a $u(1)$-auxiliary field that is needed to write the four-scalar action off shell as in Lemma \ref{lem:bb2-offshell}. If both $\mathbb{F}_i$ and $\mathbb{F}_j$ are equal to $\com$ we revert to Proposition \ref{lem:bb2-nosol} to show that there is no supersymmetric solution for $M$ and $N_{i,j}$ that satisfies the demands for $\Cw{i,j}$, $\Cw{j,i}$ and $\P_{i,j}$ from supersymmetry.

In the third case $R = -1$ on the off-diagonal representations in $\H_F$. If both $\mathbb{F}_{i,j}$ are equal to $\mathbb{R}$ or $\mathbb{H}$ then there is no $u(1)$ gauge field and thus the spectral action cannot be written off shell. If either $\mathbb{F}_i$ or $\mathbb{F}_j$ equals $\mathbb{R}$ or $\mathbb{H}$, then there is one $u(1)$-field, but the calculation for the action carries through as in Proposition \ref{lem:bb2-nosol} and there is no supersymmetric solution for $M$ and $N_{i,j}$. Finally, if both $\mathbb{F}_{i,j}$ are equal to $\com$, there are two $u(1)$-fields and the cross term as in Remark \ref{rmk:bb2-obstr} spoils supersymmetry. 

Thus, all almost-commutative geometries for which $\A_F = M_{N_i}(\mathbb{F}_i) \oplus M_{N_j}(\mathbb{F}_j)$ and that have off-diagonal representations fail to be supersymmetric off shell.
\end{proof}

The set up described in this section has the same particle content as the supersymmetric version of a single ($R = +1$) particle--antiparticle pair and corresponds in that respect to a single chiral superfield in the superfield formalism, see Example \ref{ex:intro-susy-wz} and \cite[4.3]{DGR04}. In constrast, its action is not fully supersymmetric. 
We stress however, that the scope of Proposition \ref{lem:bb2-nosol} is that of a \emph{single} building block of the second type. As was mentioned before, the expressions for many of the coefficients typically vary with the contents of the finite spectral triple and they should only be assessed for the full model. 

Another interesting difference with the superfield formalism is that a building block of the second type really requires two building blocks of the first type, describing gauginos and gauge bosons. In the superfield formalism a theory consisting of only a chiral multiplet, not having gauge interactions, is in many textbooks the first model to be considered. This underlines that noncommutative geometry inherently describes gauge theories.

There are ways to extend almost-commutative geometries by introducing new types of building blocks ---giving new possibilities for supersymmetry--- or by combining ones that we have already defined. In the next section we will cover an example of the latter situation, in which there arise interactions between two or more building blocks of the second type.

\subsection{Interaction between building blocks of the second type}\label{sec:2bb2}

In the previous section we have fully exploited the options that a finite algebra with two components over the complex numbers gave us. If we want to extend our theory, the finite algebra \eqref{eq:alg} needs to have a third summand --- say $M_{N_k}(\com)$. A building block of the first type (cf.~Section \ref{sec:bb1}) can easily be added, but then we already stumble upon severe problems:
\begin{prop}\label{prop:2bb2-obstr}
	The action \eqref{eq:spectral_action_acg_flat} of an almost-commutative geometry whose finite algebra consists of three summands $M_{N_{i,j,k}}(\com)$ over $\com$ and whose finite Hilbert space features building blocks \Bc{ij}{\pm} and \Bc{ik}{\pm} is not supersymmetric.
\end{prop}
\begin{proof}
	The inner fluctuations of the canonical Dirac operator on \rep{i}{j} and \rep{i}{k} read:
	\bas
		& \dirac + g_iA_i - g_jA_j^o + \frac{g_{B_i}}{N_i}B_i - \frac{g_{B_j}}{N_j}B_j, \nn\\
		& \dirac + g_iA_i - g_kA_k^o + \frac{g_{B_i}}{N_i}B_i - \frac{g_{B_k}}{N_k}B_k,
	\eas	
	where $A_{i,j,k} = \gamma_\mu A^{\mu}_{i, j, k}$, with $A^{\mu}_{i,j, k}(x) \in su(N_{i,j, k})$ and similarly $B^{\mu}_{i,j, k}(x) \in u(1)$. The unimodularity condition will, in the case that the representation of at least one of the two building blocks has $R = +1$, leave two of the three independent $u(1)$ fields ---say--- $B_i$ and $B_j$. The kinetic terms of the gauge bosons on both representations will then feature a cross term \eqref{eq:u1cross} of different $u(1)$ field strengths, an obstruction for supersymmetry.
\end{proof}

To resolve this, we allow ---inspired by the NCSM--- for one or more copies of the quaternions $\mathbb{H}$ in the finite algebra. If we define a building block of the first type over such a component (with the finite Hilbert space $M_2(\com)$ as a bimodule of the complexification $M_1(\mathbb{H})^\com = M_2(\com)$ of the algebra, instead of $\mathbb{H}$ itself, cf.~\cite[\S 4.1]{Bhowmick2011}, \cite{CC08}), the self-adjoint inner fluctuations of the canonical Dirac operator are already seen to be in $su(2)$ (e.g.~traceless) prior to applying the unimodularity condition. On a representation \rep{i}{j} (from a building block \Bc{ij}{\pm} of the second type), of which one of the indices comes from a component $\mathbb{H}$, only one $u(1)$ field will act.

\emph{From here on, using three or more components in the algebra, we will always assume at most two to be of the form $M_N(\com)$ and all others to be equal to $\mathbb{H}$}.  

\begin{figure}[ht]
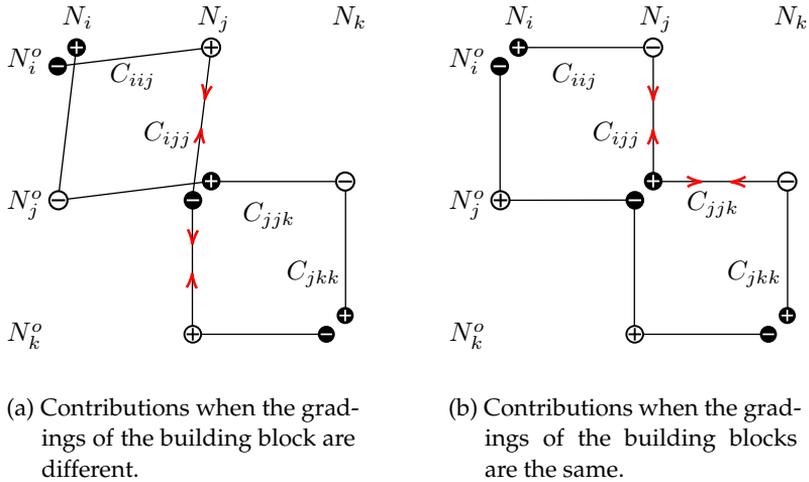

		\captionsetup{width=.9\textwidth}
	\centering
	\begin{subfigure}{.37\textwidth}
		\centering
		\def\svgwidth{\textwidth}
		\subimport{\the\svgpath}{2bb2s-different.pdf_tex}	
		\caption{Contributions when the gradings of the building block are different.}
		\label{fig:2bb2s-different}
	\end{subfigure}
	\hspace{30pt}
	\begin{subfigure}{.37\textwidth}
		\centering
		\def\svgwidth{\textwidth}
		\subimport{\the\svgpath}{2bb2s-same.pdf_tex}
		\caption{Contributions when the gradings of the building blocks are the same.}
		\label{fig:2bb2s-same}
	\end{subfigure}
\caption{In the case that there are two building blocks of the second type sharing one of their indices, there are extra interactions in the action.}
\label{fig:2bb2s}
\end{figure}

The action of an almost commutative geometry whose finite spectral triple features two building blocks of the second type sharing one of their indices (i.e.~that are in the same row or column in a Krajewski diagram) contains extra four-scalar contributions. The specific form of these terms depends on the value of the grading and of the indices appearing. When the first indices of two building blocks are the same, and they have the same grading (e.g.~\Bc{ji}{+} and \Bc{jk}{+}, cf.~Figure \ref{fig:2bb2s-different}) the resulting extra interactions are given by
\ba
	S_{ij,jk}[\sfer_{ij}, \sfer_{jk}] &= \frac{f(0)}{\pi^2} N_j \int_M |C_{ijj}\sfer_{ij}C_{jjk}\sfer_{jk}|^2 \sqrt{g}\mathrm{d}^4x.\label{eq:2bb2s-same}
\ea 
In the other case (cf.~Figure \ref{fig:2bb2s-same}) it is given by
\ba
S_{ij,jk}[\sfer_{ij}, \sfer_{jk}] &=	\frac{f(0)}{\pi^2} \int_M |C_{ijj}\sfer_{ij}|^2|C_{jjk}\sfer_{jk}|^2\sqrt{g}\mathrm{d}^4x.\label{eq:2bb2s-different}
\ea 
The paths corresponding to these contributions are depicted in Figure \ref{fig:2bb2s}.

However, to write all four-scalar interactions from the spectral action off shell in terms of the auxiliary fields $G_{i,j,k}$, one requires interactions of the form of both \eqref{eq:2bb2s-same} and \eqref{eq:2bb2s-different} to be present. The reason for this is the following. Upon writing the four-scalar part of the action of the building blocks \B{ij} and \B{jk} in terms of the auxiliary fields as in Lemma \ref{lem:bb2-offshell}, we find for the terms with $G_j$ in particular:
\bas
	- \frac{1}{2n_j}\tr_{N_j} G_j^2 - \tr_{N_j} G_j\Big(\asfer_{ij}\P_{j,i}'\sfer_{ij}\Big) - \tr_{N_j} G_j\Big(\P_{j,k}'\sfer_{jk}\asfer_{jk}\Big).
\eas
On shell, the cross terms of this expression then give the additional four-scalar interaction
\ba\label{eq:2bb2-different-aux}
	&  n_j|\P_{j,i}'^{1/2}\sfer_{ij}\P_{j,k}'^{1/2}\sfer_{jk}|^2 - \frac{n_j}{N_j}|\P_{j,i}'^{1/2}\sfer_{ij}|^2|\P_{j,i}'^{1/2}\sfer_{jk}|^2.
\ea
When the scaled counterparts \eqref{eq:bb2-scalingG} of $\P_{j,i}'$ and $\P_{j,k}'$ satisfy the constraints \eqref{eq:bb2-resultCiij} for supersymmetry, this interaction reads
\bas
	&n_jg_j^2\Big( |\sfer_{ij}\sfer_{jk}|^2 - \frac{1}{N_j}|\sfer_{ij}|^2|\sfer_{jk}|^2\Big)
\eas
after scaling the fields. When having two or more building blocks of the second type that share one of their indices, we have either \eqref{eq:2bb2s-same} or \eqref{eq:2bb2s-different} in the spectral action, while we need \eqref{eq:2bb2-different-aux} for a supersymmetric action. To possibly restore supersymmetry we need additional interactions, such as those of the next section.

\section{Third building block: extra interactions}\label{sec:bb3}

\begin{figure}
	\begin{center}
		\def\svgwidth{.4\textwidth}
		\subimport{\the\svgpath}{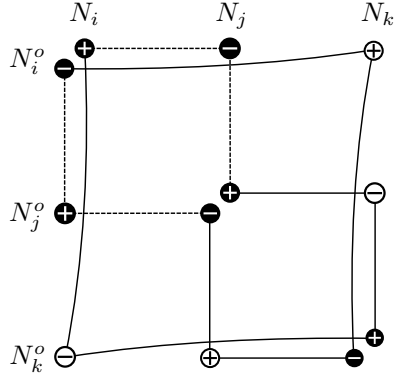}	
		\captionsetup{width=.9\textwidth}
	\caption{A situation in which all three building blocks of the second type are present whose two indices are either $i$, $j$ or $k$.} 
	\label{fig:3bb2}	
\end{center}
\end{figure}	

In a situation in which the finite algebra has three components and there are two adjacent building blocks of the second type, as depicted in Figure \ref{fig:2bb2s-same}, there is allowed a component 
		\begin{align}\label{eq:bb3-possible-comp}
		 	D_{ij}^{\phantom{ij}kj} : \rep{k}{j} \to \rep{i}{j}
		\end{align}
of the finite Dirac operator. We parametrize it with $\yuks{i}{k}$, that acts (non-trivially) on family space. Such a component satisfies the first order condition and its inner fluctuations 
	\bas
		 \sum_n a_n [\D{ij}{kj}, b_n] = \sum_n (a_i)_{n}\Big(\yuks{i}{k} (b_k)_{n} - (b_i)_{n}\yuks{i}{k}\Big)
	\eas
generate a scalar $\sfer_{ik} \in \rep{i}{k}$. Since there is no corresponding fermion $\fer{ik}$ present, a necessary condition for restoring supersymmetry is the existence of a building block \Bc{ik}{\pm} of the second type. The component \eqref{eq:bb3-possible-comp} then gives ---amongst others--- an extra fermionic contribution 
\begin{align*}
	\inpr{J_M\afer{ij}}{\gamma^5 \yuks{i}{k}\sfer_{ik}\afer{jk}}
\end{align*}
to the action. 
Using the transformations \eqref{eq:susytransforms4} and \eqref{eq:susytransforms5}, under which a building block of the second type is supersymmetric, 
we infer that this new term spoils supersymmetry. To overcome this, we need to add two extra components
\begin{align*}
\D{jk}{ik} : \rep{i}{k} & \to \rep{j}{k}, &\D{ij}{ik} : \rep{i}{k} &\to \rep{i}{j}
\end{align*}
to the finite Dirac operator, as well as their adjoints and the components that can be obtained by demanding that $[D_F,J_F] = 0$. We parametrize these two components with $\yuks{i}{j}$ and $\yuks{j}{k}$ respectively. They give extra contributions to the fermionic action that are of the form
\begin{align*}
	 \inpr{J_M\afer{jk}}{\gamma^5 \asfer_{ij}\yuks{i}{j}\fer{ik}} + \inpr{J_M\afer{ij}}{\gamma^5 \fer{ik}\asfer_{jk}\yuks{j}{k}}.
\end{align*}
Both components require the representation \rep{i}{k} to have an eigenvalue of $\gamma_F$ that is opposite to those of \rep{i}{j} and \rep{j}{k}. This is the situation as is depicted in Figure \ref{fig:3bb2}. 

\begin{figure}[ht]
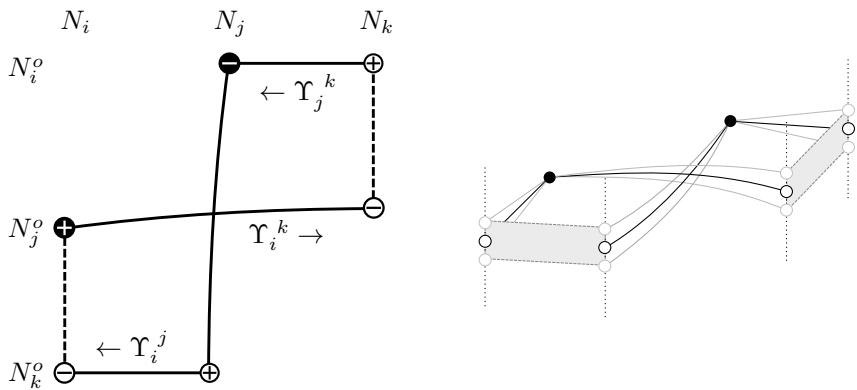

\centering
\begin{subfigure}{.4\textwidth}
\centering
		\def\svgwidth{\textwidth}
		\subimport{\the\svgpath}{bb3.pdf_tex}	
	\caption{For clarity we have omitted here the edges and vertices that stem from the building blocks of the first and second type.} 
	\label{fig:bb3-kraj}
\end{subfigure}
\hspace{30pt}
\begin{subfigure}{.4\textwidth}
		\centering
		\def\svgwidth{\textwidth}
		\subimport{\the\svgpath}{kraj_diagram_generations_bb3.pdf_tex}
		\caption{The same building block as shown on the left side but with the possible family structure of the two scalar fields with $R = 1$ being visualized.}
		\label{fig:bb3-gen}
\end{subfigure}
	\caption{A building block \B{ijk} of the third type in the language of Krajewski diagrams.}
	\label{fig:bb3}
\end{figure}

This brings us to the following definition:
\begin{defin}\label{def:bb3}
For an almost-commutative geometry in which \Bc{ij}{\pm}, \Bc{ik}{\mp} and \Bc{jk}{\pm} are present, a \emph{building block of the third type} \B{ijk} is the collection of all allowed components of the Dirac operator, mapping between the three representations \rep{i}{j}, \rep{i}{k} and \rep{k}{j} and their conjugates. Symbolically it is denoted by
\begin{align}\label{eq:bb3-comps}
	\BBB{ijk} = (0, \D{ij}{kj} + \D{jk}{ik} + \D{ij}{ik}) \in \H_F \oplus \End(\H_F).
\end{align}
\end{defin}
The Krajewski diagram corresponding to \B{ijk} is depicted in Figure \ref{fig:bb3}.

The parameters of \eqref{eq:bb3-comps} are chosen such that the sfermions $\sfer_{ij}$ and $\sfer_{jk}$ are generated by the inner fluctuations of \yuk{i}{j} and \yuk{j}{k} respectively, whereas $\sfer_{ik}$ is generated by $\yuks{i}{k}$. This is because $\sfer_{ik}$ crosses the particle/antiparticle-diagonal in the Krajewski diagram. Note that $i$, $j$, $k$ are labels, not matrix indices.

\begin{figure}[ht]
	\begin{center}
		\captionsetup{width=.9\textwidth}
		\includegraphics[width=.8\textwidth]{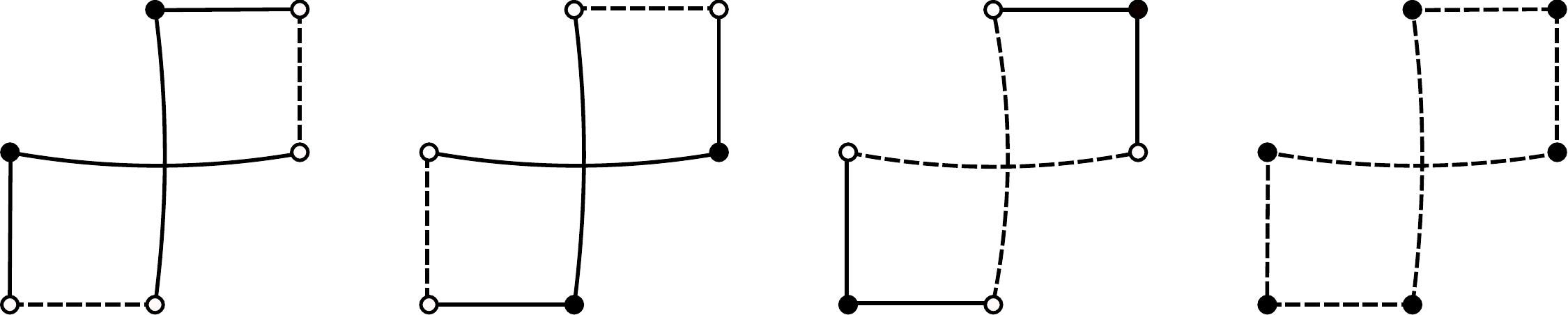}
	\caption{All possible combinations of values for the $R$-parity operator in a building block of the third type. Three of those possibilities have one representation on which $R = -1$, in the other possibility all three of them have $R = -1$. This last option essentially entails having no family structure.} 
	\label{fig:bb3Pos}	
\end{center}
\end{figure}

There are several possible values of $R$ that the vertices and edges can have. Requiring a grading that yields $-1$ on each of the diagonal vertices, all possibilities for an explicit construction of $R \in \A_F \otimes \A_F^o$ are given by $R = - P\otimes P^o$, $P = (\pm 1, \pm 1, \pm 1) \in \A_F$ where each of the three signs can vary independently. This yields 8 possibilities, but each of them appears in fact twice. Of the effectively four remaining combinations, three have one off-diagonal vertex that has $R = -1$ and in the other combination all three off-diagonal vertices have $R = -1$. These four possibilities are depicted in Figure \ref{fig:bb3Pos}. We will typically work in the case of the first image of Figure \ref{fig:bb3Pos}, as is visualised in Figure \ref{fig:bb3-gen}, and will indicate where changes might occur when working in one of the other possibilities. If in this context the $R = 1$ representations in $\H_F$ come in $M$ copies (`generations'), all components of the finite Dirac operator are in general acting non-trivially on these $M$ copies, except $C_{iij}$ and $C_{ijj}$, since they parametrize components of the finite Dirac operator mapping between $R = -1$ representations.  

Note that in the action the expressions \eqref{eq:exprN} for the pre-factors $\n_{ij}^2$, $\n_{ik}^2$ and $\n_{jk}^2$ of the sfermion kinetic terms all get an extra contribution from the new edges of the Krajewski diagram of Figure~\ref{fig:bb3}. The first of these becomes
\ba
	\n_{ij}^2 & \to\frac{f(0)}{2\pi^2}(N_iC_{iij}^*C_{iij} + N_jC_{ijj}^*C_{ijj} + N_k\yuks{i}{j}\yuk{i}{j})\label{eq:kin_term_bb3}.
\ea
The other two can be obtained replacing $N_i$, $C_{iij}$, $C_{ijj}$ and $\yuk{i}{j}$ by their respective analogues. 

The presence of a building block of the third type allows us to take a specific parametrization of the $C_{iij}$ in terms of $\yuk{i}{j}$. To this end, we introduce the shorthand notations
\ba
	q_i &:= \frac{f(0)}{\pi^2}g_i^2, & r_i &:= q_in_i, & \w{ij} &:= 1 - r_iN_i - r_jN_j,\label{eq:kintermnorm}
\ea
where we can infer from the normalization of the kinetic terms of the gauge bosons (i.e.~setting $\K_i = 1$) that $q_i$ must be rational. Then, similarly as in Proposition \ref{lem:bb2-nosol}, we write out $C_{iij}^*C_{iij}$, with $C_{iij}$ satisfying \eqref{eq:bb2-resultCiij} from supersymmetry, and insert the pre-factor \eqref{eq:kin_term_bb3} of the kinetic term. This reads 
\bas
C_{iij}^*C_{iij} &= r_i\big(N_i C_{iij}^*C_{iij} + N_jC_{ijj}^*C_{ijj} + N_k\yuks{i}{j}\yuk{i}{j}\big).
\eas
Using $r_i C_{ijj}^*C_{ijj} = r_jC_{iij}^*C_{iij}$, which can be directly obtained from the result \eqref{eq:bb2-resultCiij}, we obtain
\ba
	C_{iij}^*C_{iij} &= \frac{r_i}{\w{ij}}N_k\yuks{i}{j}\yuk{i}{j}
\ea
for the parametrization of $C_{iij}$ that satisfies \eqref{eq:bb2-resultCiij}. For future convenience we will take 
\ba\label{eq:bb3-expressionCs-sqrt}
	C_{iij} = \sgnc_{i,j} \sqrt{\frac{r_i}{\w{ij}}}\big(N_k \yuks{i}{j}\yuk{i}{j}\big)^{1/2},
\ea
with $\sgnc_{i,j} \in \{\pm\}$ the sign introduced in Theorem \ref{prop:bb2}. The other parameter, $C_{ijj}$, can be obtained by $r_i \to r_j$, $\sgnc_{i,j} \to \sgnc_{j,i}$. This yields for the pre-factor \eqref{eq:kin_term_bb3} of the kinetic term of $\sfer_{ij}$:
\ba
\n_{ij}^2&= \frac{f(0)}{2\pi^2}\bigg(N_i\frac{r_i}{\w{ij}} + N_j\frac{r_j}{\w{ij}} + 1\bigg) N_k\yuks{i}{j}\yuk{i}{j} = \frac{f(0)}{2\pi^2}\frac{1}{\w{ij}}N_k\yuks{i}{j}\yuk{i}{j}. \label{eq:bb3-exprN}
\ea
prior to the scaling \eqref{eq:bb3-scalingfields}. When $\sfer_{ij}$ has $R = 1$ and therefore does not carry a family structure (as in Figure \eqref{fig:bb3-gen}) then the trace over the representations where $\sfer_{ij}\asfer_{ij}$ and $\asfer_{ij}\sfer_{ij}$ are in, decouples from that over $M_{M}(\com)$. Consequently, the third term in \eqref{eq:kin_term_bb3} and the right hand sides of the solutions \eqref{eq:bb3-expressionCs-sqrt} and \eqref{eq:bb3-exprN} receive additional traces over family indices, i.e.~$N_k \yuks{i}{j}\yuk{i}{j} \to N_k \tr_M \yuks{i}{j}\yuk{i}{j}$. The strategy to write $C_{iij}$ in terms of parameters of building blocks of the third type works equally well when the kinetic term of $\sfer_{ij}$ gets contributions from multiple building blocks of the third type. In that case $N_k\yuks{i}{j}\yuk{i}{j}$ must be replaced by a sum of all such terms: $\sum_l N_l\yuks{i,l}{j}\yuk{i,l}{j}$ (see e.g.~Section \ref{sec:2bb3}), where the label $l$ is used to distinguish the building blocks \B{ijl} that all give a contribution to the kinetic term of $\sfer_{ij}$. 

There are several contributions to the action as a result of adding a building block of the third type. 
The action is given by
\begin{align}
	S_{ijk}[\zeta, \szeta] = S_{f, ijk}[\zeta, \szeta] + S_{b,ijk}[\szeta], \label{eq:bb3action}
\end{align}
with its fermionic part $S_{f, ijk}[\zeta, \szeta]$ reading
\begin{align}
	S_{f, ijk}[\zeta, \szeta] &= \inpr{J_M \afer{ij}}{\gamma^5\fer{ik}\asfer_{jk}\yuks{j}{k}} + \inpr{J_M \afer{ij}}{\gamma^5\yuks{i}{k}\sfer_{ik}\afer{jk}} \nn\\ 
&\qquad + \inpr{J_M\afer{jk}}{\gamma^5\asfer_{ij}\yuks{i}{j}\fer{ik}} + \inpr{J_M\afer{ik}}{\gamma^5\yuk{i}{j}\sfer_{ij}\fer{jk}} \label{eq:bb3-action-ferm}\\ 
&\qquad\qquad + \inpr{J_M\afer{ik}}{\gamma^5\fer{ij}\yuk{j}{k}\sfer_{jk}} + \inpr{J_M\fer{jk}}{\gamma^5\asfer_{ik}\yuk{i}{k}\fer{ij}}.\nn
\end{align}
 The bosonic part of the action is given by:
\ba
	S_{b,ijk}[\szeta]	&=	\frac{f(0)}{2\pi^2} \Big[N_i|\yuk{j}{k}\sfer_{jk}\asfer_{jk}\yuks{j}{k}|^2 + N_j|\yuks{i}{k}\sfer_{ik}\asfer_{ik}\yuk{i}{k}|^2 \nn\\
		&\qquad +  N_k\tr_M(\yuks{i}{j}\yuk{i}{j})^2|\sfer_{ij}\asfer_{ij}|^2\Big] \nn\\
			&\qquad\qquad + S_{b,ij,jk}[\szeta] + S_{b,ik,jk}[\szeta] + S_{b,ij,ik}[\szeta], \label{eq:bb3-boson-action}
\ea
with
\ba
		&S_{b,ij,jk}[\szeta]\nn\\
	&=	\frac{f(0)}{\pi^2} \Big[N_i |C_{iij}\sfer_{ij}\yuk{j}{k}\sfer_{jk}|^2 + N_k |\yuk{i}{j}\sfer_{ij}C_{jkk}\sfer_{jk}|^2 + |\sfer_{ij}|^2|\yuks{i}{j}\yuk{j}{k}\sfer_{jk}|^2\nn\\
		&\qquad\qquad	+ \Big(\tr \asfer_{jk}\yuks{j}{k}(\asfer_{ij}C_{iij}^*)^o(C_{iij}\sfer_{ij})^o \yuk{j}{k}\sfer_{jk} \nn\\
		&\qquad\qquad\qquad + \tr\asfer_{jk}C_{jjk}^*(\asfer_{ij}\yuks{i}{j})^o(\yuk{i}{j}\sfer_{ij})^oC_{jjk}\sfer_{jk}\nn\\
	&\qquad\qquad\qquad\qquad + \tr\asfer_{jk}\yuks{j}{k}(\asfer_{ij}C_{ijj}^*)^o(\yuk{i}{j}\sfer_{ij})^oC_{jjk}\sfer_{jk} + h.c.\Big)\Big],\label{eq:bb3-boson-action-part}
\ea
where the traces above are over $(\rep{k}{i}{})^{\oplus M}$. The fact that in this context $\sfer_{ij}$ has $R = 1$ makes it possible to separate the trace over the family-index in the last term of the first line of \eqref{eq:bb3-boson-action}. A more detailed derivation of the four-scalar action that corresponds to a building block of the third type, including the expressions for $S_{b,ik,jk}[\szeta]$ and $S_{b,ij,ik}[\szeta]$, is given in Appendix \ref{sec:bb3-calc-action}. 

The expression \eqref{eq:bb3-boson-action} contains interactions that in form we either have seen earlier (cf.~\eqref{eq:exprM1}, \eqref{eq:2bb2s-same}) or that we needed but were lacking in a set up consisting only of building blocks of the second type (cf.~\eqref{eq:2bb2s-different}, see also the discussion in Section \ref{sec:2bb2}). In addition, it features terms that we need in order to have a supersymmetric action. 

We can deduce from the transformations \eqref{eq:susytransforms4} that, for the expression \eqref{eq:bb3-action-ferm} (i.e.~the fermionic action that we have) to be part of a supersymmetric action, the bosonic action must involve terms with the auxiliary fields $F_{ij}$, $F_{ik}$ and $F_{jk}$ (that are available to us from the respective building blocks of the second type), coupled to two scalar fields. We will therefore formulate the most general action featuring these auxiliary fields and constrain its coefficients by demanding it to be supersymmetric in combination with \eqref{eq:bb3-action-ferm}. Subsequently, we will check if and when the spectral action \eqref{eq:bb3-boson-action} (after subtracting the terms that are needed for \eqref{eq:2bb2s-different}) is of the correct form to be written off shell in such a general form. This will be done for the general case in Section \ref{sec:4s-aux}.

The most general Lagrangian featuring the auxiliary fields $F_{ij}$, $F_{ik}$, $F_{jk}$ that can yield four-scalar terms is
\begin{align}
	S_{b, ijk, \mathrm{off}}[F_{ij}, F_{ik}, F_{jk}, \szeta] &= \int_M \mathcal{L}_{b, ijk, \mathrm{off}}(F_{ij}, F_{ik}, F_{jk}, \szeta)\sqrt{g}\mathrm{d}^4x, \label{eq:bb3-auxfields}
\end{align}
with
\bas
	&\mathcal{L}_{b, ijk, \mathrm{off}}(F_{ij}, F_{ik}, F_{jk}, \szeta) \nn\\&\qquad =- \tr F_{ij}^*F_{ij} + \big(\tr F_{ij}^*\beta_{ij,k}\sfer_{ik}\asfer_{jk} + h.c.\big) \nn\\
	&\qquad\qquad\qquad - \tr F_{ik}^*F_{ik} + \big(\tr F_{ik}^*\beta_{ik,j}^*\sfer_{ij}\sfer_{jk} + h.c.\big)\nn\\
			&\qquad\qquad\qquad\qquad - \tr F_{jk}^*F_{jk} + \big(\tr F_{jk}^*\beta_{jk,i}\asfer_{ij}\sfer_{ik} + h.c.\big)\nn.
\eas
Here $\beta_{ij,k}$, $\beta_{ik,j}$ and $\beta_{jk,i}$ are matrices acting on the generations and consequently the traces are performed over $\srep{j}^{\oplus M}$ (the first two terms) and $\srep{k}^{\oplus M}$ (the last four terms) respectively. Using the Euler-Lagrange equations the on shell counterpart of \eqref{eq:bb3-auxfields} is seen to be
	\begin{align*}
		S_{b, ijk, \mathrm{on}}[\szeta] = \int_M \sqrt{g}\mathrm{d}^4x\Big(|\beta_{ij,k}\sfer_{ik}\asfer_{jk}|^2 + |\beta_{ik,j}^*\sfer_{ij}\sfer_{jk}|^2 + |\beta_{jk,i}\asfer_{ij}\sfer_{ik}|^2\Big)
	\end{align*} 
	cf.~the second and third terms of \eqref{eq:bb3-boson-action}. We have the following result:


\begin{theorem}\label{thm:bb3}
	The action consisting of the sum of \eqref{eq:bb3-action-ferm} and \eqref{eq:bb3-auxfields} is supersymmetric under the transformations \eqref{eq:susytransforms4} and \eqref{eq:susytransforms5} if and only if the parameters of the finite Dirac operator are related via
\begin{align}
\yuk{j}{k}C_{jkk}^{-1} &= - (C_{ikk}^*)^{-1}\yuk{i}{k}, &
(C_{iik}^*)^{-1}\yuk{i}{k} &= - \yuk{i}{j}C_{iij}^{-1}, \nn\\
\yuk{i}{j}C_{ijj}^{-1} &= - \yuk{j}{k}C_{jjk}^{-1}.&&\label{eq:improvedUpsilons}
\end{align}
and
\ba\label{eq:bb3-susy-demand2}
\bps_{ij,k}\bp_{ij,k} &= \yukp{j}{k}\yukps{j}{k}= \yukp{i}{k}\yukps{i}{k}, &
\bps_{ik,j}\bp_{ik,j} &= \yukp{i}{j}\yukps{i}{j}= \yukp{j}{k}\yukps{j}{k}, \nn\\
\bps_{jk,i}\bp_{jk,i} &= \yukp{i}{k}\yukps{i}{k}= \yukp{i}{j}\yukps{i}{j}, &
\ea
where 
\bas
 \bp_{ij,k} &:= \n_{jk}^{-1}\beta_{ij,k}\n_{ik}^{-1},& \bp_{ik,j} &:= \n_{jk}^{-1} \beta_{ik,j} \n_{ij}^{-1},& \bp_{jk,i} &:= \n_{ij}^{-1}\beta_{jk,i}\n_{ik}^{-1}
\eas
and
\ba\label{eq:def-yukp}
	\yukp{i}{j} &:= \yuk{i}{j}\n_{ij}^{-1},& 	
	\yukp{i}{k} &:= \n_{ik}^{-1}\yuk{i}{k},& 	
	\yukp{j}{k} &:= \yuk{j}{k}\n_{jk}^{-1}, 	
\ea
denote the scaled versions of the $\beta_{ij,k}$'s and the $\yuk{i}{j}$'s respectively.
\end{theorem}
\begin{proof}
See Appendix \ref{sec:bb3-proof}.
\end{proof}

For future use we rewrite \eqref{eq:improvedUpsilons} using the parametrization \eqref{eq:bb3-expressionCs-sqrt} for the $C_{iij}$, giving 
\ba
	\sgnc_{i,j} \sqrt{\w{ij}}\,\yukw{i}{j} &= - \sgnc_{i,k}\sqrt{\w{ik}}\,\yukw{i}{k}, &
	\sgnc_{j,i} \sqrt{\w{ij}}\,\yukw{i}{j} &= - \sgnc_{j,k}\sqrt{\w{jk}}\,\yukw{j}{k}, \nn\\
	\sgnc_{k,i} \sqrt{\w{ik}}\,\yukw{i}{k} &= - \sgnc_{k,j}\sqrt{\w{jk}}\,\yukw{j}{k}, &&
\label{eq:improvedUpsilons1}
\ea
where we have written
\ba\label{eq:def-yukw}
	\yukw{i}{j} &:= \yuk{i}{j}(N_k\tr\yuks{i}{j}\yuk{i}{j})^{-1/2}, & \yukw{i}{k} &:= (N_j\yuk{i}{k}\yuks{i}{k})^{-1/2}\yuk{i}{k},\nn\\
	 \yukw{j}{k} &:= \yuk{j}{k}(N_i\yuks{j}{k}\yuk{j}{k})^{-1/2}.&&
\ea
There is a trace over the generations in the first term because the corresponding sfermion $\sfer_{ij}$ has $R = 1$ and consequently no family-index. Using these demands on the parameters, the (spectral) action from a building block of the third type becomes much more succinct. First of all it allows us to reduce all three parameters of the finite Dirac operator of Definition \ref{def:bb3} to only one, e.g.~$\yuk{}{} \equiv \yuk{i}{j}$. Second, upon using \eqref{eq:improvedUpsilons} the second and third lines of \eqref{eq:bb3-boson-action-part} are seen to cancel.\footnote{More generally, this also happens for the other combinations: the four-scalar interactions of \eqref{eq:bb3-action-2} are seen to cancel those of \eqref{eq:bb3-action-5}}
If the demands \eqref{eq:improvedUpsilons} and \eqref{eq:bb3-susy-demand2} are met, the on shell action \eqref{eq:bb3action} that arises from a building block \B{ijk} of the third type reads
\ba
S_{ijk}[\zeta, \szeta, \mathbb{A}] = g_m\sqrt{\frac{2\w{1}}{q_m}}&\Big[\inpr{J_M\afer{2}}{\gamma^5\yukw{}{}\sfer_{1}\fer{3}} + \kappa_{j}
\inpr{J_M\afer{2}}{\gamma^5\fer{1}\yukw{}{}\sfer_{3}} \nn\\
	&\qquad  + \kappa_{i}\inpr{J_M\fer{3}}{\gamma^5\asfer_{2}\yukw{}{}\fer{1}} + h.c.\Big]\nn\\
	 + g_m^2\frac{4\w{1}}{q_m}&\Big[(1 - \w{2}) |\yukw{}{}\sfer_{1}\sfer_{3}|^2 
		+ (1 - \w{1}) |\yukw{}{}\sfer_{3}\asfer_{2}|^2 \nn\\
	&\qquad	+ (1 - \w{3}) |\yukw{}{}\asfer_{2}\sfer_{1}|^2\Big].\label{eq:bb3-action-final}
\ea
Here we used the shorthand notations $ij \to 1$, $ik \to 2$, $jk \to 3$ and $\kappa_{j} = \sgnc_{j,i}\sgnc_{j,k}$, $\kappa_{i} = \sgnc_{i,j}\sgnc_{i,k}$ to avoid notational clutter as much as possible and where we have written everything in terms of $\yukw{}{} \equiv \yukw{i}{j}$ (as defined above), the parameter that corresponds to the sfermion having $R = 1$ (and consequently also multiplicity $1$). The index $m$ in $g_m$ and $q_m$ can take any of the values that appear in the model, e.g.~$i$, $j$ or $k$. As with a building block of the second type there is a sign ambiguity that stems from those of the $C_{iij}$. In addition, the terms that are not listed here but are in \eqref{eq:bb3action} give contributions to terms that already appeared in the action from building blocks of the second type. See Section \ref{sec:4s-aux} for details on this.

For notational convenience we have used two different notations for scaled variables: $\yukw{i}{j}$ from \eqref{eq:def-yukw} and $\yukp{i}{j}$ from \eqref{eq:def-yukp}. Using the expression \eqref{eq:bb3-exprN} for $\n_{ij}$ in terms of $\yuk{i}{j}$ these are related via
\ba\label{eq:yukwyukp}
	\yukp{i}{k} \equiv \n_{ik}^{-1}\yuk{i}{k} = \sqrt{\frac{2\pi^2}{f(0)}\w{ik}}\big(N_j \yuk{i}{k}\yuks{i}{k}\big)^{-1/2}\yuk{i}{k} \equiv g_l\sqrt{\frac{2\w{ik}}{q_l}}\yukw{i}{k},
\ea
assuming that $\sfer_{ik}$ has $R = -1$. The other two scaled variables give analogous expressions but the order of $\yuk{}{}$ and $\yuks{}{}$ is reversed and the sfermion with $R = 1$ gets an additional trace over family indices.

\begin{rmk}\label{rmk:bb3-relativesigns}
Note that we can use this result to say something about the signs of the $C_{iij}$ appearing in a building block of the third type. We first combine all three equations of \eqref{eq:improvedUpsilons} into one,
\bas
	\yuk{j}{k} = (-1)^3 (C_{iik}C_{ikk}^{-1})^{*} \yuk{j}{k}(C_{jjk}^{-1}C_{jkk})(C_{ijj}C_{iij}^{-1}),
\eas
when it is $C_{iij}$ and $C_{ijj}$ that do not have a family structure. All these parameters are only determined up to a sign. We will write 
\bas
	C_{iij}C_{ijj}^{-1} = s_{ij} \sqrt{\frac{n_i\K_j}{n_j\K_i}} \frac{g_i}{g_j},\qquad\text{with } s_{ij} := \sgnc_{i,j}\sgnc_{j,i} = \pm 1,
\eas
cf.~\eqref{eq:bb2-resultCiij}, etc.~which gives  
$
	\yuk{j}{k} = - s_{ij}s_{jk}s_{ki} \yuk{j}{k}
$
 for the relation above. So for consistency either one, or all three combinations of $C_{iij}$ and $C_{ijj}$ associated to a building block \B{ij} that is part of a \B{ijk} must be of opposite sign.
\end{rmk}

\begin{rmk}\label{rmk:bb3-R=1}
	If instead of $\sfer_{ij}$ it is $\sfer_{ik}$ or $\sfer_{jk}$ that has $R = 1$ (see Figure \ref{fig:bb3Pos}) the demand on the parameters $\yuk{i}{j}$, $\yuk{i}{k}$ and $\yuk{j}{k}$ is a slightly modified version of \eqref{eq:improvedUpsilons}: 
\ba
(\yuk{j}{k}C_{jkk}^{-1})^t &= - (C_{ikk}^*)^{-1}\yuk{i}{k}, &
(C_{iik}^*)^{-1}\yuk{i}{k} &= - \yuk{i}{j}C_{iij}^{-1}, \nn\\
\yuk{i}{j}C_{ijj}^{-1} &= - (\yuk{j}{k}C_{jjk}^{-1})^t,&&\label{eq:improvedUpsilons2}
\ea
where $A^t$ denotes the transpose of the matrix $A$. This result can be verified by considering Lemma \ref{lem:bb3-lem1} for these cases.
\end{rmk}

By introducing a building block of the third type we generated the interactions that we lacked in a situation with multiple building blocks of the second type. The wish for supersymmetry thus forces us to extend any model given by Figure \ref{fig:3bb2} with a building block of the third type.

\begin{figure}[ht]
\centering
	\def\svgwidth{.4\textwidth}
	\subimport{\the\svgpath}{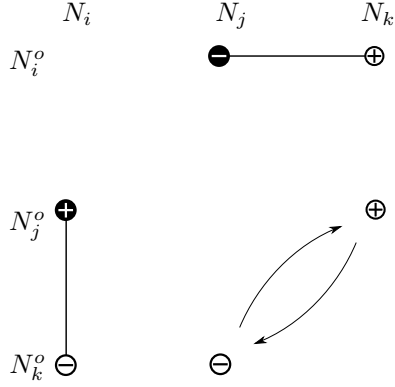}	
	\captionsetup{width=.6\textwidth}
	\caption{A set up similar to that of Figure \ref{fig:bb3}, but with the values of the grading reversed for $\rep{j}{k}$ and its opposite. Consequently, only one of the three components that characterize a building block of the first type can now be defined.}
	\label{fig:bb3_opposite_grading}
\end{figure}

If we again seek the analogy with the superfield formalism, then a building block of the third type is a Euclidean analogy of an action on a Minkowskian background that comes from a superpotential term
\ba\label{eq:bb3-superpot}
	\int \Big(\mathcal{W}(\{\Phi_m\})\Big|_F + h.c.\Big)\mathrm{d}^4 x,\quad \text{with}\quad \mathcal{W}(\{\Phi_m\}) = f_{mnp} \Phi_m\Phi_n \Phi_p,
\ea
where $\Phi_{m,n,p}$ are chiral superfields, $f_{mnp}$ is symmetric in its indices \cite[\S 5.1]{DGR04} and with $|_{F}$ we mean multiplying by $\bar\theta\bar\theta$ and integrating over superspace $\int \mathrm{d}^2\theta\mathrm{d}^2\bar\theta$. To specify this statement, we write $\Phi_{ij} = \phi_{ij} + \sqrt{2}\theta \psi_{ij} + \theta\theta F_{ij}$ for a chiral superfield. Similarly, we introduce $\Phi_{jk}$ and $\Phi_{ki}$. We then have that
\bas
&	\int_M \Big[\Phi_{ij} \Phi_{jk}\Phi_{ki}  \Big]_F + h.c. \nn\\
&\qquad =\int_M  -  \psi_{ij}\phi_{jk}\psi_{ki} -  \psi_{ij}\psi_{jk}\phi_{ki}-  \phi_{ij}\psi_{jk}\psi_{ki}\nn\\
	&\qquad\qquad\qquad + F_{ij}\phi_{jk}\phi_{ki} + \phi_{ij}\phi_{jk}F_{ki} + \phi_{ij}F_{jk}\phi_{ki} + h.c. 
\eas
This gives on shell the following contribution\footnote{On a Minkowskian background the product of a superfield and its conjugate appears in the action as $F_{ij}^*F_{ij}$, i.e.~with pre-factor $+1$ \cite[\S 4.3]{DGR04}, in contrast to \eqref{eq:bb2-auxfields}.}:
\bas
& - \int_M \Big(\psi_{ij}\phi_{jk}\psi_{ki} +  \psi_{ij}\psi_{jk}\phi_{ki}+  \phi_{ij}\psi_{jk}\psi_{ki} \nn\\
		&\qquad\qquad + \frac{1}{2}|\phi_{jk}\phi_{ki}|^2 + \frac{1}{2}|\phi_{ij}\phi_{jk}|^2 + \frac{1}{2}|\phi_{ki}\phi_{ij}|^2 + h.c.\Big), 
\eas
to be compared with \eqref{eq:bb3-action-final}. In a set up similar to that of Figure \ref{fig:3bb2}, but with the chirality of one or two of the building blocks \B{ij}, \B{jk} and \B{ik} being flipped, not all three components of $D_F$ such as in Definition \ref{def:bb3} can still be defined, see Figure \ref{fig:bb3_opposite_grading}. Interestingly, one can check that in such a case the resulting action corresponds to a superpotential that is not holomorphic, but e.g.~of the form $\Phi_{ij} \Phi_{ik}\Phi_{jk}^\dagger$ instead. To see this, we calculate the action \eqref{eq:bb3-superpot} in this case, giving 
\bas
	\int_M \Big[\Phi_{ij} \Phi_{jk}^\dagger  \Phi_{ki}\Big]_F + h.c. 
&=  \int_M -  \psi_{ij}\phi_{jk}^*\psi_{ki} + F_{ij}\phi_{jk}^*\phi_{ki} + \phi_{ij}\phi_{jk}^*F_{ki} + h.c., 
\eas
which on shell equals
\bas
- \int_M \psi_{ij} \phi_{jk}^* \psi_{ki} + \frac{1}{2}|\phi_{jk}^*\phi_{ki}|^2 + \frac{1}{2}|\phi_{ij}\phi_{jk}^*|^2 + h.c.
\eas
This is indeed analogous to the interactions that the spectral triple depicted in Figure \ref{fig:bb3_opposite_grading} (still) gives rise to.

\subsection{Interaction between building blocks of the third type}\label{sec:2bb3}
\begin{figure}
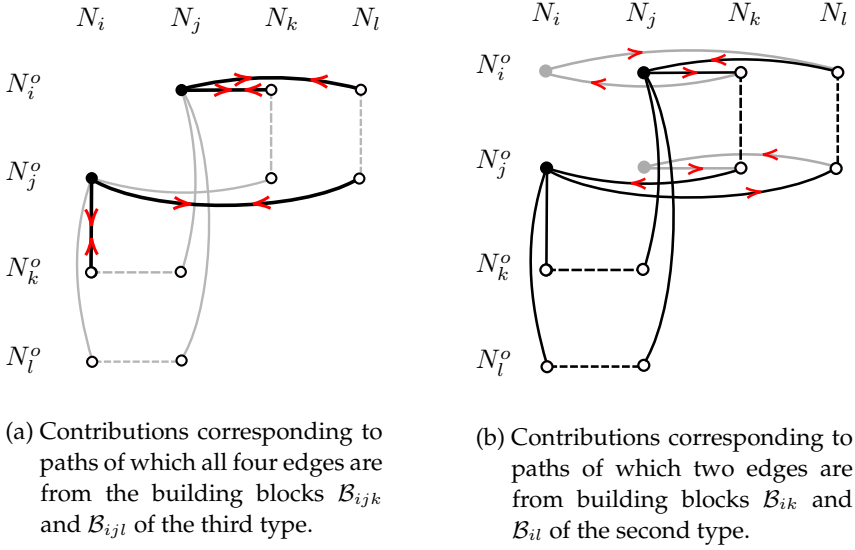

	\centering
	\begin{subfigure}{.4\textwidth}
		\centering
		\def\svgwidth{\textwidth}
		\subimport{\the\svgpath}{2bb3s-inner.pdf_tex}
		\caption{Contributions corresponding to paths of which all four edges are from the building blocks \B{ijk} and \B{ijl} of the third type.}
		\label{fig:2bb3s-inner}
	\end{subfigure}
	\hspace{30pt}
	\begin{subfigure}{.4\textwidth}
		\centering
		\def\svgwidth{\textwidth}
		\subimport{\the\svgpath}{2bb3s-outer.pdf_tex}	
		\caption{Contributions corresponding to paths of which two edges are from building blocks \B{ik} and \B{il} of the second type.}
		\label{fig:2bb3s-outer}
	\end{subfigure}
\caption{In the case that there are two building blocks of the third type sharing two of their indices, there are extra four-scalar contributions to the action. They are given by \eqref{eq:2bb3-action}.}
\label{fig:2bb3s}
\end{figure}

Suppose we have two building blocks \B{ijk} and \B{ijl} of the third type that share two of their indices, as is depicted in Figure \ref{fig:2bb3s}. This situation gives rise to the following extra terms in the action:
\ba
	&\frac{f(0)}{\pi^2}\Big[N_j|\asfer_{jk}C_{jjk}^*C_{jjl}\sfer_{jl}|^2 + N_i|\asfer_{jk}\yuks{j}{k}\yuk{j}{l}\sfer_{jl}|^2 \nn\\
	&\qquad\qquad\qquad + |\yuk{j}{k}\sfer_{jk}|^2|\yuks{i}{l}\sfer_{il}|^2\Big] + (i \leftrightarrow j)\nn\\
	&\qquad + \frac{f(0)}{\pi^2}\Big(N_i  \tr C_{iik}\sfer_{ik}\asfer_{jk}\yuks{j}{k}\yuk{j}{l}\sfer_{jl}\asfer_{il}C_{iil}^* \nn\\
	 &\qquad\qquad\qquad\qquad +  N_j\tr \yuks{i}{k}\sfer_{ik}\asfer_{jk}C_{jjk}^*C_{jjl}\sfer_{jl}\asfer_{il}\yuk{i}{l} + h.c.\Big),\label{eq:2bb3-action}
\ea
where with `$(i\leftrightarrow j)$' we mean the expression preceding it, but everywhere with $i$ and $j$ interchanged. The first line of \eqref{eq:2bb3-action} corresponds to paths within the two building blocks \B{ijk} and \B{ijl} (such as the ones depicted in Figure \ref{fig:2bb3s-inner}) and the second line corresponds to paths of which two of the edges come from the building blocks of the second type that were needed in order to define the building blocks of the third type (Figure \ref{fig:2bb3s-outer}).

If we scale the fields appearing in this expression according to \eqref{eq:bb3-scalingfields} and use the identity \eqref{eq:improvedUpsilons} for the parameters of a building block of the third type, we can write \eqref{eq:2bb3-action} more compactly as
\ba
&4n_jr_jN_jg_j^2|\asfer_{jk}\sfer_{jl}|^2 + 4\frac{g_m^2}{q_m}\w{ij}^2N_i|\asfer_{jk}\yukws{k}{}\yukw{l}{}\sfer_{jl}|^2 \nn\\
	&\qquad + 4\frac{g_m^2}{q_m}\w{ij}^2|\yukw{k}{}\sfer_{jk}|^2|\yukws{l}{}\sfer_{il}|^2 + (i \leftrightarrow j, \yuk{}{} \leftrightarrow \yuks{}{})\nn\\
	&\qquad\qquad + \kappa_{k}\kappa_{l}\, 4\frac{g_m^2}{q_m}(1 - \w{ij})\w{ij}\tr \yukw{l}{}\yukws{k}{}\sfer_{ik}\asfer_{jk}\sfer_{jl}\asfer_{il} + h.c.,\label{eq:2bb3-action-scaled}
\ea
where $\kappa_{k} = \sgnc_{k,i}\sgnc_{k,j}$, $\kappa_{l} = \sgnc_{l,i}\sgnc_{l,j} \in \{\pm 1\}$, $\yukw{k}{} \equiv \yukw{i,k}{j}$ of \B{ijk} and $\yukw{l}{} \equiv \yukw{i, l}{j}$ of \B{ijl}, as defined in \eqref{eq:def-yukw} but with contributions from two building blocks of the third type: 
\begin{subequations}
\ba
	\yukw{i,k}{j} &= \yuk{i,k}{j} (N_k \tr \yuks{i,k}{j}\yuk{i,k}{j} + N_l\tr\yuks{i,l}{j}\yuk{i,l}{j})^{-1/2}, \\
	\yukw{i,l}{j} &= \yuk{i,l}{j} (N_k \tr \yuks{i,k}{j}\yuk{i,k}{j} + N_l\tr\yuks{i,l}{j}\yuk{i,l}{j})^{-1/2}.\label{eq:def-yukw-mult}
\ea
\end{subequations}
This expression can be generalized to any number of building blocks of the third type. In addition, we have assumed that $s_{ik}s_{il} = s_{jk}s_{jl}$ for the products of the relative signs between the parameters $C_{iik}$ and $C_{ikk}$ etc.~(cf.~Remark \ref{rmk:bb3-relativesigns}). 

These new interactions must be accounted for by the auxiliary fields. The first and second terms are of the form \eqref{eq:2bb2s-same} and should therefore be covered by the auxiliary fields $G_{i,j}$. The third term is of the form \eqref{eq:2bb2s-different} and should consequently be described by the combination of $G_{i,j}$ and the $u(1)$-field $H$. The second line of \eqref{eq:2bb3-action} should be rewritten in terms of the auxiliary field $F_{ij}$. This can indeed be achieved via the off shell Lagrangian
%
%
\bas
-\tr F_{ij}^*F_{ij}	+ \big(\tr F_{ij}^*( \beta_{ij,k} \sfer_{ik}\asfer_{jk} + \beta_{ij,l}\sfer_{il}\asfer_{jl}) + h.c.\big),
\eas
which on shell gives the following cross terms:
\ba
 \tr \beta_{ij,l}^*\beta_{ij,k}\sfer_{ik}\asfer_{jk}\sfer_{jl}\asfer_{il} + h.c. \label{eq:2bb3-aux}
\ea
In form, this indeed corresponds to the second line of \eqref{eq:2bb3-action-scaled}. In Section \ref{sec:4s-aux} a more detailed version of this argument is presented. 

\begin{figure}
	\begin{center}
		\def\svgwidth{.5\textwidth}
		\subimport{\the\svgpath}{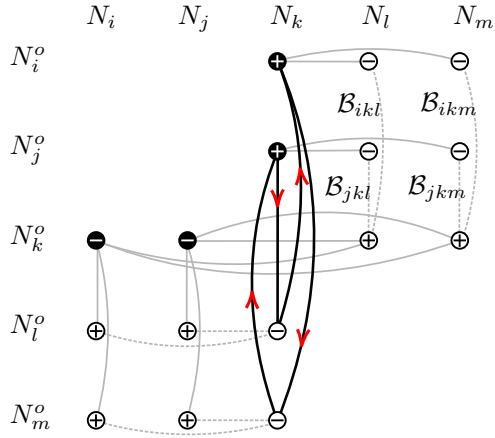}	
		\captionsetup{width=.8\textwidth}
	\caption{When four building blocks of the third kind share one common index (in this case $k$) and each pair of building blocks shares one of its two remaining indices ($i$, $j$, $l$ or $m$) with one other building block, there is an additional path that contributes to the trace of $D_F^4$ (including its inner fluctuations). The interaction is given by \eqref{eq:4bb3s}.}
	\label{fig:4bb3s}	
\end{center}
\end{figure}

Furthermore, it can be that there are four different building blocks of the third type that all share one particular index ---say \B{ikl}, \B{ikm}, \B{jkl} and \B{jkm}, sharing index $k$--- then there arises one extra interaction, that is of the form
\bas
	& N_k\frac{f(0)}{\pi^2}\big[\tr\yuks{i}{m}\sfer_{im}\asfer_{jm}\yuk{j}{m}\yuks{j}{l}\sfer_{jl}\asfer_{il}\yuk{i}{l} + h.c.\big].\nn
\eas
Scaling the fields and rewriting the parameters using \eqref{eq:improvedUpsilons1} gives
\ba
	&4\frac{g_n^2}{q_n}\w{ik}\w{jk}
 \tr \tilde \Upsilon_{l}\tilde \Upsilon_{m}^*\sfer_{im}\asfer_{jm}(\tilde\Upsilon_{l}'\tilde\Upsilon_{m}'^*)^*\sfer_{jl}\asfer_{il} + h.c., \label{eq:4bb3s}
\ea
where $g_n$ can equal any of the coupling constants that appear in the theory and we have written 
\bas 
	\tilde \Upsilon_{m} &\equiv \yuk{i,m}{k}(N_m\yuks{i,m}{k}\yuk{i,m}{k} + N_l\yuks{i,l}{k}\yuk{i,l}{k})^{-1/2},\nn\\
 \tilde \Upsilon_{m}' &\equiv \yuk{j,m}{k}(N_m\yuks{j,m}{k}\yuk{j,m}{k} + N_l\yuks{j,l}{k}\yuk{j,l}{k})^{-1/2},
\eas
and the same for $m \leftrightarrow l$. The path to which such an interaction corresponds, is given in Figure \ref{fig:4bb3s}. One can check that this interaction can only be described off shell by invoking either one or both of the auxiliary fields $F_{ij}$ and $F_{lm}$. This means that in order to have a chance at supersymmetry, the finite spectral triple that corresponds to the Krajewski diagram of Figure \ref{fig:4bb3s} requires in addition at least \B{ij} or \B{lm}.

\section{Higher degree building blocks?}

The first three building blocks that gave supersymmetric actions are characterized by one, two and three indices respectively. One might wonder whether there are building blocks of higher order, carrying four or more indices. 

Each of the elements of a finite spectral triple is characterized by one (components of the algebra, adjoint representations in the Hilbert space), two (non-adjoint representations in the Hilbert space) or three (components of the finite Dirac operator that satisfy the order-one condition) indices. For each of these elements corresponding building blocks have been identified. Any object that carries four or more different indices (e.g.~two or more off-diagonal representations, multiple components of a finite Dirac operator) must therefore be part of more than one building block of the first, second or third type. These blocks are, so to say, the irreducible ones. 

This does not imply that there are no other building blocks left to be identified. However, as we will see in the next section, they are characterized by less than four indices.

\section{Mass terms}\label{sec:mass_terms}

There is a possibility that we have not covered yet. The finite Hilbert space can contain two or more copies of one particular representation. This can happen in two slightly different ways. The first is when there is a building block $\B{11'}$ of the second type, on which the same component $\com$ of the algebra acts both on the left and on the right in the same way. For the second way it is required that there are two copies of a particular building block $\B{ij}$ of the second type. If the gradings of the representations are of opposite sign (in the first situation this is automatically the case for finite KO-dimension $6$, in the second case by construction) there is allowed a component of the Dirac operator whose inner fluctuations will not generate a field, rather the resulting term will act as a mass term. In the first case such a term is called a Majorana mass term. We will cover both of them separately.

\subsection{Fourth building block: Majorana mass terms}\label{sec:bb4}

\begin{figure}
	\begin{center}
		\def\svgwidth{.35\textwidth}
		\subimport{\the\svgpath}{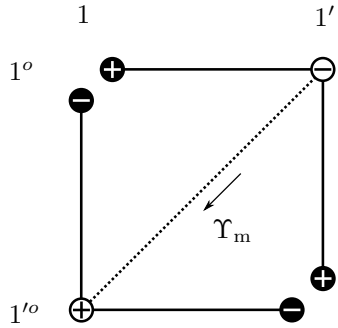}	
		\captionsetup{width=.5\textwidth}
		\caption{A component of the finite Dirac operator that acts as a Majorana mass is represented by a dotted line in a Krajewski diagram.}
		\label{fig:MajMass}	
\end{center}
\end{figure}

The finite Hilbert space can, for example due to some breaking procedure \cite{CC08, CCM07}, contain representations
\begin{align*}
	&\repl{1}{1'} \oplus \repl{1'}{1} \simeq \mathbb{C} \oplus \mathbb{C},
\end{align*} 
which are each other's antiparticles, e.g.~these representations are not in the adjoint (`diagonal') representation, but the same component $\com$ of the algebra\footnote{For a component $\mathbb{R}$ in the finite algebra this would work as well, but such a component would not give rise to gauge interactions and is therefore unfavourable.} acts on them. Then there is allowed a component $\D{1'1}{11'}$ of the Dirac operator connecting the two. It satisfies the first order condition \eqref{eq:order_one} and its inner fluctuations automatically vanish. Consequently, this component does not generate a scalar, unlike the typical component of a finite Dirac operator. Writing $(\xi, \xi') \in (\com\oplus \com)^{\oplus M}$ (where $M$ denotes the multiplicity of the representation) for the finite part of the fermions, the demand of $D_F$ to commute with $J_F$ reads
\bas
		(\D{11'}{1'1} \bar\xi, \D{1'1}{11'}\bar \xi') = \Big(\overline{\D{1'1}{11'}\xi}, \overline{\D{11'}{1'1}\xi'}\Big).
\eas
Using that $(\D{ij}{ik})^* = \D{ik}{ij}$ this teaches us that the component must be a symmetric matrix. It can be considered as a Majorana mass for the particle $\fer{11'}$ whose finite part is in the representation $\repl{1}{1'}$ (cf.~the Majorana mass for the right handed neutrino in the Standard Model \cite{CCM07}). Then we have 
\begin{defin}\label{def:bb4}
	For an almost-commutative geometry that contains a building block \B{11'} of the second type, a \emph{building block of the fourth type} \BBBB{11'} consists of a component
	\begin{align*}
		\D{1'1}{11'} : \repl{1}{1'} \to \repl{1'}{1} 
	\end{align*}
	of the finite Dirac operator. Symbolically it is denoted by 
\begin{align*}
	\BBBB{11'} = (0, \D{1'1}{11'}) \in \H_F \oplus \End(\H_F),
\end{align*}
where for the symmetric matrix that parametrizes this component we write $\maj$.
\end{defin}
In the language of Krajewski diagrams such a Majorana mass is symbolized by a dotted line, cf.~Figure \ref{fig:MajMass}.

\begin{figure}
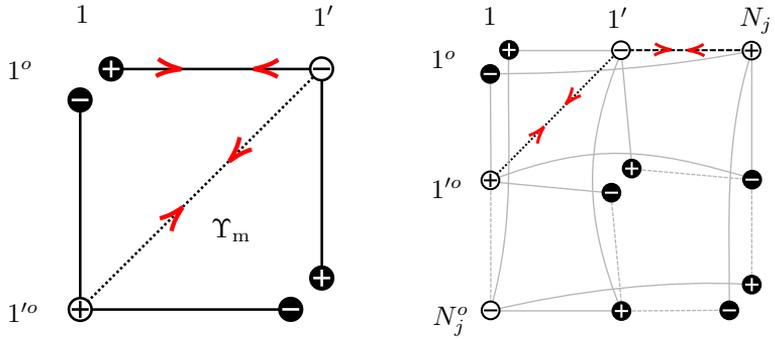
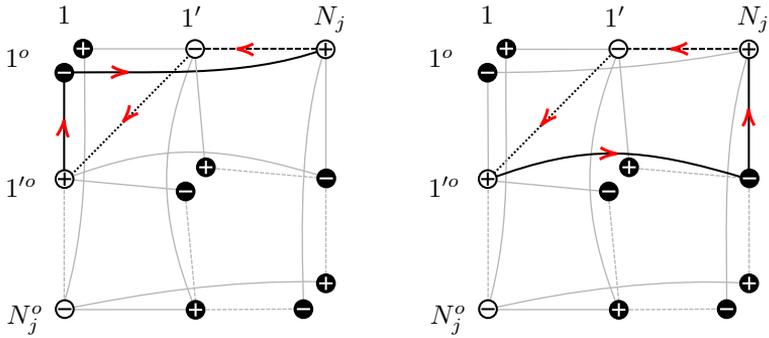

	\centering
	\begin{subfigure}{.35\textwidth}
		\centering
		\def\svgwidth{\textwidth}
		\subimport{\the\svgpath}{bb4-path1.pdf_tex}
		\caption{A path featuring edges from a building block of the second type.}
		\label{fig:bb4-path1}
	\end{subfigure}
	\hspace{30pt}
\begin{subfigure}{.35\textwidth}
		\centering
		\def\svgwidth{\textwidth}
		\subimport{\the\svgpath}{bb4-path3.pdf_tex}
		\caption{A path featuring edges from a building block of the third type.}
		\label{fig:bb4-path3}
	\end{subfigure}
	\vspace{20pt}
	
	\begin{subfigure}{.35\textwidth}
		\centering
		\def\svgwidth{\textwidth}
		\subimport{\the\svgpath}{bb4-path2.pdf_tex}
		\caption{A path featuring edges from building blocks of the second and third type.}
		\label{fig:bb4-path2}
	\end{subfigure}
	\hspace{30pt}
\begin{subfigure}{.35\textwidth}
		\centering
		\def\svgwidth{\textwidth}
		\subimport{\the\svgpath}{bb4-path4.pdf_tex}
		\caption{A second path featuring edges from a building block of the third type.}
		\label{fig:bb4-path4}
	\end{subfigure}

\captionsetup{width=.8\textwidth}
\caption{In the case that there is a building block of the fourth type, there are extra interactions in the action.}
\label{fig:bb4-paths}
\end{figure}

A \BBBB{11'} adds the following to the action \eqref{eq:spectral_action_acg_flat}:
\begin{align}
&	\frac{1}{2}\inpr{J_M\fer{11'L}}{\gamma^5 \maj^* \fer{11'L}} + \frac{1}{2}\inpr{J_M\afer{11'R}}{\gamma^5 \maj \afer{11'R}}\nn\\
&\qquad  + \frac{f(0)}{\pi^2} \Big[|\maj \asfer_{11'}C_{111'}^*|^2 + |\maj \asfer_{11'}C_{1'1'1}^*|^2 \nn\\
&\qquad\qquad\qquad\qquad + \sum_j \Big( |\maj\yuk{1'}{j}\sfer_{1'j}|^2 + |\maj^*\yuks{1}{j}\sfer_{1j}|^2\Big) \Big]\nn\\[-7pt]
& \qquad + \frac{f(0)}{\pi^2}\sum_j \Big(\tr (\asfer_{11'}C_{111'}^*)^o\maj\yuk{1'}{j}\sfer_{1'j}\asfer_{1j}C_{11j}^* \nn\\[-7pt]
&\qquad\qquad\qquad\qquad +  \tr \maj(\asfer_{11'}C_{1'1'1}^*)^oC_{1'1'j}\sfer_{1'j}\asfer_{1j}\yuk{1}{j}\nn\\
&\qquad\qquad\qquad\qquad	+  \tr \maj\yuk{1'}{j}\sfer_{1'j}(\asfer_{11'}\yuks{1}{1'})^o\asfer_{1j}\yuk{1}{j} + h.c.\Big),\label{eq:bb4-action}
\end{align}
where the traces are over $(\repl{1}{1'})^{\oplus M}$. In this expression, the first contribution comes from the inner product. The paths in the Krajewski diagram corresponding to the other contributions are depicted in Figure \ref{fig:bb4-paths}. In this set up it is $\sfer_{1'j}$ that does not have a family index. Consequently we can separate the traces over the family-index and that over \srep{j} in the penultimate term of the second line of \eqref{eq:bb4-action}. We would like to rewrite the above action in terms of $\yukw{}{} \equiv \yuk{1'}{j}$ by using the identity \eqref{eq:improvedUpsilons2}. For this we first need to rewrite the $C_{iij}$ to the $C_{ijj}$ by employing Remark \ref{rmk:bb3-relativesigns}. Writing out the family indices of the third and fourth line of \eqref{eq:bb4-action} gives
\ba
&  \tr ((\asfer_{11'}C_{111'}^*)^o\maj)_a\sfer_{1'j}\asfer_{1jc}(C_{11j}^*(\yuk{1'}{j})^t)_{ca} \nn\\
	&\qquad \qquad + \tr (\maj(\asfer_{11'}C_{1'1'1}^*)^o)_aC_{1'1'j}\sfer_{1'j}\asfer_{1jc}(\yuk{1}{j})_{ca}\nn\\
&=  \sqrt{\frac{n_1\K_j}{n_j\K_1}}\frac{g_1}{g_j}(s_{1'j} - s_{1j}s_{11'})\Big[\tr (\asfer_{11'}C_{11'1'}^*)^o\maj \yuk{1'}{j}\sfer_{1'j}\asfer_{1j}C_{1jj}^*\Big]\label{eq:bb4-idnyuks},
\ea
where $a, b, c$ are family indices, $s_{ij}$ is the product of the signs of $C_{iij}$ and $C_{ijj}$ (cf.~the notation in Remark \ref{rmk:bb3-relativesigns}) and where we have used that $\maj$ is a symmetric matrix. 

Then to make things a bit more apparent, we scale the fields in \eqref{eq:bb4-action} (with the third and fourth line replaced by \eqref{eq:bb4-idnyuks}) according to \eqref{eq:bb3-scalingfields} and put in the expressions for the $C_{ijj}$ from \eqref{eq:bb3-expressionCs-sqrt}, which gives
\ba
&\mkern-18mu 	\frac{1}{2}\inpr{J_M\fer{11'L}}{\gamma^5 \maj^* \fer{11'L}} + \frac{1}{2}\inpr{J_M\afer{11'R}}{\gamma^5 \maj \afer{11'R}}\nn\\
&\quad +  4r_1|\maj\asfer_{11'}|^2 + 2\sum_j \w{1j}\Big(|\maj\yukw{j}{}|_M^2 |\sfer_{1'j}|^2 + |\maj^*\yukws{j}{}\sfer_{1j}|^2\Big) \label{eq:bb4-action-scaled}\\
&\quad + \kappa_{1'}\kappa_j\sum_j 2g_m\sqrt{\frac{2\w{1j}}{q_m}} \Big( \tr \asfer_{11'}(r_1 + \w{1j}\yukw{j}{}\yukws{j}{})^t\maj\yukw{j}{}\sfer_{1'j}\asfer_{1j} + h.c.\Big), \nn\\
\ea
where we have written $|a|^2_M = \tr_M a^*a$ for the trace over the family-index, $\yukw{j}{} \equiv \yukw{1'}{j}$, and where $\kappa_{1'} = \sgnc_{1',j}\sgnc_{1',1}$, $\kappa_{j} = \sgnc_{j,1'}\sgnc_{j,1}\in \{\pm 1\}$. We replaced $\asfer_{11'}^o$ by $\asfer_{11'}$ since these coincide when $\asfer_{11'}$ is a gauge singlet. Consequently, the traces are now over $\mathbf{1}^{\oplus M}$. In addition we used the relation \eqref{eq:improvedUpsilons2} between $\yuk{1}{j}$, $\yuk{1'}{j}$ and $\yuk{1}{1'}$, the symmetry of $\maj$ and that $g_1 \equiv g_{1'}$ (which follows from the set up) and consequently $r_1 = r_{1'}$ and $\w{1'j} = \w{1j}$. In contrast to the previous case, not all scalar interactions that appear here can be accounted for by auxiliary fields:

\begin{lem}\label{prop:bb4}

	For a finite spectral triple that contains, in addition to building blocks of the first, second and third type, one building block of the fourth type, the only terms in the associated spectral action that can be written off shell using the available auxiliary fields are those featuring $\sfer_{11'}$ or its conjugate. 	
\end{lem}
\begin{proof}
The bosonic terms in \eqref{eq:bb4-action} must be the on shell expressions of an off shell Lagrangian that features the auxiliary fields available to us. Respecting gauge invariance, the latter must be  
\ba\label{eq:bb4-auxfields}
- \tr F_{11'}^*F_{11'} - \Big(\tr F_{11'}^*\big(\gamma_{11'} \asfer_{11'} + \sum_j \beta_{11',j}\sfer_{1j}\asfer_{1'j}\big) + h.c.\Big).
\ea
On shell this then gives the following contributions featuring $\sfer_{11'}$ and its conjugate: 
\bas
|\gamma_{11'} \asfer_{11'}|^2 
 + \sum_j 
 \big(\tr \gamma_{11'}\asfer_{11'}\sfer_{1'j}\asfer_{1j}\beta_{11',j}^* + h.c.\big),
\eas
which corresponds at least in form to all bosonic terms of \eqref{eq:bb4-action-scaled}, except the second term of the second line.
\end{proof}

We can use an argument similar to the one we used for building blocks of the third type:

\begin{lem}\label{lem:bb4-aux-susy}
	The action consisting of the fermionic terms of \eqref{eq:bb4-action-scaled} and the terms of \eqref{eq:bb4-auxfields} that do not feature $\beta_{11',j}$ or its conjugate is supersymmetric under the transformations \eqref{eq:susytransforms5} iff
	\ba\label{eq:bb4-aux-susy-demand}
			\gamma^*_{11'}\gamma_{11'} = \maj^*\maj
	\ea
	and the gauginos represented by the black vertices in Figure \ref{fig:bb4-path1} that have the same chirality are associated with each other.
\end{lem}
\begin{proof}
	See Section \ref{sec:bb4-proof}.
\end{proof}

Combining the above two Lemmas, then gives the following result.

\begin{prop}\label{cor:bb4}
The action \eqref{eq:bb4-action-scaled} of a single building block of the fourth type breaks supersymmetry only softly via
\bas
 2\sum_j \w{1j}\Big(|\maj\yukw{j}{}|_M^2|\sfer_{1'j}|^2 + |\maj^*\yukws{j}{}\sfer_{1j}|^2\Big) \nn
\eas
iff
\ba\label{eq:bb4-susy-demands}
	r_1 &= \frac{1}{4} &&\text{and} & \w{1j}\yukw{j}{}\yukws{j}{} &= \Big(- \frac{1}{4} \pm \frac{\kappa_{1'}\kappa_j}{2}\Big)\id_M,
\ea
where the latter should hold for all $j$ appearing in the sum in \eqref{eq:bb4-action}. Here $\kappa_{1'} = \sgnc_{1',j}\sgnc_{1',1}$ and $\kappa_j = \sgnc_{j,1'}\sgnc_{j,1}\in \{\pm 1\}$.
\end{prop} 
\begin{proof}
To prove this, we must match the coefficients of the contribution \eqref{eq:bb4-action-scaled} to the spectral action from a building block \B{11'} to those of the auxiliary fields \eqref{eq:bb4-auxfields}. This requires
\ba
	& \gamma_{11'} = 2\sqrt{r_1}e^{i\phi_\gamma}\maj, \nn\\
	& \kappa_{1'}\kappa_j 2g_m \sqrt{\frac{2\w{1j}}{q_m}}(r_1\id_M  + \w{1j} \yukw{j}{}\yukws{j}{})^t\maj\yukw{j}{} = \gamma_{11'}(\beta_{11',j}^*)^t\label{eq:bb4-susy-demands-interm}
\ea
for all $j$, where $e^{i\phi_\gamma}$ denotes the phase ambiguity left in $\maj$ from \eqref{eq:bb4-aux-susy-demand} and where we have used the symmetry of $\maj$. From supersymmetry $\gamma_{11'}$ is in addition constrained by \eqref{eq:bb4-aux-susy-demand}, which requires the first relation of \eqref{eq:bb4-susy-demands} to hold. For the building block \B{11'j} to have a supersymmetric action we demand
\bas	
		\beta_{11',j}^* &= g_m \sqrt{\frac{2\w{1j}}{q_m}} e^{-i\phi_{\beta_j}}(\yukw{j}{})^t,
\eas
which can be obtained by combining the demand \eqref{eq:bb3-susy-demand2} with the relation \eqref{eq:yukwyukp}, but keeping Remark \ref{rmk:bb3-R=1} in mind since it is $\sfer_{1'j}$ that does not have a family index. As is with $\maj$, the demand \eqref{eq:bb3-susy-demand2} determines $\beta_{11',j}$ only up to a phase $\phi_{\beta_j}$. Comparing this with the second demand of \eqref{eq:bb4-susy-demands-interm}, inserting \eqref{eq:bb4-aux-susy-demand} and using the symmetry of $\maj$, we must have 
\bas
	\phi_\gamma &= \phi_{\beta_j} \mod \pi, & 2(r_1\id_M + \w{1j}\yukw{j}{}\yukws{j}{}) = \pm \kappa_{1'}\kappa_j 2\sqrt{r_1}\id_M.
\eas
 Inserting the first relation of \eqref{eq:bb4-susy-demands}, its second relation follows. The second term of the second line of \eqref{eq:bb4-action-scaled} cannot be accounted for by the auxiliary fields at hand, which establishes the result.
\end{proof}

It is not per se impossible to write all of \eqref{eq:bb4-action-scaled} off shell in terms of auxiliary fields, but to avoid the obstruction from Lemma \ref{prop:bb4} at least requires the presence of mass terms for the representation $\sfer_{1j}$ and $\sfer_{1'j}$ such as the ones that are discussed in the next section.

%
%
%
%
%

\subsection{Fifth building block: `mass' terms}\label{sec:bb5}

\begin{figure}
	\begin{center}
		\def\svgwidth{.45\textwidth}
		\subimport{\the\svgpath}{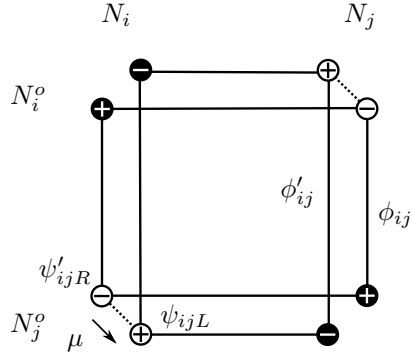}	
		\captionsetup{width=.7\textwidth}
		\caption{The case with two building blocks of the second type that have the same indices but an opposite grading; a component of the finite Dirac operator mapping between the two copies will generate a mass-term, indicated by the dotted line with the `$\mu$'.}
		\label{fig:bb5}	
\end{center}
\end{figure}

If there are two building blocks of the second type with the same indices ---say $i$ and $j$--- but with different values for the grading, we are in the situation as depicted in Figure \ref{fig:bb5}. On the basis 
\ba\label{eq:bb5-basis}
	\Big[(\rep{i}{j})_L \oplus (\rep{j}{i})_R \oplus (\rep{i}{j})_R \oplus (\rep{j}{i})_L\Big]^{\oplus M}, 
\ea
the most general finite Dirac operator that satisfies the demand of self-adjointness, the first order condition \eqref{eq:order_one} and that commutes with $J_F$ is of the form
\ba\label{eq:bb5-DF}
	D_F = \begin{pmatrix}
			0  & 0 & \mu_i + \mu_j^o  & 0 \\
		0 & 0 & 0 & (\mu_i^o)^* + \mu_j^*\\
		\mu_i^* + (\mu_j^*)^o & 0 & 0 & 0   \\
		0 &  \mu_i^o + \mu_j  & 0 & 0 
	\end{pmatrix}
\ea
with $\mu_i \in M_{N_iM}(\com)$ and $\mu_j \in M_{N_jM}(\com)$. The inner fluctuations for general such matrices $\mu_{i,j}$ will generate scalar fields in the representations $M_{N_{i,j}}(\com)$. If we want these components to result in mass terms in the action, we should restrict them both to only act non-trivially on possible generations, i.e.~for a single generation the components are equal to a complex number. We will write $\mu := \mu_i + \mu_j^o \in M_{M}(\com)$ for the restricted component. 

This gives rise to the following definition.
\begin{defin}\label{def:bb5}
For a finite spectral triple that contains building blocks \Bc{ij}{\pm} and \Bc{ij}{\mp} of the second type (both with multiplicity $M$), a \emph{building block of the fifth type} is a component of $D_F$ that runs between the representations of the two building blocks and acts only non-trivially on the $M$ copies. Symbolically:
\begin{align*}
	\B{\mathrm{mass}, ij} = (0, \D{ijL}{ijR}) \in \H_F \oplus \End(\H_F).
\end{align*}
We denote this component with $\mu \in M_M(\com)$. 
\end{defin}

\begin{figure}
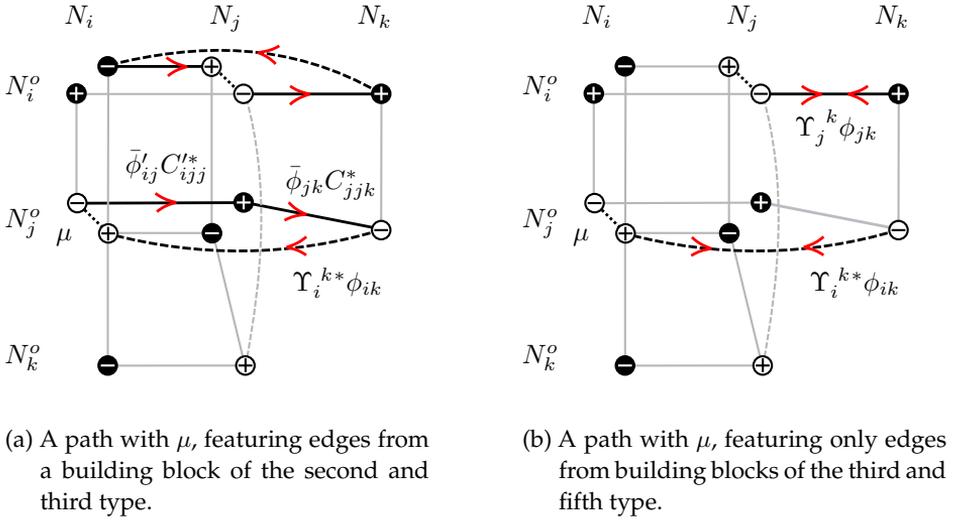

	\centering
	\begin{subfigure}{.45\textwidth}
		\centering
		\def\svgwidth{\textwidth}
		\subimport{\the\svgpath}{bb5-path1.pdf_tex}
		\caption{A path with $\mu$, featuring edges from a building block of the second and third type.}
		\label{fig:bb5-path1}
	\end{subfigure}
	\hspace{30pt}
	\begin{subfigure}{.45\textwidth}
		\centering
		\def\svgwidth{\textwidth}
		\subimport{\the\svgpath}{bb5-path2.pdf_tex}
		\caption{A path with $\mu$, featuring only edges from building blocks of the third and fifth type.}
		\label{fig:bb5-path2}
	\end{subfigure}
		\caption{In the case of a building block of the fifth type, there are various extra contributions to the action, depending on the content of the finite spectral triple.}
		\label{fig:bb5-paths}	
\end{figure}

If for convenience we restrict to the upper signs for the chiralities of the building blocks and write 
\bas
 (\fer{ijL},  \afer{ijR}, \fer{ijR}', \afer{ijL}')
\eas
for the elements of $L^2(M, S \otimes \H_F)$ on the basis \eqref{eq:bb5-basis} (where the first two fields are associated to \Bc{ij}{+} and the last two to \Bc{ij}{-}), then the contribution of \eqref{eq:bb5-DF} to the fermionic action reads
\ba
	S_{f,\textrm{mass}}[\zeta] &= \frac{1}{2}\inpr{J(\fer{ijL}, \afer{ijR}, \fer{ijR}', \afer{ijL}')}{\gamma^5 D_F (\fer{ijL}, \fer{ijR}', \afer{ijR}, \afer{ijL}')}\nn\\
&\qquad= \inpr{J_M \afer{ijR}}{\gamma^5 \mu\fer{ijR}'}
+ \inpr{J_M\afer{ijL}'}{\gamma^5 \mu^*\fer{ijL}}.\label{eq:bb5-action-ferm}
\ea
Let $\sfer$ and $\sfer'$ be the sfermions that are associated to \Bc{ij}{+} and \Bc{ij}{-} respectively, then the extra contributions to the spectral action as a result of adding this building block are given by
\begin{align}
	& S_{b,\textrm{mass}}[\szeta] \nn\\
 &= \frac{f(0)}{\pi^2}(N_i|\mu^*C_{iij}\sfer_{ij}|^2 + N_j|\mu^* C_{ijj}\sfer_{ij}|^2 + N_i|\mu C_{iij}'\sfer_{ij}'|^2 + N_j|\mu C_{ijj}'\sfer_{ij}'|^2)\nn\\
&\quad + \frac{f(0)}{\pi^2}\sum_{k}\Big[ N_i \tr \mu^*\asfer_{ij}'C_{iij}'^*C_{iik}\sfer_{ik}\asfer_{jk}\yuks{j}{k} \nn\\
		&\qquad + N_j \tr \asfer_{ij}'C_{ijj}'^*\,\mu^*\yuks{i}{k}\sfer_{ik}\asfer_{jk}C_{jjk}^* + h.c.\nn\\
&\qquad + \Big(N_j\tr_M(\mu\mu^*\yuks{i}{k}\yuk{i}{k})|\sfer_{ik}|^2 + N_i|\mu\yuk{j}{k}\sfer_{jk}|^2\Big)\Big],\label{eq:bb5-action}
\end{align}
where the second and third lines arise in a situation where for some $k$, $\B{ijk}$ is present. The paths corresponding to these expressions are depicted in Figure \ref{fig:bb5-paths}. Here, the $C_{iij}$ with a prime correspond to the components of the Dirac operator of \Bc{ij}{-}. We assume that they also satisfy \eqref{eq:bb2-resultCiij}. In this context $\sfer_{ik}$ does not have a family-index and consequently we could separate the traces in the first term of the third line of \eqref{eq:bb5-action}.

In a similar way as with the building block of the fourth type we can rewrite the second line of \eqref{eq:bb5-action} using Remarks \ref{rmk:bb3-relativesigns} and \ref{rmk:bb3-R=1}, giving
\ba
&\frac{f(0)}{\pi^2} \Big[N_i \tr (\asfer_{ij}'C_{iij}'^*)_aC_{iik}\sfer_{ik}\asfer_{jkb}(\yuks{j}{k}(\mu^*)^t)_{ba} \nn\\
&\qquad + N_j \tr (\asfer_{ij}'C_{ijj}'^*)_a\,(\mu^*\yuks{i}{k})_{ac}\sfer_{ik}\asfer_{jkb}(C_{jjk}^*)_{bc} + h.c.\Big]\nn\\
&= s_{jk}\bigg(\frac{N_ir_i + N_jr_j}{\sqrt{n_jn_k}g_jg_k}\bigg) \tr \asfer_{ij}'C_{ijj}'^*\mu^*\yuks{i}{k}\sfer_{ik}\asfer_{jk}C_{jkk}^*  + h.c.\label{eq:bb5-relativesigns}
\ea
Replacing the second line of \eqref{eq:bb5-action} with \eqref{eq:bb5-relativesigns} and then scaling the fields and rewriting $\yuk{i}{j}$ and $\yuk{j}{k}$ in terms of $\yuk{i}{k} \equiv\yuk{}{}$ using the identities \eqref{eq:improvedUpsilons2}, reduces the bosonic contribution \eqref{eq:bb5-action} to
\begin{align}
	& 2(1 - \w{ij})\big(|\mu^* \sfer_{ij}|^2 + |\mu \sfer_{ij}'|^2\big) \nn\\
	&\qquad + 2\sum_{k}\bigg[\kappa_{j} g_l (1 - \w{ij})\sqrt{\frac{2\w{ik}}{q_l}} \tr \asfer_{ij}'\mu^*\yukws{}{}\sfer_{ik}\asfer_{jk} + h.c.\nn\\
&\qquad\qquad + \w{ik}\Big(N_j|\yuk{}{}\mu|_M^2|\sfer_{ik}|^2 + N_i|\mu\yukw{}{}\sfer_{jk}|^2\Big)\bigg],\label{eq:bb5-action-scaled}
\end{align}
where we have again employed the notation $|a|_M^2 = \tr_M a^*a$ for the trace over the family-index and used that $s_{jk}\sgnc_{j,i}\sgnc_{k,j} = \sgnc_{j,i}\sgnc_{j,k} \equiv \kappa_j \in \{\pm\}$. The index $l$ can take any of the values that appear in the model. 

Here we have a similar result as in the previous section:

\begin{lem}\label{lem:bb5}

	For a finite spectral triple that contains, in addition to building blocks of the first, second and third type, one building block of the fifth type, the only terms in the associated spectral action that can be written off shell are those featuring $\sfer_{ij}$, $\sfer_{ij}'$ or their conjugates. 	
\end{lem}
\begin{proof}
	In order to rewrite the first terms of \eqref{eq:bb5-action-scaled} in terms of auxiliary fields, we must introduce an interaction featuring one auxiliary field $F$ and one sfermion. Since $\sfer_{ij}$ and $\sfer_{ij}'$ are in the same representation of the algebra, we can choose whether to couple $\sfer_{ij}$ to $F_{ij}$ (corresponding to \Bc{ij}{+}) or to $F_{ij}'$ (corresponding to \Bc{ij}{-}). The same holds for $\sfer_{ij}'$. Transforming the fermions in \eqref{eq:bb5-action-ferm} according to \eqref{eq:susytransforms4} suggests that, in order to have a chance at supersymmetry, we must couple $F_{ij}'$ to $\sfer_{ij}$ and $F_{ij}$ to $\sfer_{ij}'$. We thus write 
	\ba\label{eq:bb5-auxfields}
		- \tr F_{ij}^*F_{ij} - \tr F_{ij}'^*F_{ij}' - \big( \tr F_{ij}^*\delta_{ij}'\sfer_{ij}' + \tr F_{ij}'^*\delta_{ij} \sfer_{ij} + h.c.\big)
	\ea
with $\delta_{ij}, \delta_{ij}' \in M_M(\com)$. This yields on shell 
$
	|\delta_{ij}\sfer_{ij}|^2 + |\delta_{ij}'\sfer_{ij}'|^2
$, which is indeed of the same form as the first two terms in \eqref{eq:bb5-action-scaled}. In the case that there is a building block \B{ijk} of the third type present, the extra contributions to the action must come from the cross terms of
\bas
	\mkern-18mu	- \tr F_{ij}^*F_{ij} - \tr F_{ij}'^*F_{ij}' -  \Big[ \tr F_{ij}^*\big(\delta_{ij}'\sfer_{ij}' + \beta_{ij,k}\sfer_{ik}\asfer_{jk}\big) + \tr F_{ij}'^*\delta_{ij}\sfer_{ij} + h.c.\Big]
\eas
where the interaction with $\beta_{ij,k}$ corresponds to the second term of \eqref{eq:bb3-auxfields}. On shell this gives us the additional interaction
\ba
 \tr \asfer_{ij}'\delta_{ij}'^*\beta_{ij,k}\sfer_{ik}\asfer_{jk} + h.c.\label{eq:bb5-cross-term}
\ea
In form, this indeed coincides with the second line of \eqref{eq:bb5-action-scaled}. The last two terms of \eqref{eq:bb5-action-scaled} do not appear here and consequently they cannot be addressed using the auxiliary fields that are available to us when having only building blocks of the first, second and third type.
\end{proof}

Similar as with the previous building blocks we can check what the demands for off shell supersymmetry are.

\begin{lem}\label{prop:bb5}
	The action consisting of the fermionic action \eqref{eq:bb5-action-ferm} and the off shell action \eqref{eq:bb5-auxfields} is supersymmetric under the transformations \eqref{eq:susytransforms5} if and only if
	\ba\label{eq:bb5-constraints}
		\delta\delta^* &= \mu^*\mu,& \delta'\delta'^* &= \mu\mu^*.
	\ea
\end{lem}
\begin{proof}
	See Section \ref{sec:bb5-proof}.
\end{proof}

Combining the above lemmas gives the following result for a building block of the fifth type.

\begin{prop}\label{cor:bb5}
	For a finite spectral triple that contains, in addition to building blocks of the first, second and third type, one building block of the fifth type, the action of a single building block of the fifth type breaks supersymmetry only softly via 
\bas
&\w{ik}\Big(N_j|\yukw{}{}\mu|_M^2|\sfer_{ik}|^2 + N_i|\mu\yukw{}{}\sfer_{jk}|^2\Big)
\eas
iff
\bas
	\w{ij} &= \frac{1}{2}
\eas
and the product of the possible phases of $\delta'^*$ and $\beta_{ij,k}$ (cf.~\eqref{eq:bb5-constraints} and \eqref{eq:bb3-susy-demand2} respectively) is equal to $\sgnc_{j,i}\sgnc_{j,k}$.
\end{prop}
\begin{proof}
This follows from comparing the spectral action \eqref{eq:bb5-action-scaled} with the off shell action \eqref{eq:bb5-auxfields} and using the demands \eqref{eq:bb5-constraints} and \eqref{eq:bb3-susy-demand2}.
\end{proof}

The form of the soft breaking term suggests that, in order to let it be part of a truly supersymmetric action, we have the following necessary requirement. Each two building blocks of the second type that are connected to each other via an edge of a building block of the third type, both need to have a building block of the fifth type defined on them. In the case above this would have been $\sfer_{ik}$ and $\sfer_{jk}$. 

	\section{Conditions for a supersymmetric spectral action}\label{sec:4s-aux}

Our aim is to determine whether the total action that corresponds to an almost-commutative geometry consisting of various of the five identified building blocks, is supersymmetric. More than once we used the following strategy for that. First, we identified the off shell counterparts for the contributions of $\tr_F \Phi^4$ to the (on shell) spectral action, using the available auxiliary fields and coefficients whose values were undetermined still. Second, we derived constraints for these coefficients based on the demand of having supersymmetry for the fermionic action and this off shell action. Finally, we should check if the off shell interactions correspond on shell to the spectral action again, when their coefficients satisfy the constraints that supersymmetry puts on them. If this is the case then the action from noncommutative geometry is an on shell counterpart of an off shell action that is supersymmetric.

In the previous sections we have experienced multiple times that the pre-factors of all bosonic interactions can get additional contributions when extending the almost-commutative geometry. As was stated before, we should therefore assess whether or not the demands from supersymmetry on the coefficients are satisfied for the final model only. In this section we will present an overview of all four-scalar interactions that have appeared previously, from which building blocks their pre-factors get what contributions and which demands hold for them. We identify several such demands, thus constructing a checklist for supersymmetry.

\begin{enumerate}
\item To have supersymmetry for a building block \B{ij} of the second type, the components of the finite Dirac operator should satisfy \eqref{eq:bb2-resultCiij}, after scaling them. For a single building block of the second type this demand can only be satisfied for $N_i = N_j$ and $M = 4$ (Proposition \ref{lem:bb2-nosol}). When \B{ij} is part of a building block of the third type the demand is automatically satisfied via the solution \eqref{eq:bb3-expressionCs-sqrt}.

\item A necessary requirement to have supersymmetry for any building block \B{ijk} of the third type (Section \ref{sec:bb3}), is that the scaled parameters of the finite Dirac operator that make up such a building block satisfy 
\ba\label{eq:demand-0}
	\w{jk}\yukws{j}{k}\yukw{j}{k} &= \w{ik}\yukws{i}{k}\yukw{i}{k} =  
	\w{ij}\yukws{i}{j}\yukw{i}{j} =: \Om{ijk}^*\Om{ijk}.
\ea
This relation can be obtained from \eqref{eq:improvedUpsilons1}, multiplying each term with its conjugate. For notational convenience we have introduced the variable $\Om{ijk}^*\Om{ijk}$. 

\item Terms $\propto |\sfer_{ij}\asfer_{ij}|^2$ appear for the first time with a building block of the second type (\eqref{eq:exprM1} in Section \ref{sec:bb2}) but also get contributions from a building block \B{ijk} of the third type (first term of \eqref{eq:bb3-boson-action}). The total expression reads 
	\ba
		& \frac{f(0)}{2\pi^2} \Big[ N_i|C_{iij}^*C_{iij}\sfer_{ij} \asfer_{ij}|^2 + N_j|C_{iij}^*C_{ijj}\sfer_{ij} \asfer_{ij}|^2 \nn\\
		&\qquad\qquad + \sum_k N_k |\yuks{i,k}{j}\yuk{i,k}{j}\sfer_{ij}\asfer_{ij}|^2\Big]\nn\\
		%
		&\to 2\frac{g_i^2}{q_i}\bigg|\bigg( N_ir_i^2 + \alpha_{ij}  \sum_k N_k(\Om{ijk}^*\Om{ijk})^2\bigg)^{1/2}\sfer_{ij}\asfer_{ij}\bigg|^2\nn\\
		&\qquad + 2\frac{g_j^2}{q_j}\bigg|\bigg(N_jr_j^2 + (1 - \alpha_{ij}) \sum_k N_k (\Om{ijk}^*\Om{ijk})^2\bigg)^{1/2}\sfer_{ij}\asfer_{ij}\bigg|^2,\nn
	\ea
	upon scaling the fields. Here we have introduced a parameter $\alpha_{ij} \in \mathbb{R}$ that tells how any new contributions are divided over the initial two. Such terms can only be described off shell using the auxiliary fields $G_{i}$ and $G_{j}$ (cf.~Lemma \ref{lem:bb2-offshell}) via
\bas
	 - \frac{1}{2n_i}\tr G_i\big(G_i + 2n_i\P_i \sfer_{ij}\asfer_{ij}\big)  
	 - \frac{1}{2n_j}\tr G_j\big(G_j + 2n_j \asfer_{ij}\P_j\sfer_{ij}\big),
\eas
which on shell equals
\bas
	\frac{n_i}{2}|\P_i \sfer_{ij}\asfer_{ij}|^2 + \frac{n_j}{2}|\asfer_{ij}\P_j \sfer_{ij}|^2, 
\eas
cf.~\eqref{eq:bb2-auxterms}. Comparing this with the above expression sets the coefficients $\P_{i}$ and $\P_{j}$:
\bas
		\frac{n_i}{2}\P_i^2 &= 2\frac{g_i^2}{q_i}\bigg(N_ir_i^2 + \alpha_{ij}  \sum_k N_k (\Om{ijk}^*\Om{ijk})^2\bigg), \nn\\
		\frac{n_j}{2}\P_j^2 &= 2\frac{g_j^2}{q_i}\bigg(N_jr_j^2 + (1 - \alpha_{ij}) \sum_k N_k (\Om{ijk}^*\Om{ijk})^2\bigg),\nn
\eas
where there is an additional trace over the last terms if $\sfer_{ij}$ has no family index. If the action is supersymmetric then \eqref{eq:bb2-resultCiij} can be used with $\K_i = \K_j = 1$ and the above relations read
	\ba\label{eq:demand-1}
		\frac{r_i}{4} &= N_ir_i^2 + \alpha_{ij}  \sum_k N_k \tr[(\Om{ijk}^*\Om{ijk})^2],\nn\\
  	\frac{r_j}{4} &= N_jr_j^2 + (1 - \alpha_{ij}) \sum_k N_k \tr[(\Om{ijk}^*\Om{ijk})^2], 
	\ea
when $\sfer_{ij}$ has no family index and 
	\ba\label{eq:demand-1.1}
		\frac{r_i}{4}\id_M &= N_ir_i^2\id_M + \alpha_{ij}  \sum_k N_k (\Om{ijk}^*\Om{ijk})^2, \nn\\
  	\frac{r_j}{4}\id_M &= N_jr_j^2\id_M + (1 - \alpha_{ij}) \sum_k N_k (\Om{ijk}^*\Om{ijk})^2, 
	\ea
when it does. Here we have used that $r_i = q_in_i$. 

\item An interaction $\propto |\sfer_{ij}\sfer_{jk}|^2$ can receive contributions in two different ways; one comes from a building block \B{ijk} of the third type \eqref{eq:bb3-action-final}, the other comes from two adjacent building blocks \B{ijl} and \B{jkl} (first and second term of \eqref{eq:2bb3-action-scaled}, but occurs only for particular values of the grading): 
\bas
	&	g_m^2\frac{4\w{ij}}{q_m}(1 - \w{ik}) |\yukw{i,k}{j}\sfer_{ij}\sfer_{jk}|^2 \nn\\
	&\qquad + 4\bigg(n_jr_jN_jg_j^2|\sfer_{ij}\sfer_{jk}|^2 + \frac{g_m^2}{q_m}\w{ij}\w{jk}N_l|\yukw{i,l}{j}\sfer_{ij}\yukw{j,l}{k}\sfer_{jk}|^2\bigg).\nn
\eas
From this, however, we need to subtract the value $n_j g_j^2|\sfer_{ij}\sfer_{jk}|^2$ that is expected from the cross term
\bas
 -  \tr G_j\big(\P_{j,i}\asfer_{ij}\sfer_{ij} + \P_{j,k}\sfer_{jk}\asfer_{jk}\big),
\eas
that should already be there when the almost-commutative geometry contains \Bc{ij}{\pm} and \Bc{jk}{\mp} but nevertheless does not appear in the spectral action (see Section \ref{sec:2bb2} and the discussion above Theorem \ref{prop:bb3}). 
The remaining terms must be accounted for by 
\ba\label{eq:bb3-auxfields2}
	- \tr F_{ik}^*F_{ik} + \big(\tr F_{ik}^*\bps_{ik,j}\sfer_{ij}\sfer_{jk} + h.c.\big) 
\ea
which equals
\bas \tr \asfer_{jk}\asfer_{ij}\bp_{ik,j}\bps_{ik,j}\sfer_{ij}\sfer_{jk}
\eas
on shell. Since $\beta_{ik,j}\beta_{ik,j}^*$ is positive definite we can also write the above as
\bas
	 |(\bp_{ik,j}\bps_{ik,j})^{1/2}\sfer_{ij}\sfer_{jk}|^2.
\eas
Comparing the above relations, the off shell action \eqref{eq:bb3-auxfields2} corresponds on shell to the spectral action, iff
\bas
	& \bp_{ik,j}\bps_{ik,j} \nn\\
	&= g_m^2\frac{4\w{ij}}{q_m}(1 - \w{ik}) \yukws{i,k}{j}\yukw{i,k}{j} - n_j g_j^2\id_M\nn\\
			&\qquad + 4\bigg(n_jr_jN_j g_j^2\id_M + \frac{g_m^2}{q_m}\w{ij}\w{jk}N_l(\yukw{i,l}{j}\yukw{j,l}{k})^*(\yukw{i,l}{j}\yukw{j,l}{k})\bigg),
\eas
where we have assumed that it is $\sfer_{ij}$ not having a family structure.
Furthermore, from the demand of supersymmetry $\bp_{ik,j}$ must satisfy 
\bas
	\bp_{ik,j}\bps_{ik,j} &= g_m^2\frac{2\w{ij}}{q_m}\yukws{i}{j}\yukw{i}{j} \equiv 2\frac{g_m^2}{q_m}\Om{ijk}^*\Om{ijk}
\eas
i.e.~\eqref{eq:bb3-susy-demand2},\footnote{In fact, in \eqref{eq:bb3-susy-demand2} the variables are in reversed order compared to here but looking at \eqref{eq:bb3-constr2} ---from which the former is derived--- one sees immediately that this also holds.} but with $\yukp{}{}$ replaced by $\yukw{}{}$ using \eqref{eq:yukwyukp}. Combining the above two relations, we require that 
\bas
	&\frac{2g_m^2}{q_m}\Om{ijk}^*\Om{ijk}\nn\\
	& = 4\frac{g_m^2}{q_m}(1 - \w{ik})\Om{ijk}^*\Om{ijk} - n_j g_j^2\id_M \nn\\ &\qquad + 4\bigg( 
n_jr_jN_jg_j^2\id_M + \frac{g_m^2}{q_m}\w{ij}\w{jk}N_l(\yukw{i,l}{j}\yukw{j,l}{k})^*(\yukw{i,l}{j}\yukw{j,l}{k})\bigg),
\eas
using the notation introduced in \eqref{eq:demand-0}. Setting $m = j$ in particular, this reduces to 
\ba
	 &2(1 - 2\w{ik})\Om{ijk}^*\Om{ijk} - r_j \id_M \nn\\
	&\qquad + 4\bigg( 
N_jr_j^2\id_M + \w{ij}\w{jk}N_l(\yukw{i,l}{j}\yukw{j,l}{k})^*(\yukw{i,l}{j}\yukw{j,l}{k})\bigg) = 0.\label{eq:demand-2}
\ea


%

\item The interaction $\propto \tr\sfer_{ik}\asfer_{jk}\sfer_{jl}\asfer_{il}$ only appears in the case of two adjacent building blocks \B{ijk} and \B{ijl} of the third type (cf.~the Lagrangian \eqref{eq:2bb3-action-scaled}). Equating this term to \eqref{eq:2bb3-aux} that appears from the auxiliary field $F_{ij}$, gives
\bas
 &\kappa_{k}\kappa_{l} 4\frac{g_m^2}{q_m}(1 - \w{ij})\w{ij}\tr \yukw{l}{}\yukws{k}{}\sfer_{ik}\asfer_{jk}\sfer_{jl}\asfer_{il} + h.c. \nn\\
	&\qquad = \tr \bps_{ij,l}\bp_{ij,k}\sfer_{ik}\asfer_{jk}\sfer_{jl}\asfer_{il} + h.c.,
\eas
with $\kappa_{k} = \sgnc_{k,i}\sgnc_{k,j}, \kappa_{l} = \sgnc_{l,i}\sgnc_{l,j}$. From the demand of supersymmetry $\bps_{ij,l}$ and $\bp_{ij,k}$ should satisfy \eqref{eq:bb3-susy-demand2}. Their phases, if any, must be opposite modulo $\pi$ for the action to be real. We write $\phi_{kl}$ for the remaining sign ambiguity. Inserting these demands above and using \eqref{eq:yukwyukp} requires that $\kappa_{k}\kappa_{l} 4\w{ij}(1- \w{ij}) = 2\phi_{kl}\w{ij}$ for this interaction to be covered by the auxiliary field $F_{ij}$. This has two solutions, the only acceptable of which is 
\ba\label{eq:demand-3}
	\phi_{kl} &= \kappa_{k}\kappa_{l},& \w{ij} &= \frac{1}{2} \quad\Longrightarrow\quad
			  r_iN_i + r_jN_j = \frac{1}{2},
\ea
where we have used \eqref{eq:kintermnorm}.
\item From the spectral action interactions $\propto |\sfer_{ij}|^4$ only appear in the context of a building block of the second type as 
\bas
	\frac{f(0)}{\pi^2}|C_{iij}\sfer_{ij}|^2 |C_{ijj}\sfer_{ij}|^2 \to 4\frac{g_l^2}{q_l}r_ir_j |\sfer_{ij}|^4,
\eas
see \eqref{eq:bb2-action}. Via the auxiliary fields on the other hand they appear in two ways; from the $G_{i,j}$ and via the $u(1)$-field $H$ (see Lemma \ref{lem:bb2-offshell} for both). The latter give on shell the contributions
\bas
	\bigg(\frac{\Q_{ij}^2}{2} - n_i\frac{\P_i^2}{2N_i} - n_j\frac{\P_j^2}{2N_j}\bigg)|\sfer_{ij}|^4,
\eas		
where the minus-signs stem from the identity \eqref{eq:idn-sun-gens} between the generators $T^a_{i,j}$ of $su(N_{i,j})$. Demanding supersymmetry, $\P_i^2$ must equal $g_i^2$ and similarly $\P_j^2 = g_j^2$. In order for the interactions from the spectral action to equal the above equation, $\Q_{ij}^2$ is then set to be
	\ba\label{eq:demand-4}
		\Q_{ij}^2 = \frac{g_l^2}{q_l}\bigg(8r_ir_j + \frac{r_i}{N_i} + \frac{r_j}{N_j}\bigg).
	\ea
	In the case that $\sfer_{ij}$ has family indices, the expressions for $\P_{i,j}^2$ and $\Q_{ij}^2$ must be multiplied with the $M \times M$ identity matrix $\id_M$.

\item Interactions $\propto |\sfer_{ij}|^2|\sfer_{jk}|^2$ (having one common index $j$) appear via the spectral action in two different ways. First of all from two adjacent building blocks \B{ij} and \B{jk} of the second type (cf.~\eqref{eq:2bb2s-different}), and secondly from a building block of the third type (second line of \eqref{eq:bb3-boson-action}). This gives
\bas
& \frac{f(0)}{\pi^2}\Big( |C_{ijj}\sfer_{ij}|^2|C_{jjk}\sfer_{jk}|^2
		+ |\sfer_{ij}|^2|\yuks{i}{j}\yuk{j}{k}\sfer_{jk}|^2\Big) \nn\\
&\qquad \to 4\frac{g_l^2}{q_l}\Big( r_j^2|\sfer_{ij}|^2|\sfer_{jk}|^2
		 + \w{jk}\w{ij}|\sfer_{ij}|^2|\yukws{i}{j}\yukw{j}{k}\sfer_{jk}|^2\Big),
\eas
where we have assumed $\sfer_{ij}$ not to have a family-index. We can write this as
\bas
	& 4\frac{g_l^2}{q_l}\big|\sfer_{ij}\big|^2 \big|\big(r_j^2\id_M + \w{ij}\w{jk}(\yukws{i}{j}\yukw{j}{k})^*\yukws{i}{j}\yukw{j}{k}\big)^{1/2}\sfer_{jk}\big|^2.
\eas
From the auxiliary fields these terms can appear via $G_j$ (with coefficients $\P_{j,i}$ and $\P_{j,k}$, i.e.~as in \eqref{eq:2bb2-different-aux}) and via the $u(1)$-field $H$ with coefficients $\Q_{ij}$ and $\Q_{jk}$:
\bas
	\bigg[\Q_{ij}\Q_{jk} - n_j\frac{\P_{j,i}\P_{j,k}}{N_j}\bigg]|\sfer_{ij}|^2 |\sfer_{jk}|^2.
\eas
Equating the terms from the spectral action and those from the auxiliary fields, and 
inserting the values for the coefficients $\P_{j,i}$, $\P_{j,k}$ (from \eqref{eq:bb2-resultCiij}), $\Q_{ij}$ and $\Q_{jk}$ (from \eqref{eq:demand-4}) that we obtain from supersymmetry, we require
\ba\label{eq:demand-5}
&\bigg(2r_ir_j + \frac{r_i}{4N_i} + \frac{r_j}{4N_j}\bigg)
\bigg(2r_jr_k + \frac{r_j}{4N_j} + \frac{r_k}{4N_k}\bigg)\id_M\nn\\
&\qquad = \Big[\Big(r_j^2 + \frac{r_j}{4N_j}\Big)\id_M + \w{ij}\w{jk}(\yukws{i}{j}\yukw{j}{k})^*\yukws{i}{j}\yukw{j}{k}\Big]^2.
	\ea

\item There are interactions $\propto |\sfer_{ik}|^2|\sfer_{jl}|^2$ and $\propto |\sfer_{jk}|^2|\sfer_{il}|^2$ that arise from two adjacent building blocks \B{ijk} and \B{ijl} of the third type. The first of these is given by 
\bas
4\frac{g_m^2}{q_m}|(\w{ik}\yukw{i,j}{k}\yukws{i,j}{k})^{1/2}\sfer_{ik}|^2|(\w{jl}\yukws{j,i}{l}\yukw{j,i}{l})^{1/2}\sfer_{jl}|^2,\nn
\eas
see \eqref{eq:2bb3-action-scaled}. Since the interactions are characterized by four different indices, the auxiliary fields $G_i$ cannot account for these and consequently they should be described by the $u(1)$-field $H$:
\bas
	|\Q_{ik}^{1/2}\sfer_{ik}|^2|\Q_{jl}^{1/2}\sfer_{jl}|^2.
\eas
 In order for the spectral action to be written off shell we thus require that 
	\bas
		 \Q_{ik}\Q_{jl} &= 4\frac{g_m^2}{q_m}\Om{ijk}\Om{ijk}^*\Om{ijl}^*\Om{ijl}.\nn
	\eas
	With $\Q_{ik}$ and $\Q_{jl}$ being determined by \eqref{eq:demand-4} from the demand of supersymmetry, we can infer from this that for the squares of these expressions we must have
	\ba
		\Big(2r_ir_k + \frac{r_i}{4N_i} + \frac{r_k}{4N_k}\Big)\id_M &= \Om{ijk}\Om{ijk}^*,\nn\\ 
		\Big(2r_jr_l + \frac{r_j}{4N_j} + \frac{r_l}{4N_l}\Big)\id_M &= \Om{ijl}^*\Om{ijl}\label{eq:demand-6}.
	\ea

\item As was already covered in Section \ref{sec:bb4}, a building block \BBBB{} of the fourth type only breaks supersymmetry softly iff
\ba\label{eq:demand-7}
	r_1 &= \frac{1}{4}&&\text{and}& \w{1j}\yukw{j}{}\yukws{j}{} &= \Big(- \frac{1}{4} \pm \frac{\kappa_{1'}\kappa_{j}}{2}\Big)\id_M
\ea
(see Proposition \ref{cor:bb4}), where the latter should hold for each building block \B{11'j} of the third type. Here $\kappa_{1'}, \kappa_{j} \in \{\pm 1\}$.  

\item Covered in Section \ref{sec:bb5}, a building block \B{\textrm{mass}, ij} of the fifth type also breaks supersymmertry only softly iff
\ba\label{eq:demand-8}
	 \w{ij} &= \frac{1}{2},
\ea
see Proposition \ref{cor:bb5}. 

\end{enumerate}

To be able to say whether an almost-commutative geometry that is built out of building blocks of the first to the fifth type has a supersymmetric action then entails checking whether all the relevant relations above are satisfied. 

\subsection{Applied to a single building block of the third type}\label{sec:4s-aux-bb3}

We apply a number of the demands above to the case of a single building block of the third type (and the building blocks of the second and first type that are needed to define it) to see whether this possibly exhibits supersymmetry. We will assume that $\fer{ij}$ has $R = - 1$ (and consequently no family index), but of course we could equally well have taken one of the other two (see e.g.~Remark \ref{rmk:bb3-R=1}). 
The generalization of Remark \ref{rmk:bb2-rmk} for the expressions of the $r_i$ that results from normalizing the gauge bosons' kinetic terms is
\bas
	r_i &= \frac{3}{2N_i + N_j + MN_k}, &
	r_j &= \frac{3}{N_i + 2N_j + MN_k}, \nn\\
	r_k &= \frac{3}{M(N_i + N_j) + 2N_k}. &&
\eas

For the first of the demands of the previous section, \eqref{eq:demand-0}, one of the three terms that are equated to each other reads
\bas
	\w{ik}\yukw{i}{k}\yukws{i}{k} &\equiv \w{ik}(N_j\yuk{i}{k}\yuks{i}{k})^{-1/2}\yuk{i}{k}\yuks{i}{k}(N_j\yuk{i}{k}\yuks{i}{k})^{-1/2} \nn\\
		&= \frac{\w{ik}}{N_j}\id_M = \w{ik}\yukws{i}{k}\yukw{i}{k},
\eas
where we have used the definition \eqref{eq:def-yukw} of $\yukw{i}{k}$. Similarly,
\bas
	\w{jk}\yukws{j}{k}\yukw{j}{k} &= \frac{\w{jk}}{N_i}\id_M && 
\text{and} &
	\w{ij}\yukws{i}{j}\yukw{i}{j} &= \frac{\w{ij}}{N_k}\yuks{i}{j}\yuk{i}{j} (\tr \yuks{i}{j}\yuk{i}{j})^{-1}
\eas
for the other two. Equating these, we obtain: 
\ba\label{eq:idn-bb3-sols}
 \frac{\w{ik}}{N_j}\id_M &= \frac{\w{jk}}{N_i}\id_M = \frac{\w{ij}}{N_k}\yuks{i}{j}\yuk{i}{j} (\tr \yuks{i}{j}\yuk{i}{j})^{-1},
\ea
i.e.~$\yuk{i}{j}$ is constrained to be proportional to a unitary matrix. Taking the trace gives the demand
\ba\label{eq:demand-0-bb3}
 M\frac{\w{ik}}{N_j} &= M\frac{\w{jk}}{N_i} = \frac{\w{ij}}{N_k}.
\ea
Given the expressions for $r_{i,j,k}$ above, we can test whether this demand admits solutions. Indeed, we find
\ba\label{eq:solutions-bb3}
	N_i &= N_j = N_k \equiv N,& M &= 1 \lor 2.
\ea
In the first case we find that 
	\bas
		r_iN_i &= r_jN_j = r_k N_k = \frac{3}{4},& \w{ij} &= \w{ik} = \w{jk} = - \frac{1}{2}, 
	\eas 
	whereas in the second case we have 
	\bas
		r_iN_i &= r_jN_j = \frac{3}{5},& r_kN_k &= \frac{1}{2},& \w{ij} &= - \frac{1}{5},& \w{ik} &= \w{jk} = - \frac{1}{10}.
	\eas

Next, we have the demand \eqref{eq:demand-1} to ensure that terms of the form $|\sfer_{ij}\asfer_{ij}|^2$ can be written off shell in a supersymmetric manner. In this context it reads 
\bas
		\frac{r_i}{4} &= N_ir_i^2 + \alpha_{ij} N_k \w{ij}^2\tr[(\yukws{i}{j}\yukw{i}{j})^2],\nn\\
  	\frac{r_j}{4} &= N_jr_j^2 + \alpha_{ji} N_k \w{ij}^2\tr[(\yukws{i}{j}\yukw{i}{j})^2],\nn
\eas
for $\sfer_{ij}$ (where the trace in the last term comes from the fact that $\sfer_{ij}$ does not have family indices) and 
\bas
		\frac{r_k}{4}\id_M &= N_kr_k^2\id_M + \alpha_{kj} N_i \w{jk}^2(\yukws{j}{k}\yukw{j}{k})^2, \nn\\
		\frac{r_j}{4}\id_M &= N_jr_j^2\id_M + \alpha_{jk} N_i \w{jk}^2(\yukws{j}{k}\yukw{j}{k})^2,\nn\\
		\frac{r_k}{4}\id_M &= N_kr_k^2\id_M + \alpha_{ki} N_j \w{ik}^2(\yukw{i}{k}\yukws{i}{k})^2,\nn\\
		\frac{r_i}{4}\id_M &= N_ir_i^2\id_M + \alpha_{ik} N_j \w{ik}^2(\yukw{i}{k}\yukws{i}{k})^2,
\eas
for $\sfer_{jk}$ and $\sfer_{ik}$ respectively. Here we have written $\alpha_{ji} = 1 - \alpha_{ij}$, etc. We can remove all variables $\yukw{i}{j}$, $\yukw{i}{k}$ and $\yukw{j}{k}$ by using the squares of the expressions in \eqref{eq:idn-bb3-sols}. This gives
\bas
		\frac{N_ir_i}{4} &= (N_ir_i)^2 + \alpha_{ij} N_k \frac{\w{jk}^2}{N_i}M, &
  	\frac{N_jr_j}{4} &= (N_jr_j)^2 + \alpha_{ji} N_k \frac{\w{ik}^2}{N_j}M,\nn\\
		\frac{N_kr_k}{4} &= (N_kr_k)^2 + \alpha_{kj} N_k \frac{\w{jk}^2}{N_i}, &
		\frac{N_jr_j}{4} &= (N_jr_j)^2 + \alpha_{jk} N_i \frac{\w{ik}^2}{N_j},\nn\\
		\frac{N_kr_k}{4} &= (N_kr_k)^2 + \alpha_{ki} N_k \frac{\w{ik}^2}{N_j},&
		\frac{N_ir_i}{4} &= (N_ir_i)^2 + \alpha_{ik} N_j \frac{\w{jk}^2}{N_i},
\eas
where the $M$ in the first line above comes from taking the trace over $\id_M$. Comparing the expressions featuring the same combinations $r_{i}N_{i}$, $r_{j}N_{j}$, $r_{k}N_{k}$ and using \eqref{eq:demand-0-bb3} we must have that
\bas
	\alpha_{ij}N_kM &= \alpha_{ik}N_j,&
	(1- \alpha_{jk})N_i &= (1 - \alpha_{ik})N_j,&
	(1 - \alpha_{ij})N_kM &= \alpha_{jk}N_i.
\eas
Since both solutions \eqref{eq:solutions-bb3} to the relation \eqref{eq:demand-0-bb3} have $N_i = N_j = N_k$, this solves 
\bas
	\alpha_{ij} &= \frac{1}{2},& \alpha_{ik} &= \frac{1}{2}M, & \alpha_{jk} &= \frac{1}{2}M
\eas
and the demands above reduce to
\bas
		N_ir_i &=4(N_ir_i)^2 + 2 \w{jk}^2M, &
  	N_jr_j &=4(N_jr_j)^2 + 2 \w{ik}^2M, \nn\\
		N_kr_k &=4(N_kr_k)^2 +  \w{ik}^2(4 - 2M). &&
\eas
We can check that for neither of the two cases of \eqref{eq:solutions-bb3} these are satisfied. As a cross check of this result we will employ one more demand.

In the context of a single building block of the third type the demand \eqref{eq:demand-2} that is necessary to write terms of the form $|\sfer_{ij}\sfer_{jk}|^2$ off shell in a supersymmetric manner, reduces to
\bas
	 2(1 - 2\w{ik})\w{ik} &= r_jN_j ,&
	 2(1 - 2\w{jk})\w{jk} &= r_iN_i ,\nn\\
	 2(1 - 2\w{ij})\w{ij}\yuks{i}{j}\yuk{i}{j}  &= r_kN_k\id_M \tr \yuks{i}{j}\yuk{i}{j}. &&
\eas
We can use \eqref{eq:demand-0-bb3} to rewrite the last equation in terms of $\w{ik}$ or $\w{jk}$. In any way, the LHS are seen to be negative for all values of $\w{ij}$, $\w{ik}$ and $\w{jk}$ allowed by the solutions \eqref{eq:solutions-bb3}, whereas $r_{i}N_{i}$, $r_{j}N_{j}$ and $r_{k}N_{k}$ are necessarily positive. We thus get a contradiction. 

A single building block of the third type (together with the building blocks needed to define it) is thus not supersymmetric.



\svgpath={./gfx/}

\chapter[Soft supersymmetry breaking]{Soft supersymmetry breaking\NoCaseChange{\footnote{The contents of this chapter are based on \cite{BS13II}.}}}\label{ch:breaking}

Already shortly after the advent of supersymmetry it was realized \cite{WZ74-2} that if it is a real symmetry of nature, then the superpartners should be of equal mass. This, however, is very much not the case. If it were, we should have seen all the sfermions and gauginos that feature in the Minimal Supersymmetric Standard Model (MSSM, e.g.~\cite{DGR04}) in particle accelerators by now. In the context of the MSSM we need \cite{GH86} a supersymmetry breaking Higgs potential to get electroweak symmetry breaking and give mass to the SM particles. Somehow there should be a mechanism at play that \emph{breaks} supersymmetry. Over the years many mechanisms have been suggested that break supersymmetry and explain why the masses of superpartners should be different at low scales. Ideally this should be mediated by a \emph{spontaneous} symmetry breaking mechanism, such as $D$-term \cite{R75} or $F$-term \cite{FI74} supersymmetry breaking. But phenomenologically such schemes are disfavoured, for they require that `in each family at least one slepton/squark is lighter than the corresponding fermion' \cite[\S 9.1]{DGR04}.

	Alternatively, supersymmetry can be broken \emph{explicitly} by means of a supersymmetry breaking Lagrangian. In order for the solution to the hierarchy problem that supersymmetry provides to remain useful, the terms in this supersymmetry breaking Lagrangian should be \emph{soft} \cite{GD81}. This means that such terms have couplings of positive mass dimension, not yielding the quadratically divergent loop corrections that would spoil the solution to the hierarchy problem (the enormous sensitivity of the Higgs boson mass to perturbative corrections) that supersymmetry provides.

\section{Soft supersymmetry breaking}\label{sec:Lsoft}

Consider a simple gauge group $G$, a set of scalar fields $\{\sfer_\alpha, \alpha = 1, \ldots, N\}$, all in a representation of $G$, and gauginos $\gau{} = \gau{a} T^a$, with $T^a$ the generators of $G$. Then the most general renormalizable Lagrangian that breaks supersymmetry softly is given \cite{GG81} by
\begin{align}
	\mathcal{L}_{\mathrm{soft}} &= - \sfer_\alpha^*(m^2)_{\alpha\beta} \sfer_\beta + \bigg(\frac{1}{3!} A_{\alpha\beta\gamma} \sfer_\alpha\sfer_\beta\sfer_\gamma - \frac{1}{2}B_{\alpha\beta}\sfer_\alpha\sfer_\beta + C_\alpha\sfer_\alpha + h.c.\bigg)\nonumber \\ &\qquad 
	- \frac{1}{2}(M\gau{a} \gau{a} + h.c.)\label{eq:LsoftM},
\end{align}
where the combinations of fields should be such that each term is gauge invariant. This expression contains the following terms:
\begin{itemize}
	\item mass terms for the scalar bosons $\sfer_\alpha$. For the action to be real, the matrix $m^2$ should be self-adjoint;
	\item trilinear couplings, proportional to a symmetric tensor $A_{\alpha\beta\gamma}$ of mass dimension $1$;
	\item bilinear scalar interactions via a matrix $B_{\alpha\beta}$ of mass dimension two;
	\item for gauge singlets there can be linear couplings, with $C_\alpha \in \mathbb{C}$ having mass dimension three;
	\item gaugino mass terms, with $M \in \com$.
\end{itemize}

It is important to note that the Lagrangian \eqref{eq:LsoftM} corresponds to a theory that is defined on a Minkowskian background. Performing a Wick transformation $t \to i\tau$ for the time variable to translate it to a theory on a Euclidean background, changes all the signs in \eqref{eq:LsoftM}:
\begin{align}
	\mathcal{L}_{\mathrm{soft}}^{\mathrm{E}} &=  \sfer_\alpha^*(m^2)_{\alpha\beta} \sfer_\beta - \bigg(\frac{1}{3!} A_{\alpha\beta\gamma} \sfer_\alpha\sfer_\beta\sfer_\gamma - \frac{1}{2}B_{\alpha\beta}\sfer_\alpha\sfer_\beta + C_\alpha\sfer_\alpha + h.c.\bigg)\nonumber \\
	&\qquad + \frac{1}{2}(M\gau{a} \gau{a} + h.c.)\label{eq:LsoftE}.
\end{align}
This expression can easily be extended to the case of a direct product of simple groups, but its main purpose is to give an idea of what soft supersymmetry breaking terms typically look like.   

\section{Soft supersymmetry breaking terms from the spectral action}\label{sec:terms}

As was mentioned at the end of Section \ref{sec:gauge_action}, we have to settle with the terms in the action that the spectral action principle provides us. The question at hand is thus whether noncommutative geometry can give us terms needed to break the supersymmetry. In Chapter \ref{ch:sst} we have disregarded the second to last term ($\propto \Lambda^2$) in the expansion \eqref{eq:spectral_action_acg_flat} of the spectral action. Here we \emph{will} take this term into account.\footnote{As in Chapter \ref{ch:sst}, but on contrast to Chapter \ref{ch:constraints}, we will assume the background $M$ to be flat. This excludes (possible) supersymmetry breaking interactions from the term $\propto R\tr_F\Phi^2$ in the action \eqref{eq:spectral_action_acg}.}

In the following sections we will check for each of the terms in \eqref{eq:LsoftE} if it can also occur in our action \eqref{eq:totalaction} (with \eqref{eq:spectral_action_acg_flat} for the expansion of the spectral action) in the context of the building blocks. We will denote scalar fields generically by $\sfer_{ij} \in C^{\infty}(M, \rep{i}{j})$, fermions by $\fer{ij} \in L^2(M, S\otimes \rep{i}{j})$ and gauginos by $\gau{i} \in L^2(M, S \otimes M_{N_i}(\com))$, with $M_{N_i}(\com) \to su(N_i)$ after reducing the gaugino degrees of freedom (Section \ref{sec:equalizing}).

\subsection{Scalar masses (e.g.~Higgs masses)}\label{sec:breaking_scalar_mass}

\begin{figure}[ht]
	\begin{center}
		\def\svgwidth{.4\textwidth}
		\subimport{\the\svgpath}{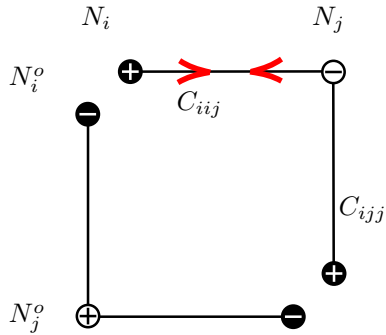}	
	\captionsetup{width=.6\textwidth}
\caption{A building block of the second type that defines a fermion--sfermion pair $(\fer{ij}, \sfer_{ij})$. Contributions to the mass term of the sfermion correspond to paths going back and forth on an edge, as is depicted on the top edge.}
	\label{fig:bb2-mass}
\end{center}
\end{figure}

Terms that describe the masses of the scalar particles such as the first term of \eqref{eq:LsoftE} are known \cite[\S 5.4]{KR97} to originate from the square of the finite Dirac operator (c.f.~\eqref{eq:spectral_action_acg_flat}). In terms of Krajewski diagrams these contributions are given by paths such as depicted in Figure \ref{fig:bb2-mass}.

Then the contribution to the action from a building block of the second type is:
\begin{align}
	- \frac{1}{2\pi^2}\Lambda^2 f_2\tr_F \Phi^2 =  - \frac{1}{2\pi^2} \Lambda^2f_2\big(4N_i|C_{iij}\sfer_{ij}|^2 + 4N_j|C_{ijj}\sfer_{ij}|^2\big)\label{eq:scalar-mass-term}
\end{align}
where $N_{i,j}$ are the dimensions of the representations $\mathbf{N_{i,j}}$ and $\sfer_{ij}$ is the field that is generated by the components of $D_F$ parametrized by $C_{iij}$ and $C_{ijj}$. Their expression depends on which building blocks are present in the spectral triple. 

In the case that there is a building block \B{ijk} of the third type present (parametrized by ---say--- $\yuk{i}{j}$, $\yuk{i}{k}$ and $\yuk{j}{k}$ acting on family-space),  we can both get the correct fermion--sfermion--gaugino interaction and a normalized kinetic term for the sfermion $\sfer_{ij}$ by on the one hand setting $C_{iij}$ and $C_{ijj}$ according to \eqref{eq:bb3-expressionCs-sqrt}.  
On the other hand we scale the sfermion according to \eqref{eq:bb3-scalingfields} with $\n_{ij}$ in that expression given by \eqref{eq:bb3-exprN}. 
There is an extra contribution from $\tr_F \Phi^2$ to $|\sfer_{ij}|^2$ compared to that of the building block of the second type. This contribution corresponds to paths going back and forth over the rightmost and bottommost edges in Figure \ref{fig:bb3}. In the parametrizations \eqref{eq:bb3-expressionCs-sqrt} and upon scaling according to \eqref{eq:bb3-scalingfields} these together yield 
\ba
& - \frac{1}{2\pi^2} \Lambda^2f_2\Big(4N_i|C_{iij}\sfer_{ij}|^2 + 4N_j|C_{ijj}\sfer_{ij}|^2 + 4N_k|\yuk{i}{j}\sfer_{ij}|^2\Big) \nn\\
&\qquad\to - 4 \Lambda^2\frac{f_2}{f(0)}|\sfer_{ij}|^2 \label{eq:scalar-mass-term3},
\ea
and similar expressions for $|\sfer_{ik}|^2$ and $|\sfer_{jk}|^2$. Interestingly, the pre-factor for this contribution is universal, i.e.~it is completely independent from the representation \rep{i}{j} the scalar resides in.

Note that, for $\Lambda \in \mathbb{R}$ and $f(x)$ a positive function (as is required for the spectral action) in both cases the scalar mass contributions are of the wrong sign, i.e.~they have the same sign as a Higgs-type scalar potential would have. The result would be a theory whose gauge group is broken maximally. We will see that, perhaps counterintuitively, we can escape this by adding gaugino-masses.

	\subsection{Gaugino masses}\label{sec:breaking_gaugino}

Having a building block of the first type, that consists of two copies of $M_N(\com)$ for a particular value of $N$, allows us to define a finite Dirac operator whose two components map between these copies, since both are of opposite grading. On the basis $\H_F = M_N(\com)_L \oplus M_N(\com)_R$ this is written as
\begin{align*}
	D_F = \begin{pmatrix} 0 & G \\ G^* & 0 \end{pmatrix},\qquad G : M_{N}(\com)_R \to M_{N}(\com)_L,
\end{align*}
since it needs to be self-adjoint. This form for $D_F$ automatically satisfies the order one condition \eqref{eq:order_one} and the demand $JD = DJ$ (see \eqref{eq:JD-anti-comm}) translates into $G = JG^*J^*$. If we want this to be a genuine mass term it should not generate any scalar field via its inner fluctuations. For this $G$ must be a multiple of the identity and consequently we write $G = M \id_{N}$, $M \in \com$. This particular pre-factor is dictated by how the term appears in \eqref{eq:LsoftE}. 

For the fermionic action we then have
\begin{align}
	\frac{1}{2}\langle J (\gau{L}, \gau{R}), \gamma^5 D_F(\gau{L}, \gau{R})\rangle =  
\frac{1}{2}M\langle J_M\gau{R}, \gamma^5 \gau{R}\rangle + \frac{1}{2}\overline{M}\langle J_M\gau{L}, \gamma^5 \gau{L}\rangle, \label{eq:gaugino-mass}
\end{align}
where $(\gau{L}, \gau{R}) \in \H^+ = L^2(S_+ \otimes M_{N}(\com)_L) \oplus L^2(S_- \otimes M_{N}(\com)_R)$, with $S_{\pm}$ the space of left- resp.~right-handed spinors. This indeed describes a gaugino mass term for a theory on a Euclidean background (cf.~\cite{CCM07}, equation 4.52).

\begin{figure}[ht]
\begin{center}
		\def\svgwidth{.4\textwidth}
		\subimport{\the\svgpath}{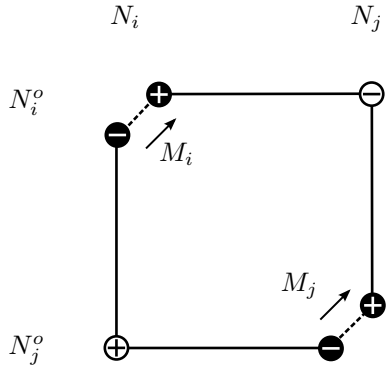}
\captionsetup{width=.9\textwidth}
\caption{A building block of the second type that defines a fermion--sfermion pair $(\fer{ij}, \sfer_{ij})$, dressed with mass terms for the corresponding gauginos (dashed edges, labeled by $M_{i,j}$).}
\label{fig:gaugMass}
\end{center}
\end{figure}

A gaugino mass term in combination with building blocks of the second type (for which two gaugino pairs are required), gives extra contributions to the spectral action. From the set up as is depicted in Figure \ref{fig:gaugMass}, one can see that $\tr D_F^4$ receives extra contributions coming from paths that traverse two edges representing a gaugino mass and two representing the scalar $\sfer_{ij}$. In detail, the extra contributions are given by:
\begin{align}
	  \frac{f(0)}{8\pi^2} \tr_F \Phi^4 &= \frac{f(0)}{\pi^2}\big(N_i|M_i|^2|C_{iij}\sfer_{ij}|^2 + N_j|M_j|^2|C_{ijj}\sfer_{ij}|^2\big)\nonumber\\
	 						&\to 2\Big(r_iN_i|M_i|^2 + r_jN_j|M_j|^2\Big)|\sfer_{ij}|^2. 
\label{eq:scalar-mass-gaugino}
\end{align}
upon scaling the fields.
 
This means that there is an extra contribution to the scalar mass terms, that is of opposite sign (i.e.~positive) as compared to the one from the previous section. When
\bas
	2r_iN_i|M_i|^2 + 2r_jN_j|M_j|^2 > 4\frac{f_2}{f(0)}\Lambda^2,
\eas 
then the mass terms of the sfermions have the correct sign, averting the problem of a maximally broken gauge group that was mentioned in the previous section. Comparing this with the expression for the Higgs mass(es) raises interesting questions about the physical interpretation of this result. In particular, if we would require the mass terms of the sfermions and Higgs boson(s) to have the correct sign already at the scale $\Lambda$ on which we perform the expansion of the spectral action, this seems to suggest that at least some gaugino masses must be very large.

Note that a gauge singlet $\fer{\mathrm{sin}} \in L^2(M, S\otimes \repl{1}{1'})$ (such as the right-handed neutrino) can be dressed with a Majorana mass matrix $\maj$ in family space (see \cite[\S 2.6]{CCM07} and Figure \ref{fig:singlet}). This yields extra supersymmetry breaking contributions: 
\begin{align}
	&\frac{f(0)}{8\pi^2}\tr \Big[4(C_{111'}\sfer_{\mathrm{sin}})\overline{\maj}(C_{111'}\sfer_{\mathrm{sin}})\overline{M} \nn\\
	&\qquad\qquad + 4(C_{11'1'}\sfer_{\mathrm{sin}})\overline{\maj}(C_{11'1'}\sfer_{\mathrm{sin}})\overline{M'}\Big] + h.c.  \nn\\
	&\qquad \to r_1(\overline{M} + \overline{M'})\tr \overline{\maj}\sfer_{\mathrm{sin}}^2 + h.c. \label{eq:singletterm}
\end{align}
where $M$ and $M'$ denote the gaugino masses of the two one-dimensional building blocks \B{1}, \B{1'} of the first type respectively and the trace is over family space. This expression is independent of whether there are building bocks of the third type present. 

Note furthermore that the gaugino masses do not give rise to mass terms for the gauge bosons. In the spectral action such terms could come from an expression featuring both $D_A = i \gamma^\mu D_\mu $ and $D_F$ twice. We do have such a term in \eqref{eq:spectral_action_acg_flat} but since it appears with a commutator between the two and since we demanded the gaugino masses to be a multiple of the identity in $M_N(\com)$, such terms vanish automatically. (In contrast, the Higgs boson does generate mass terms for the $W^{\pm}$- and $Z$-bosons, partly since the Higgs is not in the adjoint representation.)

\subsection{Linear couplings} \label{sec:breaking_lin}

The fourth term of \eqref{eq:LsoftE} can only occur for a gauge singlet, i.e.~the representation \repl{1}{1} (or, quite similarly, the representation \repl{\overline{1}}{\overline{1}}). The only situation in which such a term can arise is with a building block of the second type --- defining a fermion--sfermion pair $(\fer{\mathrm{sin}}, \sfer_{\mathrm{sin}})$ and their antiparticles (see Figure \ref{fig:singlet}). Moreover in this case a Majorana mass $\maj$ is possible, that does not generate a new field.

\begin{figure}[ht]
\begin{center}
		\def\svgwidth{.4\textwidth}
		\captionsetup{width=.9\textwidth}
		\subimport{\the\svgpath}{bb2-maj.pdf_tex}	
\caption{A building block of the second type that defines a gauge singlet fermion--sfermion pair $(\fer{\mathrm{sin}}, \sfer_{\mathrm{sin}})$. Moreover, a Majorana mass term $\maj$ is possible.}
\label{fig:singlet}
\end{center}
\end{figure}

Any such term in the spectral action must originate from a path in this Krajewski diagram consisting of either two or four steps (corresponding to the second and fourth power of the Dirac operator), ending at the same vertex at which it started (if it is to contribute to the trace) and traversing an edge labeled by $\sfer_{\mathrm{sin}}$ only once. From the diagram one readily checks that such a contribution cannot exist.

	\subsection{Bilinear couplings}\label{sec:breaking_bilin}

\begin{figure}[ht]
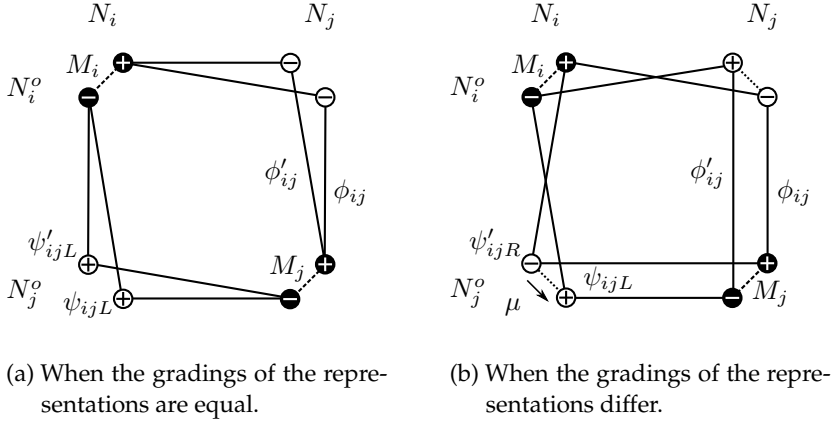

	\centering
	\begin{subfigure}{.40\textwidth}
		\centering
		\def\svgwidth{\textwidth}
		\subimport{\the\svgpath}{bilin1.pdf_tex}
		\caption{When the gradings of the representations are equal.}
		\label{fig:bilin1}
	\end{subfigure}
	\hspace{20pt}
	\begin{subfigure}{.40\textwidth}
		\centering
		\def\svgwidth{\textwidth}
		\subimport{\the\svgpath}{bilin2.pdf_tex}
		\caption{When the gradings of the representations differ.}
		\label{fig:bilin2}
	\end{subfigure}
		\captionsetup{width=.9\textwidth}
		\caption{Two building blocks of the second type defining two fermion--sfermion pairs $(\fer{ij}, \sfer_{ij})$ and $(\fer{ij}', \sfer_{ij}')$ in the same representation.} 
		\label{fig:bilin}	
\end{figure}

If a bilinear coupling (such as the third term in \eqref{eq:LsoftE}) is to be a gauge singlet, the two fields $\sfer_{ij}$ and $\sfer_{ij}'$ appearing in the expression should have opposite finite representations, e.g.~$\sfer_{ij} \in C^\infty(M, \rep{i}{j})$, $\sfer_{ij}' \in C^\infty(M, \rep{j}{i})$. We will rename $\sfer_{ij}' \to \asfer_{ij}'$ for consistency with Section \ref{sec:bb5}. The building blocks of the second type by which they are defined are depicted in Figure \ref{fig:bilin}.

The gradings of both representations are either the same (left image of Figure \ref{fig:bilin}), or they are of opposite eigenvalue (the right image). A contribution to the action that resembles the third term in \eqref{eq:LsoftE} needs to come from paths in the Krajewski diagram of Figure \ref{fig:bilin} consisting of either two or four steps, ending in the same point as where they started and traversing an edge labeled by $\sfer_{ij}$ and $\sfer_{ij}'$ only once.

One can easily check that in the left image of Figure \ref{fig:bilin} no such paths exist. In the second case (right image of Figure \ref{fig:bilin}), however, there arises the possibility of a component $\mu$ of the finite Dirac operator that maps between the vertices labeled by $\fer{ij}$ and $\fer{ij}'$ (and consequently also between $\afer{ij}$ and $\afer{ij}'$). This corresponds to a building block of the fifth type (Section \ref{sec:bb5}). There is a contribution to the action (via $\tr D_F^4$) that comes from loops traversing both an edge representing a gaugino mass and one representing $\mu$. If the component $\mu$ is parameterized by a complex number, then the contribution is 
\begin{align}	
	& \frac{f(0)}{8\pi^2} \big(8 N_i\tr M_i\asfer_{ij}C_{iij}^*\mu C_{iij}'\sfer_{ij}' + 8N_j\tr M_j \asfer_{ij}C_{ijj}^* \mu C_{ijj}'\sfer_{ij}'\big) + h.c.\nn\\
	&\qquad \to 2 \big(r_iN_iM_i + r_jN_jM_j\big)\mu \tr\asfer_{ij}\sfer_{ij}' + h.c.\label{eq:bilinear},
\end{align}
where the traces are over $\mathbf{N}_j^{\oplus M}$, with $M$ the number of copies of \rep{i}{j}. This indeed yields a bilinear term such as the third one of \eqref{eq:LsoftE}. 

\subsection{Trilinear couplings}\label{sec:breaking_trilin}

	Trilinear terms such as the second term of \eqref{eq:LsoftE} might appear in the spectral action. For that we need three fields $\sfer_{ij} \in C^{\infty}(M, \rep{i}{j})$, $\sfer_{jk} \in C^{\infty}(M, \rep{j}{k})$ and $\sfer_{ik} \in C^{\infty}(M, \rep{i}{k})$, generated by the finite Dirac operator. Such a term can only arise from the fourth power of the finite Dirac operator\footnote{Here we assume that each component of the finite Dirac operator generates only a single field, instead of ---say--- two composite ones.} which is visualized by paths in the Krajewski diagram consisting of four steps, three of which correspond to a component that generates a scalar field, the other one must be a term that does not generate inner fluctuations, e.g.~a mass term. Non-gaugino fermion mass terms were already covered in Sections \ref{sec:bb4} and \ref{sec:bb5} and were seen to generate potentially supersymmetric trilinear interactions, so the mass term must be a gaugino mass.

\begin{figure}[ht]
\begin{center}
		\def\svgwidth{.5\textwidth}
		\subimport{\the\svgpath}{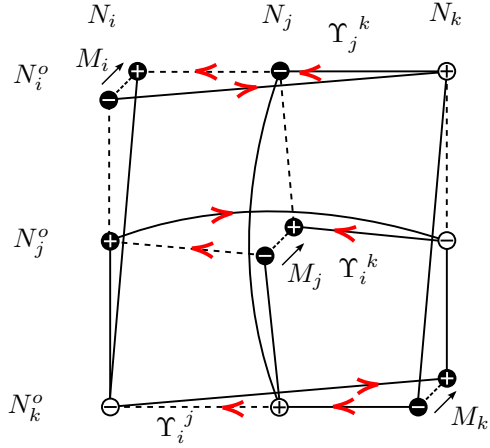}	
	\captionsetup{width=.9\textwidth}
\caption{A situation in which there are three building blocks $\mathcal{B}_{i,j,k}$ of the first type (black vertices), three building blocks $\mathcal{B}_{ij,jk,ik}$ of the second type and a building block $\mathcal{B}_{ijk}$ of the third type. Adding gaugino masses (dashed edges) gives rise to trilinear interactions, corresponding to the paths in the diagram marked by arrows.}
\label{fig:trilinear}
\end{center}
\end{figure}

If the component of the finite Dirac operator that does not generate a field is a gaugino mass term (mapping between ---say--- $M_{N_i}(\com)_R$ and $M_{N_i}(\com)_L$), then two of the three components that do generate a field must come from building blocks of the second type, since they are the only ones connecting to the adjoint representations. If we denote the non-adjoint representations from these building blocks by \rep{i}{j} and \rep{i}{k} then we can only get a contribution to $\tr D_F^4$ if there is a component of $D_F$ connecting these two representations. If $\mathbf{N_j} = \mathbf{N_k}$, such a component could yield a mass term for the fermion in the representation \rep{i}{j}, and we revert to the previous section. If $\mathbf{N_j} \ne \mathbf{N_k}$ then the remaining component of $D_F$ must be part of a building block of the third type, namely $\mathcal{B}_{ijk}$. This situation is depicted in Figure~\ref{fig:trilinear}. It gives rise to three different trilinear interactions corresponding to the paths labeled by arrows in the figure. Each of these three paths actually represents four contributions: one can traverse each path in the opposite direction, and for each path one can reflect it around the diagonal, giving another path with the same contribution to the action. 

Calculating the spectral action we get for each building block $\mathcal{B}_{ijk}$ of the third type the contributions
\begin{align}
	&\frac{f(0)}{\pi^2}\Big(N_i\overline{M_i}\tr \yuk{j}{k}\sfer_{jk}\asfer_{ik}C_{iik}^*C_{iij}\sfer_{ij} + N_j\overline{M_j}\tr C_{jjk}\sfer_{jk}\asfer_{ik}\yuk{i}{k}C_{ijj}\sfer_{ij} \nn\\
	&\qquad\qquad + N_k\overline{M_k}\tr C_{jkk}\sfer_{jk}\asfer_{ik}C_{ikk}^*\yuk{i}{j}\sfer_{ij}\Big) + h.c.\label{eq:trilinear_prior}
\end{align}
where all traces are over $\mathbf{N}_j^{\oplus M}$. A careful analysis of the demand for supersymmetry in this context (see Section \ref{sec:bb3}) requires the parameters $\yuk{i}{j}$, $\yuk{i}{k}$ and $\yuk{j}{k}$ to be related via \eqref{eq:improvedUpsilons1}, 
where $C_{iij}$ and $C_{ijj}$ act trivially on family space if $\sfer_{ij}$ is assumed to have $R =1$. From this relation we have deduced in Remark \ref{rmk:bb3-relativesigns} that $s_{ij}s_{ik}s_{jk} = -1$ for the product of the three signs $s_{ij} := \sgnc_{i,j}\sgnc_{j,i}$, $s_{ik}$, $s_{jk}$. If we replace $C_{iik} \to C_{ikk}$, $C_{iij} \to C_{ijj}$, $C_{jjk} \to C_{jkk}$ and $C_{ijj} \to C_{iij}$ in the first two terms of \eqref{eq:trilinear_prior} using \eqref{eq:bb3-expressionCs-sqrt}, employ \eqref{eq:improvedUpsilons1}, then \eqref{eq:trilinear_prior} can be written as
\bas
&\frac{f(0)}{\pi^2}\Big(N_i\overline{M_i}\frac{r_i}{r_k} + N_j\overline{M_j}\frac{r_j}{r_k}	+ N_k\overline{M_k}\Big)\tr C_{jkk}\sfer_{jk}\asfer_{ik}C_{ikk}^*\yuk{i}{j}\sfer_{ij} + h.c.
\eas
We then scale the sfermions according to \eqref{eq:bb3-scalingfields}, again using \eqref{eq:bb3-expressionCs-sqrt} for $C_{jkk}$ and $C_{ikk}^*$ to obtain 
\begin{align}
	& 2\kappa_k g_l\sqrt{2\frac{\w{ij}}{q_l}}\Big(r_iN_i\overline{M_i} + r_jN_j\overline{M_j} + r_kN_k\overline{M_k}\Big)\tr \yukw{}{}\sfer_{ij}\sfer_{jk}\asfer_{ik} + h.c.\label{eq:trilinear},
\end{align}
where we have written
\bas
	\yukw{}{} := \yuk{i}{j}(N_k\tr\yuks{i}{j}\yuk{i}{j})^{-1/2}
\eas
(cf.~\eqref{eq:def-yukw}) for the scaled version of the parameter $\yuk{i}{j}$, $\kappa_k := \sgnc_{k,j}\sgnc_{k,i}$ and the index $l$ can take any of the values that appear in the theory.



\section{Conclusion}

We have now considered all terms featuring in \eqref{eq:LsoftE}. At the same time the reader can convince himself that this exhausts all possible terms that appear via $\tr D_F^4$ and feature a gaugino mass. As for the fermionic action, a component of $D_F$ mapping between two adjoint representations can give gaugino mass terms \eqref{eq:gaugino-mass}. As for the bosonic action, any path of length two contributing to the trace and featuring a gaugino mass, cannot feature other fields. In contrast, a path of length four in a Krajewski diagram involving a gaugino mass can feature:
\begin{itemize}
	\item only that mass, as a constant term (see the comment at the end of this section);
	\item two times the scalar from a building block of the second type, when going in one direction \eqref{eq:scalar-mass-gaugino};
	\item two times the scalar from a building block of the second type, when going in two directions and when a Majorana mass is present (only possible for singlet representations, \eqref{eq:singletterm});
	\item two scalars from two different building blocks of the second type having opposite grading in combination with a building block of the fifth type \eqref{eq:bilinear}.
 	\item three scalars, partly originating from a building block of the second type and partly from one of the third type \eqref{eq:trilinear}.
\end{itemize}
Furthermore, via $\tr D_F^2$ there are contributions to the scalar masses from building blocks of the second and third type \eqref{eq:scalar-mass-term}. We can combine the main results of the previous sections into the following theorem.
\begin{theorem}
All possible terms that break supersymmetry softly and that can originate from the spectral action \eqref{eq:spectral_action_acg_flat} of an almost-commutative geometry consisting of building blocks are mass terms for scalar fields and gauginos and trilinear and bilinear couplings. More precisely, the most general Lagrangian that softly breaks supersymmetry and results from almost-commutative geometries is of the form
\begin{align}
	\mathcal{L}_{\mathrm{soft}}^{\mathrm{NCG}} &= \mathcal{L}^{(1)} + \mathcal{L}^{(2)} + \mathcal{L}^{(3)}+ \mathcal{L}^{(4)} + \mathcal{L}^{(5)}\label{eq:main-result},
\end{align}
where
\begin{subequations}
\begin{align}
			\mathcal{L}^{(1)}&= \frac{1}{2}M_i \langle J_M\gau{iR}, \gamma^5\gau{iR} \rangle + \frac{1}{2}\overline{M_i} \langle J_M\gau{iL}, \gamma^5\gau{iL} \rangle \label{eq:gauginoterm}\\
	\intertext{for each building block \ensuremath{\mathcal{B}_i} of the first type,}
	\mathcal{L}^{(2)} &= 2\Big(r_iN_i|M_i|^2 + r_jN_j|M_j|^2 - 2\frac{f_2}{f(0)}\Lambda^2\Big)|\sfer_{ij}|^2,\label{eq:massterm2}
	\intertext{for each building block \ensuremath{\mathcal{B}_{ij}} of the second type for which there is at least one building block \B{ijk} of the third type present (knowing that a single \B{ij} cannot be supersymmetric by itself, Section \ref{sec:bb2}),}
	\mathcal{L}^{(3)} &= 
	  2\kappa_k g_l\sqrt{2\frac{\w{ij}}{q_l}}\Big(r_iN_i\overline{M_i} + r_jN_j\overline{M_j} + r_kN_k\overline{M_k}\Big)\tr \yukw{}{}\sfer_{ij}\sfer_{jk}\asfer_{ik} + h.c.
\label{eq:trilinterm},
	\intertext{for each building block \ensuremath{\mathcal{B}_{ijk}} of the third type,}
\mathcal{L}^{(4)} &= r_1(\overline{M} + \overline{M'})\tr \overline{\maj}\sfer_{\mathrm{sin}}^2+ h.c. \label{eq:majterm}
	\intertext{for each building block \ensuremath{\mathcal{B}_{\mathrm{maj}}} of the fourth type (with the trace over a possible family index), and}
	\mathcal{L}^{(5)} &= 2(r_iN_iM_{i} + r_jN_jM_{j})\mu \tr\asfer_{ij}\sfer_{ij}' + h.c.\label{eq:bilinterm}
\end{align}
\end{subequations}
for each building block $\mathcal{B}_{\mathrm{mass}}$ of the fifth type.
\end{theorem}
It should be remarked that the building blocks of the fourth and fifth type typically already provide soft breaking terms of their own (see Section \ref{sec:mass_terms}).

Interestingly, all supersymmetry breaking interactions that occur are seen to be generated by the gaugino masses (except the ones coming from the trace of the square of the finite Dirac operator) and each of them can be associated to one of the five supersymmetric building blocks. Note that the gaugino masses give rise to extra contributions that are not listed in \eqref{eq:main-result}. For each gaugino mass $M_i$ there is an additional contribution 
\begin{align*}
\mathcal{L}_{M_i} &= \frac{f(0)}{4\pi^2}|M_i|^4 - \frac{f_2}{\pi^2}\Lambda^2|M_i|^2.
\end{align*}
Since such contributions do not contain fields, they are not breaking supersymmetry, but might nonetheless be interesting from a gravitational perspective.

\svgpath={./gfx/}

\chapter[The noncommutative MSSM (NCMSSM)]{The noncommutative MSSM (NCMSSM)\NoCaseChange{\footnote{The contents of this chapter are based on \cite{BS13III}.}}}\label{ch:NCMSSM}



In this chapter we turn our attention to the minimal supersymmetric Standard Model (MSSM, see Section \ref{sec:intro_MSSM}), phenomenologically the most important example of ($N = 1$) supersymmetry. 
We apply the formalism that was developed in Chapter \ref{ch:sst} to explore the possibilities for obtaining the particle content and action of the MSSM. In that chapter we already saw that the building blocks do not automatically imply that the corresponding action is also supersymmetric. We will shortly review the possible obstacles for a supersymmetric action that we have come across. These are the following:

\begin{itemize}


\item the three obstructions from Remarks \ref{rmk:bb1-obstr} and \ref{rmk:bb2-obstr} and Proposition \ref{prop:2bb2-obstr} concerning the set up of the almost-commutative geometry. The first excludes a finite algebra that is equal to $\com$ with the corresponding building block \B{1}, since it lacks gauge interactions and thus cannot be supersymmetric. The second excludes a finite algebra consisting of two summands that are both matrix algebras over $\com$ in the presence of only building blocks of the second type whose off-diagonal representations in the Hilbert space have $R$-parity equal to $-1$. The third obstruction says that for an algebra consisting of three or more summands $M_{N_{i,j,k}}(\com)$ we cannot have two building blocks \B{ij} and \B{ik} of the second type that share one of their indices. To avoid this obstruction, we can maximally have two components of the algebra that are a matrix algebra over $\com$.
%


\item to obtain the fermion--sfermion--gaugino interactions needed for a supersymmetric action, the parameters $C_{iij}$ and $C_{ijj}$ of the finite Dirac operator associated to a building block \B{ij} of the second type ---that read $\Cw{i,j}$ and $\Cw{j,i}$ after normalizing the kinetic terms of the sfermions--- should satisfy \eqref{eq:bb2-resultCiij}. 
In that condition $\sgnc_{i,j}$ and $\sgnc_{j,i}$ are signs that we are free to choose, and the $\K_{i,j}$ are the pre-factors of the kinetic terms of the gauge bosons that correspond to the building blocks \B{i,j} of the first type and should be set to $1$ to give normalized kinetic terms (the consequences of this will be reviewed at the end of Section \ref{sec:identification}). Similarly, when a building block \B{ijk} of the third type is present, its fermionic interactions can only be part of a supersymmetric action if the parameters $\yuk{i}{j}$, $\yuk{i}{k}$ and $\yuk{j}{k}$ of the finite Dirac operator satisfy \eqref{eq:improvedUpsilons} when \rep{i}{j} in $\H_F$ has $R = -1$ or \eqref{eq:improvedUpsilons2} when it is \rep{i}{k} or \rep{j}{k}. 
For any building block of the third type it is necessary that either one or all three representations \rep{i}{j}, \rep{i}{k} and \rep{j}{k} in the Hilbert space have $R$-parity $-1$.


\item for the four-scalar interactions to have an off shell counterpart that satisfies the constraints supersymmetry puts on them, the coefficients of the interactions with the auxiliary fields $G_{i}$, $H$ and $F_{ij}$ should satisfy the demands listed in Section \ref{sec:4s-aux}.

\end{itemize}
For each almost-commutative geometry that one defines in terms of the building blocks, we should explicitly check that the obstructions are avoided and the appropriate demands are satisfied.


In the next section we will list the basic properties of the almost-commutative geometry that is to give the MSSM, including the building blocks it consists of and show that this set up avoids the three possible obstructions from the first item in the list above. To confirm that we are on the right track we identify all MSSM particles and examine their properties in Section \ref{sec:identification}. Finally, in Section \ref{sec:ncmssm-checks} we will confront our model with the demands from Section \ref{sec:4s-aux}. Throughout this chapter, we will a priori allow for a number of generations other than $3$.


\section{The building blocks of the MSSM}\label{sec:NCSSM}
We start by listing the properties of the finite spectral triple that, when part of an almost-commutative geometry, should correspond to the MSSM.

\begin{figure}[ht!]
	\centering
	\begin{subfigure}{.4\textwidth}
		\centering
		\def\svgwidth{\textwidth}
		\subimport{\the\svgpath}{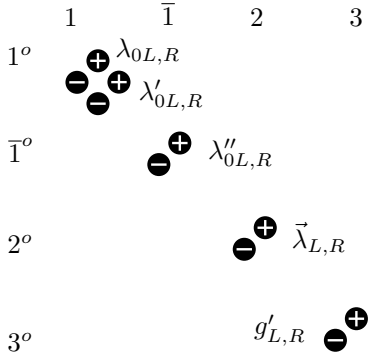}
		\caption{Blocks of the first type.\\\ \\\ }
		\label{fig:MSSM_bb1_compact}
	\end{subfigure}
	\hspace{30pt}
\begin{subfigure}{.4\textwidth}
		\centering
		\def\svgwidth{\textwidth}
		\subimport{\the\svgpath}{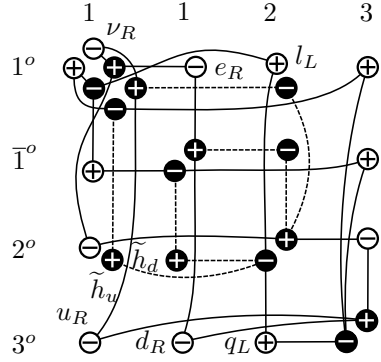}
		\caption{Blocks of the second type. Each white off-diagonal node corresponds to a SM (anti)particle.}
		\label{fig:MSSM_bb2_compact}
	\end{subfigure}
	\\[10pt]
	\begin{subfigure}{.4\textwidth}
		\centering
		\def\svgwidth{\textwidth}
		\subimport{\the\svgpath}{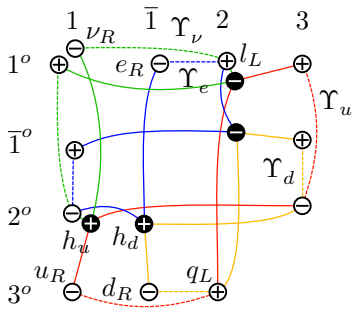}
		\caption{Blocks of the third type, parametrized by the Yukawa matrices $\Upsilon_{\nu, e, u, d}$.\\\ }
		\label{fig:MSSM_bb3_compact}
	\end{subfigure}
	\hspace{30pt}
\begin{subfigure}{.4\textwidth}
		\centering
		\def\svgwidth{\textwidth}
		\subimport{\the\svgpath}{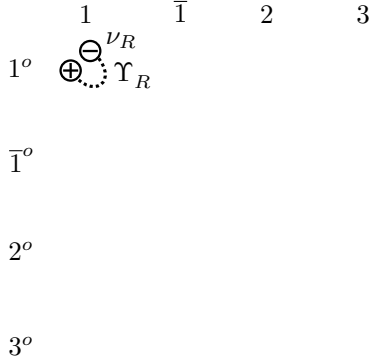}
		\caption{The block of the fourth type, representing a Majorana mass for the right-handed neutrino.}
		\label{fig:MSSM_bb4_compact}
	\end{subfigure}
\captionsetup{width=.9\textwidth}
\caption{All building blocks that together represent the particle content and interactions of the MSSM.}
\label{fig:MSSM-bbs}
\end{figure}

\begin{enumerate}


\item The gauge group of the MSSM is (up to a finite group) the same as that of the SM. In noncommutative geometry there is a strong connection between the algebra $\A$ of the almost-commutative geometry and the gauge group $\mathcal{G}$ of the corresponding theory. There is more than one algebra that may yield the correct gauge group (Lemma \ref{lem:possible_algs}) but any supersymmetric extension of the SM also contains the SM particles, which requires an algebra that has the right representations (see just below the aforementioned Lemma). This motivates us to take the Standard Model algebra:
	\begin{align}
		\A_F \equiv \A_{SM} = \com \oplus \mathbb{H} \oplus M_3(\com).\label{eq:ncmssm-alg}
	\end{align}
Note that with this choice we already avoid the third obstruction for a supersymmetric theory from the first item in the list above, since only two of the summands of this algebra are defined over $\com$.

In the derivation \cite{CCM07} of the SM from noncommutative geometry the authors first start with the `proto-algebra'
\ba
	\A_{L,R} = \com \oplus \mathbb{H}_L \oplus \mathbb{H}_R \oplus M_3(\com)\label{eq:LRalgebra}
\ea
(cf.~\cite[\S2.1]{CCM07}) that breaks into the algebra above after allowing for a Majorana mass for the right-handed neutrino \cite[\S 2.4]{CCM07}. Although we do not follow this approach here, we do mention that this algebra avoids the same obstruction too.

\item As is the case in the NCSM, we allow four inequivalent representations of the components of \eqref{eq:ncmssm-alg}: $\mathbf{1}$, $\overline{\mathbf{1}}$, $\mathbf{2}$ and $\mathbf{3}$. Here $\overline{\mathbf{1}}$ denotes the real-linear representation $\pi(\lambda)v = \bar\lambda v$, for $ v\in\overline{\mathbf{1}}$.\footnote{Keep in mind that we ensure the Hilbert space being complex by defining it as a bimodule of the complexification $\A^{\com}$ of $\A$, rather than of $\A$ itself \cite{CC08}.} This results in only three independent forces ---with coupling constants $g_1$, $g_2$ and $g_3$--- since the inner fluctuations of the canonical Dirac operator acting on the representations $\mathbf{1}$ and $\overline{\mathbf{1}}$ of $\com$ are seen to generate only a single $u(1)$ gauge field \cite[\S 3.5.2]{CCM07} (see also Section \ref{sec:ncmssm_unimod}).

\item If we want a theory that contains the superpartners of the gauge bosons, we need to define the appropriate building blocks of the first type (cf.~Section \ref{sec:bb1}). In addition, we need these building blocks to define the superpartners of the various Standard Model particles. We introduce
 \ba \label{eq:mssm-bb1s}
\B{1},\quad \B{1_R},\quad \B{\bar 1_R},\quad \B{2_L},\quad \B{3},
\ea
whose representations in $\H_F$ all have $R = -1$ to ensure that the gauginos and gauge bosons are of opposite R-parity. The Krajewski diagram that corresponds to these building blocks is given in Figure \ref{fig:MSSM_bb1_compact}. For reasons that will become clear later on, we have two building blocks featuring the representation $\mathbf{1}$, and one featuring $\overline{\mathbf{1}}$. We distinguish the first two by giving one a subscript $R$. This notation is not related to $R$-parity but instead is inspired by the derivation of the Standard Model where, in terms of the proto-algebra \eqref{eq:LRalgebra}, the component $\com$ is embedded in the component $\mathbb{H}_R$ via $\lambda \to \diag(\lambda, \bar\lambda)$. The initially two-dimensional representation $\mathbf{2}_R$ of this component (making the right-handed leptons and quarks doublets) thus breaks up into two one-dimensional representations $\mathbf{1}_R$ and $\mathbf{\bar 1}_R$ (corresponding to right-handed singlets).

At this point we thus have too many fermionic degrees of freedom, but these will be naturally identified to each other in Section \ref{sec:identification}.

\item For each of the Standard Model fermions\footnote{In the strict sense the Standard Model does not feature a right handed neutrino (nor does the MSSM), but allows for extensions that do. On the other hand the more recent derivations of the SM from noncommutative geometry naturally come with a right-handed neutrino. We will incorporate it from the outset, always having the possibility to discard it should we need to.} we define the corresponding building block of the second type:
	\begin{subequations}\label{eq:mssm-bb2s}
	\begin{align}
		\Bc{1_R1}{-} &: (\nu_R, \widetilde{\nu}_R), & 	\Bc{\bar 1_R1}{-} &: (e_R, \widetilde{e}_R), & \Bc{2_L1}{+} &: (l_L, \widetilde{l}_L), \\
		\Bc{1_R3}{-} &: (u_R, \widetilde{u}_R), & 	\Bc{\bar 1_R3}{-} &: (d_R, \widetilde{d}_R), & \Bc{2_L3}{+} &: (q_L, \widetilde{q}_L).
\intertext{Of each of the representations in the finite Hilbert space we will take $M$ copies representing the $M$ generations of particles, also leading to $M$ copies of the sfermions. We can always take $M = 3$ in particular. Each of these fermions has $R = +1$. We do the same for representations in which the SM Higgs resides:}
		\B{1_R2_L} &: (h_u, \widetilde{h}_u), & 	\B{\bar 1_R2_L} &: (h_d, \widetilde{h}_d), &  &
	\end{align} 
\end{subequations}
save that their representations in the Hilbert space have $R = -1$ and consequently we take only one copy of both. For the two Higgs/higgsino building blocks we can choose the grading still. We will set them both to be left-handed and justify that choice later. 

The Krajewski diagram that corresponds to these building blocks is given by Figure \ref{fig:MSSM_bb2_compact}. 

	The fact that there is at least one building block \B{1j}, $j = \bar 1_R, 2_L, 3$, avoids the first of the three obstructions for a supersymmetric theory mentioned in the first item of the list above. 

The building blocks introduced above fully determine the finite Hilbert space. For concreteness, it is given by
\ba
	\H_F &= \H_{F, R = +} \oplus \H_{F, R = -}, \label{eq:ncmssm_H_F}
\ea
with $\H_{F, R = \pm}$ (cf.~\eqref{eq:R-parity-grading}) reading
\bas
			\H_{F, R = +} &= \big(\mathcal{E} \oplus \mathcal{E}^o\big)^{\oplus M},& \mathcal{E} &= (\mathbf{2}_L \oplus \mathbf{1}_R \oplus \overline{\mathbf{1}}_R) \otimes (\mathbf{1} \oplus \mathbf{3})^o\nn\\
			\H_{F, R = -} &= \mathcal{F} \oplus \mathcal{F}^o, &  \mathcal{F} &= (\mathbf{1} \otimes \mathbf{1}^o)^{\oplus 2}\ \oplus\ \overline{\mathbf{1}}\otimes \overline{\mathbf{1}}^o\ \oplus\ \mathbf{2} \otimes \mathbf{2}^o\ \nn\\
		&& &\qquad\oplus\ \mathbf{3} \otimes \mathbf{3}^o \oplus\ (\mathbf{1}_R \oplus \overline{\mathbf{1}}_R) \otimes \mathbf{2}^o_L.\nn
\eas
Here $\mathcal{E}$ contains the finite part of the left- and right-handed leptons and quarks. The first four terms of $\mathcal{F}$ represent the $u(1)$, $su(2)$ and $su(3)$ gauginos and the last term the higgsinos. For the (MS)SM the number of generations $M$ is equal to $3$.

\item In terms of the `proto-algebra' \eqref{eq:LRalgebra} the operator 
\bas
	R = - (+, -, -, +) \otimes (+, -, -, +)^o
\eas
gives the right values for $R$-parity to all the fermions: $R = + 1$ for all the SM-fermions, $R = -1$ for the higgsino-representations that are in $\mathbf{2}_R\otimes \mathbf{2}_L^o$ before breaking to $(\mathbf{1}_R \oplus \overline{\mathbf{1}}_R) \otimes \mathbf{2}_L^o$. 

Since there is at least one building block of the second type whose representation in the finite Hilbert space has $R = +1$, also the second obstruction for a supersymmetric theory mentioned above is avoided. 

\item The MSSM features additional interactions, such as the Yukawa couplings of fermions with the Higgs. In the superfield formalism, these are determined by a superpotential. Its counterpart in the language of noncommutative geometry is given by the building blocks \BBB{ijk} of the third type. These should at least contain the Higgs-interactions of the Standard Model (but with the distinction between up- and down-type Higgses). The values of the grading on the representations in the finite Hilbert space are such that they allow us to extend the Higgs-interactions to the following building blocks:
\begin{align}\label{eq:mssm-bb3s}
\BBB{11_R2_L},&&	\BBB{1\bar 1_R2_L},&& \BBB{1_R2_L3},&& \BBB{\bar 1_R2_L3}. 
\end{align}
The four building blocks \BBB{ijk}, are depicted in Figure \ref{fig:MSSM_bb3_compact}. (For conciseness we have omitted here the building blocks of the first type and the components of $D_F$ from the building blocks of the second type.) 

Note that all components of $D_-$, the part of $D_F$ that anticommutes with $R$, that are allowed by the principles of NCG are in fact also non-zero now. This is in contrast with those of $D_+$, on which the (ad hoc) requirement \eqref{eq:NCSM_extra_demand} (see \cite[\S 2.6]{CCM07}). 
The reason for this is to keep the photon massless and to get the interactions of the SM. Requiring the same for the entire finite Dirac operator would forbid the majority of the components that determine the sfermions, not requiring it at all would lead to extra, non-supersymmetric interactions such as $\repl{\bar1}{1} \to \repl{3}{1}$. 
Thus, we slightly change the demand, reading 
\ba
	[D_+, \com_F] = 0.\label{eq:demandDF}
\ea
Relaxing this demand does not lead to a photon mass since it only affects the sfermions that have $R = -1$ whereas any photon mass would arise from the kinetic term of the Higgses, having $R = +1$.

At this point we can justify the choice for the grading of the up- and down-type higgsinos. If the grading of any of the two would have been of opposite sign, none of the building blocks of the third type that feature that particular higgsino could have been defined. The interactions that are still possible then cannot be combined into building blocks of the third type, which is an undesirable property. It corresponds to a superpotential that is not holomorphic (see Section \ref{sec:bb3}). 


\item Having a right-handed neutrino in $\repl{1_R}{1}$, that is a singlet of the gauge group, we are allowed to add a Majorana mass for it via 
\ba\label{eq:mssm-bb4}\BBBB{11_R}\ea 
such as in Section \ref{sec:bb4}. This is represented by the dotted diagonal line in Figure \ref{fig:MSSM_bb4_compact}. The building block is parametrized by a symmetric $M\!\times\!M$--matrix $\yuk{R}{}$. 


\end{enumerate}

Summarizing things, the finite spectral triple of the almost-commutative geometry that should yield the MSSM then reads
\ba
	& \B{1} \oplus \B{1_R} \oplus \B{\bar1_R} \oplus \B{2} \oplus \B{3} \oplus \Bc{1_R2_L}{+} \oplus \Bc{\bar1_R2_L}{+}\nn\\
	&\qquad\oplus \Bc{1_R1}{-} \oplus \Bc{\bar1_R1}{-} \oplus \Bc{2_L1}{+} \oplus \Bc{1_R3}{-} \oplus \Bc{\bar1_R3}{-} \oplus \Bc{2_L3}{+}\nn\\
	&\qquad\qquad\oplus \B{11_R2_L} \oplus \B{1\bar1_R2_L} \oplus \B{1_R2_L3} \oplus \B{\bar1_R2_L3} \oplus \B{\textrm{maj}}\label{eq:ncmssm-acg}
\ea
One of its properties is that all components that are not forbidden by the principles of NCG and the additional demand \eqref{eq:demandDF} are in fact also non-zero, save for the supersymmetry-breaking gaugino masses of Chapter \ref{ch:breaking}, that we will not cover here. 

\begin{figure}[ht]
	\centering
	\centering
	\def\svgwidth{.4\textwidth}
		\subimport{\the\svgpath}{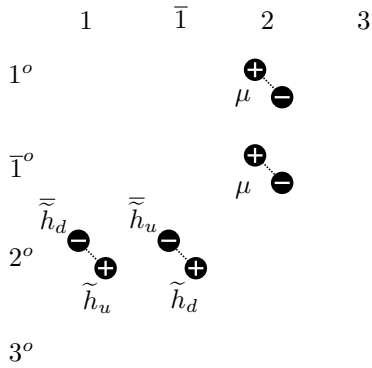}
		\captionsetup{width=.6\textwidth}
		\caption{The extra building blocks of the second type featuring a Higgs/higgsino-pair and the building blocks of the fifth type that are consequently possible.}
		\label{fig:MSSM_bb5_compact}
\end{figure}

\begin{rmk}\label{rmk:extra-higgsinos}
	Running ahead of things a bit already we note that there is an important difference with the MSSM. In the superfield-formalism there is an interaction that reads
\ba\label{eq:mu-term}
	\mu H_d \cdot H_u,
\ea
where $H_{u,d}$ represent the up-/down-type Higgs/higgsino superfields \cite[\S 8.3]{DGR04}. Suppose that \Bc{1_R2_L}{+} and \Bc{\bar1_R2_L}{+} indeed describe the up- and down-type Higgses and higgsinos. Because their vertices are on different places in the Krajewski diagram and in addition they have the same value for the grading, there is no building block of the fifth type possible that would be the equivalent of \eqref{eq:mu-term}. Moreover, in the MSSM there is a soft supersymmetry-breaking interaction
\bas
	B\mu h_d\cdot h_u + h.c.
\eas
In this framework also such an interaction can only be generated via a building block of the fifth type (in combination with gaugino masses, see Section \ref{sec:breaking_bilin}). Not having these interactions would at least leave several of the tree-level mass-eigenstates that involve the Higgses massless \cite[\S 10.3]{DGR04}. We can overcome this problem by adding two more building blocks \B{1_R2_L} and \B{\bar1_R2_L} of the second type whose values of the grading are opposite to the ones previously defined. With these values no additional components for the finite Dirac operator are possible, except for two building blocks of the fifth type that run between the representations of \Bc{1_R2_L}{\pm} and between those of \Bc{\bar1_R2_L}{\pm}. If we then identify the degrees of freedom of \Bc{1_R2_L}{+} to those of \Bc{\bar1_R2_L}{-} and those of \Bc{\bar1_R2_L}{+} to those of \Bc{1_R2_L}{-}, this would give us the interactions that correspond to the term \eqref{eq:mu-term}. The additions to the finite spectral triple \eqref{eq:ncmssm-acg} that correspond to these steps are given by
\ba\label{eq:ncmssm-acg-ext}
	\Bc{1_R2_L}{-} \oplus \Bc{\bar 1_R2_L}{-} \oplus \B{\mathrm{mass}, 1_R2_L} \oplus \B{\mathrm{mass}, \bar1_R2_L}.
\ea
This situation is depicted in Figure \ref{fig:MSSM_bb5_compact}.
\end{rmk}

We proceed by ensuring that we are indeed talking about the noncommutative counterpart of the MSSM by identifying the MSSM particles and checking that the number of fermionic and bosonic degrees of freedom are the same.

\section{Identification of particles and sparticles}\label{sec:identification}

\subsection{The gauge group and hypercharges}

To justify the nomenclature we have been using in the previous section we need to test the properties of the new particles by examining how they transform under the gauge group (e.g.~\cite[\S 7.1]{V06}). We do this by transforming elements of the finite Hilbert space and finite Dirac operator under the gauge group according to \eqref{eq:fermion_transf}, 
but using for the gauge group the definition \eqref{eq:gauge_group_new}, featuring the $R$-parity operator. 
Since we have $\H_{F, R = +} = \H_{F, SM}$, the space that describes the SM fermions, this determinant gives
\ba\label{eq:sm_gauge_group}
	SU(\A_{SM}) = \{(\lambda, q, m) \in U(1) \times SU(2) \times U(3), [\lambda\det(m)]^{4M} = 1\}.
\ea
The factor $M$ again represents the number of particle generations and stems from the fact that the algebra acts trivially on family-space. Unitary quaternions $q$ automatically have determinant $1$ and consequently all contributions to the determinant come from 
\bas
	\mathcal{E}^o = (\mathbf{1} \oplus \mathbf{3})\otimes (\mathbf{2}_L \oplus \mathbf{1}_R \oplus \overline{\mathbf{1}}_R)^o
\eas 
defined above, instead of from $\mathcal{E}$. The power $4 = 2 + 1 + 1$ above comes from the second part of the tensor product on which the unitary elements $U(\A)$ act trivially. From \eqref{eq:sm_gauge_group} we infer that the $U(1)$-part of $SU(\A_{SM})$ (the part that commutes with all other elements) is given by 
\begin{align}
	\{(\lambda, 1, \lambda^{-1/3}1_3), \lambda \in U(1)\} \subset SU(\A_{SM}).\label{eq:U1factor}
\end{align}
This part determines the hypercharges of the particles; these are given by the power with which $\lambda$ acts on the corresponding representations. This result makes the identification of the fermions that have $R = +1$ exactly the same as in the case of the SM (\cite[\S 2.5]{CCM07}).
Applying it to the gaugino and higgsino sectors of the Hilbert space, we find that:
\begin{itemize}
	\item there are the gauginos $\widetilde{g} \in \repl{3}{3}$ whose traceless part transforms as $\widetilde{g} \to  \bar v\widetilde{g} v^t$, with $\bar v\in SU(3)$ (i.e.~it is in the adjoint representation of $SU(3)$) and whose trace part transforms trivially;
	\item there are the gauginos $\widetilde{W} \in \repl{2}{2}$ whose traceless part transforms according to $\widetilde{W} \to q\widetilde{W}q^*$ with $q \in SU(2)$ (i.e.~the adjoint representation of $SU(2)$) and whose trace part transforms trivially;
	\item the higgsinos in \repl{1_R}{2_L} and \repl{\bar1_R}{2_L} transform in the representation $\mathbf{2}$ of $SU(2)$ and have hypercharge $+1$ and $-1$ respectively;
	\item the gauginos in \repl{1}{1}, \repl{2}{2} and \repl{3}{3} all have zero hypercharge.
\end{itemize}
The new scalars, parametrized by the finite Dirac operator, generically transform as $\Phi \to U\Phi U^*$. In particular, we separately consider the elements $U = uJuJ^*$ with $u = (\lambda, 1, \lambda^{-1/3}1_3)$, $(1, q, 1)$ and $(1, 1, \bar v)$. This gives the following:  
\begin{itemize}
\item with $u = (\lambda, 1, \lambda^{-1/3}1_3)$ we find for the hypercharges of the various sfermions: 
\begin{align*}
	\widetilde{q}_L&:\quad \tfrac{1}{3}, & \widetilde{u}_R &:\quad \tfrac{4}{3},& \widetilde{d}_R &:\quad -\tfrac{2}{3},\\
	\widetilde{l}_L&:\quad -1, & \widetilde{\nu}_R &:\quad 0,& \widetilde{e}_R &:\quad -2.
\end{align*}
The conjugates are found to carry the opposite charge.
\item with $u = (1, q, 1)$ we find the following sfermions that transform non-trivially: $\widetilde{q}_L$ and $\widetilde{l}_L$, each coming in $M$ generations.
\item with $u = (1, 1, \overline{v})$ we find the following sfermions that transform in the fundamental representation of $SU(3)$: $\widetilde{q}_L$, $\widetilde{u}_R$ and $\widetilde{d}_R$, each coming in $M$ generations. 
\end{itemize}

This completes the identification of the new elements in the theory with the gauginos, higgsinos and sfermions of the MSSM. 

\subsection{Unimodularity in the MSSM}\label{sec:ncmssm_unimod}

Having identified the particles there is one other thing to check; that the number of bosonic and fermionic degrees of freedom are indeed the same. We can quite easily see that at least initially this is not the case for the following reason. In order to be able to define the building blocks \Bc{\overline{1}_R1}{-}, \Bc{\overline{1}_R3}{-} and \Bc{\overline{1}_R2_L}{+} of the second type (describing the right-handed (s)electron and (s)quark and down-type Higgs/higgsino respectively), we defined the building blocks \B{\overline{1}} and \B{1_R} of the first type. Each provides extra $u(1)$ fermionic degrees of freedom, but no bosonic ones (see below). In addition, the gaugino $\widetilde{W}$ contains a trace part, whereas the corresponding gauge boson does not.

We will employ the unimodularity condition \eqref{eq:unimod_new} 
to reduce the bosonic degrees of freedom on the one hand and see what its consequences are, using the supersymmetry transformations.

First of all, we note that the inner fluctuations on the $\mathbf{1}$ and $\overline{\mathbf{1}}$ give rise to only one $u(1)$ gauge field (cf.~\cite[\S 15.4]{CCM07}). Initially there are
\bas
	\Lambda &= i\gamma^\mu\sum_j  \lambda_j\partial_\mu \lambda_j',&& \text{and} &
	\Lambda' &= i\gamma^\mu \sum_j \bar\lambda_j\partial_\mu \bar\lambda_j',
\eas
but since $\Lambda$ must be self-adjoint (as $\dirac$ is), $\Lambda_\mu = i\sum_j  \lambda_j\partial_\mu \lambda_j'$ is real-valued. Consequently $\Lambda'_\mu(x) = - \Lambda_\mu(x)$ and they indeed generate the same gauge field. But via the supersymmetry transformations this also means that 
\bas
	\delta \Lambda \propto \delta \Lambda',
\eas 
i.e.~the corresponding gauginos whose finite parts are in $\mathbf{1} \otimes \mathbf{1}^o$ and $\overline{\mathbf{1}} \otimes \overline{\mathbf{1}}^o$ should be associated to each other. 

Second, the inner fluctuations of the quaternions $\mathbb{H}$ generate an $su(2)$-valued gauge field. This can be seen as follows. The quaternions form a real algebra, spanned by $\{1_2, i\sigma^a\}$, with $\sigma^a$ the Pauli matrices. Since $\dirac$ commutes with the basis elements, the inner fluctuations
\bas
	\sum_j q_j [\dirac, q_j'],\qquad\ q_j, q_j' \in C^{\infty}(M, \mathbb{H})
\eas 
can again be written as a quaternion-valued function, i.e.~of the form
\bas
	 \sum_j f_{j0}[\dirac f_{j0}'] + f_{ja} [\dirac, if'_{ja}\sigma^a]
\eas
for certain $f_{j0}, f_{j0}', f_{ja}, f_{ja}' \in C^\infty(M, \mathbb{R})$.
Using that $[\dirac, x]^* = - [\dirac, x^*]$, only the second term above, which we will denote with $Q$, 
is seen to satisfy the demand of self-adjointness for the Dirac operator. Since the Pauli matrices are traceless, the self-adjoint inner fluctuations of $\mathbb{H}$ are automatically traceless as well.

Using the supersymmetry transformations on the gauge field $Q$, we demand that $\tr \delta Q = 0$, which sets the trace of the corresponding gaugino and auxiliary field equal to zero.  

Third, the inner fluctuations of the component $M_3(\com)$ of the algebra generate a gauge field
\bas
	V' = \sum_j m_j [\slashed{\partial}, m_j'],\qquad m_j, m_j' \in M_3(\com).
\eas
Because $D_A$ is self-adjoint $V'$ must be too and hence $V'(x) \in u(3)$. We can employ the unimodularity condition \eqref{eq:unimod_new}, which for $\H_F$ given by \eqref{eq:ncmssm_H_F} reads 
\bas
	4M(\Lambda + \tr V') = 0.
\eas
The contributions to this expression again only come from $\mathcal{E}^o$ and the factor $4 = 2 + 1 + 1$ arises from the gauge fields acting trivially on the second part of its tensor product. The inner fluctuations of the quaternions do not appear in this expression, since they are traceless. A solution to the demand above is
	\ba 
		V' = - V - \frac{1}{3}\Lambda \id_3,\label{eq:ident_V}
	\ea 
with $V(x) \in su(3)$. The sign of $V$ is chosen such that the interactions match those of the Standard Model \cite[\S 3.5]{CCM07}.

In order to introduce coupling constants into the theory, we have to redefine the fields at hand:
\bas
	\Lambda_\mu &\equiv g_1B_\mu, &
	Q_\mu &\equiv g_2W_\mu, &
	V_\mu &\equiv g_3g_\mu.
\eas
Note that we parametrize the gauge fields differently than in \cite{CCM07}. 
Then looking at the supersymmetry transformation of $V'$, we infer that its superpartner, the $u(3)$ `gluino' $\gluino{L,R}'$ and corresponding auxiliary field $G_3'$ can also be separated into a trace part and a traceless part. We parametrize them similarly as
\ba
	\gluino{L,R}' &= \gluino{L,R} - \frac{1}{3}\bino{L,R} \id_3, & G_3' &= G_3 - \frac{1}{3} G_1 \id_3,\label{eq:su3-ident}
\ea
with $\bino{L,R}$ the superpartner of $B_\mu$ and $G_1$ the associated auxiliary field. 

The unimodularity condition reduced a bosonic degree of freedom. Employing it in combination with the supersymmetry transformations allowed us to reduce fermionic and auxiliary degrees of freedom as well. A similar result comes from $\mathbf{1}$ and $\overline{\mathbf{1}}$ generating the same gauge field. All in all we are left with three gauge fields, gauginos and corresponding auxiliary fields:
\bas
	\photon_\mu &\in C^{\infty}(M, u(1)),&  
\bino{L,R} &\in L^2(M, S\otimes u(1)), &
	G_1 &\in C^{\infty}(M, u(1)),\\
	\weak_\mu &\in C^{\infty}(M, su(2)),&  
	\wino{L,R} &\in L^2(M, S\otimes su(2)),& 	
	G_2 &\in C^{\infty}(M, su(2)),\\
	\gluon_\mu &\in C^{\infty}(M, su(3)),& 
	\gluino{L,R} &\in L^2(M, S\otimes su(3)), &
	G_3 &\in C^{\infty}(M, su(3)),
\eas
exactly as in the MSSM. 

With the finite Hilbert space being determined by the building blocks of the first and second type, we can also obtain the relation between the coupling constants $g_1$, $g_2$ and $g_3$ that results from normalizing the kinetic terms of the gauge bosons, appearing in \eqref{eq:spectral_action_acg_flat}. The latter are of the form (see Section \ref{sec:gut}):
\ba
	\frac{1}{4}\K_{j} \int_M F^{j\, a}_{\mu\nu} F^{j\,a\,\mu\nu},\qquad \K_j &= \frac{f(0)}{3\pi^2}g_j^2n_j \Big(2N_j + \sum_k M_{jk} N_k\Big)\nn\\
			&\equiv \frac{r_j}{3}\Big(2N_j + \sum_k M_{jk} N_k\Big),\label{eq:ncmssm-ident-gauge-kin}
\ea
where the label $j$ denotes the type (i.e.~$u(1)$, $su(2)$ or $su(3)$) of gauge field and the index $a$ runs over the generators of the corresponding gauge group. The expressions for $\K_j$ include a factor $2$ that comes from summing over both particles and anti-particles. Its first term stems from a building block \B{j} of the first type and the other terms come from the building blocks \B{jk} of the second type, having multiplicity $M_{jk}$. The symbol $n_j$ comes from the normalization 
\bas
	\tr T^a_{j} T^b_{j} = n_{j}\delta^{ab}
\eas
of the gauge group generators $T^a_j$. For $su(2)$ and $su(3)$ these have the value $n_{2,3} = \tfrac{1}{2}$, for $u(1)$ we have $n_1 = 1$. In addition, each contribution to the kinetic term of the $u(1)$ gauge boson must be multiplied with the square of the hypercharge of the building block the contribution comes from. The contributions from each representation to each kinetic term appearing in the MSSM are given in Table \ref{tab:contr_G}.
\begin{table}[ht]
\begin{tabularx}{\textwidth}{XlllllX}
\toprule
& Particle			& Representation	& $\K_1$ 									&$\K_2$	& $\K_3$	& \\[0pt]
\midrule
& \bino{L,R} 		& \repl{1}{1}			& 0 											& 0 		& 0 & \\[0pt]
& \wino{L,R} 		& \repl{2}{2}			& 0 											& 4			& 0 & \\[0pt]
& \gluino{L,R}	& \repl{3}{3} 		& 0 											& 0			& 6& \\[0pt]
\midrule
& $\nu_R$ 			& \repl{1}{1}			& $0$ 										& 0			& 0 & \\[0pt]
& $e_R$					& \repl{1}{\bar 1}& $4M$ 										& 0			& 0 & \\[0pt]
& $l_L$				& \repl{1}{2}			& $2M$ 										& $M$	& 0 & \\[0pt]
& $d_R$ 				& \repl{\bar 1}{3}&	$3(-1+\frac{1}{3})^2M$	& 0			& $M$ & \\[0pt]
& $u_R$					& \repl{1}{3}			&	$3(1+\frac{1}{3})^2M$		& 0			& $M$ & \\[0pt]
& $q_L$					& \repl{2}{3}			&	$6(\frac{1}{3})^2M$		& $3M$	& $2M$ & \\[0pt]
\midrule
& $\hd$ 				& \repl{\bar 1}{2}& 2 											& 1			& 0 & \\[0pt] 
& $\hu$ 				& \repl{1}{2}			& 2 											& 1			& 0 & \\[0pt]
\midrule
& Total 				& 								& $4+ 120M/9$			& $6 + 4M$ & $6 + 4M$ &\\[0pt]
\bottomrule
\end{tabularx}
\caption[The contributions to the prefactors of the gauge bosons]{The contributions to the pre-factors \eqref{eq:ncmssm-ident-gauge-kin} of the gauge bosons' kinetic terms for all of the representations of the MSSM. The number of generations is denoted by $M$. The contents of this table can be obtained from those of Table \ref{tab:coeff}, inserting $a/b =1$ and multiplying the contributions to all fermions that have $R = 1$ by $M$.}
\label{tab:contr_G}
\end{table}
Summing all contributions, we find  
\bas
	\K_1 &= \frac{f(0)}{3\pi^2}n_1g_1^2(4 + 120M/9) \equiv \frac{r_1}{3}(4 + 120M/9),\nn\\ 
	\K_2 &= \frac{f(0)}{3\pi^2}n_2 g_2^2(6 + 4M) \equiv \frac{r_2}{3}(6 + 4M),  \nn\\
	\K_3 &= \frac{f(0)}{3\pi^2}n_3 g_3^2(6 + 4M) \equiv \frac{r_3}{3}(6 + 4M),
\eas
for the coefficients of the gauge bosons' kinetic terms. We have to insert an extra factor $\tfrac{1}{4}$ into $\K_1$, since we must divide the hypercharges by two to compare with \cite{CCM07}, that has a different parametrization of the gauge fields (see also Section \ref{sec:gut}).
Normalizing these kinetic term by setting $\K_{1,2,3} = 1$, we obtain for the $r_i$ (defined in \eqref{eq:ncmssm-ident-gauge-kin}):
\ba\label{eq:ncmssm-rs}
	r_3 &= r_2 = \frac{3}{6 + 4M},& r_1 &= \frac{9}{3 + 10M}.
\ea
Consequently, we find for the coefficients
\ba\label{eq:defw}
		\w{ij} := 1 - r_iN_i - r_jN_j
\ea
the following values:
\bas
	\w{11} &= \frac{10M - 15}{10M + 3\phantom{1}},&
	\w{12} &= \frac{20 M^2 - 12M -27}{20 M^2 + 36M + 9\phantom{1}},\\
	\w{13} &= \frac{40 M^2 - 54 M - 63}{40 M^2 + 72 M + 18}, &
	\w{23} &= \frac{4M - 9}{4M + 6}.
\eas

From \eqref{eq:ncmssm-rs} it is immediate that, upon taking $M = 3$ and inserting the values of $n_{1,2,3}$, the three coupling constants are related by
\ba
	g_3^2 = g_2^2 = \frac{11}{9} g_1^2.\label{eq:mssm-gut}
\ea
This is different than for the SM \cite[\S 4.2]{CCM07}, where it is the well-known $g_2^2 = g_3^2 = \tfrac{5}{3}g_1^2$. For this value of $M$, the $\w{ij}$ have the following values:
\ba
	\w{11} &= \frac{5}{11},&
	\w{12} &= \frac{13}{33}, &
	\w{13} &= \frac{5}{22}, &
	\w{23} &= \frac{1}{6}.\label{eq:values-omega-3gen}
\ea

\begin{rmk}
	In Remark \ref{rmk:extra-higgsinos} we have suggested to add one extra copy of the two building blocks that describe the Higgses and higgsinos, to match the interactions of the MSSM. Such an extension gives extra contributions to the kinetic terms of the $su(2)$ and $u(1)$ gauge bosons, leading to
	\ba\label{eq:ncmssm-rs2}
		r_3 &= \frac{3}{6 + 4M},&
		r_2 &= \frac{3}{8 + 4M},&
		r_1 &= \frac{9}{6 + 10M}.&
	\ea
Consequently,
\bas
	\w{11} &= \frac{5M - 6}{5M + 3},&
	\w{12} &= \frac{10 M^2 + 2M -15}{2 (2 + M) (3 + 5 M)},\\
	\w{13} &= \frac{20 M^2 - 21 M - 36}{2 (3 + 2 M) (3 + 5 M)}, & 
	\w{23} &= \frac{4 M^2 - M - 15}{2(2 + M) (3 + 2 M)}
\eas
for the parameters $\w{ij}$. From the ratios of the $r_1$, $r_2$ and $r_3$ we derive for the coupling constants when $M = 3$:
\bas
	g_3^2 &= \frac{10}{9}g_2^2 = \frac{4}{3} g_1^2.
\eas
The $\w{ij}$ then read
\bas
	\w{11} &= \frac{1}{2},&
	\w{12} &= \frac{9}{20},&
	\w{13} &= \frac{1}{4}, & 
	\w{23} &= \frac{1}{5}.
\eas

\end{rmk}

	\section{Supersymmetry of the action} \label{sec:ncmssm-checks}

Even though the three obstructions mentioned at the beginning of Section \ref{sec:NCSSM} are avoided and the particle content of this theory coincides with that of the MSSM, we do not know if the action associated to it is in fact supersymmetric. In this section we check this by examining the requirements from the list in Section \ref{sec:4s-aux}. We will not cover all of them here, however.

Before we get to that, we note that each of the fields $\sfer_{ij}$ appears at least once in one of the building blocks of the third type. This can easily be seen by taking all combinations $(i,j)$, $(i,k)$ and $(j,k)$ of the indices $i,j,k$ of each of the building blocks of the third type that we have. Put differently, there is at least one horizontal line between each two `columns' in the Krajewski diagram of Figure~\ref{fig:MSSM_bb3_compact}. This means that for each sfermion field $\sfer_{ij}$ of the MSSM that is defined via the building block \B{ij}, we can meet the demand \eqref{eq:bb2-resultCiij} on the parameters $C_{iij}$, $C_{ijj}$ that supersymmetry sets on them. We do this by setting them to be of the form \eqref{eq:bb3-expressionCs-sqrt}, where necessary with a sum over multiple building blocks of the third type. 
With the right choice of the signs $\sgnc_{i,j}, \sgnc_{j,i}$ occurring in \eqref{eq:bb3-expressionCs-sqrt}, the fermion--sfermion--gaugino interactions that come from the building blocks of the second type coincide with those of the MSSM. 
\begin{itemize}
\item 
For each of the four building blocks \B{11_R2_L}, \B{1_R2_L3}, \B{1\bar1_R2_L} and \B{\bar1_R2_L3} of the third type that we have, there is the necessary requirement \eqref{eq:improvedUpsilons} for supersymmetry (or \eqref{eq:improvedUpsilons1} in the parametrization \eqref{eq:bb3-expressionCs-sqrt} of the $C_{iij}$), but with a sum over building blocks of the third type where necessary.
To connect with the notation of the noncommutative Standard Model (Section \ref{sec:intro_NCSM}), we will write 
\bas 
 \yuk{\nu}{} &:= \yuk{1_R,1}{2_L},& \yuk{u}{} &:= \yuk{1_R,3}{2_L}
\eas
for the parameters of the building blocks \B{11_R2_L} and \B{1_R2_L3} that generate the up-type Higgs fields and
\bas
 \yuk{e}{} &:= \yuk{\bar1_R,1}{2_L}, & \yuk{d}{} &:= \yuk{\bar 1_R,3}{2_L}
\eas
for those of \B{1\bar1_R2_L} and \B{\bar1_R2_L3} that generate the down-type Higgs fields. Furthermore, we write
\bas 
	a_u &= \tr_M \big(\yuks{\nu}{}\yuk{\nu}{} + 3 \yuks{u}{}\yuk{u}{}\big), & a_d &= \tr_M\big(\yuks{e}{}\yuk{e}{} + 3 \yuks{d}{}\yuk{d}{}\big)
\eas
for the expressions that we encounter in the kinetic terms of the Higgses:
\bas
	&\n_{1_R2_L}^2\int_M |D_\mu \hu|^2,\qquad \n_{1_R2_L}^2 = \frac{f(0)}{2\pi^2}\frac{1}{\w{12}}a_u\nn\\
\intertext{and}
	&\n_{\bar1_R2_L}^2\int_M |D_\mu \hd|^2,\qquad \n_{\bar1_R2_L}^2 = \frac{f(0)}{2\pi^2}\frac{1}{\w{12}}a_d
\eas
respectively. (Here, the parametrization as in \eqref{eq:bb3-exprN} is used). The factors $3$ above come from the dimension of the representation $\mathbf{3}$ of $M_3(\com)$. Inserting the expressions for the $\yukw{i}{j}$ the identity \eqref{eq:improvedUpsilons1} reads for the building block \B{11_R2_L}:
\bas
		&- \sqrt{\w{12}}\Big(\yuk{2,1}{1}\yuks{2,1}{1} + \yuk{2,\bar1}{1}\yuks{2,\bar1}{1}\Big)^{-1/2}\yuk{2,1}{1} \nn\\
		&=  \sgnc_{1,2_L}\sgnc_{1,1_R} \sqrt{\w{11}} \,\yuk{1,2}{1}\Big(2 \yuks{1,2}{1}\yuk{1,2}{1}\Big)^{-1/2} 
			=  \sgnc_{2_L,1}\sgnc_{2_L,1_R} \sqrt{\w{12}} \frac{\yuk{\nu}{}{}^t}{\sqrt{a_u}}.
\eas
For \B{1\bar1_R2_L}, \B{1_R2_L3}, \B{\bar1_R2_L3} it reads 
\bas
		&- \sqrt{\w{12}}\Big(\yuk{2,1}{1}\yuks{2,1}{1} + \yuk{2,\bar1}{1}\yuks{2,\bar1}{1}\Big)^{-1/2}\yuk{2,\bar 1}{1} \nn\\
		&=  \sgnc_{1,2_L}\sgnc_{1,\bar1_R} \sqrt{\w{11}}\yuk{\bar 1,2}{1}\Big(2 \yuks{\bar 1,2}{1}\yuk{\bar 1,2}{1}\Big)^{-1/2} 
			=  \sgnc_{2_L,1}\sgnc_{2_L,\bar1_R}\sqrt{\w{12}} \frac{\yuk{e}{}{}^t}{\sqrt{a_d}},\\
		&- \sqrt{\w{23}}\, \yuk{2,1}{3}\Big(\yuks{2,1}{3}\yuk{2,1}{3} + \yuks{2,\bar1}{3}\yuk{2,\bar1}{3}\Big)^{-1/2} \nn\\
		&=  \sgnc_{3,2_L}\sgnc_{3,1_R} \sqrt{\w{13}}\Big(2 \yuk{1,2}{3}\yuks{1,2}{3}\Big)^{-1/2}\yuk{1,2}{3} 
			=  \sgnc_{2_L,3}\sgnc_{2_L,1_R} \sqrt{\w{12}} \frac{\yuk{u}{}}{\sqrt{a_u}},\\
\intertext{and}
		&- \sqrt{\w{23}}\, \yuk{2,\bar 1}{3}\Big(\yuks{2,1}{3}\yuk{2,1}{3} + \yuks{2,\bar1}{3}\yuk{2,\bar1}{3}\Big)^{-1/2} \nn\\
		&=  \sgnc_{3,2_L}\sgnc_{3,\bar1_R}\sqrt{\w{13}}\Big(2 \yuk{\bar 1,2}{3}\yuks{\bar 1,2}{3}\Big)^{-1/2}\yuk{\bar 1,2}{3} 
			=  \sgnc_{2_L,3}\sgnc_{2_L,\bar1_R}\sqrt{\w{12}} \frac{\yuk{d}{}}{\sqrt{a_d}}
\eas
respectively. We have suppressed the subscripts $L$ and $R$ here for notational convenience and used Remark \ref{rmk:bb3-R=1} for the identities associated to \B{11_R2_L} and \B{1\bar1_R2_L}, giving rise to the transposes of the matrices \yuk{\nu}{} and \yuk{e}{} above. Not only do these identities help to write some expressions appearing in the action more compactly, it also gives rise to some additional relations between the parameters. Taking the second equality of each of the four groups, multiplying each side with its conjugate and taking the trace, this gives 
\begin{subequations}
\ba
	\frac{M}{2}\w{11}a_u &= \w{12}\tr_M \yuks{\nu}{}\yuk{\nu}{}, &
	\frac{M}{2}\w{11}a_d &= \w{12}\tr_M \yuks{e}{}\yuk{e}{},\label{eq:ncmssm-rel1-line1} \\
	\frac{M}{2}\w{13}a_u &= \w{12}\tr_M \yuks{u}{}\yuk{u}{}, &
	\frac{M}{2}\w{13}a_d &= \w{12}\tr_M \yuks{d}{}\yuk{d}{},\label{eq:ncmssm-rel1-line2}
\ea
\end{subequations}
where on the LHS there is a factor $M$ coming from the identity on family-space. Summing the first and three times the third equality (or, equivalently, the second and three times the fourth), we obtain 
\ba\label{eq:MSSM-condition1.1}
	\w{11} +
	3\w{13} &= \frac{2}{M}\w{12}.
\ea
Similarly, we can equate the first and last terms of each of the four groups of equalities, multiply each side with its conjugate and subsequently sum the first two (or last two) of the resulting equations. This gives
\begin{subequations}
\ba
	\id_M &= \frac{\yuk{\nu}{}^t(\yuks{\nu}{})^t}{a_u} + \frac{\yuk{e}{}^t(\yuks{e}{})^t}{a_d} \label{eq:ncmssm-rel2-line1}\\
\intertext{and} 
	\frac{\w{23}}{\w{12}}\id_M &= \frac{\yuks{u}{}\yuk{u}{}}{a_u} + \frac{\yuks{d}{}\yuk{d}{}}{a_d}\label{eq:ncmssm-rel2-line2}
\ea
\end{subequations}
respectively. By adding the first relation to three times the second relation and taking the trace on both sides, we get
\ba\label{eq:MSSM-condition1.2}
	\w{12} = \frac{3M}{2-M}\w{23}.
\ea	
We combine both results in the following way. We add the relations of \eqref{eq:ncmssm-rel1-line1} and insert \eqref{eq:ncmssm-rel2-line1} to obtain
\bas
	\frac{M}{2}\w{11} +	\frac{M}{2}\w{11} &= \w{12}\bigg(\tr_M \frac{\yuks{\nu}{}\yuk{\nu}{}}{a_u} + \tr_M \frac{\yuks{e}{}\yuk{e}{}}{a_d}\bigg) = \w{12}M,
\eas
i.e.~
\ba\label{eq:ncmssm-condition5}
	\w{11} &= \w{12}. 
\ea
Similarly, we add the relations of \eqref{eq:ncmssm-rel1-line2}, insert \eqref{eq:ncmssm-rel2-line2} and get
\ba\label{eq:ncmssm-condition6}
	\w{13}M &= \w{12}\tr_M \bigg(\frac{\w{23}}{\w{12}}\id_M\bigg), && \text{or} & \w{13} &= \w{23}.
\ea

\item We have four combinations of two building blocks \B{ijk} and \B{ijl} of the third type that share two of their indices (Section \ref{sec:2bb3}). Together, these give two extra conditions from the demand for supersymmetry, i.e.~that $\w{ij}$ (as defined in \eqref{eq:defw}) must equal $\tfrac{1}{2}$ (Section \ref{sec:4s-aux}):
	\begin{subequations}\label{eq:ncmssm-condition4}
		\ba
			\B{11_R2_L}\ \&\ \B{1_R2_L3}: &\quad \w{12} = \frac{1}{2},\\
			\B{1_R2_L3}\ \&\ \B{\bar1_R2_L3}: &\quad \w{23} = \frac{1}{2}.
		\ea
	\end{subequations}
		The other two combinations, \B{1\bar1_R2_L}\ \&\ \B{\bar1_R2_L3} and \B{11_R2_L}\ \&\ \B{1\bar1_R2_L}, both give the first condition again.

%

\end{itemize}

Combining the conditions \eqref{eq:ncmssm-condition5}, \eqref{eq:ncmssm-condition6} and \eqref{eq:ncmssm-condition4} we at least need that
\bas
	\w{11} &= \w{12} = \w{13} = \w{23} = \frac{1}{2}
\eas
for supersymmetry. However, if we combine this result with \eqref{eq:MSSM-condition1.1} and \eqref{eq:MSSM-condition1.2} it requires
\ba\label{eq:ncmssm-doesnotwork}
	2 - M &= 3M &&\text{and}& 4 &= \frac{2}{M}\quad \Longrightarrow\quad M = \frac{1}{2}.
\ea
%

We draw the following conclusion:
\begin{theorem}
	There is no number of particle generations for which the action \eqref{eq:totalaction} associated to the almost-commutative geometry determined by \eqref{eq:ncmssm-acg}, which corresponds to the particle-content and superpotential of the MSSM, is supersymmetric.
\end{theorem}

Since the extension \eqref{eq:ncmssm-acg-ext} of the finite spectral triple with extra Higgs/higgsino copies does not have an effect on which building blocks of the third type can be defined, the calculations presented in this section and hence also the conclusion above are unaffected by this.



%

Does this mean that all is lost? Suppose we focus on further extensions of the MSSM, such as that of Theorem \ref{thm:constraints_mssm_ext}. Since such extensions have extra representations in $\H_F$, this also creates the possibility of additional components for $D_F$. Which components these are exactly, depends on the particular values of the gradings $\gamma_F$ and $R$ on the representations. However, for the extension of Theorem \ref{thm:constraints_mssm_ext} in particular, we can check that for all combinations of values, the permitted components can never all be combined into building blocks of the third type, thus obstructing supersymmetry.

In general, any other extension might allow for extra building blocks of the third type, making the results \eqref{eq:MSSM-condition1.1} and \eqref{eq:MSSM-condition1.2} subject to change. The demands \eqref{eq:ncmssm-condition4} that follow from adjacent building blocks of the third type remain, however. If we add a building block of the fourth type for the right-handed neutrino, this requires $r_1 = \tfrac{1}{4}$ (see \eqref{eq:bb4-susy-demands}). This can only hold simultaneously with \eqref{eq:ncmssm-condition4} if
\bas
	r_1 &= \frac{1}{4}, & r_2 &= \frac{1}{8}, & r_3 &= \frac{1}{12}.
\eas
Enticingly, for $M \leq 3$ these required values are all smaller than or equal to the actual ones of \eqref{eq:ncmssm-rs} and \eqref{eq:ncmssm-rs2}, implying that there might indeed be extensions of $\H_F$ for which they coincide.

\chapter{Conclusions \& Outlook}

The question whether noncommutative geometry and supersymmetry go well together has been open for quite a while now. Previous attempts to reconcile both have not led to conclusive results. At the same time, all of these can be characterized by the wish to create spectral triples based on superalgebras, rather than on `ordinary' algebras, in analogy with the superfield formalism for supersymmetry. It is certainly not illogical to believe that if an algebra is the noncommutative geometrist's way to characterize spaces, then superalgebras are the device to describe superspaces with. However, in the light of the noncommutative Standard Model (NCSM) and its successes we have chosen to take a more hands on approach by taking a better look at possible supersymmetric theories arising from almost-commutative geometries (ACGs), the class of models of which the NCSM is an example. Indeed, we circumvented the need for resorting to (non-physical) superalgebras or superspaces via the introduction of various \emph{building blocks}.

These building blocks consist of representations in the finite Hilbert space and components of the finite Dirac operator. They are irreducible from a supersymmetric point of view in the sense that these are the minimally necessary ingredients that one must add to retain a supersymmetric action. They are seen to correspond to elements in the superfield formalism. This at least suggests that these are the right objects to look at and it shows that almost-commutative geometries provide a natural context to describe theories with a supersymmetric particle content. We have identified the five building blocks that are listed in Table \ref{tab:bbs}.

\begin{table}[ht]
	\setlength{\extrarowheight}{3pt}
\begin{tabularx}{\textwidth}{XllllX}
\toprule	
	& \textbf{Block} & \textbf{Role} & \textbf{Required} &  \textbf{Superfield} & \\ 
\midrule
	& \B{i}	(\S\ref{sec:bb1})					& Gaugino, g.~boson 		& ---													& Vector multiplet & \\
	& \BB{ij}	(\S\ref{sec:bb2})				& Fermion, sfermion 		& \B{i}, \B{j}								& Chiral multiplet & \\
	& \BBB{ijk}	(\S\ref{sec:bb3})			& Scalar interactions 	& \BBB{ij}, \BBB{ik}, \BBB{jk}& Superpotential & \\
	& \BBBB{11'}	(\S\ref{sec:bb4})		& Majorana mass 				& \B{11'}											& Majorana mass & \\
	& \B{\mathrm{mass}}	(\S\ref{sec:bb5}) & Adds a mass 								& \B{ij}											& A mass for $\fer{ij}, \sfer_{ij}$& \\
\bottomrule	
\end{tabularx}
\caption{The building blocks of a supersymmetric spectral triple. In the last column we have listed their counterparts in the common superfield formalism.}
\label{tab:bbs}
\end{table}

What this at least shows is that, \emph{given the action functional} \eqref{eq:totalaction}, any generic approach of doing supersymmetry via superalgebras would have been deemed to fail, since far from all spectral triples that have a supersymmetric particle content, also have a supersymmetric action. (For the reasons mentioned in Section \ref{sec:motivation}, ``supersymmetric'' in fact means ``supersymmetric, at most softly broken''.)

In addition to this we have shown that a soft supersymmetry-breaking Lagrangian appears naturally in his context. Gaugino masses are seen to correspond to components of the finite Dirac operator that run between the two copies of a representation that characterize a building block of the first type. Moreover, the presence of gaugino masses generates appropriate soft breaking terms for each building block; a sfermion mass in the case of a building block of the second type, a trilinear interaction in the case of a building block of the third type, and a cross term $\propto \sfer_{ij}\sfer_{ij}'$ in the case of a building block of the fifth type. For one, this yields a vastly reduced number of free parameters compared to the soft breaking sector of a similar model in the superfield formalism. 

Building a model that in addition to a supersymmetric particle content possibly also has a supersymmetric action then boils down to taking the following steps, as we have done for the case of the MSSM in Chapter \ref{ch:NCMSSM}:
\begin{enumerate}
	\item Specify the algebra $\A_F$ of the finite part of the almost commutative geometry. Check that it avoids the obstruction from Proposition \ref{prop:2bb2-obstr}.
	\item Add to the finite Hilbert space the gauginos corresponding to the components of $\A_F$. 
	\item Add the off-diagonal representations to the Hilbert space, with the appropriate multiplicity. This both allows for the components of the finite Dirac operator that describe the superpartner of the fermion, and sets the values of the $r_i$ that come from the normalisation of the kinetic terms of the gauge bosons;
		\bas
			r_i = \frac{3}{2N_i + \sum_k M_{ik}N_k}
		\eas 
		where $M_{ik}$ denotes the multiplicity of the representation \rep{i}{k}. Check that the finite Hilbert space avoids the obstruction from Remark \ref{rmk:bb2-obstr}. With the finite Hilbert space determined now, employ the unimodularity condition to both the gauge bosons and ---via the supersymmetry transformations--- the gauginos to check whether the number of bosonic and fermionic degrees of freedom indeed match.
	\item Check whether the values of the grading for all representations are such that all allowed components of the finite Dirac operator together combine into building blocks of the third type (or devise a principle that forbids them, such as the demands \eqref{eq:NCSM_extra_demand} or \eqref{eq:demandDF}).
	\item Check if the resulting spectral action is supersymmetric, i.e.~whether the $C_{iij}$ of \Bc{ij}{\pm} and the $\yuk{i,j}{k}$ of \B{ijk} satisfy the demands in the list of Section \ref{sec:4s-aux}.
	\item Add possible building blocks of the fourth or fifth type, knowing that they generate soft supersymmetry breaking terms.
	\item Add gaugino masses that generate the typical soft supersymmetry-breaking Lagrangian.
\end{enumerate}

In this way the only explicit examples of ACGs that exhibit a supersymmetric action are the super-Yang Mills models of \S \ref{sec:bb1} (in the superfield formalism: vector multiplets) consisting of a single building block of the first type. The action of a single building block of the second type (\S \ref{sec:bb2}, with the required building blocks of the first type), however, was proved not to be supersymmetric in Proposition \ref{lem:bb2-nosol}. The same holds (see Section \ref{sec:4s-aux-bb3}) for an ACG consising of a single building block of the third type (together with the necessary building blocks of the first and second type). We have also defined the almost-commutative geometry whose particle content and fermion--sfermion--gaugino interactions exactly coincide with those of the MSSM. We were forced to conclude, however, that also here the four-scalar terms that arise from this almost-commutative geometry prevent the associated action from being supersymmetric. The logical next question to ask ---whether this theory provides falsifiable phenomenological predictions--- then still remains a premature one. 

Since the demands above are not automatically met, this setup is possibly not the most natural one to do supersymmetry in. However, the question whether this is an \emph{economical} way of obtaining supersymmetric theories or not, is hardly relevant as long as the hunt for theories that exhibit supersymmetry is still on experimentally. With this we mean that possibly noncommutative geometry might point us towards the right supersymmetric theory once supersymmetry would indeed be verified experimentally, precisely \emph{because} supersymmetry is far from automatic.

This entire exposition hinges on the spectral action principle. Certainly, the success of describing the \emph{complete} action of the Standard Model makes it a favourable way to associate an action to a spectral triple, but this does not mean that there are no other ways of doing something similar that might be more natural in the context of supersymmetry. Indeed, in \cite{Chamseddine1998} the observation was made that the action functional \eqref{eq:totalaction} inherently treats bosons and fermions in a different way. The suggestion \cite{Chamseddine1998} to change the fermionic part of the functional to address the difference, or attempts such as \cite{AKL11} to generate the bosonic action from the fermionic one are very interesting, particularly from a supersymmetric point of view. 

\begin{figure}[h!]
\centering
	\includegraphics[width=.6\textwidth]{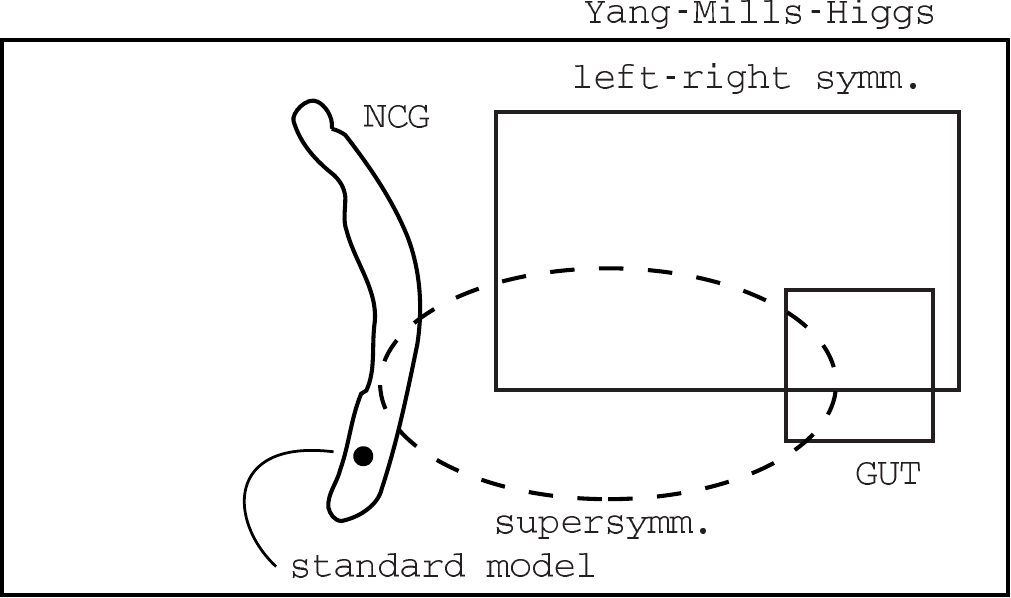}
	\caption{The intersection between supersymmetric and noncommutative theories.}
	\label{fig:schucker-modified}
\end{figure}

For long, the figure that was presented in \cite{Schucker2007} was believed to be correctly describing the (empty) intersection between noncommutative geometry and supersymmetry. In our view this should be slightly modified, to yield Figure \ref{fig:schucker-modified}.
The question that is still open is how big that intersection actually is, and exactly how `far' the MSSM is away from it, i.e.~if there are models resembling the MSSM that do exhibit a supersymmetric action. This calls for a constructive approach generating a list of (experimentally viable) supersymmetric spectral triples.

\appendix
\cleardoublepage

\chapter{Appendix}
	\section{Minkowksi versus Euclidean signature}\label{sec:mink_eucl}

Typically, Quantum Field Theory is written down in Minkowski space, having a diagonal metric on which one of the coordinates has the role of time (i.e.~it is an example of a pseudo-Riemannian manifold). This results in some expressions differing from those defined on a Riemannian (Euclidean) background as is used in the NCSM. We list in this section the relevant (sign) differences between the Minkowskian and Euclidean cases.

\begin{itemize}
	\item \emph{Wick rotation} \\
Let $\eta_{\mu\nu} = \diag(-, +, +, +)$ be the Minkowski-metric in the `mostly plus'-convention. To go from a Minkowskian time variable ($t_M$) to a Euclidean one ($t_E$) we perform a Wick rotation:
\begin{align*}
	t_E &:= i t_M,
\end{align*}
with the other variables remaining unchanged. Consequentially,
\bas
\partial_{t_E} &= -i \partial_{t_M}, &\mathrm{d}t_E &= i \mathrm{d}t_M.
\eas
Then under the Wick rotation this yields
\bas
	\eta_{\mu\nu}\mathrm{d}x^\mu_M\mathrm{d}x^\nu_M = \delta_{\mu\nu}\mathrm{d}x^\mu_E\mathrm{d}x^\nu_E.
\eas



	\item \emph{Spinors and $\gamma$-matrices}

In Minkowski space, we have (e.g.~\cite[5.4]{W05-1}) for the $\gamma$-matrices:
\bas
	\{\gamma^\mu_M, \gamma^\nu_M\} = 2\eta^{\mu\nu} \quad \Rightarrow\quad (- \gamma^0_M)^2 = 1 = (\gamma^k_M)^2, k = 1, 2, 3
\eas
and $\gamma^5_M = i \gamma^0_M\gamma^1_M\gamma^2_M\gamma^3_M$. We set
\bas
 \gamma_E^4 &:= i\gamma_M^0,&	\gamma_E^k &:= \gamma^k_M, k = 1, 2, 3, 
\eas
in order for
$
	\{\gamma^\mu_E, \gamma^\nu_E\} = 2\delta^{\mu\nu}
$. 
 Then for the grading in Euclidean space we find
\bas
	\gamma^5_E &:= - \gamma^1_E\gamma^2_E\gamma^3_E\gamma^4_E\\
					&= - i\gamma^1_M\gamma^2_M\gamma^3_M\gamma^0_M\\
					&= -(-)^3 i\gamma^0_M\gamma^1_M\gamma^2_M\gamma^3_M\\
					&= \gamma^5_M,
\eas
where in the first step we have used the expression from Example \ref{ex:canon_real_even}. Furthermore, in Minkowski space the conjugate $\bar\psi$ is given by $\bar\psi= \psi^*\gamma^0$, with\footnote{If it were not for the $\gamma^0$ in $\bar\psi$, $\bar\psi\psi$ and $\bar\psi i\slashed{\partial}\psi$ would not be Lorentz invariant.} $\psi^*$ the Hermitian conjugate of $\psi$, whereas in Euclidean space it is just $\bar\psi = \psi^*$. In addition $\psi$ and $\bar\psi$ should really be considered as independent variables, e.g.~\cite[\S5.2]{C85}, \cite{NW96}. 


	\item \emph{The action}

	In Minkowski space, the exponential appearing in the path integral is
	\bas
		\exp(i S_M),\qquad S_M = \int \mathrm{d}t_M \mathrm{d}^3x \mathcal{L}_M(\phi, \psi, \ldots).
	\eas
	Setting $t_M = -i t_E$, $t_E = x^4$, then in Euclidean space this equals 
	\bas 
		\exp(- S_E)\qquad S_E = \int_M \mathrm{d}^4 x \mathcal{L}_E(\phi, \psi, \ldots)
	\eas
	with $\mathcal{L}_E(\phi, \psi, \ldots) = - \mathcal{L}_M(\phi, \psi, \ldots)$. See Table \ref{tab:signs} for details.


\begin{table}
\begin{tabularx}{\textwidth}{XlllllllX}
	\toprule
& Term & \multicolumn{2}{l}{M., m.m.} & \multicolumn{2}{l}{M., m.p.} & \multicolumn{2}{l}{E. (C)} &\\
\midrule
& $|D_\mu\phi|^2$ 												&$+$ &&	$-$ && $+$ & & \\
& $m^2|\phi|^2$ 													&$-$ &&	$-$ && $+$ & & \\
& $\lambda|\phi|^4$ 											&$-$ &&	$-$ && $+$	&& \\
& $m^2|h|^2$ 															&$+$ &&$+$	&& $-$ & & \\
& $\bar\psi \gamma^\mu \partial_\mu \psi$ &$i$ &&	$i$ &&	$-i$&& \\
& $m\bar\psi\psi$ 												&$-$ &&	$-$ && $+i$& & \\
& $B_{\mu\nu}B^{\mu\nu}$ 									&$-$ &&	$-$ && $+$ & & \\
& $\exp(\alpha S), \alpha =	$							&$i$ &&$i$	&& $-$	&& \\
\bottomrule
\end{tabularx}	
\caption{Coefficients for various terms in the action. Here, M. = Minkowski, E. = Euclidean, m.m. = `mostly minus'-metric and m.p. = `mostly plus'. References are Zee \cite{Z03} for `mostly minus', Srednicki \cite{SR07} for `mostly plus' and Coleman \cite{C85} for Euclidean. Differences between m.m and m.p. are a minus sign per contracted index (except the ones in which $\gamma$-matrices appear).
Differences between M., m.p. and E. are an overall minus sign, except for the fermion mass terms, but these are only defined upto an overall phase anyway.} 
\label{tab:signs}
\end{table}
\end{itemize}	 

	\section{Expansion of the spectral action}\label{sec:spectral_action}

In this section we go a bit more into detail on how to calculate the spectral action \eqref{eq:spectral_action}. We apply this to two specific cases in the subsequent sections. What follows, heavily relies on \cite{GIL84} and \cite{CM07}.

Let $V$ be a \gls{vector bundle} on a compact Riemannian manifold $(M, g)$. For a second-order elliptic differential operator $P:\Gamma^{\infty}(V)\to \Gamma^{\infty}(V)$ of the form
\begin{equation}
 P = - \big(g^{\mu\nu}\partial_{\mu}\partial_{\nu} + K^{\mu}\partial_{\mu} + L)\label{eq:elliptic} 
\end{equation}
with $K^{\mu}, L \in \Gamma^{\infty}(\End(V))$, then there exist a unique \gls{connection} $\nabla$ and an \gls{endomorphism} $E$ on $V$ such that
\begin{align*}
  P = \nabla\nabla^* - E.
\end{align*}
Explicitly, we write locally $\nabla_{\mu} = \partial_{\mu} + \omega'_{\mu}$, where $\omega_\mu' = \frac{1}{2}(g_{\mu\nu}K^\nu + g_{\mu\nu}g^{\rho\sigma}\Gamma_{\rho\sigma}^{\nu})$, with $\Gamma_{\rho\sigma}^{\nu}$ the Christoffel symbols. Using this $\omega'_\mu$ and $L$ we find $E \in \Gamma^{\infty}(\End(V))$ and compute for the curvature $\Omega_{\mu\nu}$ of $\nabla$:
\begin{align}
  E &:= L - g^{\mu\nu}\partial_{\nu}(\omega_{\mu}') - g^{\mu\nu}\omega_\mu'\omega_\nu' + g^{\mu\nu}\omega_\rho'\Gamma_{\mu\nu}^\rho,\nn \\
 \Omega_{\mu\nu} &:= \partial_{\mu}(\omega'_{\nu}) - \partial_{\nu}(\omega'_{\mu}) + [\omega'_{\mu},\omega'_{\nu}].\label{eq:EOmega}
\end{align}
In this case one can make an asymptotic (`heat kernel') expansion (as $t \to 0^+$) of the trace of the operator $e^{-tP}$ in powers of $t$:
\begin{align}
\tr e^{-tP} &\sim \sum_{n \geq 0}t^{(n-m)/2}\int_Ma_n(x, P)\sqrt{g}\mathrm{d}^mx,\qquad a_n(x, P) := \tr e_n(x, P),\label{eq:app_heat_kernel}
\end{align}
where $m$ is the dimension of $M$, $\sqrt{g}\mathrm{d}^mx$ (with $g \equiv \det g$) its \emph{volume form} and the coefficients $\glslink{a024}{a_n(x, P)}$ are called the \emph{Seeley--DeWitt coefficients}. One finds \cite[\S 1.7]{GIL84} that for $n$ odd, $e_n(x, P) = 0$ and the first three even coefficients are given \cite[\S 4.8]{GIL84} by
\begin{subequations}
	\label{eq:gilkey}
	\begin{align}
		e_0(x, P) &= (4\pi)^{-m/2}(\id)\label{eq:gilkey1},\\
		e_2(x, P) &= (4\pi)^{-m/2}(-R/6\,\id + E)\label{eq:gilkey2},\\
		e_4(x, P) &= (4\pi)^{-m/2}\frac{1}{360}\big(-12R_{;\mu}^{\phantom{;\mu}\mu} + 5R^2 - 2R_{\mu\nu}R^{\mu\nu} \label{eq:gilkey3} \\
 &\qquad+ 2R_{\mu\nu\rho\sigma}R^{\mu\nu\rho\sigma} - 60RE + 180E^2 +60 E_{;\mu}^{\phantom{;\mu}\mu} + 30\Omega_{\mu\nu}\Omega^{\mu\nu}\big) \nonumber ,
	\end{align}
\end{subequations}
where 
\bas
	\glslink{Riemm}{R_{\mu\nu\rho}^{\sigma}} := \partial_\rho \Gamma^\sigma_{\mu\nu} - \partial_\nu \Gamma^\sigma_{\mu\rho} + \Gamma^{\lambda}_{\mu\nu}\Gamma^{\sigma}_{\rho\lambda} - \Gamma^{\lambda}_{\mu\rho}\Gamma^{\sigma}_{\nu\lambda}
\eas
is the Riemann curvature tensor, $\gls{Rmunu} := R_{\mu\rho\nu}^{\rho}$ the Ricci curvature tensor and $\gls{R} := g^{\mu\nu}R_{\mu\nu}$ the scalar curvature of $M$. Furthermore, $R_{;\mu}^{\phantom{;\mu}\mu} := \nabla^{\mu}\nabla_{\mu}R$ and the same for $E$.

Now suppose $D_A^2$ of a spectral triple is of the form \eqref{eq:elliptic}. Upon taking for the spectral action an even function $f$ that is a Laplace transform of some other function $g$,
\bas
	f(x) = \int_0^\infty g(t)e^{-t x^2}\mathrm{d}t,
\eas
we get for the spectral action
\bas
	\tr f(D_A/\Lambda) &=\int_0^\infty g(t)\tr e^{-t D_A^2/\Lambda^2}\mathrm{d}t,
\eas
for which we can use the expansion \eqref{eq:app_heat_kernel}:
\ba
	\tr f(D_A/\Lambda) &\sim \sum_{n \geq 0}\Lambda^{m-n}\int_M a_n(x, D_A^2)\sqrt{g}\mathrm{d}^mx\int_0^\infty g(t) t^{(n-m)/2} \mathrm{d}t,\label{eq:sp_action_interm}
\ea
In order to rewrite the last terms back into $f$, we can use that for $(n - m)/2 < 0$ \cite[\S 11]{CM07}, 
\bas
	t^{(n - m)/2} = \frac{2}{\Gamma((m -n)/2)} \int_0^\infty e^{-tx^2} x^{(m - n) -1}\mathrm{d}x,
\eas
which gives
\bas
\int_0^\infty g(t) t^{(n-m)/2} \mathrm{d}t &= \frac{2}{\Gamma((m -n)/2)} \int_0^\infty g(t) \int_0^{\infty} e^{-tx^2} x^{m - n -1}\mathrm{d}x\mathrm{d}t.
\eas
Interchanging the order of integration, 
we can then rewrite $g$ back into $f$:
\bas
\int_0^\infty g(t) t^{(n-m)/2} \mathrm{d}t &= \frac{2}{\Gamma((m -n)/2)} \int_0^{\infty} f(x) x^{m - n -1}\mathrm{d}x \\&\equiv \frac{2}{\Gamma((m - n)/2)}f_{m-n},
\eas
where $f_i$ denotes the $(i-1)$th moment of $f$:
\begin{align*}
	f_i := \int_0^{\infty} f(x) x^{i - 1}\mathrm{d}x\qquad (i > 0).
\end{align*}
For $n = m$ we can rewrite the last part of \eqref{eq:sp_action_interm} using 
\bas
 \int_0^\infty g(t) \mathrm{d}t = \lim{}_{x \to 0} \int_0^\infty g(t)e^{-tx^2}\mathrm{d}t = \lim{}_{x \to 0} f(x) = f(0).
\eas
Taking $m = \dim M = 4$ we can use these tricks to get for the first three contributions for the second term of \eqref{eq:totalaction}:
\begin{align}
	\tr f(D_A/\Lambda) &\sim  2\Lambda^4 f_4 a_0(D^2_A) + 2f_2\Lambda^2 a_2(D_A^2) + f(0) a_4(D_A^2) + \mathcal{O}(\Lambda^{-2}).\label{eq:total_action_exp}
\end{align}

\subsection{The spectral action for a canonical spectral triple}\label{sec:sp_act_canon}

As a relatively simple application, consider a $4$-dimensional canonical spectral triple (Examples \ref{ex:canon} and \ref{ex:canon_real_even}), whose background $M$ does not have a boundary. In this case the canonical Dirac operator $\dirac$ \eqref{eq:dirac_canon} does not receive inner fluctuations, see the comment in Section \ref{sec:gauge_action}. For its square we find 
\bas
	\dirac^2 = - \big[g^{\mu\nu}\partial_\mu \partial_\nu + (2\omega^\mu - \Gamma^\mu)\partial_\mu + \partial^\mu \omega_\mu + \omega^\mu \omega_\mu - \Gamma^\mu \omega_\mu + \tfrac{1}{4}R\big]
\eas
where $\Gamma^\mu = \Gamma^{\mu}_{\nu\lambda}\gamma^\nu\gamma^\lambda$ with $\Gamma^{\mu}_{\nu\lambda}$ the Christoffel symbols and $R$ is the scalar curvature of $M$. From this we can determine that 
\bas
E &= \frac{1}{4}R, & \Omega_{\mu\nu} &\equiv \partial_\mu \omega_\nu - \partial_\nu\omega_\mu + [\omega_\mu, \omega_\nu] = \frac{1}{4} R^{ab}_{\mu\nu} \gamma_{a}\gamma_{b},
\eas
where $R^{ab}_{\mu\nu}$ is the Riemann tensor and the latin indices $a$, $b$ indicate the use of a frame field $\gamma_\mu  = \gls{h}^a_\mu \gamma_a$. This sets
\bas
	a_0(\dirac^2) &= \frac{1}{16\pi^2}\int_M \tr\id\sqrt{g} \mathrm{d}^4x = \frac{1}{4\pi^2}\int_M \vol,\\
	a_2(\dirac^2) &= \frac{1}{16\pi^2}\int_M \frac{1}{12} R \tr\id\sqrt{g}\mathrm{d}^4 x = \frac{1}{48\pi^2} \int_M R \vol,\\
	a_4(\dirac^2) &= \frac{1}{16\pi^2}\frac{1}{360}\int_M \bigg[\Big(5 - \frac{60}{4} + \frac{180}{16}\Big)R^2\tr \id - 2R_{\mu\nu}R^{\mu\nu}\tr\id \\
			&\qquad + 2R_{\mu\nu\rho\sigma}R^{\mu\nu\rho\sigma}\tr\id + \frac{15}{8} R^{ab}_{\mu\nu}R^{cd\mu\nu}\tr\gamma_{a}\gamma_{b}\gamma_{c}\gamma_{d}\big)\bigg]\vol\\ 
	 			&= \frac{1}{16\pi^2}\frac{1}{360}\int_M \big(5R^2 - 8R_{\mu\nu}R^{\mu\nu} - 7R_{\mu\nu\rho\sigma}R^{\mu\nu\rho\sigma}\big) \vol, 
\eas
where the traces are over $L^2(S)$ and in the last step we have used that 
\bas
	\tr\gamma_{a}\gamma_{b}\gamma_{c}\gamma_{d} = 4[\delta_{ab}\delta_{bc} - \delta_{ac}\delta_{bd} + \delta_{ad}\delta_{bc}]
\eas
and $R_{abcd} =- R_{bacd}$, $R_{abcd} = - R_{abdc}$. Collecting terms, we have for the total action \eqref{eq:totalaction}: 
\bas
	S[\psi, g] &\sim \int_M\bigg(\frac{\Lambda^4 f_4}{2\pi^2} + \frac{f_2\Lambda^2 }{24\pi^2} R 
	 			+ \frac{f(0)}{16\pi^2}\frac{1}{360} \big(5R^2 - 8R_{\mu\nu}R^{\mu\nu} \\
			&\qquad\qquad - 7R_{\mu\nu\rho\sigma}R^{\mu\nu\rho\sigma}\big)\bigg) + S_f[\psi],
\eas
where the last term is equal to the first one in \eqref{eq:totalaction}. Comparing this with the Einstein-Hilbert action, we can identify the last term with the `matter Lagrangian density' that curves space. The coefficients of the first two terms are connected to the cosmological constant and gravitational constant $G$ respectively. The term $\propto f(0)$ has an interpretation of its own \cite[\S 11.4]{CM07}, but we will not go into that here. Note that this description holds at some, a priori unknown, energy scale and that at this energy the values of the coefficients might differ substantially from what we measure them to be.

\subsection{For an almost-commutative spectral triple}

For a general real and even almost-commutative geometry (Example \ref{eq:acg}) on a four-dimensional background, the expression for the contributions to the heat kernel expansion are somewhat different. In contrast to the case above, the canonical Dirac operator in general receives inner fluctuations (Section \ref{sec:gauge_action}) leading to gauge fields. Secondly, there might be a finite Dirac operator possible on the finite Hilbert space $\H_F$. The general expressions for the operators $E$ and $\Omega_{\mu\nu}$ (c.f.~\eqref{eq:EOmega}) when taking in \eqref{eq:elliptic} $D_A^2$ for $P$, are
\begin{align}
 E&= \frac{1}{4}R \otimes \id_{\H_F} - \frac{1}{2}\gamma^{\mu}\gamma^{\nu}\mathbb{F}_{\mu\nu} - \id_{L^2(S)}\otimes \Phi^2 + \{\can_A, \gamma^5 \otimes \Phi\},\label{eq:spectral-action-acg-E}\\
\Omega_{\mu\nu} &= \frac{1}{4} R_{\mu\nu}^{ab}\gamma_{a}\gamma_{b}\otimes \id_{\H_F} + \id_{L^2(S)}\otimes \mathbb{F}_{\mu\nu}\label{eq:spectral-action-acg-Omega},
\end{align}
where $\mathbb{F}_{\mu\nu}$ is the (skew-Hermitian) field strength (or curvature) of $\mathbb{A}_\mu$ and with $\Phi$ we mean $D_F$ plus its inner fluctuations \eqref{eq:inner_flucts}. Considering the fact that some terms are orthogonal in the Clifford algebra, we get the following expressions for the traces:
\begin{align*}
	\mkern-18mu \tr E^2 
		&= \frac{1}{4}R^2 \mathcal{N}(F) - 2\tr_F\mathbb{F}_{\mu\nu}\mathbb{F}^{\mu\nu} + 4\tr_F\Phi^4 + 4\tr_F [D_\mu, \Phi]^2 - 2R\tr_F\Phi^2,\\
\mkern-18mu	\tr RE 
	 			&= R^2 \mathcal{N}(F)  - 4R\tr_F \Phi^2,
\end{align*}
where we have written $\tr_F := \tr_{\H_F}$ and $\mathcal{N}(F) := \dim \H_F$ and have used \eqref{eq:commutatorExpr}. Then the first three Seeley-DeWitt coefficients $a_{0,2,4}(D_A^2)$ read:
\ba
a_0(D_A^2) &= \mathcal{N}(F)a_0(\dirac^2),\label{eq:sw_acg1}\\
a_2(D_A^2) &= \mathcal{N}(F)a_2(\dirac^2) - \frac{1}{4\pi^2}\int_M\tr_F\Phi^2\vol,\label{eq:sw_acg2}\\
a_4(D_A^2) &= 
	 \frac{1}{8\pi^2}\int_M\bigg[- \frac{1}{3}\tr_F\mathbb{F}_{\mu\nu}\mathbb{F}^{\mu\nu} + \tr_F \Phi^4 + \tr_F [D_\mu, \Phi]^2\nn\\
	&\qquad\qquad\qquad - \frac{1}{6}R\tr_F \Phi^2 \bigg]\vol+ \mathcal{N}(F) a_4(\dirac^2),\label{eq:sw_acg3}
\ea
where we have used that $\mathbb{F}_{\mu\nu}$ is antisymmetric in its indices, and that $\tr \gamma^5\gamma^\mu = 0$. The total spectral action for a general almost-commutative geometry on a 4-dimensional Riemannian spin-manifold without boundary is given by the expression \eqref{eq:spectral_action_acg} in Section \ref{sec:introNCG}. In Table \ref{tab:action_contrib} the various contributions to the action (including the fermionic contributions) are listed, along with the interactions they correspond to.  

\begin{table}[ht]
\begin{tabularx}{\textwidth}{Xp{.12\textwidth}p{.15\textwidth}p{.55\textwidth}X}
\toprule
	& \textbf{Object} & \text{Term} & \textbf{Accounts for} & \\ 
\midrule
	& $a_0(D_A^2)$ & $\id$& Cosmological constant & \\
	&	$a_2(D_A^2)$ & $R$ 	& Einstein-Hilbert action & \\
	&							 & $E$ 	& Scalar masses & \\
	& $a_4(D_A^2)$ & $RE$ & Gravitational coupling scalars & \\			
	&  						 & $E^2$& Gauge kinetic terms, scalar kinetic terms, scalar potential, gravitational terms & \\			
	&				 & $\Omega^{\mu\nu}\Omega^{\mu\nu}$ & Gauge kinetic terms, gravitational terms & \\			
	& \multicolumn{2}{l}{ $\tfrac{1}{2}\inpr{J\zeta}{(\dirac + i\mathbb{A})\zeta}$} & Kinetic terms and gauge interactions of fermions. & \\
	& \multicolumn{2}{l}{ $\tfrac{1}{2}\inpr{J\zeta}{\gamma^5 \otimes \Phi\zeta}$} & fermion--fermion--scalar interactions. & \\
\bottomrule	
\end{tabularx}
	\caption{The various objects that appear in the action \eqref{eq:totalaction}, the terms they are comprised of and the particle interactions they correspond to.}
	\label{tab:action_contrib}
\end{table}

	\graphicspath{{./gfx/}}
	\svgpath={./gfx/}
	\section{The action from a building block of the third type}\label{sec:bb3-calc-action}

In this section we derive in detail the action that comes from a building block \B{ijk} of the third type (cf.~Section \ref{sec:bb3}), such as that of Figure \ref{fig:bb3}. If we constrain ourselves for now to the off-diagonal part of the finite Hilbert space, then on the basis 
\bas
	\H_{F,\mathrm{off}} &= (\rep{i}{j})_L\, \oplus\, (\rep{i}{k})_R\, \oplus\, (\rep{j}{k})_L\nn\\
		&\qquad  \oplus\, (\rep{j}{i})_R\, \oplus\, (\rep{k}{i})_L\, \oplus\, (\rep{k}{j})_R
\eas
the most general allowed finite Dirac operator is of the form
\ba
	D_F= & \begin{pmatrix}
			0 &   \yuks{j}{k\,o}& 0 & 0 & 0 &   \yuks{i}{k}\\
			  \yuk{j}{k\,o}	& 0 &   \yuk{i}{j} & 0 & 0 & 0 \\
			0 &   \yuks{i}{j} & 0  &   \yuks{i}{k\,o}& 0 & 0 \\
			0 & 0 &   \yuk{i}{k\,o}& 0 &   \yuk{j}{k} & 0 \\	
			0 & 0 & 0 &   \yuks{j}{k} & 0 &   \yuks{i}{j\,o}\\	
			  \yuk{i}{k} & 0 & 0 & 0 &   \yuk{i}{j\,o}& 0 
	\end{pmatrix}\label{eq:bb3-DF}
\ea
We write for a generic element $\zeta$ of $\frac{1}{2} (1 + \gamma) L^2(S \otimes \H_{F, \mathrm{off}})$ 
\bas
	\zeta = (\fer{ijL}, \fer{ikR}, \fer{jkL}, \afer{ijR}, \afer{ikL}, \afer{jkR})
\eas
where $\afer{ijR} \in L^2( S_- \otimes \rep{j}{i})$, etc. Applying the matrix \eqref{eq:bb3-DF} to this element yields
\bas
	\gamma^5 D_F \zeta &= \gamma^5\Big(\fer{ikR}\asfer_{jk}\yuks{j}{k} + \yuks{i}{k}\sfer_{ik}\afer{jkR}, \fer{ijL}\yuk{j}{k}\sfer_{jk} + \yuk{i}{j}\sfer_{ij}\fer{jkL}, \\
						&\qquad \asfer_{ij}\yuks{i}{j}\fer{ikR} + \afer{ijR}\yuks{i}{k}\sfer_{ik}, \fer{jkL}\asfer_{ik}\yuk{i}{k} + \yuk{j}{k}\sfer_{jk}\afer{ikL}, \\
				&\qquad \asfer_{jk}\yuks{j}{k}\afer{ijR} + \afer{jkR}\asfer_{ij}\yuks{i}{j}, \afer{ikL}\yuk{i}{j}\sfer_{ij} + \asfer_{ik}\yuk{i}{k}\fer{ijL}\Big).
\eas
Notice that for the pairs $(i,j)$ and $(j,k)$ we always encounter $\sfer_{ij}$ in combination with $\yuk{i}{j}$, whereas for $(i,k)$ it is the combination $\sfer_{ik}$ and $\yuks{i}{k}$. This has to do with the fact that the sfermion $\sfer_{ik}$ crosses the particle/antiparticle-diagonal in the Krajewski diagram. Since 
\bas
	J\zeta &= J(\fer{ijL}, \fer{ikR}, \fer{jkL}, \afer{ijR}, \afer{ikL}, \afer{jkR}) \\
				&= (J_M \afer{ijR}, J_M\afer{ikL}, J_M\afer{jkR}, J_M\fer{ijL}, J_M\fer{ikR}, J_M\fer{jkL}),
\eas
the extra contributions to the inner product are written as
\bas
\mkern-18mu	\frac{1}{2}\inpr{J\zeta}{\gamma^5 D_F\zeta} 
&= \frac{1}{2}\inpr{J_M \afer{ijR}}{\gamma^5(\fer{ikR}\asfer_{jk}\yuks{j}{k} + \sfer_{ik}\yuks{i}{k}\afer{jkR})}\nn\\
&\quad + \frac{1}{2}\inpr{J_M\afer{ikL}}{\gamma^5(\fer{ijL}\yuk{j}{k}\sfer_{jk} + \yuk{i}{j}\sfer_{ij}\fer{jkL})} \\ 
&\qquad + \frac{1}{2}\inpr{J_M\afer{jkR}}{\gamma^5(\asfer_{ij}\yuks{i}{j}\fer{ikR} + \afer{ijR}\yuks{i}{k}\sfer_{ik})} \nn\\ 
&\quad\qquad+ \frac{1}{2}\inpr{J_M\fer{ijL}}{\gamma^5(\fer{jkL}\asfer_{ik}\yuk{i}{k} + \yuk{j}{k}\sfer_{jk}\afer{ikL})} \nn\\ 
&\qquad \qquad + \frac{1}{2}\inpr{J_M\fer{ikR}}{\gamma^5(\asfer_{jk}\yuks{j}{k}\afer{ijR} + \afer{jkR}\asfer_{ij}\yuks{i}{j})}\nn\\
&\quad\qquad\qquad + \frac{1}{2}\inpr{J_M\fer{jkL}}{\gamma^5(\afer{ikL}\yuk{i}{j}\sfer_{ij} + \asfer_{ik}\yuk{i}{k}\fer{ijL})}.
\eas
Using the symmetry properties \eqref{eq:identitySymJ} of the inner product, this equals
\bas
&\inpr{J_M \afer{ijR}}{\gamma^5\fer{ikR}\asfer_{jk}\yuks{j}{k}} + \inpr{J_M \afer{ijR}}{\gamma^5\yuks{i}{k}\sfer_{ik}\afer{jkR}} \nn\\
&\qquad + \inpr{J_M\afer{ikL}}{\gamma^5\fer{ijL}\yuk{j}{k}\sfer_{jk}} + \inpr{J_M\afer{ikL}}{\gamma^5\yuk{i}{j}\sfer_{ij}\fer{jkL}} \nn\\ 
&\qquad\qquad + \inpr{J_M\afer{jkR}}{\gamma^5\asfer_{ij}\yuks{i}{j}\fer{ikR}} + \inpr{J_M\fer{jkL}}{\gamma^5\asfer_{ik}\yuk{i}{k}\fer{ijL}}.
\eas
We drop the subscripts $L$ and $R$, keeping in mind the chirality of each field, and for brevity we replace $ij \to 1$, $ik \to 2$, $jk \to 3$:
\ba
 S_{123,F}[\zeta, \szeta] &= \inpr{J_M \afer{1}}{\gamma^5\fer{2}\asfer_{3}\yuks{3}{}} + \inpr{J_M \afer{1}}{\gamma^5\yuks{2}{}\sfer_{2}\afer{3}} \nn\\
&\qquad + \inpr{J_M\afer{2}}{\gamma^5\fer{1}\yuk{3}{}\sfer_{3}} + \inpr{J_M\afer{2}}{\gamma^5\yuk{1}{}\sfer_{1}\fer{3}} \nn\\
&\qquad\qquad + \inpr{J_M\afer{3}}{\gamma^5\asfer_{1}\yuks{1}{}\fer{2}} + \inpr{J_M\fer{3}}{\gamma^5\asfer_{2}\yuk{2}{}\fer{1}}\label{eq:bb3-action-ferm-detail}.
\ea

The spectral action gives rise to some new interactions compared to those coming from building blocks of the second type. They arise from the trace of the fourth power of the finite Dirac operator and are given by the following list.

\begin{figure}
\begin{center}
	\def\svgwidth{.8\textwidth}
	\subimport{\the\svgpath}{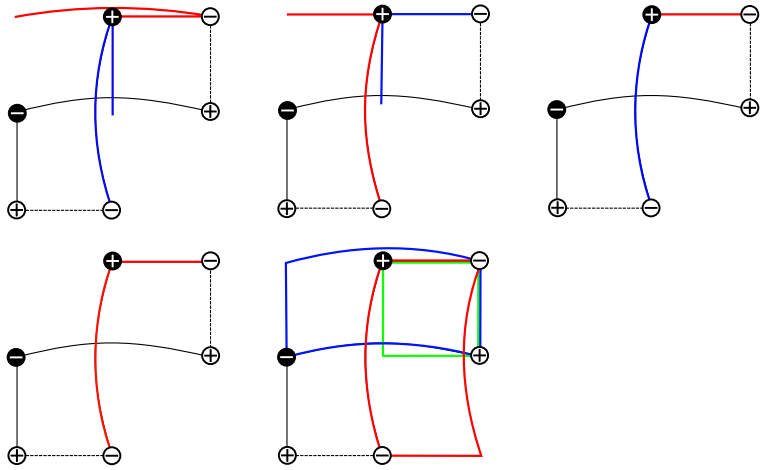}	
	\caption{The various contributions to $\tr D_F^4$ in the language of Krajewski diagrams corresponding to a building block \B{ijk} of the third type.}
	\label{fig:bb3-contributions}
\end{center}
\end{figure}

\begin{itemize}
	\item From paths of the type such as the one in the upper left corner of Figure \ref{fig:bb3-contributions} the contribution is 
	\ba
			&\mkern-36mu 8\Big[N_i|C_{iij}\sfer_{ij}\yuk{j}{k}\sfer_{jk}|^2  
			 + N_k|\yuk{i}{j}\sfer_{ij}C_{jkk}\sfer_{jk}|^2  
			 + N_j|\asfer_{ij}C_{ijj}^*\yuks{i}{k}\sfer_{ik}|^2\label{eq:bb3-action-1}\\
			 &\mkern-18mu +  N_k|\asfer_{ij}\yuks{i}{j}C_{ikk}\sfer_{ik}|^2
		 + N_i|\yuk{j}{k}\sfer_{jk}\asfer_{ik}C_{iik}^*|^2	
		+ N_j|C_{jjk}\sfer_{jk}\asfer_{ik}\yuk{i}{k}|^2	\Big].\nn
	\ea
	Here the multiplicity $8 = 2(1 + 1+ 2)$ comes from the fact that there are three vertices involved in each path, on each of which the path can start. In the case of the `middle' vertices the path can be traversed in two distinct orders. Furthermore a factor two comes from that each path occurs twice; also mirrored along the diagonal of the diagram.
	\item From paths such as the upper middle one in Figure \ref{fig:bb3-contributions} the contribution is:
	\ba
			&\mkern-36mu 8\Big[ \tr (C_{iij}\sfer_{ij})^o \yuk{j}{k}\sfer_{jk}\asfer_{jk}\yuks{j}{k}(\asfer_{ij}C_{iij}^*)^o\nn\\
			&		+ \tr(\yuk{i}{j}\sfer_{ij})^oC_{jjk}\sfer_{jk}\asfer_{jk}C_{jjk}^*(\asfer_{ij}\yuks{i}{j})^o\nn\\
			&\qquad		+ \tr(\asfer_{ij}C_{iij}^*)^o\yuks{i}{k}\sfer_{ik}\asfer_{ik}\yuk{i}{k}(C_{iij}\sfer_{ij})^o\nn\\
			&\qquad\qquad +  \tr(\asfer_{ij}\yuks{i}{j})^oC_{iik}\sfer_{ik}\asfer_{ik}C_{iik}^*(\yuk{i}{j}\sfer_{ij})^o\nn\\
			&\qquad\qquad\qquad +  \tr(\asfer_{ik}C_{ikk}^*)^o\yuk{j}{k}\sfer_{jk}\asfer_{jk}\yuks{j}{k}(C_{ikk}\sfer_{ik})^o\nn\\
			&\qquad\qquad\qquad\qquad  +  \tr(\asfer_{ik}\yuk{i}{k})^oC_{jjk}\sfer_{jk}\asfer_{jk}C_{jkk}^*(\yuks{i}{k}\sfer_{ik})^o	\Big],\label{eq:bb3-action-2}
	\ea
		where the arguments for determining the multiplicity are the same as for the previous contribution.
	\item From paths such as the upper right one in Figure \ref{fig:bb3-contributions}, going back and forth along the same edge twice, the contribution is:
	\ba
	\mkern-36mu	4\Big[N_i|\yuk{j}{k}\sfer_{jk}\asfer_{jk}\yuks{j}{k}|^2 + N_j|\yuks{i}{k}\sfer_{ik}\asfer_{ik}\yuk{i}{k}|^2 + N_k|\yuk{i}{j}\sfer_{ij}\asfer_{ij}\yuks{i}{j}|^2\Big]. \label{eq:bb3-action-3}
	\ea
	The multiplicity arises from $2$ vertices on which the path can start and each such path occurs again reflected.

	\item From paths such as the lower left one in Figure \ref{fig:bb3-contributions} the contribution is:
	\ba
		&\mkern-36mu 8 \Big[|\sfer_{ij}|^2|\yuk{i}{j}\yuks{i}{k}\sfer_{ik}|^2 + |\sfer_{ij}|^2|\yuks{i}{j}\yuk{j}{k}\sfer_{jk}|^2 + |\yuks{i}{k}\sfer_{ik}|^2|\yuk{j}{k}\sfer_{jk}|^2 \Big]. \label{eq:bb3-action-4}	
	\ea

	\item From paths such as the lower right one in Figure \ref{fig:bb3-contributions} the contribution is:
	\ba
			&\mkern-36mu 8 \Big[ \tr (\asfer_{ik}C_{iik}^*(\yuk{i}{j}\sfer_{ij})^o(\asfer_{ij} C_{iij}^*)^o \yuks{i}{k}\sfer_{ik}) \nn\\
		&\quad + \tr(\asfer_{jk}\yuks{j}{k}(\asfer_{ij}C_{ijj}^*)^o(\yuk{i}{j}\sfer_{ij})^oC_{jjk}\sfer_{jk}) \nn\\
		&\quad\qquad + \tr((\asfer_{ik}\yuk{i}{k})^oC_{jkk}\sfer_{jk}\asfer_{jk}\yuks{j}{k}(C_{ikk}\sfer_{ik})^o)  + h.c. \Big],\label{eq:bb3-action-5}
	\ea
	corresponding with the blue, green and red paths respectively. The multiplicity arises from the fact that any such path has four vertices on which it can start and also occurs reflected around the diagonal. Besides, each path can also be traversed in the opposite direction, hence the `h.c.'.
\end{itemize}
Adding \eqref{eq:bb3-action-1}, \eqref{eq:bb3-action-2}, \eqref{eq:bb3-action-3}, \eqref{eq:bb3-action-4} and \eqref{eq:bb3-action-5} the total \emph{extra} contribution to $\tr D_F^4$ from adding a building block \B{ijk} of the third type, is given by \eqref{eq:bb3-boson-action}.

\section{Proofs}\label{sec:AppSusySTAction}

In this section we give the actual proofs and calculations of the Lemmas and Theorems presented in the text. First we introduce some notation. With $(.,.)_{\mathcal{S}} : \Gamma^{\infty}(\mathcal{S}) \times \Gamma^{\infty}(\mathcal{S}) \to C^{\infty}(M)$ we mean the $C^{\infty}(M)$-valued Hermitian structure on $\Gamma^{\infty}(\mathcal{S})$. 
The Hermitian form on $\Gamma^{\infty}(\mathcal{S})$ is to be distinguished from the $C^{\infty}(M)$-valued form on $\H \equiv L^2(M, S\otimes \H_F)$:
\begin{align*}
		(., .)_{\H} : \Gamma(\mathcal{S} \otimes \H_F) \times \Gamma(\mathcal{S} \otimes \H_F) \to C^{\infty}(M)
\end{align*}
given by
\begin{align*}
		(\fer{1}, \fer{2})_{\H}  := (\zeta_1,\zeta_2)_{\mathcal{S}}\langle m_1, m_2 \rangle_{F},\qquad \psi_{1,2} = \zeta_{1,2} \otimes m_{1,2},
\end{align*}
where $\inpr{\,.\,}{\,.\,}_F$ denotes the inner product on the finite Hilbert space $\H_F$. The inner product on the full Hilbert space $\H$ is then obtained by integrating over the manifold $M$: 
\begin{align*}
	\langle \psi_1, \psi_2\rangle_{\H} := \int_M  (\psi_1, \psi_2)_{\H}\,\sqrt{g}\mathrm{d}^4x.
\end{align*}
If no confusion is likely to arise between $(.,.)_{\mathcal{S}}$ and $(.,.)_{\H}$, we omit the subscript.

In the proofs there appear a number of a priori unknown constants. To avoid confusion: capital letters always refer to parameters of the Dirac operator, lowercase letters always refer to proportionality constants for the superfield transformations. For the latter the number of indices determines what field they belong to: constants with one index belong to a gauge boson--gaugino pair, constants with two indices belong to a fermion--sfermion pair. 

\subsection{First building block}\label{sec:SYM}

This section forms the proof of Theorem \ref{prop:bb1}. In this case the action is given by \eqref{eq:SYM}. Its constituents are the ---flat--- metric metric $g$, the gauge field $A^j \in \End(\Gamma(\cS) \otimes su(N_j))$ and spinor $\gau{j} \in L^2(M, S \otimes su(N_j))$, both in the adjoint representation and the spinor after reducing its degrees of freedom (see Section {\ref{sec:equalizing}}). 

Now for $\epsilon \equiv (\eL, \eR) \in L^2(M, S)$, decomposed into Weyl spinors that vanish covariantly (i.e.~$\nabla^S\epsilon = 0$), we define 
\begin{subequations}\label{eq:susytransforms2}
\begin{align}
		\delta A_j &= c_{j}\gamma^\mu\big[(J_M\eR, \gamma_\mu\gau{jL})_\cS +  (J_M\eL, \gamma_\mu\gau{jR})_\cS\big] \equiv \gamma^\mu (\delta A_{\mu j +} + \delta A_{\mu j -})\label{eq:transforms2.1},\\
		\delta \gau{jL,R} &= (c_{j}' F^j + c_{G_j}'G_j)\eLR,\qquad F^j  \equiv \gamma^\mu\gamma^\nu F_{\mu\nu}^{j}\label{eq:transforms2.2},\\ 
		\delta G_j &= c_{G_j}\big[(J_M\eL, \can_A\gau{jR})_{\cS} + (J_M\eR, \can_A\gau{jL})_{\cS}\big]\label{eq:transforms2.3},
\end{align}
\end{subequations}
where the coefficients $c_j, c_j', c_{G_j}, c_{G_j}'$ are yet to be determined. In the rest of this section we will drop the index $j$ for notational convenience and discard the factor $n_j$ from the normalization of the gauge group generators, since it appears in the same way for each term. 


\begin{itemize}
\item The fermionic part of the Lagrangian, upon transforming the fields, equals:
\begin{align}
	&\langle J_M\gau{L}, \can_A\gau{R}\rangle \nn\\&\to \int_M (J_M[c'F + c_{G}'G]\eL, \can_A\gau{R})_{\H} + (J_M\gau{L}, \can_A[c'F + c_{G}'G]\eR)_{\H}\nn\\
		&\qquad + gc(J_M\gau{L}, \gamma^\mu\ad[(J_M\eL, \gamma_\mu\gau{R})_{\cS} + (J_M\eR, \gamma_\mu\gau{L})_{\cS}]\gau{R})_{\H}\label{eq:bb1-transf1}.
\end{align}
Here we mean with $\ad(X)$ the adjoint: $\ad(X)Y := [X, Y]$.

\item The kinetic terms for the gauge bosons transform to:
\begin{align}
 & \frac{1}{4}\K \int_M \tr_N F^{\mu\nu}F_{\mu\nu} \nn\\
	&\to  c\frac{\K}{2} \int_M \tr_{N} F^{\mu\nu}\Big(\partial_{[\mu} \big[(J_M\eR, \gamma_{\nu]}\gau{L})_\cS +  (J_M\eL, \gamma_{\nu]}\gau{R})_\cS\big] \nn\\
	&\qquad\qquad -ig[(J_M\eR, \gamma_{\mu}\gau{L})_\cS +  (J_M\eL, \gamma_{\mu}\gau{R})_\cS, A_\nu]\label{eq:bb1-transf2}\\
	&\qquad\qquad\qquad  -ig[A_\mu, (J_M\eR, \gamma_{\nu}\gau{L})_\cS +  (J_M\eL, \gamma_{\nu}\gau{R})_\cS]\Big)\sqrt{g}\mathrm{d}^4x.\nn
\end{align}
where $A_{[\mu}B_{\nu]} \equiv A_\mu B_\nu - A_\nu B_\mu$.
\item And finally the term for the auxiliary fields transforms to
\begin{align}
	- \frac{1}{2}\int_M \tr_{N} G^2 &\to - c_{G}\int_M \tr_{N} G\big[ (J_M\eR, \can_A\gau{L})_{\cS} + (J_M\eL, \can_A\gau{R})_{\cS}\big].\label{eq:bb1-transf3}
\end{align}
\end{itemize}

If we collect the terms of \eqref{eq:bb1-transf1}, \eqref{eq:bb1-transf2} and \eqref{eq:bb1-transf3} containing the same field content, we get three groups of terms that separately need to vanish in order to have a supersymmetric theory. These groups are:
\begin{itemize}
\item one consisting of only one term with four fermionic fields (coming from the second line of \eqref{eq:bb1-transf1}):
\begin{align}
	gc(J_M\gau{L}, \gamma^\mu\ad(J_M\eL, \gamma_\mu\gau{R})_{\cS}\gau{R})_{\H}.\label{eq:bb1-group1}
\end{align}
There is a second such term with $\eL \to \eR$ and $\gau{R} \to \gau{L}$ that is obtained via $(J_M\eL, \gamma_\mu\gau{R})_{\cS} \to (J_M\eR, \gamma_\mu\gau{L})_{\cS}$.

\item one consisting of a gaugino and two or three gauge fields:
\begin{align}
	& \int_M \bigg[c' (J_M\gau{L}, \can_AF\eR)_{\H} + c\frac{\K}{2} \tr_{N} F^{\mu\nu}\Big(\partial_{[\mu}(J_M\eR, \gamma_{\nu]}\gau{L})_\cS \nn\\
	&\quad\qquad -ig\big[(J_M\eR, \gamma_\mu\gau{L})_\cS, A_\nu\big] -ig\big[A_\mu, (J_M\eR, \gamma_{\nu}\gau{L})_\cS\big]\Big)\bigg]\label{eq:bb1-group2}
\end{align}
featuring the third term of \eqref{eq:bb1-transf1} and the terms of \eqref{eq:bb1-transf2} featuring $\gau{L}$. There is another such group with $\eR \to \eL$ and $\gau{L} \to \gau{R}$ consisting of the first term of \eqref{eq:bb1-transf1} and the other terms of \eqref{eq:bb1-transf2}.

\item one consisting of the auxiliary field $G$, a gauge field and a gaugino:
\begin{align}
	 \int_M \Big[ c_{G}'(J_M\gau{L},\can_AG\eR)_{\H}- c_{G} \tr_{N} G(J_M\eR, \can_A\gau{L})_{\cS} \Big]\label{eq:bb1-group3}
\end{align}
featuring the second part of the third term of \eqref{eq:bb1-transf1} and the first term of \eqref{eq:bb1-transf3}. There is another such group with $\eR\to\eL$ and $\gau{L} \to \gau{R}$.
\end{itemize}
We will tackle each of these groups separately in the following Lemmas.


\begin{lem}\label{lem:bb1-group1}
The term \eqref{eq:bb1-group1} equals zero.
\end{lem}
\begin{proof}
Evaluating \eqref{eq:bb1-group1} point-wise, applying the finite inner product and using the normalization for the generators of the gauge group, yields up to a constant factor
\begin{align}
	f^{abc} (J_M\gau{L}^a, \gamma^\mu \gau{R}^b)_\cS(J_M\eL, \gamma_\mu \gau{R}^c)_\cS\label{eq:fierz_start}.
\end{align}
Here the $f^{abc}$ are the structure constants of the Lie algebra $SU(N)$. We employ a Fierz transformation (See Appendix \ref{sec:fierz}), using $C_{10} = - C_{14} = 4$, $C_{11} =  C_{13} = -2$, $C_{12} = 0$, to rewrite \eqref{eq:fierz_start} as
\begin{align*}
	&\mkern-18mu f^{abc} (J_M\gau{L}^a, \gamma^\mu \gau{R}^b)_\cS(J_M\eL, \gamma_\mu \gau{R}^c)_\cS = - \frac{1}{4}f^{abc}\Big[4(J_M\eL, \gau{R}^b)_\cS(J_M\gau{L}^a, \gau{R}^c)_\cS\\
& - 2 (J_M\eL, \gamma_\mu \gau{R}^b)_\cS					(J_M\gau{L}^a, \gamma^\mu \gau{R}^c)_\cS  - 2 (J_M\eL, \gamma_\mu \gamma^5 \gau{R}^b)_\cS(J_M\gau{L}^a, \gamma^\mu \gamma^5 \gau{R}^c)_\cS\\
&\quad - 4 (J_M\eL, \gamma^5 \gau{R}^b)_\cS						(J_M\gau{L}^a, \gamma^5 \gau{R}^c)_\cS\Big].
\end{align*}
The first and last terms on the right hand side of this expression are seen to cancel each other, whereas the second and third term add. We retain 
\begin{align*}
	& f^{abc} (J_M\gau{L}^a, \gamma^\mu \gau{R}^b)_\cS(J_M\eL, \gamma_\mu \gau{R}^c)_\cS =  f^{abc} (J_M\gau{L}^a, \gamma^\mu \gau{R}^c)_\cS (J_M\eL, \gamma_\mu \gau{R}^b)_\cS	. 
\end{align*}
Since $f^{abc}$ is fully antisymmetric in its indices, this expression equals zero.
\end{proof}


\begin{lem}\label{lem:bb1-group2}
The term \eqref{eq:bb1-group2} equals zero iff
\begin{align}
	2ic' &= - c\K \label{bb1-constr2}.
\end{align}
\end{lem}
\begin{proof}
If we use that the spin connection is Hermitian and employ \eqref{eq:idnNablaS}, this yields:
\begin{align*}
	\partial_\mu \delta A_{\nu\, +} &= c (J_M\eR, \gamma_\nu \nabla^S_\mu\gau{L}).
\end{align*}
Here we have used that $[\nabla^S_\mu, J_M] = 0$, that we have a flat metric and that $\nabla^S\eLR = 0$. Now using that $A_\mu(J_M\eR, \gamma_\nu\gau{L})_\cS = (J_M\eR, A_\mu \gamma_\nu \gau{L})_\cS$ and inserting these results into the second part of \eqref{eq:bb1-group2} gives
\begin{align*}
 	&  c\frac{\mathcal{K}}{2}\int_M \tr_{N} F^{\mu\nu}(J_M\eR, D_{[\mu}\gamma_{\nu]}\gau{L})_{\cS},\quad D_\mu = \nabla^S_\mu - ig \ad(A_\mu).
\end{align*}
Using Lemma \ref{lem:pullScalar} and employing the antisymmetry of $F_{\mu\nu}$ we get
\begin{align*}
 	&  c\mathcal{K}\int_M (J_MF^{\mu\nu}\eR, D_{\mu}\gamma_{\nu}\gau{L})_{\H}. 
\end{align*}
We take the first term of \eqref{eq:bb1-group2}
and write out the expression $\can_AF = i\gamma^\mu D_\mu \gamma^\nu\gamma^\lambda F_{\nu\lambda}$. We can commute the $D_\mu$ through the $\gamma^\nu\gamma^\lambda$-combination since the metric is flat. Employing the identity
\begin{align}
	\gamma^\mu\gamma^\nu\gamma^\lambda = g^{\mu\nu}\gamma^\lambda + g^{\nu\lambda}\gamma^\mu - g^{\mu\lambda}\gamma^\nu +\epsilon^{\sigma\mu\nu\lambda}\gamma^5\gamma_\sigma
\end{align}
yields
\begin{align*}
\can_A F = i\big(2g^{\mu\nu}\gamma^\lambda + \epsilon^{\sigma\mu\nu\lambda}\gamma^5\gamma_\sigma) D_\mu F_{\nu\lambda}.
\end{align*}
Applying this operator to $\eR$ gives
\begin{align*}
\can_A F \eR = 2ig^{\mu\nu}\gamma^\lambda D_\mu F_{\nu\lambda} \eR = 2i\gamma_\lambda D_\mu F^{\mu\lambda} \eR, 
\end{align*}
for the other term cancels via the Bianchi identity and the fact that $\nabla^S\eR = 0$. 
With the above results, \eqref{eq:bb1-group2} is seen to be equal to 
\begin{align}
2ic'\langle J\gau{L}, \gamma_\nu D_\mu F^{\mu\nu} \eR\rangle + c\mathcal{K}\int_M (J_MF^{\mu\nu}\eR, D_{\mu}\gamma_{\nu}\gau{L})_{\H}.
\end{align}
Using the symmetry of the inner product, the result follows.
\end{proof}


\begin{lem}\label{lem:bb1-group3}
The term \eqref{eq:bb1-group3} equals zero if and only if 
\begin{align}
	c_G = - c_G'.\label{eq:bb1-constr3}
\end{align}
\end{lem}
\begin{proof}
Using the cyclicity of the trace, the symmetry property \eqref{eq:identitySymJ2} of the inner product and Lemma \ref{lem:pullScalar}, the second term of \eqref{eq:bb1-group3} can be rewritten to 
\begin{align*}
 c_{G} \int_M (J_M\gau{L}, \can_A G\eR)_{\H}
\end{align*}
from which the result immediately follows. 
\end{proof}


By combining the above three lemmas we can prove Theorem \ref{thm:bb1}:
\begin{prop}\label{prop:bb1}
	A spectral triple whose finite part consists of a building block of the first type (Def.~\ref{def:bb1}) has a supersymmetric action \eqref{eq:SYM} under the transformations \eqref{eq:susytransforms2} iff
	\begin{align*}
			2ic' &= - c\K, & c_G = - c_G'.
	\end{align*}
\end{prop}

\subsection{Second building block}\label{sec:bb2-proof}

We apply the transformations \eqref{eq:bb1-transforms2}, \eqref{eq:susytransforms4} and \eqref{eq:susytransforms5} to the terms in the action \emph{that appear for the first time}\footnote{We add this explicitly since we do not need the terms in the Yang-Mills action for together they were already supersymmetric.} as a result of the new content of the spectral triple, i.e.~\eqref{eq:bb2-action-offshell}. In the fermionic part of the action, the second and fourth terms transform under \eqref{eq:susytransforms4} to
\begin{align}
	& \langle J_M\afer{R}, \gamma^5\gau{iR}\Cw{i,j}\sfer \rangle \nn\\
	&\qquad \to 
	\langle J_M c_{ij}'^*\gamma^5[\can_A,\asfer]\epsilon_L, \gamma^5\gau{iR}\Cw{i,j}\sfer \rangle +
	\langle J_M d_{ij}'^*F_{ij}^*\epsilon_R, \gamma^5\gau{iR}\Cw{i,j}\sfer \rangle \nn\\
	&\qquad\qquad + c_i' \langle J_M\afer{R}, \gamma^5F_i\Cw{i,j}\sfer\epsilon_R \rangle 
+ c_{G_i}' \langle J_M\afer{R}, \gamma^5G_i\Cw{i,j}\sfer\epsilon_R \rangle\nn \\
		&\qquad\qquad\qquad  + \langle J_M\afer{R}, \gamma^5\gau{iR}\Cw{i,j}c_{ij}(J_M\epsilon_L, \gamma^5 \fer{L})\rangle \label{eq:bb2-transf1}
\end{align}
and
\begin{align}
	& \langle J_M\fer{L}, \gamma^5\asfer \Cw{i,j}^*\gau{iL}\rangle \nn\\
	&\qquad \to c_{ij}'	\langle J_M\gamma^5[\can_A, \sfer]\epsilon_R, \gamma^5\asfer \Cw{i,j}^*\gau{iL}\rangle +
d_{ij}'\langle J_M F_{ij}\epsilon_L, \gamma^5\asfer \Cw{i,j}^*\gau{iL}\rangle \nn\\
&\qquad\qquad  
+ c_i'\langle J_M\fer{L}, \gamma^5\gamma^\mu\gamma^\nu\asfer \Cw{i,j}^*F_{i\mu\nu}\epsilon_L\rangle + c_{G_i}'\langle J_M\fer{L}, \gamma^5\asfer \Cw{i,j}^*G_i\epsilon_L\rangle\nn\\
	&\qquad\qquad\qquad+ \langle J_M\fer{L}, \gamma^5c_{ij}^*(J_M\epsilon_R, \gamma^5\afer{R}) \Cw{i,j}^* \gau{iL}\rangle\label{eq:bb2-transf2}
\end{align}
respectively. 
We omit the terms with $\gau{jL,R}$ instead of $\gau{iL,R}$; transformation of these yield essentially the same terms. For the kinetic term of the $R = 1$ fermions (the first term of \eqref{eq:bb2-action-ferm}) we have under the same transformations:
\begin{align}
	 \langle J_M \afer{R}, \can_A \fer{L}\rangle &\to \langle J_M c_{ij}'^*\gamma^5[\can_A, \asfer]\epsilon_L, \can_A \fer{L}\rangle \nn\\
	&\qquad + g_ic_i\langle J_M \afer{R}, \gamma^\mu [ (J_M\epsilon_L, \gamma_\mu \gau{iR}) + (J_M\epsilon_R, \gamma_\mu \gau{iL})] \fer{L}\rangle \nn\\
	&\qquad\qquad + \langle J_M \afer{R}, \can_A c_{ij}'\gamma^5[\can_A \sfer]\eR\rangle 
 + \langle J_Md_{ij}'^*F_{ij}^*\epsilon_R, \can_A \fer{L}\rangle \nn\\
 &\qquad\qquad\qquad + \langle J_M \afer{R}, \can_A d_{ij}'F_{ij}\epsilon_L\rangle. 
\label{eq:bb2-transf3}
\end{align}
As with the previous contributions to the action, we omit the terms $\delta A_j$ (instead of $\delta A_i$) for brevity. In the bosonic action, we have the kinetic terms of the sfermions, transforming to
\begin{align}
	\tr_{N_j} D^\mu\asfer  D_\mu\sfer 
&\to i g_ic_i\tr_{N_j}\big( \asfer[ (J_M\epsilon_L, \gamma_\mu \gau{iR}) + (J_M\epsilon_R, \gamma_\mu \gau{iL})] D^\mu \sfer\big)\nn\\
		&\quad\qquad  -i g_ic_i\tr_{N_j}\big( D_\mu \asfer[ (J_M\epsilon_L, \gamma^\mu \gau{iR}) + (J_M\epsilon_R, \gamma^\mu \gau{iL})] \sfer\big)\nn\\
		&\quad\qquad\qquad  + \tr_{N_j}\big( D_\mu c_{ij}^*(J_M\epsilon_R, \gamma^5 \afer{R}) D^\mu \sfer\big) \nn\\
	&\quad\qquad\qquad\qquad	+ \tr_{N_j}\big( D_\mu \asfer  D^\mu c_{ij}(J_M\epsilon_L, \gamma^5 \fer{L})\big)\label{eq:bb2-transf4}
\end{align}
(and terms with $\gau{j}$ instead of $\gau{i}$) and from the terms with the auxiliary fields we have 
\begin{align}
	\tr_{N_i} \P_{i}\sfer\asfer G_i & \to \tr_{N_i} \P_{i}c_{ij}(J_M\epsilon_L, \gamma^5\fer{L})\asfer G_i + \tr_{N_i} \P_{i}\sfer c_{ij}^*(J_M\epsilon_R, \gamma^5\afer{R}) G_i\nn \\
	&\qquad + c_{G_i}\tr_{N_i} \P_{i}\sfer\asfer[ (J_M\epsilon_L, \can_A\gau{iR}) + (J_M\epsilon_R, \can_A\gau{iL})]\label{eq:bb2-transf5}.
\end{align}
And finally we have the kinetic terms of the auxiliary fields $F_{ij}$, $F_{ij}^*$ that transform to 
\begin{align}
	\tr F_{ij}^*F_{ij} &\to \tr F_{ij}^*\Big[ d_{ij}(J_M\epsilon_R, \can_A\fer{L})_{S} + d_{ij,i}(J_M\epsilon_R, \gamma^5\gau{iR}\sfer)_{\cS}\nn\\
	&\qquad\qquad	 - d_{ij,j}(J_M\epsilon_R, \gamma^5\sfer\gau{jR})_{\cS}\Big] \nn\\
		&\qquad + \tr\Big[d_{ij}^*(J_M\epsilon_L, \can_A\afer{R})_{S}  + d_{ij,i}^*(J_M\epsilon_L, \gamma^5\asfer\gau{iL})_{\cS} \nn\\
	&\qquad\qquad	- d_{ij,j}^*(J_M\epsilon_L, \gamma^5\gau{jL}\asfer)_{\cS}\Big]F_{ij},\label{eq:bb2-transf6}
\end{align}
where the traces are over $\mathbf{N}_j^{\oplus M}$. Analyzing the result of this, we can put them in groups of terms featuring the very same fields. Each of these groups should separately give zero in order to have a supersymmetric action. We have:
\begin{itemize}
\item Terms with four fermionic fields; the fifth term of \eqref{eq:bb2-transf1}, and part of the second term of \eqref{eq:bb2-transf3}: 
\begin{align}
	& \langle J_M\afer{R}, \gamma^5\gau{iR}\Cw{i,j}c_{ij}(J_M\epsilon_L, \gamma^5 \fer{L})\rangle + 
	 g_ic_i\langle J_M \afer{R}, \gamma^\mu(J_M\epsilon_L, \gamma_\mu \gau{iR}) \fer{L}\rangle \label{eq:bb2-group1}.
\end{align}
 The third term of \eqref{eq:bb2-transf2} and the other part of the second term of \eqref{eq:bb2-transf3} give a similar contribution but with $\epsilon_L \to \epsilon_R$, $\gau{iL} \to \gau{iR}$.

\item Terms with one gaugino and two sfermions, consisting of the first term of \eqref{eq:bb2-transf1}, part of the first and second terms of \eqref{eq:bb2-transf4}, and part of the third term of \eqref{eq:bb2-transf5}:
\begin{align}
	& \langle J_Mc_{ij}'^*\gamma^5[\can_A,\asfer]\epsilon_L, \gamma^5\gau{iR}\Cw{i,j}\sfer \rangle 
  +i g_ic_i\int \tr_{N_j}\big( \asfer (J_M\epsilon_L, \gamma_\mu \gau{iR}) D^\mu \sfer\big)\nn\\
		&\qquad -i g_ic_i \int \tr_{N_j}\big( D_\mu \asfer (J_M\epsilon_L, \gamma^\mu \gau{iR}) \sfer\big)\nn\\
&\qquad\qquad - c_{G_i}\int \tr_{N_i} \P_i\sfer\asfer (J_M\epsilon_L, \can_A\gau{iR}) \label{eq:bb2-group2}.
\end{align}
The first term of \eqref{eq:bb2-transf2}, the other parts of the first and second terms of \eqref{eq:bb2-transf4} and the other part of the third term of \eqref{eq:bb2-transf5} give similar terms but with $\eL \to \eR$, $\gau{iR} \to \gau{iL}$.

\item Terms with two gauge fields, a fermion and a sfermion, consisting of the third term of \eqref{eq:bb2-transf1}, the third term of \eqref{eq:bb2-transf4} and the third term of \eqref{eq:bb2-transf3}:
\begin{align}
& c_i' \langle J_M\afer{R}, \gamma^5F_i\Cw{i,j}\sfer\epsilon_R \rangle 
		 + \int \tr_{N_j}\big( D_\mu c_{ij}^*(J_M\epsilon_R, \gamma^5 \afer{R}) D^\mu \sfer\big) \nn\\ 
	&\qquad	+  \langle J_M \afer{R}, \can_A \gamma^5c_{ij}'[\can_A, \sfer]\epsilon_R\rangle \label{eq:bb2-group3}
\end{align}
The fourth term of \eqref{eq:bb2-transf2}, the first term of \eqref{eq:bb2-transf3} and the fourth term of \eqref{eq:bb2-transf4} make up a similar group but with $\epsilon_R \to \epsilon_L$ and $\afer{R} \to \fer{L}$.

\item Terms with the auxiliary field $G_i$, consisting of the fourth term of \eqref{eq:bb2-transf1} and the second term of \eqref{eq:bb2-transf5}:
\begin{align}
 & c_{G_i}' \langle J_M\afer{R}, \gamma^5G_i\Cw{i,j}\sfer\epsilon_R \rangle - \int \tr_{N_i} \P_i\sfer c_{ij}^*(J_M\epsilon_R, \gamma^5\afer{R}) G_i \label{eq:bb2-group4}
\end{align}
The fifth term of \eqref{eq:bb2-transf2} and the first term of \eqref{eq:bb2-transf5} make up another such group but with $\epsilon_R \to \epsilon_L$ and $\afer{R} \to \fer{L}$.

\item And finally all terms with either $F_{ij}$ or $F_{ij}^*$, consisting of the second term of \eqref{eq:bb2-transf1}, the second term of \eqref{eq:bb2-transf2}, the fourth and fifth terms of \eqref{eq:bb2-transf3} and the terms of \eqref{eq:bb2-transf6} (of which we have omitted the terms with $\gau{j}$ for now):
\begin{align}
	& \langle J_M d_{ij}'^*F_{ij}^*\epsilon_R, \gamma^5\gau{iR}\Cw{i,j}\sfer \rangle + \langle J_Md_{ij}'^*F_{ij}^*\epsilon_R, \can_A \fer{L}\rangle  \nn\\
 &\qquad - \int \tr_{N_j} F_{ij}^*\big[ d_{ij}(J_M\epsilon_R, \can_A\fer{L})_{S} + d_{ij,i}(J_M\epsilon_R, \gamma^5\gau{iR}\sfer)_{\cS}\big] \label{eq:bb2-group5}
\end{align}
and 
\begin{align*}
& \langle J_M F_{ij}d_{ij}' \epsilon_L, \gamma^5\asfer \Cw{i,j}^*\gau{iL}\rangle + \langle J_M \afer{R}, \can_A d_{ij}'F_{ij}\epsilon_L\rangle   \\
		&\qquad - \int \tr_{N_j}\big[d_{ij}^*(J_M\epsilon_L, \can_A\afer{R})_{S}  + d_{ij,i}^*(J_M\epsilon_L, \gamma^5\asfer\gau{iL})_{\cS} \big]F_{ij}.
\end{align*}

\end{itemize}


We will tackle each of these five groups in the next five lemmas. For the first group we have:
\begin{lem}\label{lem:bb2-group1}
	The expression \eqref{eq:bb2-group1} vanishes, provided that
	\begin{align}
\frac{1}{2}\Cw{i,j} c_{ij} &= - c_{i}g_i\label{eq:bb2-group1-constr}
	\end{align}
\end{lem}
\begin{proof}
	Since the expression contains only fermionic terms, we need to prove this via a Fierz transformation, which is valid only point-wise. We will write
\begin{align*}
	\gau{i} &= \gau{}^a \otimes T^a \in L^2(S_- \otimes su(N_i)_R),\nn\\
	 \fer{L} &= \fer{mn} \otimes e_{i,m}\otimes \bar e_{j,n} \in L^2(S_+ \otimes \rep{i}{j}),\nn\\
	 \afer{R} &= \afer{rs} \otimes e_{j,r} \otimes \bar e_{i,s} \in L^2(S_- \otimes \rep{j}{i}),
\end{align*}
where a sum over $a$, $m$, $n$, $r$ and $s$ is implied, to avoid a clash of notation. Here the $T^a$ are the generators of $su(N_i)$. Using this notation, \eqref{eq:bb2-group1} is point-wise seen to be equivalent to
\begin{align*}
	& (J_M\afer{jk}, \gamma^5\gau{}^a)(J_M\epsilon_L, \gamma^5 \Cw{i,j}c_{ij}\fer{ij}) T^a_{ki}\nn\\
	&\qquad\qquad + g_ic_i(J_M \afer{jk}, \gamma^\mu \fer{ij})(J_M\epsilon_L, \gamma_\mu \gau{}^a)T^a_{ki},
\end{align*}
after relabeling indices. Since it appears in both expressions, we may simply omit $T^a_{ki}$ from our considerations. For brevity we will omit the subscripts of the fermions from here on. We then apply a Fierz transformation (see Appendix \ref{sec:fierz}) for the first term, giving:
\begin{align*}
&\mkern-18mu	( J_M\afer{}, \gamma^5\gau{}^a)(J_M\epsilon_L, \gamma^5 \fer{}) = - \frac{C_{40}}{4}(J_M\afer{}, \fer{})(J_M\epsilon_L, \gau{}^a)\nn\\
& - \frac{C_{41}}{4}(J_M\afer{}, \gamma^\mu\fer{})(J_M\epsilon_L, \gamma_\mu \gau{}^a) - \frac{C_{42}}{4}(J_M\afer{}, \gamma^\mu\gamma^\nu\fer{})(J_M\epsilon_L, \gamma_\mu\gamma_\nu \gau{}^a)\nn\\
&\qquad - \frac{C_{43}}{4}(J_M\afer{}, \gamma^\mu\gamma^5\fer{})(J_M\epsilon_L, \gamma_\mu\gamma^5 \gau{}^a) - \frac{C_{44}}{4}(J_M\afer{}, \gamma^5\fer{})(J_M\epsilon_L, \gamma^5 \gau{}^a). 
\end{align*}
(Note that the sum in the third term on the RHS runs over $\mu < \nu$, see Example \ref{exmpl:dim4}.) We calculate: $C_{40} = C_{43} = C_{44} = -C_{41} = -C_{42}  = 1$ and use that $\fer{}$ and $\afer{}$ are of opposite parity, as are $\fer{}$ and $\gau{}^a$, to arrive at
\begin{align*}
 ( J_M\afer{}, \gamma^5\gau{}^a)(J_M\epsilon_L, \gamma^5 \fer{})
 &= \frac{1}{4}(J_M\afer{}, \gamma^\mu\fer{})(J_M\epsilon_L, \gamma_\mu \gau{}^a) \nn\\
	&\qquad\qquad	- \frac{1}{4}(J_M\afer{}, \gamma^\mu\gamma^5\fer{})(J_M\epsilon_L, \gamma_\mu\gamma^5 \gau{}^a)\\
	&= \frac{1}{2}(J_M\afer{}, \gamma^\mu \fer{})(J_M\epsilon_L, \gamma_\mu\gau{}^a).
\end{align*}

\end{proof}

\begin{rmk}\label{rmk:bb2-group1-constr1.1}
	From the action there in fact arises also a similar group of terms as \eqref{eq:bb2-group1}, that reads
\ba
	& \langle J_M\afer{R}, \gamma^5\Cw{j,i}c_{ij}(J_M\epsilon_L, \gamma^5 \fer{L})\gau{jR}\rangle 
	 - g_jc_j\langle J_M \afer{R}, \gamma^\mu \fer{L}(J_M\epsilon_L, \gamma_\mu \gau{jR})\rangle \label{eq:bb2-group1.1},
\ea
where the minus sign comes from the one in \eqref{eq:fluctDfull}. Performing the same calculations, we find 
 \ba
\frac{1}{2}\Cw{j,i}c_{ij} &= c_{j}g_j\label{eq:bb2-group1-constr1.1}
\ea
here.
\end{rmk}


\begin{lem}\label{lem:bb2-group2}
	The term \eqref{eq:bb2-group2} vanishes provided that
	\begin{align}
		 \frac{1}{2}c_{ij}'^*\Cw{i,j} &= - g_ic_i = \mathcal{P}_i c_{G_i}\label{eq:bb2-group2-constr}.
	\end{align}
\end{lem}
\begin{proof}
	Using that $[J_M, \gamma^5] = 0$, $(\gamma^5)^* = \gamma^5$ and $(\gamma^5)^2 = 1$, the first term of \eqref{eq:bb2-group2} can be rewritten as
	\begin{align*}
		c_{ij}'^* \langle J_M[\can_A,\asfer]\epsilon_L, \gau{iR}\Cw{i,j}\sfer \rangle = c_{ij}'^*\inpr{J_M \asfer \epsilon_L}{\can_A\gau{iR}\Cw{i,j}\sfer},
	\end{align*}
	where we have used the self-adjointness of $\can_A$. The third term of \eqref{eq:bb2-group2} can be written as
\begin{align}
	& g_ic_i\langle J_M\asfer\epsilon_L, \can_A\gau{iR}\sfer \rangle \label{eq:bb2-group2.1}
\end{align}
where we have used that $\slashed{\partial}\epsilon_L = 0$. On the other hand, the second and fourth terms of \eqref{eq:bb2-group2} can be rewritten to yield
\begin{align}
	&  +i g_ic_i\int \tr_{N_j}\big( \asfer (J_M\epsilon_L, \gamma_\mu \gau{iR}) D^\mu \sfer\big)
 -  c_{G_i} \tr_{N_i} \P_i\sfer\asfer (J_M\epsilon_L, \can_A\gau{iR})_{\cS}\nn \\
 & =  g_ic_i \langle J_M\asfer \epsilon_L, \can_A\gau{iR}\sfer \rangle \label{eq:bb2-group2.2}
\end{align} 
provided that $g_ic_i = - \P_i c_{G_i}$. Then the two terms \eqref{eq:bb2-group2.1} and \eqref{eq:bb2-group2.2} cancel, provided that
\begin{align*}
c_{ij}'^*\Cw{i,j} + 2g_ic_i &= 0.
\end{align*}
\end{proof}


\begin{lem}\label{lem:bb2-group3}
	The expression \eqref{eq:bb2-group3} vanishes, provided that
 	\begin{align}
 			c_{ij}^* =  c_{ij}' = - 2ic_i'\Cw{i,j}g_i^{-1} = 2ic_j'\Cw{j,i}g_j^{-1}\label{eq:bb2-group3-constr}.
	\end{align}
\end{lem}
\begin{proof}
We start with \eqref{eq:bb2-group3}:	
\begin{align*}
 &c_i' \langle J_M\afer{R}, \gamma^5F_i\Cw{i,j}\sfer\epsilon_R \rangle 
		 + c_{ij}^* \int \tr_{N_j}\big( D_\mu (J_M\epsilon_R, \gamma^5 \afer{R}) D^\mu \sfer\big)  \nn\\
	&\qquad\qquad	-   c_{ij}'\langle J_M \afer{R}, \gamma^5\can_A[\can_A, \sfer]\epsilon_R\rangle,
\end{align*}
where we have used that $\{\gamma^5, \can_A\} = 0$. Note that the second term in this expression can be rewritten as
\begin{align*}
	- c_{ij}^*\langle J_M \afer{R}, \gamma^5D_\mu D^\mu  \sfer\epsilon_R\rangle 
\end{align*}	
by using the cyclicity of the trace, the Leibniz rule for the partial derivative and Lemma \ref{lem:symmJ}. (We have discarded a boundary term here.) Together, the three terms can thus be written as
\begin{align*}
	\langle J_M \afer{R}, \gamma^5 \mathcal{O} \sfer\epsilon_R), \quad\mathcal{O} = c_i' \Cw{i,j} F_i -  c_{ij}^*D_\mu D^\mu - c_{ij}' \can_A^2,
\end{align*}
where we have used that $\slashed{\partial}\epsilon_{R} = 0$. We must show that the above expression can equal zero. Using Lemma \ref{lem:Dsquared} we have, on a flat background:
\begin{align*}
	\can_A^2 + D_\mu D^\mu &= - \frac{1}{2}\gamma^\mu\gamma^\nu \mathbb{F}_{\mu\nu} = \frac{i}{2} \gamma^\mu\gamma^\nu (g_i F^i_{\mu\nu} - g_j F^{j\,o}_{\mu\nu})
\end{align*}
since $\mathbb{A}_\mu = -ig_i\mathrm{ad}(A_\mu)$. Comparing the above equation with the expression for $\mathcal{O}$ we see that if $- c_{ij}^* = - c_{ij}' = 2ic_i' \Cw{i,j}g_i^{-1}$, the operator $\mathcal{O}$ ---applied to $\sfer\epsilon_R$--- indeed equals zero. From transforming the fermionic action we also obtain the term
\bas
	c_j'\inpr{J_M\afer{R}}{\gamma^5 \Cw{i,j}\sfer_{}F_j\eR}
\eas
	from which we infer the last equality of \eqref{eq:bb2-group3-constr}
\end{proof}


\begin{lem}\label{lem:bb2-group4}
	The expression \eqref{eq:bb2-group4} vanishes, provided that
	\begin{align}
		c_{ij}^*\P_i &= c_{G_i}'\Cw{i,j}\label{eq:bb2-group4-constr}
	\end{align}
\end{lem}
\begin{proof}
	The second term of \eqref{eq:bb2-group4} is rewritten using Lemmas \ref{lem:symmJ}, \ref{lem:pullScalar} and \ref{lem:moveScalar} to give
	\begin{align*}
  - c_{ij}^*\langle J_M\afer{R}, \gamma^5G_i\P_i\sfer \eR\rangle 
	\end{align*}
	establishing the result.
\end{proof}


Then finally for the last group of terms we have:
\begin{lem}\label{lem:bb2-group5}
	The expression \eqref{eq:bb2-group5} vanishes, provided that
	\begin{align}
		d_{ij} &= d_{ij}'^*, & d_{ij,i} &= d_{ij}'^*\Cw{i,j}, & d_{ij,j} &= - d_{ij}'^*\Cw{j,i}.\label{eq:bb2-group5-constr}
	\end{align}
\end{lem}
\begin{proof}
	The first two identities of \eqref{eq:bb2-group5-constr} are immediate. The third follows from the term that we have omitted in {\eqref{eq:bb2-group5}}, which is equal to the other term except that $\gau{iR}\sfer_{} \to \sfer_{}\gau{jR}$, $\Cw{i,j} \to \Cw{j,i}$ and $d_{ij,i} \to - d_{ij,j}$.
\end{proof}


Combining the five lemmas above, we complete the proof of Theorem \ref{prop:bb2} with the following proposition: 


\begin{prop}\label{thm:bb2}
	A supersymmetric action remains supersymmetric $\mathcal{O}(\Lambda^{0})$ after adding a `building block of the second type' to the spectral triple if the scaled parameters in the finite Dirac operator are given by
\begin{align}
\Cw{i,j} &= \sgnc_{i,j} \sqrt{\frac{2}{\K_i}} g_i \id_M, &
\Cw{j,i} &= \sgnc_{j,i} \sqrt{\frac{2}{\K_j}} g_j \id_M \label{eq:bb2-results1}
\end{align}
and if 
\begin{subequations}\label{eq:bb2-results2}
	\begin{align}
		c_{ij}' &=  c_{ij}^* = \sgnc_{i,j} \sqrt{2\K_i}c_i = - \sgnc_{j,i} \sqrt{2\K_j}c_j, \label{eq:bb2-results2a}\\
		 d_{ij} &= d_{ij}'^* = \sgnc_{i,j} \sqrt{\frac{\K_i}{2}} \frac{d_{ij,i}}{g_i}  = - \sgnc_{j,i} \sqrt{\frac{\K_j}{2}} \frac{d_{ij,j}}{g_j},\label{eq:bb2-results2b}\\
\mathcal{P}_i^2 &= g_i^2\mathcal{K}_i^{-1}, \label{eq:bb2-results2d} \\
		c_{G_i} &= \sgnc_{i} \sqrt{\K_i} c_i, \label{eq:bb2-results2c}
	\end{align}
\end{subequations}
	with $\sgnc_{ij}, \sgnc_{ji}, \sgnc_{i} \in \{\pm\}$.
\end{prop}
\begin{proof}
	Using Lemmas \ref{lem:bb2-group1}, \ref{lem:bb2-group2}, \ref{lem:bb2-group3}, \ref{lem:bb2-group4} and \ref{lem:bb2-group5}, the action is seen to be fully supersymmetric if the relations \eqref{eq:bb2-group1-constr}, \eqref{eq:bb2-group2-constr}, \eqref{eq:bb2-group3-constr}, \eqref{eq:bb2-group4-constr} and \eqref{eq:bb2-group5-constr} can simultaneously be met. We can combine \eqref{eq:bb2-group1-constr} and the second equality of \eqref{eq:bb2-group3-constr} to yield 
	\begin{align*}
		ic_i'\Cw{i,j}^*\Cw{i,j} = g_i^2c_i^*\qquad\Longrightarrow \qquad  \Cw{i,j}^*\Cw{i,j}c_i = - \frac{2g_i^2}{\K_i}c_i^*,
	\end{align*}
	where in the last step we have used the relation \eqref{eq:bb1-constr-final} between $c_i$ and $c_i'$. Inserting the expression for $\Cw{i,j}$ from \eqref{eq:bb2-scalingG} and assuming that $c_i \in i \mathbb{R}$ to ensure the reality of $\Cw{i,j}$, we find the first relation of \eqref{eq:bb2-results1}. The other parameter, $\Cw{j,i}$, can be obtained by invoking Remark \ref{rmk:bb2-group1-constr1.1} and using \eqref{eq:bb2-group3-constr}, leading to the second relation of \eqref{eq:bb2-results1}. Plugging the former result into \eqref{eq:bb2-group3-constr} and \eqref{eq:bb2-group5-constr} (and invoking \eqref{eq:bb1-constr-final}) gives the second equality in \eqref{eq:bb2-results2a} and those of \eqref{eq:bb2-results2b} respectively. Combining \eqref{eq:bb2-group4-constr}, \eqref{eq:bb2-results1} and the second equality of \eqref{eq:bb2-results2a}, we find 
	\ba \label{eq:bb2-proof-interm} c_{G_i} = - g_i^{-1} \K_i \P_i c_i. \ea The combination of the second equality of \eqref{eq:bb2-group2-constr} with \eqref{eq:bb2-proof-interm} yields \eqref{eq:bb2-results2d}. Finally, plugging this result back into \eqref{eq:bb2-proof-interm} gives \eqref{eq:bb2-results2c}.
\end{proof}

Note that upon setting $\K_i \equiv 1$ (as should be done in the end) we recover the well known results for both the supersymmetry transformation constants and the parameters of the fermion--sfermion--gaugino interaction.

\subsection{Third building block}\label{sec:bb3-proof}

The off shell counterparts of the \emph{new interactions} that we get in the four-scalar action, are of the form (c.f.~\eqref{eq:bb3-action-1})  
\ba
&S_{123,B}[\zeta, \szeta, F] \nn\\
&= \int_M \Big[\tr F_{ij}^*(\beta_{ij,k}\sfer_{ik}\asfer_{jk})+ \tr(\sfer_{jk}\asfer_{ik}\beta_{ij,k}^*)F_{ij} + 
\tr F_{ik}^*(\beta_{ik,j}^*\sfer_{ij}\sfer_{jk})+ \nn \\
&\qquad\qquad+\tr(\asfer_{jk}\asfer_{ij}\beta_{ik,j})F_{ik}  + \tr(\beta_{jk,i}\asfer_{ij}\sfer_{ik})F_{jk}^* + \tr(\asfer_{ik}\sfer_{ij}\beta_{jk,i}^*)F_{jk}\Big]\nn\\
&\equiv \int_M \Big[\tr F_{1}^*(\beta_{1}\sfer_{2}\asfer_{3}) 
+ \tr(\asfer_{3}\asfer_{1}\beta_{2})F_{2} + \tr(\beta_{3}\asfer_{1}\sfer_{2})F_{3}^* + h.c.\Big] \nn\\
&\to \int_M \Big[\tr F_{1}^*(\beta_{1}'\sfer_{2}\asfer_{3}) 
+ \tr(\asfer_{3}\asfer_{1}\bp_{2})F_{2} + \tr(\bp_{3}\asfer_{1}\sfer_{2})F_{3}^* + h.c.\Big] \label{eq:bb3-action}.
\ea
Here we have already scaled the fields according to \eqref{eq:bb3-scalingfields} and have written 
\ba\label{eq:def-scale-beta} \bp_1 &:= \n_3^{-1}\beta_1\n_2^{-1},& \bp_2 &:= \n_3^{-1} \beta_2 \n_1^{-1},& \bp_3 &:= \n_1^{-1}\beta_3\n_2^{-1}.\ea We apply the transformations \eqref{eq:susytransforms4} and \eqref{eq:susytransforms5} to the first term of \eqref{eq:bb3-action} above, giving:
\ba
 \mkern-18mu\tr F_{1}^*(\bp_{1}\sfer_{2}\asfer_{3}) &\to 
 \tr \Big[\Big(d_1^*(J_M\eL, \can_A\afer{1}) + d_{1,i}^*(J_M\eL,\gamma^5\asfer_{1}\gau{iL}) \nn\\
	&\quad\qquad\qquad- d_{1,j}^*(J_M\eL, \gamma^5\gau{jL}\asfer_{1})\Big)(\bp_{1}\sfer_{2}\asfer_{3})\label{eq:bb3-transform1}\\
	&\quad  +  \tr F_{1}^*\bp_{1}c_2(J_M\eR, \gamma^5\fer{2})\asfer_{3} + \tr F_{1}^*\bp_{1}\sfer_{2}(J_M\eR, \gamma^5\afer{3})c_3^*\Big]\nn,
\ea
where $c_{1,2,3}$ should not be confused with the transformation parameter $c_i$ of the building blocks of the first type. We have two more terms that can be obtained from the above ones by interchanging the indices $1$, $2$ and $3$:
\ba
 \mkern-18mu\tr(\asfer_{3}\asfer_{1}\bp_2)F_{2} &\to \tr\Big[(\asfer_{3}\asfer_{1}\bp_2)\Big(d_{2}(J_M\eL, \can_A\fer{2}) + d_{2,i}(J_M\eL, \gamma^5\gau{iL}\sfer_{2})_{\cS}\nn\\
	&\quad\qquad\qquad - d_{2,k}(J_M\eL, \gamma^5\sfer_{2}\gau{kL})\Big)\label{eq:bb3-transform2}\\
			&\quad + \tr c_3^*(J_M\eR, \gamma^5\afer{3})\asfer_{1}\bp_2F_2 + \tr \asfer_3c_1^*(J_M\eR, \gamma^5\afer{1})\bp_2F_2\Big]\nn
\ea
and
\ba
  \mkern-18mu\tr F_{3}^*(\bp_{3}\asfer_{1}\sfer_{2}) &\to  \tr \Big[\Big(d_{3}^*(J_M\eL, \can_A\afer{3})_{S} + d_{3,j}^*(J_M\eL, \gamma^5\asfer_{3}\gau{jL})_{\cS} \nn\\
		&\quad\qquad\qquad- d_{3,k}^*(J_M\eL, \gamma^5\gau{kL}\asfer_{3})\Big)(\bp_{3}\asfer_1\sfer_2)\label{eq:bb3-transform3} \\
	&\quad + \tr F_{3}^*\bp_{3}c_1^*(J_M\eR, \gamma^5\afer{1})\sfer_{2} +  \tr F_{3}^*\bp_{3}\asfer_{1}c_2(J_M\eR, \gamma^5\fer{2})\Big].\nn
\ea	
We can omit the other half of the terms in \eqref{eq:bb3-action} from our considerations. 

We introduce the notation 
\ba\label{eq:def-scale-yuk}
	\yukp{1}{} &:= \yuk{1}{}\n_1^{-1},& 	
	\yukp{2}{} &:= \n_2^{-1}\yuk{2}{},& 	
	\yukp{3}{} &:= \yuk{3}{}\n_3^{-1},& 	
\ea
for the scaled version of the parameters. Then for three of the fermionic terms of \eqref{eq:bb3-action-ferm-detail}, after scaling the fields, we get:
\ba
\inpr{J_M \afer{1}}{\gamma^5\fer{2}\asfer_{3}\yukps{3}{}} &\to 
\inpr{J_M ( c_{1}'^*\gamma^5 [\can_A, \asfer_{1}]\eL +d_{1}'^* F_{1}^*\eR)}{\gamma^5\fer{2}\asfer_{3}\yukps{3}{}}\nn\\ 
&\qquad	 +\inpr{J_M \afer{1}}{\gamma^5\fer{2} c_{3}^*(J_M\eR, \gamma^5 \afer{3})\yukps{3}{}} \nn\\
&\qquad  +\inpr{J_M \afer{1}}{\gamma^5(c_{2}' \gamma^5 [\can_A, \sfer_{2}]\eL + d_{2}'F_{2}\eR)\asfer_{3}\yukps{3}{}} ,\label{eq:bb3-transform4}\\
\inpr{J_M \afer{1}}{\gamma^5\yukps{2}{}\sfer_{2}\afer{3}} &\to 
\inpr{J_M ( c_{1}'^*\gamma^5 [\can_A, \asfer_{1}]\eL + F_{1}^*d_{1}'^*\eR)}{\gamma^5\yukps{2}{}\sfer_{2}\afer{3}} \nn\\
&\qquad + \inpr{J_M \afer{1}}{\gamma^5\yukps{2}{}c_{2}(J_M\eR, \gamma^5 \fer{2})\afer{3}}\nn\\
&\qquad + \inpr{J_M \afer{1}}{\gamma^5\yukps{2}{}\sfer_{2}( c_{3}'^*\gamma^5 [\can_A, \asfer_{3}]\eL + d_{3}'^*F_{3}^*\eR)},\label{eq:bb3-transform5} \\
& \text{and} \nn\\
\inpr{J_M\afer{3}}{\gamma^5\asfer_{1}\yukps{1}{}\fer{2}} &\to 
\inpr{J_M( c_{3}'^*\gamma^5 [\can_A, \asfer_{3}]\eL + d_{3}'^*F_{3}^*\eR)}{\gamma^5\asfer_{1}\yukps{1}{}\fer{2}}\nn\\
&\qquad + \inpr{J_M\afer{3}}{\gamma^5 c_{1}^*(J_M\eR, \gamma^5 \afer{1})\yukps{1}{}\fer{2}}\nn\\
&\qquad + \inpr{J_M\afer{3}}{\gamma^5\asfer_{1}\yukps{1}{}( \gamma^5 [\can_A, c_{2}'\sfer_{2}]\eL + d_{2}'F_{2}\eR)}.\label{eq:bb3-transform6}
\ea
We can safely omit the other terms of the fermionic action \eqref{eq:bb3-action-ferm-detail}.

Collecting the terms from \eqref{eq:bb3-transform1} -- \eqref{eq:bb3-transform6} containing the same variables, we obtain the following groups of terms:
\begin{itemize}

	\item a group with three fermionic terms:
	\ba
	& \inpr{J_M \afer{1}}{\gamma^5\fer{2} c_{3}^*(J_M\eR, \gamma^5 \afer{3})\yukps{3}{}} + \inpr{J_M \afer{1}}{\gamma^5\yukps{2}{}c_{2}(J_M\eR, \gamma^5 \fer{2})\afer{3}} \nn\\
	&\qquad + \inpr{J_M\afer{3}}{\gamma^5 c_{1}^*(J_M\eR, \gamma^5 \afer{1})\yukps{1}{}\fer{2}}\nn\\
	&= \inpr{J_M \afer{1}}{\gamma^5\fer{2a} c_{3}^*(J_M\eR, \gamma^5 \afer{3b})}(\yukps{3}{})_{ba} \nn\\
	&\qquad + \inpr{J_M \afer{1}}{\gamma^5c_{2}(J_M\eR, \gamma^5 \fer{2a})\afer{3b}}(\yukps{2}{})_{ba} \nn\\
	&\qquad\qquad + \inpr{J_M\afer{3b}}{\gamma^5 c_{1}^*(J_M\eR, \gamma^5 \afer{1})\fer{2a}}(\yukps{1}{})_{ba}\label{eq:bb3-group1},
	\ea
	consisting of part of the second term of \eqref{eq:bb3-transform4}, the second term of \eqref{eq:bb3-transform5} and the second term of \eqref{eq:bb3-transform6}. Here we have explicitly written possible family indices and have assumed that it is $\sfer_{ij}$ and $\fer{ij}$ that lack these.

	\item Three similar groups containing all terms with the auxiliary fields $F_1^*$, $F_2$ and $F_3^*$ respectively:
	\begin{subequations}\label{eq:bb3-group2}
	\ba
			 &\inpr{J_M d_{1}'^*F_{1}^*\eR}{\gamma^5\fer{2}\asfer_{3}\yukps{3}{}} +  \inpr{J_M d_{1}'^*F_{1}^*\eR}{\gamma^5\yukps{2}{}\sfer_{2}\afer{3}}\nn \\ 
			 &\qquad\qquad + \int_M  \tr F_{1}^*\bp_{1}c_2(J_M\eR, \gamma^5\fer{2})\asfer_{3} + \tr F_{1}^*\bp_{1}\sfer_{2}c_3^*(J_M\eR, \gamma^5\afer{3}) \label{eq:bb3-group2.1},\\
 &  \inpr{J_M \afer{1}}{\gamma^5d_{2}'F_{2}\asfer_{3}\yukps{3}{}\eR} + \inpr{J_M\afer{3}}{\gamma^5\asfer_{1}\yukps{1}{}d_{2}'F_{2}\eR)}\nn\\
 &\qquad\qquad + \int_M  \tr \asfer_3c_1^*(J_M\eR, \gamma^5\afer{1})\bp_{2}F_2 + \tr c_3^*(J_M\eR, \gamma^5\afer{3})\asfer_{1}\bp_{2}F_2 \label{eq:bb3-group2.2}\\
	&\qquad \text{and}\nn\\
		& \inpr{J_M \afer{1}}{\gamma^5\yukps{2}{}\sfer_{2} d_{3}'^*F_{3}^*\eR} + \inpr{J_Md_{3}'^*F_{3}^*\eR}{\gamma^5\asfer_{1}\yukps{1}{}\fer{2}}\nn\\
		&\qquad\qquad + \int_M \tr F_{3}^*\bp_{3}c_1^*(J_M\eR, \gamma^5\afer{1})\sfer_{2} +  \tr F_{3}^*\bp_{3}\asfer_{1}c_2(J_M\eR, \gamma^5\fer{2}) \label{eq:bb3-group2.3},
	\ea
	\end{subequations}
	where, for example, the first group comes from parts of the first terms of \eqref{eq:bb3-transform4} and of \eqref{eq:bb3-transform5} and from the last two terms of \eqref{eq:bb3-transform1}.

	\item A group with the gauginos $\gau{iL}$, $\gau{jL}$:
	\ba
		 \int_M &\tr\big[ d_{1,i}^*(J_M\eL,\gamma^5\asfer_{1}\gau{iL}) - d_{1,j}^*(J_M\eL, \gamma^5\gau{jL}\asfer_{1})\big](\bp_1\sfer_{2}\asfer_{3})\nn\\
				&\qquad + \tr(\asfer_{3}\asfer_{1}\bp_2)\big[d_{2,i}(J_M\eL, \gamma^5\gau{iL}\sfer_{2}) - d_{2,k}(J_M\eL, \gamma^5\sfer_{2}\gau{kL})\big]\nn\\
 		&\qquad 	+  \tr\big[ d_{3,j}^*(J_M\eL, \gamma^5\asfer_{3}\gau{jL}) - d_{3,k}^*(J_M\eL, \gamma^5\gau{kL}\asfer_{3})\big](\bp_3\asfer_1\sfer_2)\label{eq:bb3-group3},
	\ea
coming from the second and third terms of \eqref{eq:bb3-transform1}, \eqref{eq:bb3-transform2} and \eqref{eq:bb3-transform3} respectively.

	\item And finally three groups of terms containing the Dirac operator $\can_A$:
\begin{subequations}\label{eq:bb3-group4}  
	\ba
&\inpr{J_M \afer{1}}{c_{2}'[\can_A, \sfer_{2}]\asfer_{3}\yukps{3}{}\eL} +
\inpr{J_M \afer{1}}{ \yukps{2}{}\sfer_{2} c_{3}'^*[\can_A, \asfer_{3}]\eL}\nn\\
	&\qquad\qquad +	\int_M \tr  d_1^*(J_M\eL, \can_A\afer{1})\bp_1\sfer_{2}\asfer_{3}\label{eq:bb3-group4.1}  
	,\\
&\inpr{J_M c_{1}'^*[\can_A, \asfer_{1}] \eL}{\fer{2}\asfer_{3}\yukps{3}{}}
+ \inpr{J_Mc_{3}'^*[\can_A, \asfer_{3}] \eL}{\asfer_{1}\yukps{1}{}\fer{2}}\nn\\
	&\qquad\qquad
 + \int_M\tr\asfer_{3}\asfer_{1}\bp_2d_{2}(J_M\eL, \can_A\fer{2}), \label{eq:bb3-group4.2}  
	\\&\qquad\qquad\text{and}\nn\\
&\inpr{J_M c_{1}'^* [\can_A, \asfer_{1}]\eL}{ \yukps{2}{}\sfer_{2}\afer{3}} + 
\inpr{J_M\afer{3}}{\asfer_{1}  \yukps{1}{} c_{2}'[\can_A, \sfer_{2}]\eL}\nn\\
	&\qquad\qquad
+ \int_M \tr  d_{3}^* (J_M\eL, \can_A\afer{3})\bp_3 \asfer_1\sfer_2 \label{eq:bb3-group4.3}, 
	\ea
	\end{subequations}
coming from parts of the first and third terms of \eqref{eq:bb3-transform4} -- \eqref{eq:bb3-transform6} and from the first terms of \eqref{eq:bb3-transform1} -- \eqref{eq:bb3-transform3}.
\end{itemize}

\begin{lem}\label{lem:bb3-lem1}
	The group \eqref{eq:bb3-group1} vanishes, provided that
	\ba
		c_3^*\yukps{3}{} = c_2\yukps{2}{} = c_1^*\yukps{1}{}	\label{eq:bb3-constr1}
	\ea
\end{lem}
\begin{proof}
	Since the terms contain four fermions, we must employ a Fierz transformation (Appendix \ref{sec:fierz}). Point-wise, we have for the first term of \eqref{eq:bb3-group1} (omitting its pre-factor for now):
	\bas
		&\rinpr{J_M\afer{1}}{\gamma^5\fer{2}}\rinpr{J_M\eR}{\gamma^5\afer{3}}\nn\\ &= - \frac{C_{40}}{4}\rinpr{J_M\afer{1}}{\afer{3}}\rinpr{J_M\eR}{\fer{2}}
- \frac{C_{41}}{4}\rinpr{J_M\afer{1}}{\gamma^\mu\afer{3}}\rinpr{J_M\eR}{\gamma_\mu\fer{2}}\\
&\qquad - \frac{C_{42}}{4}\rinpr{J_M\afer{1}}{\gamma^\mu\gamma^\nu\afer{3}}\rinpr{J_M\eR}{\gamma_\mu\gamma_\nu\fer{2}}\nn\\
&\qquad\qquad - \frac{C_{43}}{4}\rinpr{J_M\afer{1}}{\gamma^\mu\gamma^5\afer{3}}\rinpr{J_M\eR}{\gamma_\mu\gamma^5\fer{2}}\\
&\qquad\qquad\qquad - \frac{C_{44}}{4}\rinpr{J_M\afer{1}}{\gamma^5\afer{3}}\rinpr{J_M\eR}{\gamma^5\fer{2}}\\
	&= - \frac{1}{2}\rinpr{J_M\afer{1}}{\gamma^5\afer{3}}\rinpr{J_M\eR}{\gamma^5\fer{2}}
+ \frac{1}{4}\rinpr{J_M\afer{1}}{\gamma^\mu\gamma^\nu\afer{3}}\rinpr{J_M\eR}{\gamma_\mu\gamma_\nu\fer{2}},
	\eas 
where we have used that $C_{40} = C_{44} = -C_{42} = 1$ and that all fermions are of the same chirality. (Note that the sum in the last term runs over $\mu < \nu$, see Example \ref{exmpl:dim4}.) Similarly, we can take the third term of \eqref{eq:bb3-group1}, use the symmetries of the inner product for both terms, and apply the same transformation. This yields
\ba
&\rinpr{J_M\afer{3}}{\gamma^5 (J_M\eR, \gamma^5 \afer{1})\fer{2}} \nn\\
&= \rinpr{J_M\fer{2}}{\gamma^5 \afer{3}}\rinpr{J_M\afer{1}}{\gamma^5\eR}\nn\\
	&= 	- \frac{1}{2}\rinpr{J_M\fer{2}}{\gamma^5\eR}\rinpr{J_M\afer{1}}{\gamma^5 \afer{3}}
	 	+ \frac{1}{4}\rinpr{J_M\fer{2}}{\gamma^\mu\gamma^\nu\eR}\rinpr{J_M\afer{1}}{\gamma_\mu\gamma_\nu \afer{3}}\nn\\
	&= 	-  \frac{1}{2}\rinpr{J_M\eR}{\gamma^5\fer{2}}\rinpr{J_M\afer{1}}{\gamma^5 \afer{3}}
	 	- \frac{1}{4}\rinpr{J_M\eR}{\gamma^\mu\gamma^\nu\fer{2}}\rinpr{J_M\afer{1}}{\gamma_\mu\gamma_\nu \afer{3}}\label{eq:bb3-fierz1},
\ea
where we have used the symmetries \eqref{eq:identitySymJ} for the second inner product in each of the two terms of \eqref{eq:bb3-fierz1}. We can add the two results, yielding
\bas
&\rinpr{J_M\afer{1}}{\gamma^5\fer{2}}\rinpr{J_M\eR}{\gamma^5\afer{3}c_3^*\yukps{3}{}} + \rinpr{J_M\afer{3}}{\gamma^5 (J_M\eR, \gamma^5 \afer{1})c_1^*\yukps{1}{}\fer{2}}\nn\\
&= - \frac{1}{2}(c_1^*\yukps{1}{}
 +  c_3^*\yukps{3}{} )_{ba} \rinpr{J_M\afer{1}}{\gamma^5\afer{3b}}\rinpr{J_M\eR}{\gamma^5\fer{2a}} \nn\\
&\qquad + \frac{1}{4} (c_3^*\yukps{3}{}
 - c_1^*\yukps{1}{} )_{ba} \rinpr{J_M\eR}{\gamma^\mu\gamma^\nu\fer{2a}}\rinpr{J_M\afer{1}}{\gamma^\mu\gamma^\nu \afer{3b}}
\eas
When $c_3^*\yukps{3}{} = c_1^*\yukps{1}{} = c_2\yukps{2}{}$, this result is seen to cancel the remaining term in \eqref{eq:bb3-group1}.
\end{proof}


\begin{lem}
	The groups of terms \eqref{eq:bb3-group2} vanish, provided that
	\ba
		c_2\bp_1  &= - d_1'^*\yukps{3}{}, &c_3^*\bp_1  &= - d_1'^*\yukps{2}{},& c_3^*\bp_2  &= - d_2'\yukps{1}{},\nn\\
		c_1^*\bp_2  &= - d_2'\yukps{3}{},& c_1^*\bp_3  &= - d_3'^*\yukps{2}{}, & c_2 \bp_3&= - d_3'^*\yukps{1}{}.\label{eq:bb3-constr2}
	\ea
\end{lem}
\begin{proof}
	This can readily be seen upon using Lemma \ref{lem:symmJ}, the cyclicity of the trace and Lemma \ref{lem:pullScalar}.
\end{proof}

\begin{lem}\label{lem:bb3-group3}
	The group of terms \eqref{eq:bb3-group3} vanishes, provided that
	\ba
		d_{1,i}^*\bp_1 &= -d_{2,i}\bp_2, & d_{1,j}^*\bp_1	&= d_{3,j}^*\bp_3, & d_{2,k}\bp_2 	&= -d_{3,k}^*\bp_3.\label{eq:bb3-constr3}
	\ea
\end{lem}
\begin{proof}
	This can readily be seen upon using the cyclicity of the trace and Lemma \ref{lem:pullScalar}.
\end{proof}

\begin{lem}\label{lem:bb3-group4}
	The three groups of terms \eqref{eq:bb3-group4} vanish, provided that
	\ba
		\yukps{3}{}c_2' &= c_3'^*\yukps{2}{} = - d_1^* \bp_1,&
		\yukps{3}{}c_1'^* &= c_3'^*\yukps{1}{} = - \bp_2d_2, \nn\\
		c_1'^*\yukps{2}{} &= \yukps{1}{}c_2' = - d_3^*\bp_3. &\label{eq:bb3-constr4}
	\ea
\end{lem}
\begin{proof}
	This can be checked quite easily using the symmetry \eqref{eq:identitySymJ}, the Leibniz rule for $\can_A$ and the fact that it is self-adjoint, that $\epsilon_{L,R}$ vanish covariantly and Lemmas \ref{lem:pullScalar} and \ref{lem:moveScalar}.
\end{proof}

Combining the above lemmas, we get:

\begin{prop}\label{prop:bb3}
	The extra action as a result of adding a building block \B{ijk} of the third type is supersymmetric if and only if the coefficients $\yuk{i}{j}$, $\yuk{i}{k}$ and $\yuk{j}{k}$ are related to each other via 	
	\ba
	\yuk{i}{j}C_{iij}^{-1} &= -(C_{iik}^*)^{-1}\yuk{i}{k},  &
	\yuk{i}{j}C_{ijj}^{-1} &= - \yuk{j}{k}C_{jjk}^{-1},			\nn\\ 
	(C_{ikk}^*)^{-1}\yuk{i}{k} &= - \yuk{j}{k}C_{jkk}^{-1}, && \label{eq:bb3-result3}
	\ea
 the constants of the transformations satisfy
	\ba
	|d_1|^2 &= 
	|d_2|^2 = 
	|d_3|^2 = |c_1|^2 =  |c_2|^2 = |c_3|^2\label{eq:bb3-result2}
	\ea
and the coefficients $\bp_{ij}$ are given by
	\ba
\bps_1\bp_1 &= \bps_2\bp_2 = \bps_3\bp_3 = \yukp{1}{}\yukps{1}{}=  \yukp{2}{}\yukps{2}{} = \yukp{3}{}\yukps{3}{}.\label{eq:bb3-result1}
	\ea
\end{prop}
\begin{proof}
	First of all, we plug the intermediate result \eqref{eq:bb2-group1-constr} for $\Cw{i,j}$ as given by \eqref{eq:bb2-scalingG} (but keeping in mind the results of Remark \ref{rmk:bb2-group1-constr1.1}) into the Hermitian conjugate of the result \eqref{eq:bb3-constr1} such that pairwise the same combination $c_ig_i$ appears on both sides. This yields 
\bas
	\mkern-18mu \yuk{i}{j}(- 2c_ig_i)C_{iij}^{-1} &= (-2c_ig_iC_{iik}^{-1})^*\yuk{i}{k}, &
	\yuk{i}{j}(2c_jg_jC_{ijj}^{-1}) &= \yuk{j}{k}(-2c_jg_j)C_{jjk}^{-1},\nn\\
	\mkern-18mu(2c_kg_kC_{ikk}^{-1})^*\yuk{i}{k} &= \yuk{j}{k}(2c_kg_k)C_{jkk}^{-1}. &&\nn
\eas
Using that the $c_{i,j,k}$ are purely imaginary (cf.~Theorem \ref{thm:bb2}), we obtain \eqref{eq:bb3-result3}. Secondly, comparing the relations \eqref{eq:bb3-constr2} with \eqref{eq:bb3-constr4} gives
	\bas
	d_1d_1' &= (c_2c_2')^* = c_3c_3', &
	(d_2d_2')^* &= c_1c_1' = c_3c_3', &
	d_3d_3' &= c_1c_1' = (c_2c_2')^*.
	\eas
Using the relations \eqref{eq:bb2-results2a} and \eqref{eq:bb2-results2b} between the constraints, \eqref{eq:bb3-result2} follows. Plugging the relations from \eqref{eq:bb3-result2} into those of \eqref{eq:bb3-constr2}, we obtain
\bas
\bps_1\bp_1 &= \yukp{3}{}\yukps{3}{}= \yukp{2}{}\yukps{2}{}, &
\bps_2\bp_2 &= \yukp{1}{}\yukps{1}{}= \yukp{3}{}\yukps{3}{}, \nn\\
\bps_3\bp_3 &= \yukp{2}{}\yukps{2}{}= \yukp{1}{}\yukps{1}{}, &&
\eas
from which \eqref{eq:bb3-result1} directly follows.
\end{proof}

N.B.~Using \eqref{eq:def-scale-beta} and \eqref{eq:def-scale-yuk} we can phrase the identities \eqref{eq:bb3-result1} in terms of the unscaled quantities $\beta_{1,2,3}$ and $\yuk{1,2,3}{}$ as
\bas
	\n_3^{-1}\beta_2 &= \beta_3\n_2^{-1} = \yuks{1}{},&
	\n_3^{-1}\beta_1 &= \beta_3\n_1^{-1} = \yuks{2}{},\nn\\ 
	\n_1^{-1}\beta_2 &= \beta_1\n_2^{-1} = \yuks{3}{},&&
\eas
where we have used that $\n_{1} \in \mathbb{R}$ since $\sfer_{1}$ has $R = 1$ (and consequently multiplicity 1).

\subsection{Fourth building block}\label{sec:bb4-proof}

Phrased in terms of the auxiliary field $F_{11'} =: F$, a building block of the fourth type induces the following action:
\bas
&	\frac{1}{2}\inpr{J_M\fer{}}{\gamma^5 \maj^* \fer{}} + \frac{1}{2}\inpr{J_M\afer{}}{\gamma^5 \maj \afer{}} - \tr \Big(F^* \gamma \asfer + h.c.\Big). 
\eas
Here we have written $\fer{} := \fer{11'L}$, $\afer{} := \afer{11'R}$ and $\sfer := \sfer_{11'}$ for conciseness. Transforming the fields that appear in the above action, we have the following.
\begin{itemize}
\item From the first term:
\bas
	& \frac{1}{2}\inpr{J_M(c^*\gamma^5[\can_A, \sfer_{}]\eR + d^* F\eL)}{\gamma^5 \maj^* \fer{}} \nn\\ &\qquad  + 
	\frac{1}{2}\inpr{J_M\fer{}}{\gamma^5 \maj^* (c^*\gamma^5[\can_A, \sfer_{}]\eR + d^* F\eL)}.
\eas

\item From the second term:
\bas
	& \frac{1}{2}\inpr{J_M(c\gamma^5[\can_A, \asfer_{}]\eL + d F^*\eR)}{\gamma^5 \maj \afer{}} \nn\\ &\qquad + 
	\frac{1}{2}\inpr{J_M\afer{}}{\gamma^5 \maj (c\gamma^5[\can_A, \asfer_{}]\eL + d F^*\eR)}.
\eas

\item From the terms with the auxiliary fields:
\bas
	& - \tr \Big[d^*(J_M\eL, \can_A\afer{}) + d'^*(J_M\eL, \gamma^5\asfer\gau{1L}) - d''^*(J_M\eL, \gamma^5\gau{1'L}\asfer)\Big] \gamma \asfer \nn\\
	&\qquad - c^* \tr F^* \gamma \rinpr{J_M\eR}{\gamma^5 \afer{}} 
\eas
and 
\bas 
	& - \tr  \sfer{}\gamma^*\Big[d(J_M\eR, \can_A\fer{}) + d'(J_M\eR, \gamma^5\gau{1R}\sfer) - d''(J_M\eR, \gamma^5\sfer\gau{1'R})\Big]\nn\\
	&\qquad - c\tr (J_M\eL, \gamma^5\fer{})\gamma^* F.	
\eas
\end{itemize}
Here we have written $c := c_{ij}$, $d := d_{ij}$ (where we have expressed $c_{ij}'^*$ as $c_{ij}$ and $d_{ij}'^*$ as $d_{ij}$ using \eqref{eq:bb2-results2a} and \eqref{eq:bb2-results2b}) and $d' := d_{11',1}$, $d'' := d_{11',1'}$. We group all terms according to the fields that appear in them, leaving essentially the following three.
\begin{itemize}

\item The group consisting of all terms with $F^*$ and $\afer{}$:
\bas
	&\frac{1}{2}\inpr{J_M d F^*\eR}{\gamma^5 \maj \afer{}} + 
	\frac{1}{2}\inpr{J_M\afer{}}{\gamma^5 \maj  d F^*\eR}\nn\\
	 &\qquad\qquad - c^*\int_M \tr F^* \gamma \rinpr{J_M\eR}{\gamma^5 \afer{}} \\
	&= \inpr{J_M F^*\eR}{\gamma^5(d \maj - c^*\gamma)\afer{}} 
\eas
where we have used the symmetry of the inner product from Lemma \ref{lem:symmJ} and Lemma \ref{lem:pullScalar}. This group thus only vanishes if
\ba\label{eq:bb4-group1-constr}
	d \maj &= c^*\gamma.
\ea
There is also a group of terms featuring $F$ and $\fer{}$, but this is of the same form as the one above.

\item A group of three terms with $\fer{}$ and $\sfer{}$:
\bas
	& \frac{1}{2}\inpr{J_Mc^*\gamma^5[\can_A, \sfer_{}]\eR }{\gamma^5 \maj^* \fer{}} + 
	\frac{1}{2}\inpr{J_M\fer{}}{\gamma^5 \maj^* c^*\gamma^5[\can_A, \sfer_{}]\eR}\nn\\
	 &\qquad - \int_M \tr  \sfer{}\gamma^*d(J_M\eR, \can_A\fer{}) \\
	&= \inpr{J_Mc^*\gamma^5[\can_A, \sfer_{}]\eR }{\gamma^5 \maj^* \fer{}}
	 -   \inpr{J_M\sfer{}\eR}{\can_A\gamma^*d\fer{}},
\eas
where also here we have used Lemmas \ref{lem:symmJ} and \ref{lem:pullScalar}. Using the self-adjointness of $\can_A$ this is only seen to vanish if
\ba\label{eq:bb4-group2-constr}
	c^*\maj^* &= \gamma^* d.
\ea
There is also a group of terms featuring $\afer{}$ and $\asfer{}$ but these are seen to be of the same form as the terms above.
\item Finally, there are terms that feature gauginos:
\bas
	& -\int_M \Big[ \tr d'^*(J_M\eL, \gamma^5\asfer\gau{1L}) - d''^*(J_M\eL, \gamma^5\gau{1'L}\asfer)\Big] \gamma \asfer \nn\\
	&\qquad  - \int_M \tr \sfer{}\gamma^*\Big[d'(J_M\eR, \gamma^5\gau{1R}\sfer) - d''(J_M\eR, \gamma^5\sfer\gau{1'R})\Big].
\eas
This expression is immediately seen to vanish when 
\bas
d'^*\gau{1L}  &= d''^*\gau{1'L}, &
d'\gau{1R} &= d''\gau{1'R}.
\eas
For this to happen we need that the gauginos are associated to each other and that $d' = d''$.

\end{itemize}

Combining the demands \eqref{eq:bb4-group1-constr} and \eqref{eq:bb4-group2-constr} we obtain
\bas
	\maj^*\maj = \frac{|c|^2}{|d|^2} \gamma^*\gamma = \frac{|d|^2}{|c|^2}\gamma^*\gamma
\eas
i.e.
\bas
	\maj^*\maj &=  \gamma^*\gamma, & |d|^2 &= |c|^2.
\eas
\subsection{Fifth building block}\label{sec:bb5-proof}

We transform the fields that appear in the action according to \eqref{eq:susytransforms4} and \eqref{eq:susytransforms5}. We suppress the indices $i$ and $j$ as much as possible, writing $c \equiv c_{ij}, d \equiv d_{ij}$ for the transformation coefficients \eqref{eq:susytransforms5} of the building block \Bc{ij}{+} of the second type. We eliminate $c_{ij}'$ and $d_{ij}'$ in these transformations using the first relations of \eqref{eq:bb2-results2a} and \eqref{eq:bb2-results2b} so that we can write $c', d'$ for those associated to \Bc{ij}{-}. 

The first fermionic term of \eqref{eq:bb5-action-scaled} transforms as
\bas
 \inpr{J_M \afer{R}}{\gamma^5 \mu\fer{R}'} &\to
 \inpr{J_M (\gamma^5 c[\can_A, \asfer]\eL + dF^*\eR)}{\gamma^5 \mu\fer{R}'}\\
	&\qquad + \inpr{J_M \afer{R}}{\gamma^5 \mu(c'^* \gamma^5 [\can_A, \sfer']\eL + d'^*F'\eR)} 
\eas

The second fermionic term of \eqref{eq:bb5-action-scaled} transforms as
\bas
 \inpr{J_M\afer{L}'}{\gamma^5 \mu^*\fer{L}} &\to 
 \inpr{J_M(c'\gamma^5 [\can_A, \asfer']\eR + d'F'^*\eL)}{\gamma^5 \mu^*\fer{L}} \\ 
 &\qquad + \inpr{J_M\afer{L}'}{\gamma^5 \mu^*(c^* \gamma^5 [\can_A, \sfer]\eR + d^*F\eL)}
\eas

The four terms in \eqref{eq:bb5-auxfields} transform as
\bas
	\mkern-18mu - \int_M \tr F'^* \delta \sfer  &\to - \int_M \Big(\tr \big[d'^*(J_M\eR, \can_A\afer{L}') + d_{ij,i}'^*(J_M\eR, \gamma^5\asfer'\gau{iR}) \nn\\
		&\quad\qquad - d_{ij,j}'^*(J_M\eR, \gamma^5\gau{jR}\asfer')\big] \delta \sfer  + \tr F'^* \delta c(J_M\eL, \gamma^5 \fer{L})\Big) \nn,\\
	\mkern-18mu - \int_M  \tr F^* \delta' \sfer' &\to - \int_M \Big(	\tr \big[d^*(J_M\eL, \can_A\afer{R}) + d_{ij,i}^*(J_M\eL, \gamma^5\asfer\gau{iL})\nn\\
			&\quad\qquad  - d_{ij,j}^*(J_M\eL, \gamma^5\gau{jL}\asfer)\big] \delta' \sfer'  + \tr F^* \delta' c'(J_M\eR, \gamma^5 \fer{R}') \Big)\nn,\\
	\mkern-18mu - \int_M  \tr \asfer \delta^* F' &\to - \int_M \Big( \tr c^*(J_M\eR, \gamma^5 \afer{R}) \delta^* F' + \tr \asfer \delta^* \big[d'(J_M\eL, \can_A\fer{R}') \nn\\
	&\quad\qquad+ d_{ij,i}'(J_M\eL, \gamma^5\gau{iL}\sfer') - d_{ij,j}'(J_M\eL, \gamma^5\sfer'\gau{jL})\big]\Big)\nn
\intertext{and }
	\mkern-18mu - \int_M \tr \asfer' \delta'^* F &\to - \int_M \Big(\tr c'^*(J_M\eL, \gamma^5 \afer{L}') \delta'^* F  + \tr \asfer' \delta'^* \big[ d(J_M\eR, \can_A\fer{L}) \nn\\
		&\quad\qquad + d_{ij,i}(J_M\eR, \gamma^5\gau{iR}\sfer) - d_{ij,j}(J_M\eR, \gamma^5\sfer\gau{jR})\big] \Big).
\eas

We group all terms that feature the same fields, which gives
\begin{itemize}
\item  a group with $F$ and $F'$:
\bas
 & d'^*\inpr{J_M \afer{R}}{\gamma^5 \mu F'\eR} +  d^*\inpr{J_M\afer{L}'}{\gamma^5 \mu^*F\eL}\nn\\
	&\qquad\qquad - \int_M\Big( \tr c^*(J_M\eR, \gamma^5 \afer{R}) \delta^* F' + \tr c'^*(J_M\eL, \gamma^5 \afer{L}') \delta'^* F\Big). \nn
\eas
Using Lemmas \ref{lem:moveScalar} and \ref{lem:pullScalar} and employing the symmetries of the inner product (Lemma \ref{lem:symmJ}), this is seen to equal
\bas
 & d'^*\inpr{J_M \afer{R}}{\gamma^5 \mu F'\eR} +  d^*\inpr{J_M\afer{L}'}{\gamma^5 \mu^*F\eL}\nn\\
	&\qquad\qquad -  c^*\inpr{J_M\afer{R}}{\gamma^5  \delta^* F'\eR} -  c'^*\inpr{J_M \afer{L}'}{\gamma^5 \delta'^* F\eL} \nn\\
 &= \inpr{J_M \afer{R}}{\gamma^5 \big[ d'^*\mu -  c^* \delta^*\big] F'\eR} +  \inpr{J_M\afer{L}'}{\gamma^5 \big[ d^*\mu^* - c'^*\delta'^*\big]F\eL}\nn
\eas
This only vanishes if
\ba\label{eq:bb5-constr1}
 d'^*\mu &=  c^* \delta^*, & d^*\mu^* &= c'^*\delta'^*.
\ea

\item a group with $F^*$ and $F'^*$, that vanishes automatically if and only if \eqref{eq:bb5-constr1} is satisfied.

\item a group featuring $\fer{R}'$ and $\fer{L}$:
\bas
 & \inpr{J_M  c[\can_A, \asfer]\eL}{ \mu\fer{R}'} + c'\inpr{J_M [\can_A, \asfer']\eR }{ \mu^*\fer{L}} \nn\\
	&\qquad\qquad - \int_M \Big( \tr \asfer \delta^*d'(J_M\eL, \can_A\fer{R}') + \tr \asfer' \delta'^* d(J_M\eR, \can_A\fer{L})  \Big).
\eas
Employing Lemmas \ref{lem:pullScalar} and \ref{lem:moveScalar} this is seen to equal
\bas
 & \inpr{J_M  c[\can_A, \asfer]\eL}{ \mu\fer{R}'} +  c'\inpr{J_M [\can_A, \asfer']\eR }{ \mu^*\fer{L}} \nn\\
	&\qquad\qquad - d'\inpr{J_M\asfer \delta^*\eL}{\can_A\fer{R}'} -  d\inpr{J_M\asfer' \delta'^*\eR}{\can_A\fer{L}} 
\eas
Using the self-adjointness of $\can_A$, that $[\mu, \can_A] = 0$ and the symmetries of the inner product, this reads
\bas
 & \inpr{J_M  \asfer\eL}{ \big[c\mu - d'\delta^*\big]\can_A\fer{R}'} +  \inpr{J_M \asfer'\eR }{\big[c' \mu^* - d \delta'^*\big]\can_A\fer{L}} \nn.
\eas
We thus require that 
\ba\label{eq:bb5-constr2}
c\mu &= d'\delta^*, & c' \mu^* &= d \delta'^*
\ea
for this to vanish.

\item a group with $\afer{R}$ and $\afer{L}'$ that vanishes if and only if \eqref{eq:bb5-constr2} is satisfied. 
%

	\item a group with the left-handed gauginos: 
\bas
	&- \int_M \Big( \tr \big[d_{ij,i}^*(J_M\eL, \gamma^5\asfer\gau{iL}) - d_{ij,j}^*(J_M\eL, \gamma^5\gau{jL}\asfer)\big] \delta' \sfer' \nn\\
	&\qquad\qquad + \tr \asfer \delta^* \big[ d_{ij,i}'(J_M\eL, \gamma^5\gau{iL}\sfer') - d_{ij,j}'(J_M\eL, \gamma^5\sfer'\gau{jL})\big]\Big)\nn\\
	& =  - \inpr{J_M \big(d_{ij,i}^* \delta'\sfer'\asfer  + d_{ij,i}'\sfer'\asfer \delta^*\big)\eL}{\gamma^5\gau{iL}} \nn\\
	&\qquad\qquad + \inpr{J_M\big(d_{ij,j}^*\asfer \delta' \sfer' + d_{ij,j}'\asfer \delta^* \sfer'\big)\eL}{\gamma^5\gau{jL}},  
\eas
where we have used Lemmas \ref{lem:moveScalar} and \ref{lem:pullScalar}. For this to vanish, we require that 
\bas
	 d_{ij,i}^* \delta' &= - d_{ij,i}' \delta^*, & d_{ij,j}^* \delta' &= - d_{ij,j}' \delta^* . 
\eas
Inserting \eqref{eq:bb5-constr2} above this is equivalent to
\bas
d_{ij,i}^*\frac{c'^*}{d^*} &= - d_{ij,i}'\frac{c}{d'}, &
d_{ij,j}^*\frac{c'^*}{d^*} &= - d_{ij,j}'\frac{c}{d'}.
\eas

\item A group with the right-handed gauginos 
\bas
	  &- \int_M \tr \big[ d_{ij,i}'^*(J_M\eR, \gamma^5\asfer'\gau{iR})  - d_{ij,j}'^*(J_M\eR, \gamma^5\gau{jR}\asfer')\big] \delta \sfer \nn\\
	 &\qquad - \int_M \tr  \asfer'\delta'^*\big[  d_{ij,i}(J_M\eR, \gamma^5\gau{iR}\sfer) - d_{ij,j}(J_M\eR, \gamma^5\sfer\gau{jR})\big] \nn\\
	  &= -  \inpr{J_M\big(d_{ij,i}'^* \delta \sfer \asfer' + d_{ij,i}\sfer\asfer'\delta'^*\big)\eR}{ \gamma^5\gau{iR}} \nn\\
		&\qquad\qquad + \inpr{J_M\big(d_{ij,j}'^*\asfer'\delta \sfer + d_{ij,j}\asfer' \delta'^*\sfer\big)\eR}{ \gamma^5\gau{jR}},
\eas
which vanishes iff
\bas
	  d_{ij,i}'^* \delta &= - d_{ij,i}\delta'^* ,& d_{ij,j}'^*\delta&= - d_{ij,j} \delta'^*.
\eas

\end{itemize}

Combining all relations, above, we require that 
\bas
	|c|^2 &= |d'|^2,& |c'|^2 &= |d|^2,& |d_{ij,i}|^2 &= |d_{ij,i}'|^2,& |d_{ij,j}|^2 &= |d_{ij,j}'|^2,
\eas
for the transformation constants and 
\bas
	\delta\delta^* &= \mu^*\mu,  & \delta'\delta'^* &= \mu\mu^*
\eas
for the parameters in the off shell action.

\section{Auxiliary lemmas and identities}

In this section we provide some auxiliary lemmas and identities that are used in and throughout the previous proofs. 


\begin{lem}
For the spin-connection $\nabla^S :\Gamma(S) \to \mathcal{A}^1(M) \otimes_{C^\infty(M)} \Gamma(S)$ on a flat manifold we have:
\begin{align}
	[\nabla^S, \gamma^\mu]	&= 0.\label{eq:idnNablaS}
\end{align}
\end{lem}
\begin{proof}
	The spin-connection is the unique connection compatible with the Levi-Civita connection on $T^*M$, which means that it satisfies
	\begin{align*}
		\nabla^Sc(\alpha) = c(\alpha)\nabla^S + c(\nabla^g \alpha)	
	\end{align*}
	for any $\alpha \in \Gamma^{\infty}(\mathbb{C}l(M))$. Here $c : \Gamma(\mathbb{C}l(M)) \to \Gamma(\End(S))$ is the \emph{spin homomorphism}. Taking in particular $\alpha = \mathrm{d}x^\mu$, writing $\gamma^\mu = c(\mathrm{d}x^\mu)$ and using that $\nabla^g\mathrm{d}x^\mu = - \Gamma^\mu_{\nu\lambda} \mathrm{d}x^\nu \otimes \mathrm{d}x^\lambda = 0$ for a flat manifold, we have
	\begin{align*}
		\nabla^S\gamma^\mu &= \gamma^\mu\nabla^S. 
	\end{align*}
	Here we have used that $c(\mathrm{d}x^\mu \otimes \mathrm{d}x^\nu) = c(\mathrm{d}x^\mu \wedge \mathrm{d}x^\nu) + c(\{\mathrm{d}x^\mu,\mathrm{d}x^\nu\})$.
\end{proof}

\begin{lem}\label{lem:Dsquared}
Let
$
	\can_A = -i c \circ (\nabla^S + \mathbb{A})
$
 and $D_\mu = (\nabla^S + \mathbb{A})_\mu$. For a flat manifold, we have locally:
\begin{align*}
	\can_A^2 + D_\mu D^\mu = - \frac{1}{2} \gamma^\mu\gamma^\nu \mathbb{F}_{\mu\nu}.
\end{align*}
\end{lem}
\begin{proof}
Locally we write 
\begin{align*}
	\can_A = -i c(\mathrm{d}x^\mu) (\nabla^S_\mu + \mathbb{A}_\mu)
\end{align*}
where $\mathbb{A}_\mu$ is skew-Hermitian in order for $\can_A$ to be self-adjoint. Now for the square of this, we have
\begin{align*}
	 \can_A^2  &= - c(\mathrm{d}x^\mu) (\nabla^S_\mu + \mathbb{A}_\mu)c(\mathrm{d}x^\nu) (\nabla^S_\nu + \mathbb{A}_\nu)\\
					&= - c(\mathrm{d}x^\mu) c(\mathrm{d}x^\nu)(\nabla^S_\mu + \mathbb{A}_\mu)(\nabla^S_\nu + \mathbb{A}_\nu) 
					- c(\mathrm{d}x^\mu) c(\nabla^g_{\partial_\mu}\mathrm{d}x^\nu)(\nabla^S_\nu + \mathbb{A}_\nu), 
\end{align*}
of which the last term vanishes for a flat manifold. Here we have employed that the spin connection is the unique connection compatible with the Levi-Civita connection. We write:
\begin{align*}
	c(\mathrm{d}x^\mu) c(\mathrm{d}x^\nu) &= \frac{1}{2}\{c(\mathrm{d}x^\mu), c(\mathrm{d}x^\nu)\} + \frac{1}{2}[c(\mathrm{d}x^\mu), c(\mathrm{d}x^\nu)]\nn\\
			&= g^{\mu\nu} + \frac{1}{2}[c(\mathrm{d}x^\mu), c(\mathrm{d}x^\nu)]
\end{align*}
to arrive at
\begin{align*}
 \can_A^2 &= - (\nabla^{S} + \mathbb{A})^\mu(\nabla^S + \mathbb{A})_\mu - \frac{1}{2}c(\mathrm{d}x^\mu)c(\mathrm{d}x^\nu)[\nabla^S_\mu + \mathbb{A}_\mu, \nabla^S_\nu + \mathbb{A}_\nu] 
\end{align*}
obtaining the result.
\end{proof}

\begin{cor}\label{cor:4bcurvs}
By applying the previous result, we have for $\szeta_{ik} \in C^{\infty}(M, \mathbf{N_i}\otimes \mathbf{N_k})$, $\epsilon \in L^2(M, S)$
\begin{align*}
	(\can_A [\can_A, \szeta_{ik}]\epsilon + D_\mu [D^\mu, \szeta_{ik}])\epsilon = \frac{1}{2}[\mathbb{F}, \szeta_{ik}]\epsilon + [D^\mu, \szeta_{ik}]\nabla^S_\mu\epsilon + [\can_A, \szeta_{ik}]\slashed{\partial}\epsilon,
\end{align*}
where the term with $R$ vanished due to the commutator.
\end{cor}


\begin{lem}\label{lem:symmJ}
Let $M$ be a four-dimensional Riemannian spin manifold and $\inpr{\,.\,}{\,.\,} : L^2(S) \times L^2(S) \to \mathbb{C}$ the inner product on sections of the spinor bundle. For $\mathcal{P}$ a basis element of $\Gamma(\com l(M))$, we have the following identities:
\bas
	\inpr{J_M\zeta_1}{\mathcal{P}\zeta_2} = \pi_{\mathcal{P}}\inpr{J_M\zeta_2}{\mathcal{P}\zeta_1},\qquad \pi_{\mathcal{P}} \in \{\pm\},
\eas 
for any $\zeta_{1,2}$, the Grassmann variables corresponding to $\zeta_{1,2}' \in L^2(S)$. The signs $\pi_{\P}$ are given by 
\begin{align}
	\pi_{\id} &= 1, & \pi_{\gamma^\mu} &= - 1, & \pi_{\gamma^\mu\gamma^\nu} &= - 1\quad (\mu < \nu), \nn\\
 	\pi_{\gamma^\mu\gamma^5} &=  1, & \pi_{\gamma^5} &=  1.\label{eq:identitySymJ}
\end{align} 
\end{lem}
\begin{proof}
Using that $J_M^2 = -1$ and $\inpr{J_M\zeta_1'}{J_M\zeta_2'} = \inpr{\zeta_2'}{\zeta_1'}$, we have
\bas
	\inpr{J_M\zeta_1'}{\mathcal{P}\zeta_2'} = - \inpr{J_M\zeta_1'}{J_M^2\mathcal{P}\zeta_2'} = - \inpr{J_M\mathcal{P}\zeta_2'}{\zeta_1'}.
\eas
When considering Grassmann variables, we obtain an extra minus sign (see the discussion in \cite[\S 4.2.6]{DS12}). From $J_M\gamma^\mu = - \gamma^\mu J_M$, $(\gamma^\mu)^* = \gamma^\mu$ and $\gamma^\mu\gamma^\nu = - \gamma^\nu\gamma^\mu$ for $\mu \ne \nu$, we obtain the result.
\end{proof}

\begin{cor}\label{cor:symmInnerProd}
	Similarly (\cite[\S 4]{CCM07}) we find by using that $\dirac^* = \dirac$ and $J_M\dirac = \dirac J_M$, that
	\ba
		\inpr{J_M\zeta_1}{\dirac\zeta_2} =  \inpr{J_M\zeta_2}{\dirac\zeta_1}\label{eq:identitySymJ2}
	\ea
	for the Grassmann variables corresponding to any two $\zeta_{1, 2}' \in L^2(S)$.
\end{cor}

\begin{lem}\label{lem:pullScalar}
	For any $\sfer \in C^{\infty}(M, \rep{i}{j})$, $\fer{} \in L^2(S \otimes \rep{j}{i})$ and $\epsilon \in L^2(S)$ we have
	\begin{align*}
		\tr_{N_i} \sfer(J_M\epsilon, \fer{})_{\cS} = (J\sfer\epsilon, \fer{})_{\H}.
	\end{align*}
\end{lem}
\begin{proof}
This can be seen easily by writing out the elements in full detail:
	 \begin{align*}
		\szeta &= f \otimes e \otimes \bar e', & \fer{} &= \zeta \otimes \eta \otimes \bar \eta',\qquad f \in C^{\infty}(M, \com), \zeta \in L^2(S).
		\end{align*}
\end{proof}


\begin{lem}\label{lem:moveScalar}
	Let $\fer{1} \in L^2(S\otimes \rep{i}{j})$, $\fer{2} \in L^2(S \otimes \rep{k}{i})$, $\afer{2} \in L^2(S \otimes \rep{j}{k})$, $\sfer \in C^{\infty}(M, \rep{j}{k})$ and $\sfer' \in C^{\infty}(M, \rep{k}{i})$, then
\ba
	\inpr{J\fer{1}\sfer}{\fer{2}} &= \inpr{J\fer{1}}{\sfer\fer{2}}& &\text{and}& \inpr{J\fer{1}}{\fer{2}\sfer'} &= \inpr{J\sfer'\fer{1}}{\fer{2}}.\label{eq:moveScalar}
\ea
\end{lem}
\begin{proof}
	This can simply be proven by using that the right action is implemented via $J$ and that $J$ is an anti-isometry with $J^2 = \pm $.
\end{proof}

\subsection{Fierz transformations}\label{sec:fierz}

Details for the Fierz transformation in this context can be found in the Appendix of \cite{BS10} but we list the main result here.

\begin{defin}[Orthonormal Clifford basis] Let $Cl(V)$ be the Clifford algebra over a vector space $V$ of dimension $n$. Then $\gamma_K := \gamma_{k_1}\cdots\gamma_{k_r}$ for all strictly ordered sets $K = \{k_1 < \ldots < k_r\} \subseteq \{1, \ldots, n\}$ form a basis for $Cl(V)$. If $\gamma_K$ is as above, we denote with $\gamma^K$ the element $\gamma^{k_1}\cdots\gamma^{k_r}$. The basis spanned by the $\gamma_K$ is said to be \emph{orthonormal} if $\tr\gamma_K\gamma_L = nn_K\delta_{KL}\ \forall\ K, L$. Here $n_K := (-1)^{r(r-1)/2}$, where $r$ denotes the cardinality of the set $K$ and with $\delta_{KL}$ we mean 
\begin{align}
  \delta_{KL} = \left\{\begin{array}{ll} 1\quad \text{if}\ K = L\\
                         0 \quad \text{else}\\
                       \end{array}.
                \right.
\end{align}
\end{defin}

\begin{exmpl}\label{exmpl:dim4} Take $V = \mathbb{R}^4$ and let $Cl(4, 0)$ be the Euclidean Clifford algebra [i.e. with signature \mbox{($+$\ $+$\ $+$\ $+$)}]. Its basis are the sixteen matrices
\begin{align*}
 & 1                                                      & & \nonumber\\
 & \gamma_\mu                                             & & \text{(4 elements)} \nonumber\\
 & \gamma_\mu\gamma_\nu\quad          \quad (\mu < \nu)   &        & \text{(6 elements)}\nonumber\\
 & \gamma_\mu\gamma_\nu\gamma_\lambda \quad (\mu < \nu < \lambda)& & \text{(4 elements)} \nonumber\\
 &\gamma_1\gamma_2\gamma_3\gamma_4=:\gamma_5.             &  &\nonumber
\end{align*}
We can identify
\begin{align}\label{eq:mink_eucl}
 \gamma_1\gamma_2\gamma_3 &= \gamma_4\gamma_5, & \gamma_1\gamma_3\gamma_4 &= \gamma_2\gamma_5 &
 \gamma_1\gamma_2\gamma_4 &= - \gamma_3\gamma_5, & \gamma_2\gamma_3\gamma_4 &= -\gamma_1\gamma_5,
\end{align}
establishing a connection with the basis most commonly used by physicists.
\end{exmpl}

We then have the following result:
\begin{prop}[(Generalized) Fierz identity]\label{prop:fierz} If for any two strictly ordered sets $K, L$ there exists a third strictly ordered set $M$ and $c \in \mathbb{N}$ such that $\gamma_K\gamma_L = c\,\gamma_M$, we have for any $\psi_1, \ldots, \psi_4$ in the $n$-dimensional spin representation of the Clifford algebra 
\begin{align}
   \inpr{\psi_1}{\gamma^K\psi_2}\inpr{\psi_3}{\gamma_K\psi_4} &= -\frac{1}{n}\sum_L C_{KL}\inpr{\psi_3}{\gamma^L\psi_2}\inpr{\psi_1}{\gamma_{L}\psi_4}\label{eq:fierzf},
\end{align}
where the constants $C_{LK} \equiv n_Lf_{LK}$, $f_{LK}\in \mathbb{N}$ are
defined via 
$
	  \gamma^K\gamma^L\gamma_K = f_{KL}\gamma^L$ (no sum over $L$).
Here we have denoted by $\inpr{.}{.}$ the inner product on the spinor representation.
\end{prop}

%
%
%
%
%

	\graphicspath{{./gfx/}}
	\svgpath={./gfx/}

\printglossary[type=symbolslist,style=mcolindex]
\printglossary[type=main,style=long]

\newpage
\manualmark
\markboth{\spacedlowsmallcaps{\bibname}}{\spacedlowsmallcaps{\bibname}} 
\refstepcounter{dummy}
\addtocontents{toc}{\protect\vspace{\beforebibskip}} 
\addcontentsline{toc}{chapter}{\tocEntry{\bibname}}
\bibliographystyle{plainnat}
\label{app:bibliography} 
\bibliography{thesis.bbl}

\providecommand{\noopsort}[1]{}
\begin{thebibliography}{102}
\providecommand{\natexlab}[1]{#1}
\providecommand{\url}[1]{\texttt{#1}}
\expandafter\ifx\csname urlstyle\endcsname\relax
  \providecommand{\doi}[1]{doi: #1}\else
  \providecommand{\doi}{doi: \begingroup \urlstyle{rm}\Url}\fi

\bibitem[Alvarez et~al.(1995)Alvarez, Gracia-Bond\'ia, and
  Mart\'in]{Alvarez1995}
E.~Alvarez, J.M. Gracia-Bond\'ia, and C.P. Mart\'in.
\newblock Anomaly cancellation and gauge group of the standard model in {NCG}.
\newblock \emph{Phys. Lett. B}, 364:\penalty0 33--40, 1995.

\bibitem[Andrianov et~al.(2011)Andrianov, Kurkov, and Lizzi]{AKL11}
A.A. Andrianov, M.A. Kurkov, and F.~Lizzi.
\newblock Spectral action, {W}eyl anomaly and the {H}iggs-dilaton potential.
\newblock \emph{Journal of High Energy Physics}, 2011-10:\penalty0 1, 2011.

\bibitem[{\noopsort{atlas}}ATLAS
  Collaboration(2013{\natexlab{a}})]{ATLAS-CONF-2013-034}
{\noopsort{atlas}}ATLAS Collaboration.
\newblock {Combined coupling measurements of the Higgs-like boson with the
  ATLAS detector using up to 25 fb$^{-1}$ of proton-proton collision data}.
\newblock Technical Report ATLAS-CONF-2013-034, CERN, Geneva,
  2013{\natexlab{a}}.

\bibitem[{\noopsort{atlas}}ATLAS
  Collaboration(2013{\natexlab{b}})]{ATLAS-CONF-2013-040}
{\noopsort{atlas}}ATLAS Collaboration.
\newblock {Study of the spin of the new boson with up to 25~fb$^{-1}$ of ATLAS
  data}.
\newblock Technical Report ATLAS-CONF-2013-040, CERN, Geneva,
  2013{\natexlab{b}}.

\bibitem[Baez and Huerta(2010)]{BH10}
J.~Baez and J.~Huerta.
\newblock The algebra of grand unified theories.
\newblock \emph{Bull.~Amer.~Math.~Soc.}, 47:\penalty0 483--552, 2010.

\bibitem[Bamba et~al.(2012)Bamba, Capozziello, Nojiri, and Odintsov]{BCNO12}
K.~Bamba, S.~Capozziello, S.~Nojiri, and S.D. Odintsov.
\newblock Dark energy cosmology: the equivalent description via different
  theoretical models and cosmography tests.
\newblock \emph{Astrophys. Space Sci.}, 342:\penalty0 155--228, 2012.

\bibitem[Beenakker et~al.(2014{\natexlab{a}})Beenakker, van~den Broek, and van
  Suijlekom]{BS13I}
W.~Beenakker, T.~van~den Broek, and W.D. van Suijlekom.
\newblock Noncommutative geometry and supersymmetry. {P}art \rnum{1}:
  {S}upersymmetric almost-commutative geometries.
\newblock 2014{\natexlab{a}}.

\bibitem[Beenakker et~al.(2014{\natexlab{b}})Beenakker, van~den Broek, and van
  Suijlekom]{BS13II}
W.~Beenakker, T.~van~den Broek, and W.D. van Suijlekom.
\newblock Noncommutative geometry and supersymmetry. {P}art \rnum{2}:
  {S}upersymmetry breaking.
\newblock 2014{\natexlab{b}}.

\bibitem[Beenakker et~al.(2014{\natexlab{c}})Beenakker, van~den Broek, and van
  Suijlekom]{BS13III}
W.~Beenakker, T.~van~den Broek, and W.D. van Suijlekom.
\newblock Noncommutative geometry and supersymmetry. {P}art \rnum{3}: {T}he
  noncommutative supersymmetric {S}tandard {M}odel.
\newblock 2014{\natexlab{c}}.

\bibitem[Bertlmann(1996)]{BERT96}
R.A. Bertlmann.
\newblock \emph{Anomalies in Quantum Field Theory}.
\newblock Clarendon Press (Oxford), 1996.

\bibitem[Bhowmick et~al.(2011)Bhowmick, D'Andrea, Das, and
  D\k{a}browski]{Bhowmick2011}
J.~Bhowmick, F.~D'Andrea, B.~Das, and L.~D\k{a}browski.
\newblock Quantum gauge symmetries in noncommutative geometry.
\newblock \emph{arXiv:1112.3622}, 2011.

\bibitem[Bilal(2007)]{Bilal2007}
A.~Bilal.
\newblock Introduction to supersymmetry.
\newblock \emph{hep-th/0101055}, 2007.

\bibitem[Born and Jordan(1925)]{BJ25}
M.~Born and P.~Jordan.
\newblock Zur {Q}uantenmechanik.
\newblock \emph{Z. Phys. C}, 34:\penalty0 858--888, 1925.

\bibitem[Brazzale(2013)]{ATL-PHYS-PROC-2013-339}
S.~Brazzale.
\newblock Overview of {SUSY} results from the {ATLAS} experiment.
\newblock Technical Report ATL-PHYS-PROC-2013-339, CERN, Geneva, 2013.

\bibitem[{\noopsort{broek}}van~den Broek and van Suijlekom(2010)]{BS10}
T.~{\noopsort{broek}}van~den Broek and W.D. van Suijlekom.
\newblock Supersymmetric {QCD} and noncommutative geometry.
\newblock \emph{Comm. Math. Phys.}, 303\penalty0 (1):\penalty0 149--173, 2010.

\bibitem[{\noopsort{broek}}van~den Broek and van Suijlekom(2013)]{BS12}
T.~{\noopsort{broek}}van~den Broek and W.D. van Suijlekom.
\newblock Going beyond the {S}tandard {M}odel with noncommutative geometry.
\newblock \emph{J. High Energy Phys.}, 3:\penalty0 112, 2013.

\bibitem[Chamseddine(1994)]{Cha94}
A.H. Chamseddine.
\newblock Connection between space-time supersymmetry and noncommutative
  geometry.
\newblock \emph{Phys. Lett. B}, B332:\penalty0 349--357, 1994.

\bibitem[Chamseddine(1998)]{Chamseddine1998}
A.H. Chamseddine.
\newblock Remarks on the spectral action principle.
\newblock \emph{Phys. Lett. B}, 436:\penalty0 84--90, 1998.

\bibitem[Chamseddine and Connes(1996)]{CC96}
A.H. Chamseddine and A.~Connes.
\newblock Universal formula for noncommutative geometry actions: {U}nifications
  of gravity and the standard model.
\newblock \emph{Phys. Rev. Lett.}, 77:\penalty0 4868--4871, 1996.

\bibitem[Chamseddine and Connes(1997)]{CC97}
A.H. Chamseddine and A.~Connes.
\newblock The spectral action principle.
\newblock \emph{Comm. Math. Phys.}, 186:\penalty0 731--750, 1997.

\bibitem[Chamseddine and Connes(2008)]{CC08}
A.H. Chamseddine and A.~Connes.
\newblock Why the {S}tandard {M}odel.
\newblock \emph{J. Geom. Phys.}, 58:\penalty0 38--47, 2008.

\bibitem[Chamseddine and Connes(2012)]{CC12}
A.H. Chamseddine and A.~Connes.
\newblock Resilience of the {S}pectral {S}tandard {M}odel.
\newblock \emph{J. High Energy Phys.}, 1209:\penalty0 104, 2012.

\bibitem[Chamseddine et~al.(2007)Chamseddine, Connes, and Marcolli]{CCM07}
A.H. Chamseddine, A.~Connes, and M.~Marcolli.
\newblock Gravity and the standard model with neutrino mixing.
\newblock \emph{Adv. Theor. Math. Phys.}, 11:\penalty0 991--1089, 2007.

\bibitem[Chanowitz et~al.(1977)Chanowitz, Ellis, and Gaillard]{CEG77}
M.S. Chanowitz, J.~Ellis, and M.K. Gaillard.
\newblock The price of natural flavour conservation in neutral weak
  interactions.
\newblock \emph{Nuclear Phys. B Proc. Suppl.}, 128:\penalty0 506--536, 1977.

\bibitem[Chung et~al.(2005)Chung, Everett, Kane, King, Lykken, and
  Wang]{CEKKLW05}
D.J.H. Chung, L.L. Everett, G.L. Kane, S.F. King, J.~Lykken, and L.T. Wang.
\newblock The soft supersymmetry-breaking {L}agrangian: Theory and
  applications.
\newblock \emph{Phys. Rep.}, 407:\penalty0 1--203, 2005.

\bibitem[Colafrancesco(2010)]{Colafrancesco2010}
S.~Colafrancesco.
\newblock Dark {M}atter in {M}odern {C}osmology.
\newblock \emph{arXiv:1004.3869}, 2010.

\bibitem[Coleman(1985)]{C85}
S.~Coleman.
\newblock \emph{Aspects of Symmetry}.
\newblock Cambridge University Press, 1985.

\bibitem[Coleman and Mandula(1967)]{CoMa67}
S.~Coleman and J.~Mandula.
\newblock All possible symmetries of the ${S}$-matrix.
\newblock \emph{Phys. Rev. D (3)}, 159\penalty0 (5):\penalty0 1251--1256, 1967.

\bibitem[{C}ollaboration.(2013)]{PlanckCollaboration2013}
Planck {C}ollaboration.
\newblock Planck 2013 results. \rnum{16}. {C}osmological parameters.
\newblock \emph{arXiv:1303.5076}, 2013.

\bibitem[{C}ollaboration(2013)]{PlanckCollaboration2013a}
Planck {C}ollaboration.
\newblock Planck 2013 results. \rnum{1}. {O}verview of products and scientific
  results.
\newblock \emph{arXiv:1303.5062}, 2013.

\bibitem[Connes(1989)]{C89}
A.~Connes.
\newblock Compact metric spaces, fredholm modules, and hyperfiniteness.
\newblock \emph{Ergodic Theory Dynam. Systems}, 9:\penalty0 207--220, 1989.

\bibitem[Connes(1994)]{C94}
A.~Connes.
\newblock \emph{Noncommutative geometry}.
\newblock Academic Press, 1994.

\bibitem[Connes(1996)]{C96}
A.~Connes.
\newblock Gravity coupled with matter and the foundation of noncommutative
  geometry.
\newblock \emph{Commun. Math. Phys.}, 182:\penalty0 155--176, 1996.

\bibitem[Connes(2007)]{C00}
A.~Connes.
\newblock Noncommutative geometry year 2000.
\newblock \emph{math/0011193}, 2007.

\bibitem[Connes and Lott(1991)]{CL89}
A.~Connes and J.~Lott.
\newblock Particle models and noncommutative geometry.
\newblock \emph{Nuclear Phys. B Proc. Suppl.}, 18:\penalty0 29--47, 1991.

\bibitem[Connes and Marcolli(2007)]{CM07}
A.~Connes and M.~Marcolli.
\newblock \emph{Noncommutative Geometry, Quantum Fields and Motives}.
\newblock American Mathematical Society, 2007.

\bibitem[Dirac(1928)]{D28}
P.A.M. Dirac.
\newblock The quantum theory of the electron.
\newblock \emph{Proc. Roy. Soc. London}, 117\penalty0 (778):\penalty0 610--624,
  1928.

\bibitem[D\k{a}browski and Dossena(2010)]{Dabrowski2010}
L.~D\k{a}browski and G.~Dossena.
\newblock Product of real spectral triples.
\newblock \emph{Int. J. Geom. Methods Mod. Phys.}, 8\penalty0 (8):\penalty0
  1833--1848, 2010.

\bibitem[Drees et~al.(2004)Drees, Godbole, and Roy]{DGR04}
M.~Drees, R.~Godbole, and P.~Roy.
\newblock \emph{Theory and phenomenology of Sparticles}.
\newblock World Scientific Publishing Co., 2004.

\bibitem[Drewes(2013)]{D13}
M.~Drewes.
\newblock The {P}henomenology of {R}ight {H}anded {N}eutrinos.
\newblock \emph{Internat. J. Modern Phys. E}, 22:\penalty0 1330019, 2013.

\bibitem[{\noopsort{dungen}}van~den Dungen and van Suijlekom(2012)]{DS12}
K.~{\noopsort{dungen}}van~den Dungen and W.D. van Suijlekom.
\newblock Particle physics from almost-commutative spacetimes.
\newblock \emph{Rev. Math. Phys.}, 24:\penalty0 1230004, 2012.

\bibitem[{\noopsort{dungen}}van~den Dungen and van Suijlekom(2013)]{DS11}
K.~{\noopsort{dungen}}van~den Dungen and W.D. van Suijlekom.
\newblock Electrodynamics from noncommutative geometry.
\newblock \emph{J. Noncommut. Geom.}, 7:\penalty0 433--456, 2013.

\bibitem[Eguchi et~al.(1980)Eguchi, Gilkey, and Hanson]{EGH80}
T.~Eguchi, P.B. Gilkey, and A.J. Hanson.
\newblock Gravitation, gauge theories and differential geometry.
\newblock \emph{Phys. Rep.}, 66\penalty0 (6):\penalty0 213--393, 1980.

\bibitem[et~al. (Particle Data~Group)(2012, and 2013 partial update for the
  2014 edition.)]{B12}
J.~Beringer et~al. (Particle Data~Group).
\newblock The {R}eview of {P}article {P}hysics.
\newblock \emph{Phys. Rev. D}, 86:\penalty0 010001, 2012, and 2013 partial
  update for the 2014 edition.
\newblock URL \url{http://pdg.lbl.gov/}.

\bibitem[Fayet and Iliopoulos(1974)]{FI74}
P.~Fayet and J.~Iliopoulos.
\newblock Spontaneously broken supergauge symmetries and {G}oldstone spinors.
\newblock \emph{Phys.~Lett.~B}, 51:\penalty0 461--464, 1974.

\bibitem[Feng(2010)]{Feng2010}
J.L. Feng.
\newblock Dark matter candidates from particle physics and methods of
  detection.
\newblock \emph{Annu. rev. astron. astrophys.}, 48:\penalty0 495, 2010.

\bibitem[Freed et~al.(2006)Freed, Morrison, and Singer]{freed2006}
D.S. Freed, D.R. Morrison, and I.~Singer.
\newblock \emph{Quantum Field Theory, Supersymmetry \& Enumerative geometry}.
\newblock American Mathematical Society, 2006.

\bibitem[Freedman et~al.(1976)Freedman, van Nieuwenhuizen, and Ferrara]{FNF76}
D.Z. Freedman, P.~van Nieuwenhuizen, and S.~Ferrara.
\newblock Progress toward a theory of supergravity.
\newblock \emph{Phys. Rev. D}, 13:\penalty0 3214--3218, 1976.

\bibitem[Gelfand and Neumark(1943)]{GelNai43}
I.~Gelfand and M.~Neumark.
\newblock On the imbedding of normed rings into the ring of operators in
  {H}ilbert space.
\newblock \emph{Mat. Sb.}, 12\penalty0 (2):\penalty0 197--217, 1943.

\bibitem[Gell-Mann et~al.(1979)Gell-Mann, Ramond, and Slansky]{GRS79}
M.~Gell-Mann, P.~Ramond, and R.~Slansky.
\newblock Supergravity.
\newblock In F.~van Nieuwenhuizen and D.Z. Freedman, editors,
  \emph{Supergravity}, page 315. North-Holland, Amsterdam, 1979.

\bibitem[Georgi and Dimopoulos(1981)]{GD81}
H.~Georgi and S.~Dimopoulos.
\newblock Softly broken supersymmetry and ${SU(5)}$.
\newblock \emph{Nucl.~Phys.~B}, 193:\penalty0 150 -- 162, 1981.

\bibitem[Gilkey(1984)]{GIL84}
P.B. Gilkey.
\newblock \emph{Invariance theory, the heat equation and the {A}tiyah-{S}inger
  index theorem}, volume~11 of \emph{Mathematics Lecture Series}.
\newblock Publish or Perish, Wilmington, DE, 1984.

\bibitem[Girardello and Grisaru(1981)]{GG81}
L.~Girardello and M.T. Grisaru.
\newblock Soft breaking of supersymmetry.
\newblock \emph{Nuclear Phys. B Proc. Suppl.}, 194:\penalty0 65 -- 76, 1981.

\bibitem[Gracia-Bond\'ia et~al.(2000)Gracia-Bond\'ia, V\'arilly, and
  Figueroa]{GVF00}
J.M. Gracia-Bond\'ia, J.C. V\'arilly, and H.~Figueroa.
\newblock \emph{Elements of Noncommutative Geometry}.
\newblock Birkh\"auser {A}dvanced {T}exts, 2000.

\bibitem[Gunion and HABER(1986)]{GH86}
J.F. Gunion and H.E. HABER.
\newblock Higgs bosons in supersymmetric models (l).
\newblock \emph{Nuclear Phys. B}, 272:\penalty0 1, 1986.

\bibitem[Haag et~al.(1975)Haag, opusa\'{n}ski, and Sohnius]{Haag1975}
R.~Haag, J.T.~\L opusa\'{n}ski, and M.~Sohnius.
\newblock All possible generators of supersymmetries of the ${S}$-matrix.
\newblock \emph{Nuclear Phys. B Proc. Suppl.}, 88:\penalty0 257--274, 1975.

\bibitem[Heisenberg and Pauli(1930)]{HP30}
W.~Heisenberg and W.~Pauli.
\newblock Zur {Q}uantentheorie der {W}ellenfelder \rnum{2}.
\newblock \emph{Z. Phys. C}, 59\penalty0 (3--4):\penalty0 168--190, 1930.

\bibitem[Hussain and Thompson(1991{\natexlab{a}})]{HT91a}
F.~Hussain and G.~Thompson.
\newblock Noncommutative geometry and supersymmetry.
\newblock \emph{Phys. Lett. B}, 260:\penalty0 359--364, 1991{\natexlab{a}}.

\bibitem[Hussain and Thompson(1991{\natexlab{b}})]{HT91b}
F.~Hussain and G.~Thompson.
\newblock Noncommutative geometry and supersymmetry 2.
\newblock \emph{Phys. Lett. B}, 265:\penalty0 307--310, 1991{\natexlab{b}}.

\bibitem[Iochum et~al.(2004)Iochum, Sch\"ucker, and Stephan]{ISS03}
B.~Iochum, T.~Sch\"ucker, and C.~Stephan.
\newblock On a {C}lassification of {I}rreducible {A}lmost {C}ommutative
  {G}eometries.
\newblock \emph{J. Math. Phys.}, 45:\penalty0 5003--5041, 2004.

\bibitem[Jureit and Stephan(2005)]{JS05}
J.~Jureit and C.~Stephan.
\newblock On a classification of irreducible almost commutative geometries, a
  second helping.
\newblock \emph{J.~Math.~Phys.}, 46:\penalty0 043512, 2005.

\bibitem[Jureit and Stephan(2008)]{JS08}
J.~Jureit and C.~Stephan.
\newblock On a classification of irreducible almost-commutative geometries
  \rnum{4}.
\newblock \emph{J.~Math.~Phys.}, 49:\penalty0 033502, 2008.
\newblock arXiv:hep-th/0610040.

\bibitem[Jureit and Stephan(2009)]{JS09}
J.~Jureit and C.~Stephan.
\newblock On a classification of irreducible almost-commutative geometries
  \rnum{5}.
\newblock \emph{J.~Math.~Phys.}, 50:\penalty0 072301, 2009.
\newblock arXiv:0901.3214.

\bibitem[Jureit et~al.(2005)Jureit, Schucker, and Stephan]{JSS05}
J.~Jureit, T.~Schucker, and C.~Stephan.
\newblock On a classification of irreducible almost commutative geometries
  \rnum{3}.
\newblock \emph{J.~Math.~Phys.}, 46:\penalty0 072303, 2005.
\newblock arXiv:hep-th/0503190.

\bibitem[Kalau and Walze(1997)]{KW96}
W.~Kalau and M.~Walze.
\newblock Supersymmetry and noncommutative geometry.
\newblock \emph{J. Geom. Phys.}, 22:\penalty0 77--102, 1997.

\bibitem[Kazakov(2009)]{Kazakov2009}
D.I. Kazakov.
\newblock Beyond the {S}tandard {M}odel (in {S}earch of {S}upersymmetry).
\newblock \emph{hep-ph/0012288}, 2009.

\bibitem[Krajewski(1998)]{KR97}
T.~Krajewski.
\newblock Classification of finite spectral triples.
\newblock \emph{J. Geom. Phys.}, 28:\penalty0 1--30, 1998.

\bibitem[Landi(1998)]{Landi2008}
G.~Landi.
\newblock \emph{An {I}ntroduction to {N}oncommutative {S}paces and {T}heir
  {G}eometries ({L}ecture {N}otes in {P}hysics {M}onographs)}.
\newblock Springer, 1998.

\bibitem[Langacker(1981)]{PL81}
P.~Langacker.
\newblock Grand unified theories and proton decay.
\newblock \emph{Phys.~Rep.}, 72:\penalty0 185--385, 1981.

\bibitem[Langacker(2003)]{PL03}
P.~Langacker.
\newblock Structure of the {S}tandard {M}odel.
\newblock \emph{hep-ph/0304186}, 2003.

\bibitem[Lawson and Michelsohn(1989)]{LM89}
H.B. Lawson and M.L. Michelsohn.
\newblock \emph{Spin Geometry}.
\newblock Princeton University Press, 1989.

\bibitem[Lizzi et~al.(1997)Lizzi, Mangano, Miele, and Sparano]{LMMS97}
F.~Lizzi, G.~Mangano, G.~Miele, and G.~Sparano.
\newblock Fermion {H}ilbert space and fermion doubling in the noncommutative
  geometry approach to gauge theories.
\newblock \emph{Phys. Rev. D}, 55:\penalty0 6357--6366, 1997.

\bibitem[Lykken(2007)]{Lykken2007}
J.D. Lykken.
\newblock Introduction to supersymmetry.
\newblock \emph{hep-th/9612114}, 2007.

\bibitem[Martin(2011)]{Martin2011}
S.P. Martin.
\newblock A supersymmetry primer.
\newblock \emph{hep-ph/9709356}, 2011.

\bibitem[Misner et~al.(1973)Misner, Thorne, and Wheeler]{misner}
C.W. Misner, K.S. Thorne, and J.A. Wheeler.
\newblock \emph{Gravitation}.
\newblock W.H.~Freeman and company, 1973.

\bibitem[Mohapatra and Pal(2004)]{MP04}
R.N. Mohapatra and P.B. Pal.
\newblock \emph{Massive {N}eutrinos in {P}hysics and {A}strophysics}.
\newblock World Scienctific Publishing, 2004.

\bibitem[Nakahara(1990)]{NAK90}
M.~Nakahara.
\newblock \emph{Geometry, Topology and Physics}.
\newblock IOP Publishing (Bristol), 1990.

\bibitem[{\noopsort{nieuwenhuizen}}van Nieuwenhuizen and Waldron(1996)]{NW96}
P.~{\noopsort{nieuwenhuizen}}van Nieuwenhuizen and A.~Waldron.
\newblock On euclidean spinors and wick rotations.
\newblock \emph{Phys. Lett. B}, 389:\penalty0 29--36, 1996.
\newblock arXiv:hep-th/9608174.

\bibitem[\noopsort{suijlekom}van Suijlekom(2012)]{S12}
W.D. \noopsort{suijlekom}van Suijlekom.
\newblock Renormalizability conditions for almost-commutative geometries.
\newblock \emph{arXiv:1204.4070}, 2012.

\bibitem[O'Raifeartaigh(1975)]{R75}
L.~O'Raifeartaigh.
\newblock Spontaneous symmetry breaking for chiral scalar superfields.
\newblock \emph{Nuclear Phys. B Proc. Suppl.}, 96:\penalty0 331--352, 1975.

\bibitem[Osterwalder and Schrader(1973)]{OS1}
K.~Osterwalder and R.~Schrader.
\newblock Axioms for {E}uclidean {G}reen's functions \rnum{1}.
\newblock \emph{Comm. Math. Phys.}, 31:\penalty0 83--112, 1973.

\bibitem[Osterwalder and Schrader(1975)]{OS2}
K.~Osterwalder and R.~Schrader.
\newblock Axioms for {E}uclidean {G}reen's functions \rnum{2}.
\newblock \emph{Commun. Math. Phys.}, 42:\penalty0 281--305, 1975.

\bibitem[Paschke and Sitarz(1998)]{PS96}
M.~Paschke and A.~Sitarz.
\newblock Discrete spectral triples and their symmetries.
\newblock \emph{J. Math. Phys.}, 39:\penalty0 6191, 1998.

\bibitem[Peskin and Schroeder(1995)]{PS95}
M.E. Peskin and D.V. Schroeder.
\newblock \emph{An {I}ntroduction to {Q}uantum {F}ield {T}heory}.
\newblock Westview Press, 1995.

\bibitem[Plymen(1986)]{P86}
R.J. Plymen.
\newblock Strong {M}orita equivalence, spinors and symplectic spinors.
\newblock \emph{J. Operator Theory}, 16:\penalty0 305--324, 1986.

\bibitem[Reed and Simon(1980)]{RS80I}
M.~Reed and B.~Simon.
\newblock \emph{Methods of {M}odern {M}athematical {P}hysis}, volume \rnum{1}:
  {F}unctional {A}nalysis.
\newblock Academic Press, 1980.

\bibitem[Schucker(2007)]{Schucker2007}
T.~Schucker.
\newblock Forces from noncommutative geometry.
\newblock \emph{hep-th/0110068}, 2007.

\bibitem[Spiesberger et~al.(2007)Spiesberger, Spira, and Zerwas]{SSZ02}
H.~Spiesberger, M.~Spira, and P.M. Zerwas.
\newblock The {S}tandard {M}odel: {P}hysical {B}asis and {S}cattering
  {E}xperiments.
\newblock \emph{hep-ph/0011255}, 2007.

\bibitem[Srednicki(2007)]{SR07}
M.~Srednicki.
\newblock \emph{Quantum {F}ield {T}heory}.
\newblock Cambridge {U}niversity {P}ress, 2007.

\bibitem[Stephan(2006)]{ST06}
C.A. Stephan.
\newblock Almost-commutative geometry, massive neutrinos and the orientability
  axiom in ko-dimension 6.
\newblock \emph{hep-th/0610097}, 2006.

\bibitem[Vanhecke(2007)]{Vanhecke2007}
F.J. Vanhecke.
\newblock On the product of real spectral triples.
\newblock \emph{Lett. Math. Phys.}, 50:\penalty0 157--162, 2007.

\bibitem[V\'arilly(2006)]{V06}
J.C. V\'arilly.
\newblock \emph{An Introduction to Noncommutative Geometry}.
\newblock European Mathematical Society, 2006.

\bibitem[Wegge-Olsen(1993)]{WO93}
N.E. Wegge-Olsen.
\newblock \emph{{$K$}-theory and ${C}^*$-algebras ---a friendly approach}.
\newblock Oxford univ.~press, 1993.

\bibitem[Weinberg(2005{\natexlab{a}})]{W05}
S.~Weinberg.
\newblock \emph{The quantum theory of fields, Volume \rnum{2}}.
\newblock Cambridge University Press, 2005{\natexlab{a}}.

\bibitem[Weinberg(2005{\natexlab{b}})]{W05-1}
S.~Weinberg.
\newblock \emph{The quantum theory of fields, Volume \rnum{1}}.
\newblock Cambridge University Press, 2005{\natexlab{b}}.

\bibitem[Wess and Bagger(1992)]{wessbagger1992}
J.~Wess and J.~Bagger.
\newblock \emph{Supersymmetry and Supergravity}.
\newblock Princeton University Press, 1992.

\bibitem[Wess and Zumino(1974{\natexlab{a}})]{WZ74}
J.~Wess and B.~Zumino.
\newblock Supergauge transformations in four dimensions.
\newblock \emph{Nuclear Phys. B Proc. Suppl.}, 70:\penalty0 39--50,
  1974{\natexlab{a}}.

\bibitem[Wess and Zumino(1974{\natexlab{b}})]{WZ74-2}
J.~Wess and B.~Zumino.
\newblock Supergauge invariant extension of quantum electrodynamics.
\newblock \emph{Nuclear Phys. B Proc. Suppl.}, 78:\penalty0 1--13,
  1974{\natexlab{b}}.

\bibitem[Wigner(1939)]{W39}
E.P. Wigner.
\newblock On unitary representations of the inhomogeneous {L}orentz group.
\newblock \emph{Ann. of Math. (2)}, 40\penalty0 (1):\penalty0 149--204, 1939.

\bibitem[Yanagida(1979)]{Y79}
T.~Yanagida.
\newblock Workshop on {U}nified {T}heory and {B}aryon number.
\newblock In O.~Sawada and A.~Sugamoto, editors, \emph{Workshop on {U}nified
  {T}heory and {B}aryon number}, page~95. KEK, KEK, 1979.

\bibitem[Zee(2003)]{Z03}
A.~Zee.
\newblock \emph{Quantum {F}ield {T}heory in a {N}utshell}.
\newblock Princeton {U}niversity {P}ress, 2003.

\bibitem[Zinn-Justin(2002)]{ZJ02}
J.~Zinn-Justin.
\newblock \emph{Quantum Field Theory and Critical Phenomena}.
\newblock Clarendon Press (Oxford), 2002.

\end{thebibliography}

\begingroup

\begingroup
\let\cleardoublepage\clearpage

\selectlanguage{american}

\phantomsection
\addcontentsline{toc}{chapter}{\tocEntry{Summary}}
\chapter*{Summary}
\markboth{\spacedlowsmallcaps{Summary}}{\spacedlowsmallcaps{Summary}}
\label{ch:summary}
%
%

Today we know so much more about the world ---including ourselves--- and what it is made of than even a couple of decades ago. The Standard Model of elementary particles (SM) describes with remarkable precision how the elementary particles that we observe interact with each other. The experimental verification of the Higgs boson is the last piece of the puzzle, and the crown on the SM. Yet, many fundamental questions remain to captivate physicists. For instance:
	\begin{itemize}
			\item Is there a theory that properly combines quantum field theory with gravity?
			\item Why do we observe precisely the SM-particles, and why not more? Or less?
			\item What accounts for the dark matter that astrophysicists observe?
			\item Why do all fermions appear in three copies that are identical, save for their masses?
			\item What keeps the Higgs boson mass stable when considering loop corrections? 
			\item Why is the mass of the neutrinos so small compared to that of ---say--- the top quark?
	\end{itemize}

In the past decades the academic community has witnessed the birth of a plethora of theories that address one or more of the above questions. Some of them entail only minor modifications to the SM, others require us to radically reconsider the origin of the laws of nature. The hope is that there will scientific progress in the upcoming years via the falsification of many such theories by the results of the Large Hadron Collider. We live in fascinating times indeed!

In this thesis we focus on an alternative way to obtain particle theories such as the SM on the one hand, and on a particular extension of the SM on the other. In fact, it is the combination of both that we are after. The first of these comes from the field of noncommutative geometry (NCG). Historically, this is a branch of mathematics, but it has applications in physics. From the latter point of view it can be considered as a generalization of Einstein's theory of General Relativity in the sense that it admits spaces to exhibit some notion of noncommutativity. A particular class of noncommutative geometries ---called almost-commutative geometries (ACGs)--- does a marvelous job at describing gauge theories, of which the SM is an example. These ACGs are constructed by combining a commutative geometry, consisting of a curved space(time) on which there `live' fermions, with a so-called finite noncommutative geometry. The latter has three ingredients:
\begin{itemize}
	\item a finite algebra, closely related to the gauge group --- a notion common in particle theories;
	\item a finite Hilbert space, that gives the aforementioned fermions an internal structure;
	\item a matrix that acts on the finite Hilbert space, which is called the finite Dirac operator. It is the finite counterpart of the operator that appears in the Dirac equation. 
\end{itemize}  
The constraints that are imposed on these ACGs by the axioms of NCG translate to properties of particle theories that are actually observed in experiments. NCG thus provides us new ways, of geometrical nature, to \emph{understand} theories such as the SM. The latter in fact comes out as very natural in this context.

The aforementioned extension of the Standard Model encompasses \emph{supersymmetry}. This line of thought was once devised to `get the most' out of quantum field theories, by using all its possible symmetries. Applying it to the SM in particular requires extending it with a set of new particles, one for each particle that we have currently observed. This leads to a theory that provides an answer to some of the fundamental open questions, raised above. The theory is called the Minimal Supersymmetric Standard Model (MSSM). At the LHC, experimenters vigorously look for signals that hint at its validity. To date, however, these have not been observed. Despite this lack of experimental success, the MSSM remains one of the prime candidates for a `beyond the Standard Model' theory. 

A natural question to ask is then if noncommutative geometry and supersymmetry go well together, i.e.~if the framework of NCG admits models that exhibit supersymmetry. This question has already been around for some time, but despite several previous attempts by others, its answer was still inconclusive.

This PhD thesis is devoted mainly to address this subject and combining NCG and supersymmetry. We have restricted ourselves to the class of almost-commutative geometries, in combination with the spectral action principle, a combination that was of immense value in obtaining the SM from NCG. In addition, we have restricted our analysis to finite KO-dimension $6$, that allows us to solve the fermion doubling problem in $4$ space-time dimensions. 

We have first turned to general extensions of the SM in NCG, supersymmetric or not. We have translated several physical demands (anomaly-freedom, correct hypercharges) and properties (the existence of a GUT-point) into constraints on the multiplicities of particles. The SM only satisfies these constraints when three right-handed neutrinos are added to the particle content. Although the MSSM particle content is anomaly free, it does not yield a GUT-point nor does it give the correct hypercharges. This last problem can be solved by introducing the notion of R-parity ---one that is characteristic for supersymmetry--- in the context of almost-commutative geometries and modifying some expressions accordingly. 

In order to answer the question of whether a certain ACG exhibits supersymmetry or not, a distinction must be made between the almost-commutative geometry itself and its associated action functional. Necessary for supersymmetry is the equality of fermionic and bosonic degrees of freedom. At the level of the ACGs, this leads us to the identification of \emph{supersymmetric building blocks} and a diagrammatic approach to manage calculations. These are additions to the ACG (consisting of components of the finite Hilbert space and Dirac operator) that yield degrees of freedom eligible for supersymmetry. Since this demand must hold both on shell and off shell, we are forced to introduce (non-physical) auxiliary fields by hand, since the spectral action is interpreted as the on shell action. The requirement for the total action to actually be supersymmetric (i.e.~its variation under the supersymmetry transformations vanishes) then depends on the value of the components of the finite Dirac operator. In total, we have identified five such building blocks. For each of them, the action that results corresponds in form to a term in the superfield method, in which supersymmetry is most often phrased. In somewhat more detail, we have the following building blocks.
\begin{itemize}
	\item A building block consisting of two copies of an adjoint representation in the finite Hilbert space that have opposite handedness. This gives rise to a gaugino and a gauge boson and an action that is similar to that of a vector multiplet in the superfield formalism.
	\item Given two different building blocks of the above type, we can define a second type of building block. It consists of a non-adjoint representation in the finite Hilbert space ---along with its conjugate--- and all components of the finite Dirac operator that are consequentially possible. These are the components that map the non-adjoint representations to the adjoint ones, and vice versa. The non-adjoint representation gives rise to a fermion and the components of the finite Dirac operator generate a scalar field. Both are seen to automatically be in the same representation of the gauge group. This corresponds to a chiral multiplet in the parlance of superfields.
	\item Given three building blocks of the second type whose representations in the Hilbert space have the appropriate handedness, one can define another building block by considering all (six) components of the finite Dirac operator that this set up gives rise to. The extra interactions in the action correspond to those of a superpotential term that is the product of three chiral superfields. 
	\item The remaining two types of building blocks introduce mass terms. Both consist of a component of the finite Dirac operator that does not generate any scalar fields. One of these types is a Majorana mass term, which requires a building block of the second type that describes a gauge singlet (e.g.~a right-handed neutrino). The other type is a mass-like term which requires two building blocks of the second type that are the same, save for their opposite handedness.
\end{itemize}
In the process, all formal properties of and demands on almost-commutative geometries are respected.  

Characteristic for this approach is that each new addition to an ACG provides extra contributions to the pre-factors of terms that were previously already present in the action. This requires reassessing all interactions with each newly added building block. To manage this, we have set up a list with all possible terms that occur in the action and all possible contributions to them from each building block. The action is then supersymmetric if for a particular set of building blocks all the pre-factors of the terms that occur in the action can be equated to the value required for supersymmetry. At least for the most straightforward situations (a single building block of the second type, a single building block of the third type) this is not the case; the set up turns out to be over-constrained. An interesting phenomenon occurring is that in some cases the demand of a supersymmetric action puts constraints on the number of particle generations.

Inseparable from supersymmetry is its breaking, required to give (realistic) masses to the particles appearing in the theory. We observe that soft supersymmetry breaking interactions appear automatically in the spectral action. Hence, NCG provides a new soft supersymmetry breaking mechanism. There are in fact only two supersymmetry breaking sources: the trace of the finite Dirac operator squared, which yields mass-like terms for the scalars, and gaugino masses. The second is the most prominent one. Interestingly, the gaugino masses provide a cascade of other soft breaking interactions, each of them associated to one of the five building blocks. In particular, they also give contributions to the scalar mass terms. These are of opposite sign with respect to those from the trace of the finite Dirac operator squared. This is required for the scalar mass terms to have the right sign needed to prevent them from maximally breaking the gauge group.

This sets the stage for answering the central question, concerning a noncommutative version of the MSSM. There exists a set of building blocks whose particle content corresponds to that of the MSSM and whose fermionic interactions coincide with those of the MSSM. However, the relevant constraints on the four-scalar interactions that were mentioned above can only be satisfied for a non-integer number of particle generations. Thus, \emph{the almost-commutative geometry whose particle content is equal to that of the MSSM has a spectral action that is not supersymmetric.} 

Properties of this theory hint at possible extensions of the MSSM that do satisfy all constraints, but to find it (or any other positive example of a supersymmetric NCG for that matter) requires a more constructive, and possibly automated, approach. If such a search would yield one or more positive results, these will ---due to the stringency of this approach--- at least enjoy a very special status.


\selectlanguage{dutch}
\phantomsection
\chapter*{Samenvatting}\label{ch:samenvatting}
\addcontentsline{toc}{chapter}{\tocEntry{Samenvatting}}
\markboth{\spacedlowsmallcaps{Samenvatting}}{\spacedlowsmallcaps{Samenvatting}}
%

Vandaag de dag weten we zoveel meer over de wereld om ons heen ---inclusief onszelf--- en waar deze uit bestaat, dan zelfs maar een maar decennia geleden. Het zogenoemde Standaard Model van de elementaire deeltjes (SM) beschrijft met uitzonderlijke precisie welke elementaire deeltjes er zijn en hoe zij met elkaar interageren. De vondst van het Higgs boson bij CERN in Gen\`eve in 2013 vormde het tot dan toe laatste ontbrekende puzzelstukje voor het SM. Desondanks zijn er nog talloze fundamentele vragen die bij natuurkundigen tot hoofdbrekens leiden. Een paar daarvan zijn de volgende.
\begin{itemize}
	\item Bestaat er \'e\'en enkele theorie die op een succesvolle manier all fundamentele natuurkrachten weet te verenigen?
	\item Waarom nemen wij precies de deeltjes van het SM experimenteel waar? En waarom niet meer? Of minder?
	\item Waaruit bestaat de donkere materie die sterrenkundigen waarnemen?
	\item Waarom zijn er van elk fermion drie kopie\"en, die identiek zijn op hun massa na?
	\item Is er iets dat de massa van het Higgs boson stabiel houdt, als we hogere-orde luscorrecties meenemen in de berekeningen?
	\item Waarom is er zo'n enorm verschil in massa's tussen de neutrino's en bijvoorbeeld het top quark?
\end{itemize}

In de afgelopen decennia kende de academische gemeenschap de opkomst van vele nieuwe theorie\"en die \'e\'en of meer van bovenstaande vragen probeerden te beantwoorden. Sommige van die theorie\"en zijn slechts kleine aanpassingen aan het Standaard Model. Andere, daarentegen, werpen zelfs de manier waarop we tot nu toe over natuurwetten dachten omver. Men hoopt de komende jaren wetenschappelijk gezien grote vooruitgang te boeken doordat nieuwe resultaten bij CERN een groot aantal van die theorie\"en weerleggen. We leven in werkelijk interessante tijden!

In dit proefschrift leggen we ons toe op twee dergelijke mogelijke verbeteringen van het SM. Sterker nog, we proberen beiden te combineren. De eerste van die twee komt voort uit een vakgebied genaamd niet-commutatieve meetkunde (noncommutative geometry, NCG). Hoewel van origine een wiskundig vakgebied, heeft het toepassingen in de natuurkunde. Vanuit natuurkundig oogpunt kan het gezien worden als een veralgemenisering van Einsteins Algemene Relativiteitstheorie, in de zin dat voor de gekromde ruimtes die deze beschrijft een vorm van niet-commutativiteit wordt toegestaan. Een specifieke klasse van zulke niet-commutatieve meetkundes, welke we bijna-commutatieve meetkundes (almost-commutative geometries, ACGs) noemen, toont zich bijzonder succesvol in het beschrijven van ijktheorie\"en, waarvan het SM er eentje van is. De wiskundige axioma's uit de niet-commutatieve meetkunde laten zich hierin vertalen naar eigenschappen van deeltjestheorie\"en welke daadwerkelijk waargenomen worden in experimenten. NCG reikt ons zodoende nieuwe manieren (van meetkundige aard) aan om theorie\"en zoals het SM te \emph{begrijpen}. Het SM in het bijzonder komt hier als heel natuurlijke oplossing uit de bus.

De tweede mogelijke verbetering van het SM behelst supersymmetrie. Dit is een gedachtegang die aanvankelijk tot doel had om alle symmetrie\"en van kwantumveldentheorie\"en te benutten. Echter, uit het toepassen ervan op het SM volgt een model dat antwoord biedt op sommige van de fundamentele open vragen hierboven. Dit model wordt de Minimale Supersymmetrische uitbreiding van het Standaard Model (MSSM) genoemd. Op CERN houden wetenschappers zich koortsachtig bezig met het vinden van experimentele bewijzen voor dit model. En hoewel er tot op heden nog niets is gevonden dat wijst op het bestaan van het MSSM, blijft het \'e\'en van de meest prominente kandidaten voor een succesvolle uitbreiding van het SM.

Een vraag die zich dan vrij snel aandient, is of niet-commutatieve meetkunde en supersymmetrie zich goed tot elkaar verhouden. Dat wil zeggen, of er niet-commutatieve meetkundige modellen bestaan die tegelijkertijd supersymmetrisch zijn. Deze vraag is al een tijdje geleden voor het eerst gesteld, maar geen enkele van de onderzoeken die zich hieraan gewijd hebben, kwam met waterdicht bewijs.

Dit proefschrift is hoofdzakelijk aan deze vraag gewijd. We hebben ons erin beperkt tot de klasse van bijna-commutatieve meetkundes, in combinatie met het zogenoemde spectrale actie principe. Deze combinatie is van immens belang gebleken bij het afleiden van het SM vanuit de NCG. 

Eerst hebben we ons echter gewijd aan algemene (dus niet per se supersymmetrische) uitbreidingen van het SM in de context van de niet-commutatieve meetkunde. We hebben een aantal natuurkundige eisen (de afwezigheid van \emph{anomalie\"en}, de juiste \emph{hyperladingen} voor de SM-deeltjes) en eigenschappen (het bestaan van een GUT-punt, waar de koppelingsconstanten van de drie fundamentele krachten samenkomen) vertaald in eisen aan welke deeltjes in een niet-commutatieve theorie voor kunnen komen, en hoe vaak. Het blijkt dat het SM alleen aan die eisen voldoet wanneer we deze uitbreiden met drie rechtshandige neutrino's. De deeltjes-inhoud van het MSSM is weliswaar vrij van anomalie\"en, maar voldoet niet aan de andere twee eisen. Aan die van de hyperladingen kan echter tegemoet gekomen worden door in deze context het begrip R-pariteit (karakteristiek voor supersymmetrie) te introduceren en daarnaast een aantal van de gebruikte uitdrukkingen op basis daarvan aan te passen. 

Om de centrale vraag te kunnen beantwoorden, moeten we een onderscheid maken tussen de deeltjes-inhoud van een ACG enerzijds en de ermee geassocieerde actie anderzijds. Een noodzakelijke eis voor elk supersymmetrisch model is dat het aantal fermionische en bosonische vrijheidsgraden gelijk aan elkaar zijn. Op het niveau van de ACG leidt dit ons tot de identificatie van verschillende \emph{bouwblokken}. Dit zijn toevoegingen aan een ACG (bestaande uit representaties in de eindige Hilbertruimte en componenten van de eindige Dirac-operator) welke de deeltjes-inhoud geschikt maken (of houden) voor supersymmetrie. De eis van een gelijk aantal vrijheidsgraden moet gelden ongeacht of we vereisen dat de bewegingsvergelijkingen voor de deeltjes gelden of niet. Dit noopt ons om handmatig (niet-fysische) \emph{hulpvelden} te introduceren. De spectrale actie voorziet daar immers niet in. De eis van feitelijke supersymmetrie van de actie vertaalt zich dan in eisen aan de componenten van de eindige Dirac-operator. Al met al zijn zo vijf verschillende \emph{bouwblokken} te identificeren. Voor elk van hen correspondeert de bijdrage aan de actie (althans in vorm) met een term bekend uit het superveld-formalisme, waarin supersymmetrie veelal in geformuleerd wordt. Om iets meer in detail te gaan, hebben we de volgende bouwblokken:
\begin{itemize}
	\item een bouwblok dat bestaat uit twee kopie\"en van een geadjungeerde representatie in de eindige Hilbertruimte, welke van tegenovergestelde handigheid zijn. Dit geeft aanleiding tot een \emph{gaugino} en een ijkboson en een actie die erg lijkt op die van een vectormultiplet in het superveld-formalisme. 
\item Gegeven twee bouwblokken van bovenstaand type, kunnen we een tweede type bouwblok defini\"eren. Deze bestaat uit een niet-geadjungeerde representatie in de eindige Hilbertruimte ---samen met zijn geconjugeerde--- en alle componenten van de eindige Dirac-operator die als gevolg daarvan toegestaan zijn. Dit geeft aanleiding tot een scalair deeltje en een fermion, welke ---in de taal van het superveld-formalisme--- corresponderen met een chiraal multiplet. 
\item Gegeven drie bouwblokken van het tweede type waarvan de representaties in de Hilbertruimte de juiste handigheid hebben, is nog een bouwblok te defini\"eren. Deze bestaat uit alle (zes) de componenten van de eindige Dirac-operator die mogelijk zijn in deze situatie. De extra interacties die als gevolg hiervan in de actie verschijnen, komen overeen met die van een superpotentiaal-term welke het product is van drie chirale supervelden. 
\item Dan zijn er nog twee typen bouwblokken, die zich gedragen als massa-termen. Beiden bestaan uit een component van de eindige Dirac-operator welke geen scalaire velden voortbrengt. Het eerste van die twee is een Majorana massa-term. Deze vereist een bouwblok van het tweede type dat een ijksinglet beschrijft (zoals het rechtshandige neutrino). Het tweede type is een term die op een massa-term lijkt. Dit vereist twee bouwblokken van het tweede type die hetzelfde zijn, op hun handigheid na.
\end{itemize}
Deze bouwblokken voldoen aan alle eisen voor correcte bijna-commutatieve meetkundes.

Eigen aan deze aanpak is dat met elk extra bouwblok dat aan een ACG wordt toegevoegd, er extra bijdrages komen aan voor-factoren van termen die daarv\'o\'or al in de actie voor kwamen. Dit vereist het herzien van alle voorkomende interacties bij het introduceren van elk nieuw bouwblok. Om hierin structuur aan te brengen, hebben we een lijst opgesteld waarin alle termen staan die in de actie voorkomen en alle mogelijke bijdragen eraan van elk van de typen bouwblokken. De actie is vervolgens supersymmetrisch wanneer voor een bepaalde verzameling bouwblokken alle voor-factoren van de termen die in de actie voorkomen, overeenkomen met de waarde die vereist wordt vanuit supersymmetrie. In elk geval voor de meest voor de hand liggende situaties (een enkel bouwblok van het tweede of derde type) wordt hieraan niet voldaan. Er namelijk zijn teveel eisen aan de modellen, die samen alle oplossingen uitsluiten. Een interessant fenomeen dat zich voordoet is dat in sommige gevallen de eis van een supersymmetrische actie voorwaarden oplegt in termen van het aantal generaties van deeltjes.

Supersymmetrie en het breken ervan zijn onlosmakelijk met elkaar verbonden. Dat breken is vereist om de deeltjes die in de theorie voorkomen een realistische massa te kunnen laten hebben. We zien dat de spectrale actie automatisch termen genereert die supersymmetrie breken. Niet-commutatieve meetkunde voorziet zo in een nieuwe mechanisme om supersymmetrie op de gewenste manier te kunnen breken. Er zijn feitelijk slechts twee bronnen die de breking veroorzaken: de eerste is het spoor van het kwadraat van de eindige Dirac-operator (welke massa-achtige termen voor scalairen geeft) en massa-termen voor de gaugino's. Deze laatste is ook de meest prominente. Een interessante observatie is dat die gaugino massa's een sneeuwbaleffect veroorzaken in de zin dat ze andere brekingstermen tot gevolg hebben. Die termen kunnen elk geassocieerd worden met \'e\'en van de vijf typen bouwblokken. In het bijzonder zijn er bijdragen aan de massa's van de scalaire deeltjes. Deze zijn van tegenovergesteld teken dan de eerdergenoemde termen die voortkomen uit het spoort van het kwadraat van de eindige Dirac-operator. Die bijdragen zijn vereist om te voorkomen dat deze temen een dusdanig teken hebben dat zij de ijkgroep volledig breken.

Daarmee is het voorwerk gedaan dat nodig is om de centrale vraag van dit proefschrift te kunnen beantwoorden, namelijk of er een niet-commutatieve versie van het MSSM bestaat. Er bestaat inderdaad een verzameling van bouwblokken wier deeltjes-inhoud en fermionische interacties overeenkomen met die van het MSSM. Echter, er is alleen te voldoen aan de relevante vereisten die aan de vier-scalar interacties worden gesteld, met een niet-geheeltallig aantal generaties van deeltjes. Daaruit volgt het centrale resultaat van dit proefschrift, namelijk dat \emph{de bijna-commutatieve meetkunde waarvan de deeltjes-inhoud overeenkomt met die van het MSSM, geen supersymmetrische actie heeft.}

Eigenschappen van dit model suggereren het bestaan van uitbreidingen van het MSSM die w\'el aan alle vereisten voldoen. Om deze te vinden (of welke andere supersymmetrische ACG dan ook), vereist een meer opbouwende aanpak, wellicht ook eentje die meer geautomatiseerd is. Vanwege de stringente eisen die gesteld worden, staat in elk geval vast dat, als een dergelijke zoektocht \'e\'en of meer positieve resultaten oplevert, dit dan ook meteen oplossingen betreft die een erg bijzondere status genieten.


\selectlanguage{american}
\chapter*{Curriculum Vitae}
\addcontentsline{toc}{chapter}{\tocEntry{Curriculum Vitae}}
\markboth{\spacedlowsmallcaps{Curriculum Vitae}}{\spacedlowsmallcaps{Curriculum Vitae}}
\label{ch:cv}
Thijs van den Broek was born on the 23th of January 1983, in Eindhoven, the Netherlands. He attended primary school at the Sinte Lucij in Steensel and secondary school at the Rythovius College in Eersel. In 2001 he started a bachelor in Physics and Astrophysics at the Radboud University Nijmegen, the Netherlands. This he finished in 2006. 

In that same year he started a master in Physics and Astronomy at the Radboud University. His first experience with noncommutative geometry was during his masters' thesis project. This led to his thesis, titled 'Supersymmetric gauge theories in noncommutative geometry'. In 2009 he got his masters' degree, graduating cum laude. 

During his bachelors' and masters' studies he was active in the student community. He was board member and chairman of the physics students' association `Marie Curie' in 2003--2005 and student assessor of the Faculty of Science in 2005--2006. Afterwards he participated actively in the students' union AKKU in Nijmegen and the Dutch national students' union `LSVb'. 

Late 2009 he started his PhD in Theoretical High Energy Physics, continuing on the combination of noncommutative geometry and supersymmetry. His thesis advisor was Prof. dr. Ronald Kleiss and he was supervised by Dr.~Wim Beenakker and Dr.~Walter van Suijlekom. Parallel to doing research, he taught several courses via exercise sessions and supervised students doing their Bachelor thesis. In October 2013 he co-organized the workshop `Noncommutative geometry and particle physics' at the Lorentz Center in Leiden. 

Since 2010 he is politically active. 

\endgroup


\cleardoublepage

\chapter*{Acknowledgments}
\addcontentsline{toc}{chapter}{\tocEntry{Acknowledgments}}
\markboth{\spacedlowsmallcaps{Acknowledgments}}{\spacedlowsmallcaps{Acknowledgments}}

\selectlanguage{american}











First of all I would like to thank my advisor Ronald Kleiss, and both of my daily supervisors Wim Beenakker and Walter van Suijlekom, in the first place for having loads of patience with me, for being open for discussion anytime and stimulating me to keep on pushing the envelope. You are not the only ones that made the HEP department such a great place to work, though. The secretariat was a welcome place anytime, every time. Thank you Marjo, Gemma and Annelies. Admittedly, the atmosphere at the department was one of the reasons to choose this job. At many times it was more of a group of friends than an office. Many people deserve to be thanked for that: Antonia, Antonio, Ben, Folkert, Gijs, GJ, Guus, Harm, Irene, Jos\'e, Jari, Jins, Jonas, Koen, Magda, Marie, Margot, Marcel, Matthias, Melvin, Stefan J, Stefan G, Vince, Yiannis and many others. I've had wonderful times with you during but even more so after work. Thanks a lot for making the time at the department what it was. 

Being part of the TPP-program taught me that there are cool physicists outside Nijmegen as well. Chiara, Ivano, Jacob, Jan, Jordy, Lisa, Pablo, Reinier, Rob, Sander, Tiago, Wilco and again many more. I enjoyed the Fridays at Nikhef and the school in Driebergen almost as much as going to the IJ-brouwerij, the Opposites concert and Brazil. Thank you.

But honestly, I doubt whether I would have coped doing a PhD if it wasn't for all those great and valuable things to keep it in balance with. The `zwammen'-group for instance, with Anco, Frank en Galina, Kimberly, Koen, Maaike and Maarten. Whether we were having the serious conversations that have broadened my academic perspective, or the lighter ones, I enjoyed them equally well.


After all that mental labor one craves for some physical exercise once in a while. Luckily I was surrounded with many people that like cycling as much as I do (or more...). Whether it was in the vicinity of Nijmegen, on one of the NFC-weekends or even on a vacation somewhere high up in the Alps, it was always fun. Thank you, Geerten, Jeroen, LJ, Vincent and Youri.

Then there are the people from GroenLinks. Although I got to know you in the first place because of our shared ideals, many of you made me feel at home here. Special thanks goes out to April, Cilia, Giedo, Joep, Lisa, Louis, Matthijs, Nynke, Pepijn O.~and Wouter. 


An then there is a group of friends that I got to know while doing physics in Nijmegen, from AKKU (the AKKU trainingsbureau in particular) or elsewhere. You definitely deserve to be mentioned by name, but the list is long... Again, you know who you are! Bernard and Floor deserve a special reference here, though, and not only for all those Sunday evenings.

Perhaps not in the most direct sense, but I wouldn't have got here without the love and support from family. Of course also my `new' family ---Cor, Roelie, Gerda, Hanna, Linda \& Wieger--- but even more so Nanneke, Josien \& Mathijs and of course my parents. Even though you probably had no clue about what I was doing, you believed in me and supported me. And, above all, thanks for all those wonderful times. 

And last, but most certainly not least, my loving girlfriend Petra. Thank you so much for the time that we are together, for your patience, understanding and support, for all those beautiful moments or for just being there. I love you!

All of you made me who I am and brought me to where I am, and I am immensely grateful for that.


\endgroup
\end{document}